\documentclass[a4paper,11pt,twoside]{book}
\usepackage{caption}
\usepackage{bm}       
\usepackage{comment}
\usepackage{setspace}
\usepackage{graphicx}
\usepackage{grffile}
\usepackage{amssymb}
\usepackage{amsbsy}   
\usepackage{amsmath}
\usepackage{url}
\usepackage{lscape}
\usepackage{natbib}
\usepackage{aastex_hack}
\usepackage{deluxetable}
\usepackage{fancyhdr}
\usepackage[avantgarde]{quotchap}
\usepackage[calcwidth]{titlesec}
\usepackage{swinburnetitle}
\usepackage{subfigure}
\usepackage{pdflscape}
\usepackage{enumitem}
\usepackage{pdfpages}

\usepackage[bookmarks=true,breaklinks=true,hypertexnames=false,bookmarksnumbered=true]{hyperref}
 \hypersetup{
pdfauthor = {Themiya Nanayakkara},
pdftitle = {MOSFIRE Spectroscopy of Galaxies in Cosmic Noon},
pdfsubject = {Astronomy},
pdfkeywords = {--},
pdfcreator = {LaTeX with hyperref package},
pdfproducer = {dvips + ps2pdf}}

\fancypagestyle{plain}{
\fancyhf{}
\fancyfoot[CO]{\thepage}

}

\makeatletter
\def\cleardoublepage{\clearpage\if@twoside \ifodd\c@page\else
	\hbox{}
	\vspace*{\fill}
	\thispagestyle{empty}
	\newpage
	\if@twocolumn\hbox{}\newpage\fi\fi\fi}
\makeatother

\renewcommand\chapterheadstartvskip{\vspace*{-5\baselineskip}}

\titleformat{\section}[hang]{\sffamily\bfseries}
{\Large\thesection}{12pt}{\Large}[{\titlerule[0.5pt]}]

\newcommand{\Ks}{K$_{\rm s}$}
\newcommand{\Halpha}{H$\alpha$}
\newcommand{\Hbeta}{H$\beta$}

\newcommand{\msol}{M$_\odot$}

\newcommand{\SII}{[\hbox{{\rm S}\kern 0.1em{\sc ii}}]}
\newcommand{\AlIII}{\hbox{{\rm Al}\kern 0.1em{\sc iii}}}
\newcommand{\NII}{[\hbox{{\rm N}\kern 0.1em{\sc ii}}]}
\newcommand{\OII}{[\hbox{{\rm O}\kern 0.1em{\sc ii}}]}
\newcommand{\OIII}{[\hbox{{\rm O}\kern 0.1em{\sc iii}}]}
\newcommand{\MgII}{\hbox{{\rm Mg}\kern 0.1em{\sc ii}}}
\newcommand{\MgI}{\hbox{{\rm Mg}\kern 0.1em{\sc i}}}
\newcommand{\FeII}{\hbox{{\rm Fe}\kern 0.1em{\sc ii}}}
\newcommand{\CIII}{\hbox{{\rm C}\kern 0.1em{\sc iii}}}
\newcommand{\CIV}{\hbox{{\rm C}\kern 0.1em{\sc iv}}}
\newcommand{\HII}{\hbox{{\rm H}\kern 0.1em{\sc ii}}}
\newcommand{\HI}{\hbox{{\rm H}\kern 0.1em{\sc i}}}

\newcommand{\CII}{\hbox{{\rm C}\kern 0.1em{\sc ii}}}
\newcommand{\OI}{\hbox{{\rm O}\kern 0.1em{\sc i}}}
\newcommand{\NeIII}{[\hbox{{\rm Ne}\kern 0.1em{\sc iii}}] }
\newcommand{\NeII}{[\hbox{{\rm Ne}\kern 0.1em{\sc ii}}] }
\newcommand{\NaI}{\hbox{{\rm Na}\kern 0.1em{\sc i}} }
\newcommand{\CaII}{\hbox{{\rm Ca}\kern 0.1em{\sc ii}} }
\newcommand{\around}{$\sim$}
\newcommand{\AvSED}{$\mathrm{Av_{\mathrm{SED}}}$}
\newcommand{\NMAD}{$\sigma_{\mathrm{NMAD}}$}
\newcommand{\zspec}{$z_{\mathrm{spec}}$}
\newcommand{\zgrism}{$z_{\mathrm{grism}}$}
\newcommand{\zphoto}{$z_{\mathrm{photo}}$}
\newcommand{\mass}{M$_*$/M$_\odot$}

\newcommand{\AvZFIRE}{$\mathrm{A_{v}(ZF)}$}
\newcommand{\Av}{$\mathrm{A_v}$}
\newcommand{\gr}{$\mathrm{(g-r)_{0.1}}$}
\newcommand{\boxfil}{$\mathrm{[340]-[550]}$}
\newcommand{\dustfil}{$\mathrm{[150]-[260]}$}
\newcommand{\salpeter}{$\Gamma=-1.35$}
\newcommand{\vini}{$v_{ini}$}
\newcommand{\vcrit}{$v_{crit}$}
\newcommand{\zsol}{Z$_\odot$}

\newcommand{\sample}{ZFIRE-SP sample}
\newcommand{\nlimits}{56}
\newcommand{\ndetections}{46}

\newcommand{\logmass}{$\mathrm{log_{10}(M_*/M_\odot)}$}

\DeclareFontFamily{OT1}{pzc}{} 
\DeclareFontShape{OT1}{pzc}{m}{it}{<->[1.5] pzcmi8t}{}
\DeclareMathAlphabet{\mathpzc}{OT1}{pzc}{m}{it}

\begin{document}

\frontmatter
\author{ Themiya Nanayakkara }
\title{ MOSFIRE Spectroscopy of Galaxies in Cosmic Noon }
\date{ 2017 }
\maketitle

\addcontentsline{toc}{chapter}{Epigraph}
\newpage
\vspace*{2.5in}
\begin{center}\parbox{11cm}{\begin{center}
\textit{
	``Come, Kalamas. Do not go upon what has been acquired by repeated hearing; nor upon tradition; nor upon rumour; nor upon what is in a scripture; nor upon surmise; nor upon an axiom; nor upon specious reasoning; nor upon a bias towards a notion that has been pondered over; nor upon another's seeming ability; nor upon the consideration, `The monk is our teacher'. Kalamas, when you yourselves know: `These things are good; these things are not blamable; these things are praised by the wise; undertaken and observed, these things lead to benefit and happiness,' enter on and abide in them.''\\
	Buddha to the Kalama community 
}
\end{center} } \end{center}

\hfill
\addcontentsline{toc}{chapter}{Abstract}
\Huge \noindent Abstract
\normalsize
\\

The recent development of sensitive, multiplexed near infra-red instruments has presented astronomers the unique opportunity to survey mass/magnitude complete samples of galaxies at \emph{Cosmic Noon}, a time period where $\sim80\%$ of the observed baryonic mass is generated and galaxies are actively star-forming and evolving rapidly.  
This thesis takes advantage of the recently commissioned MOSFIRE spectrograph on Keck, to conduct a survey (ZFIRE) of galaxies at $1.5<z<2.5$ to measure accurate spectroscopic redshifts and basic galaxy properties derived from multiple emission lines. \\

The majority of the thesis work involved survey planning, observing, data reduction, and catalogue preparation of the ZFIRE survey and is described in detail in this thesis. 
Using the ZFIRE spectroscopic redshifts, I show why spectroscopy is instrumental to determine  fundamental galaxy properties via SED fitting techniques and to probe gravitationally bound structures in the early universe. 
The thesis further presents basic properties of the ZFIRE data products publicly released for the benefit of the astronomy community. \\

The high mass-completeness of the ZFIRE spectroscopic data at $z\sim2$ makes it ideal to study fundamental galaxy properties such as, star formation rates, metallicities, inter-stellar medium properties, galaxy kinematics, and the stellar initial mass functions in unbiased star-forming galaxies. 
This thesis focuses on one such aspect, the IMF. 
By using a mass-complete ($\mathrm{log_{10}(M_*/M_\odot)\sim9.3}$)  sample of 102 galaxies at $z=2.1$ in the COSMOS field from ZFIRE, I investigate the IMF of star-forming galaxies by revisiting the classical \citet{Kennicutt1983} technique of using the \Halpha\ equivalent widths  and rest-frame optical colours.
I present a thorough analysis of stellar population properties of the ZFIRE sample via multiple synthetic stellar population models and stellar libraries.\\

Due to an excess of high \Halpha-EW galaxies that are up to 0.3--0.5 dex above the Salpeter locus, the \Halpha-EW distribution is much broader (10-500\AA) than can be explained by a simple monotonic SFH with a standard Salpeter-slope IMF. This result is robust against uncertainties in dust correction and observational bias, and no single IMF (i.e. non-Salpeter slope) can explain the distribution.  
Starburst models cannot explain the \Halpha-EW distribution because: 1) spectral stacking still shows an excess \Halpha-EW in composite populations and 2) Monte Carlo burst models show that the timescale for high \Halpha-EW is too short to explain their abundance in the ZFIRE sample.  
Other possible physical mechanisms that could produce excess ionising photons for a given star-formation rate, and hence high equivalent widths, including models with variations in stellar rotation, binary star evolution, metallicity, and upper mass cutoff of the IMF are investigated and ruled out. \\

IMF variation is one possible explanation for the high \Halpha-EWs.  However, the highest \Halpha-EW values would require very shallow slopes ($\Gamma>-1.0$) and no single IMF change can explain the large variation in \Halpha-EWs.  Instead the IMF would have to vary stochastically. 
Therefore, currently there is no simple physical model to explain the large variation in \Halpha-EWs at $z\sim2$, but the distinct differences of the $z\sim2$ sample with that of local galaxies are found to be intriguing. 
Further study is required to fully constrain the stellar population parameters of actively star-forming galaxies at the epoch of maximum star-formation. Probing multiple rest-frame UV and optical features of galaxies simultaneously along with galaxy dynamical studies via integral field spectroscopy will be vital to understand stellar and ionized gas properties of these galaxies. 
Furthermore, low-$z$ analogues of galaxies at $z\sim2$ will provide vital clues to constrain galaxy evolution models aided by the ability to probe galaxies in high resolution to low surface brightness limits. \\

\noindent {\bf Advisers:}\\
\emph{Primary:} Prof. Karl Glazebrook\\
\emph{Coordinating:} Dr. Glenn Kacprzak\\
\emph{Coordinating:} Dr. David Fisher\\


\hfill
\addcontentsline{toc}{chapter}{Acknowledgements}
\Huge \noindent Acknowledgements
\normalsize
\\

This thesis has only been possible due to immense help from a large group of people in many different roles.

First of all, the amazing guidance of my supervisors, Karl, Glenn, and David throughout the years has been instrumental to reach this milestone. Karl - for all the quick email replies (even on the weekends!), the continuous encouragement to find the answers myself, and finding all the cool unimaginable things to do always kept me motivated and enthused. Glenn - for always being there and showing me the way to become an astronomer, from the observing runs to reading the drafts, your help has been tremendous. David - the continuous flow of ideas and feedback on my work has been greatly helpful.

Everyone in the ZFOURGE, and ZFIRE groups:  thanks for all the encouragement and patience. Its been a pleasure to work with such an encouraging collaboration. Special thanks for Vy for everything - from my first observing runs to all the workshops, science, beach, and food. You've always made me feel work is fun. 
Caroline - thanks for all the help with ZFOURGE stuff, you've been a great to work with, travel, and befriend.

I wish to thank the Centre for Astrophysics and Supercomputing community at Swinburne for being an inclusive and diverse bunch of lovely people who always made working enjoyable. Special thanks to Liz - for always being there and helping me through all the administrative matters throughout these years.  Luca, Paola, Rob, Rebecca, Tyler- you've been great to work around with. 
Jonathan, Dany, and Colin - your help to improve my codes has been an tremendous help. 
George - thanks for helping me install stuff day through night and always telling me why my code is really bad (and also why astronomers can never write good programs)! Pierluigi - thanks for all the discussions and research advice throughout the years.

Thanks to my parents and family for whom I owe everything I've achieved so far. Mum, your constant love and encouragement to pursue what I like, Dad, being the first inspiration in life to seek the wonders of the night sky has always been the driving force. Aunty Nelum, thanks for teaching me from the first letters to the complex mathematics, and putting me ahead in you life, you've been exemplary. 
Finally, Manisha for your love and for being who you are.


\hfill
\addcontentsline{toc}{chapter}{Declaration}
\Huge \noindent Declaration
\normalsize
\\

\noindent
The work presented in this thesis has been carried out in the Centre for
Astrophysics \& Supercomputing at Swinburne University of Technology between
2013 and 2016. This thesis contains no material that has been accepted for the
award of any other degree or diploma. To the best of my knowledge, this thesis
contains no material previously published or written by another author, except
where due reference is made in the text of the thesis. The content of the
chapters listed below has appeared in refereed journals. Minor alterations have
been made to the published papers in order to maintain argument continuity and
consistency of spelling and style.

\begin{itemize}
\item Chapter \ref{chap:zfire_survey} has been published in The Astrophysical journal as \emph{ZFIRE: A KECK/MOSFIRE Spectroscopic Survey of Galaxies in Rich Environments at $z\sim2$} \citep{Nanayakkara2016}.
\item Chapter \ref{chap:spec_analysis} has been published in The Astrophysical journal as \emph{ZFIRE: A KECK/MOSFIRE Spectroscopic Survey of Galaxies in Rich Environments at $z\sim2$} \citep{Nanayakkara2016}.
\item Chapter \ref{chap:imf_observations} has been accepted for Monthly Notices of the Royal Astronomical Society for peer reviewed publication as \emph{ZFIRE: Using \Halpha\ Equivalent Widths to Investigate the
  In Situ Initial Mass Function at $z\sim2$}. 
\item Chapter \ref{chap:imf_analysis} has been accepted for Monthly Notices of the Royal Astronomical Society for peer reviewed publication as \emph{ZFIRE: Using \Halpha\ Equivalent Widths to Investigate the
  In Situ Initial Mass Function at $z\sim2$}. 
\end{itemize}

The work presented in this thesis has further contributed directly to the following publications.

\begin{enumerate}

\item KECK/MOSFIRE spectroscopic confirmation of a Virgo-like cluster ancestor at z=2.095 \citep{Yuan2014}.

\item The Absence of an Environmental Dependence in the Mass-Metallicity Relation at $z = 2$ \citep{Kacprzak2015}.

\item ZFIRE: Galaxy cluster kinematics, \Halpha\ star formation rates, and gas-phase metallicities of XMM-LSS J02182-05102 at zcl = 1.6232 \citep{Tran2015}.

\item ZFIRE: The Kinematics of Star-Forming Galaxies as a Function of Environment at $z\sim2$ \citep{Alcorn2016}.

\item Large scale structure around a $z=2.1$ cluster \citep{Hung2016}.

\item Cold-mode Accretion: Driving the fundamental mass-metallicity relation at $z\sim2$ \citep{Kacprzak2016}.

\item ZFIRE: ISM properties of the z = 2.095 COSMOS Cluster \citep{Kewley2016}.

\item ZFIRE: Similar stellar growth in \Halpha-emitting cluster and field galaxies at $z\sim2$ (Tran et al., in press)

\end{enumerate}

I have contributed to the ZFOURGE data release by testing ZFOURGE photometry, SED fits, and ancillary data products. Furthermore, through my contributions to the data catalogues and actively engaging and providing comments to the ZFOURGE group research, I have been able to be a part of the following publications.

\begin{enumerate}[resume]

\item Exploring the $z = 3-4$ Massive Galaxy Population with ZFOURGE: The Prevalence of Dusty and Quiescent Galaxies \citep{Spitler2014}.

\item The Distribution of Satellites around Massive Galaxies at $1 < z < 3$ in ZFOURGE/CANDELS: Dependence on Star Formation Activity \citep{Kawinwanichakij2014}.

\item ZFOURGE/CANDELS: On the Evolution of M* Galaxy Progenitors from $z = 3$ to 0.5 \citep{Papovich2015}.

\item The Differential Size Growth of Field and Cluster Galaxies at $z = 2.1$ Using the ZFOURGE Survey \citep{Allen2015}.

\item The Sizes of Massive Quiescent and Star-forming Galaxies at $z\sim4$ with ZFOURGE and CANDELS \citep{Straatman2015}.

\item Satellite Quenching and Galactic Conformity at $0.3 < z < 2.5$ \citep{Kawinwanichakij2016}.

\item Radio galaxies in ZFOURGE/NMBS: no difference in the properties of massive galaxies with and without radio-AGN out to $z = 2.25$ \citep{Rees2016}.

\item The SFR-M* Relation and Empirical Star-Formation Histories from ZFOURGE* at $0.5 < z < 4$ \citep{Tomczak2016}.

\item UV to IR Luminosities and Dust Attenuation Determined from $\sim$4000 K-selected Galaxies at $1 < z < 3$ in the ZFOURGE Survey \citep{Forrest2016}.

\item ZFOURGE catalogue of AGN candidates: an enhancement of 160-μm-derived star formation rates in active galaxies to $z = 3.2$ \citep{Cowley2016}.

\item Differences in the structural properties and star-formation rates of field and cluster galaxies at $z\sim1$ \citep{Allen2016}.

\item The FourStar Galaxy Evolution Survey (ZFOURGE): Ultraviolet to Far-infrared Catalogs, Medium-bandwidth Photometric Redshifts with Improved Accuracy, Stellar Masses, and Confirmation of Quiescent Galaxies to $z\sim3.5$ \citep{Straatman2016}.

\end{enumerate}

\vspace{1.0cm}
\begin{flushright}
\textit{ Themiya Nanayakkara }

\textit{Melbourne, Victoria, Australia}
\textit{2016}
\end{flushright}

\begin{flushright}
\textit{ Karl Glazebrook }
\textit{2016}
\end{flushright}

\begin{flushright}
\textit{ Glenn Kacprzak }
\textit{2016}
\end{flushright}

\begin{flushright}
\textit{ David Fisher }
\textit{2016}
\end{flushright}

\hfill
\newpage
\vspace*{3.5in}
\begin{center}\parbox{11cm}{\begin{center}
\textit{Dedicated to all those who seek}
\end{center} } \end{center}

\addcontentsline{toc}{chapter}{Glossary}
\newpage
\Huge \noindent Glossary
\normalsize
\\

\Large \noindent Acronyms
\normalsize
\\

AGN: Active Galactic Nuclei

ALMA: Atacama Large Millimetre/submillimetre Array 

BPASS: Binary Population and Spectral Synthesis

CANDLES: Cosmic Assembly Near-infrared Deep Extragalactic Legacy Survey

CDFS: Chandra Deep Field South

CDM: Cold Dark Matter

CMB: Cosmic Microwave Background

COSMOS: Cosmic Evolution Survey

DRP: Data Reduction Pipeline

DYNAMO:  DYnamics of Newly-Assembled Massive Objects

EAZY: Easy Accurate $z_\mathrm{phot}$ from Yale

ESA: European Space Agency

ETG: Early Type Galaxy 

EW: Equivalent Width

FAST: Fitting and Assessment of Synthetic Templates

FMOS: Fibre Multi Object Spectrograph

FUV: Far Ultra-Violet

FWHM: Full Width Half Maximum

GDDS  Gemini Deep Deep Survey

GNIRS: Gemini Near-Infrared Spectrograph

HAWK- I: High Acuity Wide field K-band Imager

HST: Hubble Space Telescope

IMF: Initial Mass Function

ISM: Inter-Stellar Medium

JWST: James Webb Space Telescope

KCWI: Keck Cosmic Web Imager

LSB: Low surface brightness 

MAGMA: MOSFIRE Automatic GUI-based Mask Application

MOIRCS: Multi-Object Infrared Camera and Spectrograph

MOSDEF: MOSFIRE Deep Evolution Field 

MOSFIRE: Multi-object Spectrometer for infrared Exploration 

MZR: Mass Metallicity (Z) Relation 

NASA: National Aeronautics and Space Administration

NEWFIRM: NOAO Extremely Wide Field Infrared Imager

NMAD: Normalized Median Absolute Deviation

NMBS: NEWFIRM medium-band Survey

NOAO: National Optical Astronomy Observatory

NIR: Near Infra-Red 

PEGASE: Projet d'Étude des GAlaxies par Synthèse Évolutive

S99: Starburst99

SDDS: Sloan Digital Sky Survey

SED: spectral energy distribution 

SFH: Star-Formation History 

SFR: Star-Formation Rate

SNR: Signal to Noise Resolution

sSFR : specific Star-Formation Rate 

SSP: Synthetic Stellar Population

UDS: Ultra Deep Survey

UKIDSS : UKIRT Infrared Deep Sky Survey

UKIRT: United Kingdom InfraRed Telescope

UV: Ultra-Violet

VIMOS: Visible Multi Object Spectrograph

VIRCAM: VISTA InfraRed CAMera

VISTA: Visible and Infrared Survey Telescope for Astronomy

VVDS: VIMOS VLT Deep Survey

WFC3: Wide Field Camera 3

W-R: Wolf–Rayet

\newpage

\Large \noindent Abbreviations
\normalsize
\\

\NMAD = normalized absolute median deviation 

\Av: Extinction

\AvZFIRE: Extinction derived from the ZFIRE method

\AvSED: Extinction derived via SED fitting

$f$: The fractional difference in attenuation between the continuum at 6564.61\AA\ and the \Halpha\ nebular emission line.

$f_m$: The mass fraction of a galaxy generated during a star-burst. The mass fraction is generally expressed compared to the total mass at $t=3.1$ Gyr ($z\sim2$), unless otherwise stated.

$P(z)$: photometric redshift likelihood functions

Z: Metallicity

$z$: Redshift

ZFIRE-SP: ZFIRE Stellar population

\zphoto: Photometric redshift

\zspec: Spectroscopic redshift

 
\tableofcontents
\addcontentsline{toc}{chapter}{List of Figures}
\listoffigures
\addcontentsline{toc}{chapter}{List of Tables}
\listoftables

\mainmatter
\chapter[Introduction]{Introduction}
\label{chap:Introduction}

\begin{figure}
\centering
\includegraphics[width=0.75\textwidth, draft=True]{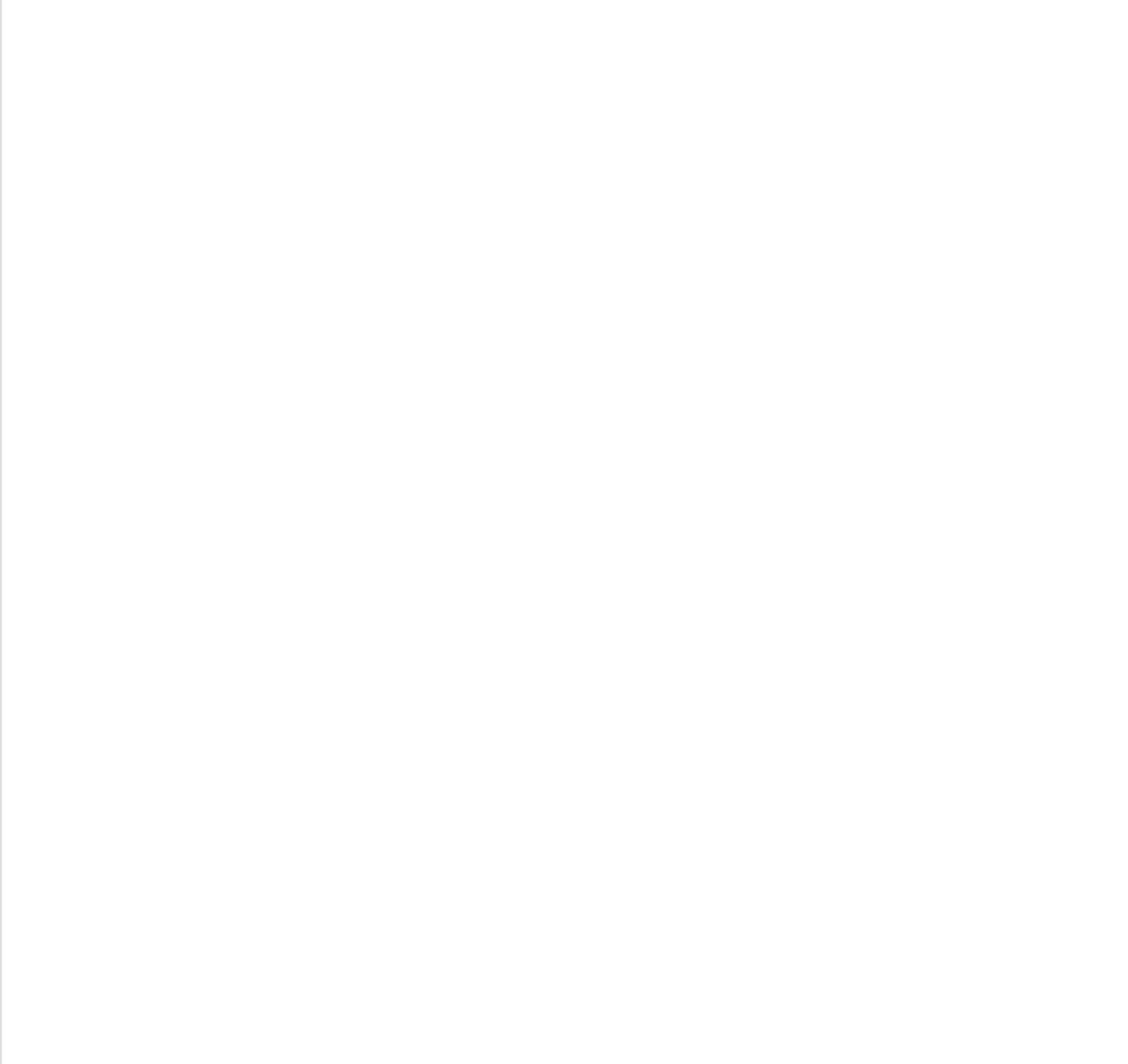} 
\includegraphics[width=0.75\textwidth, draft=True]{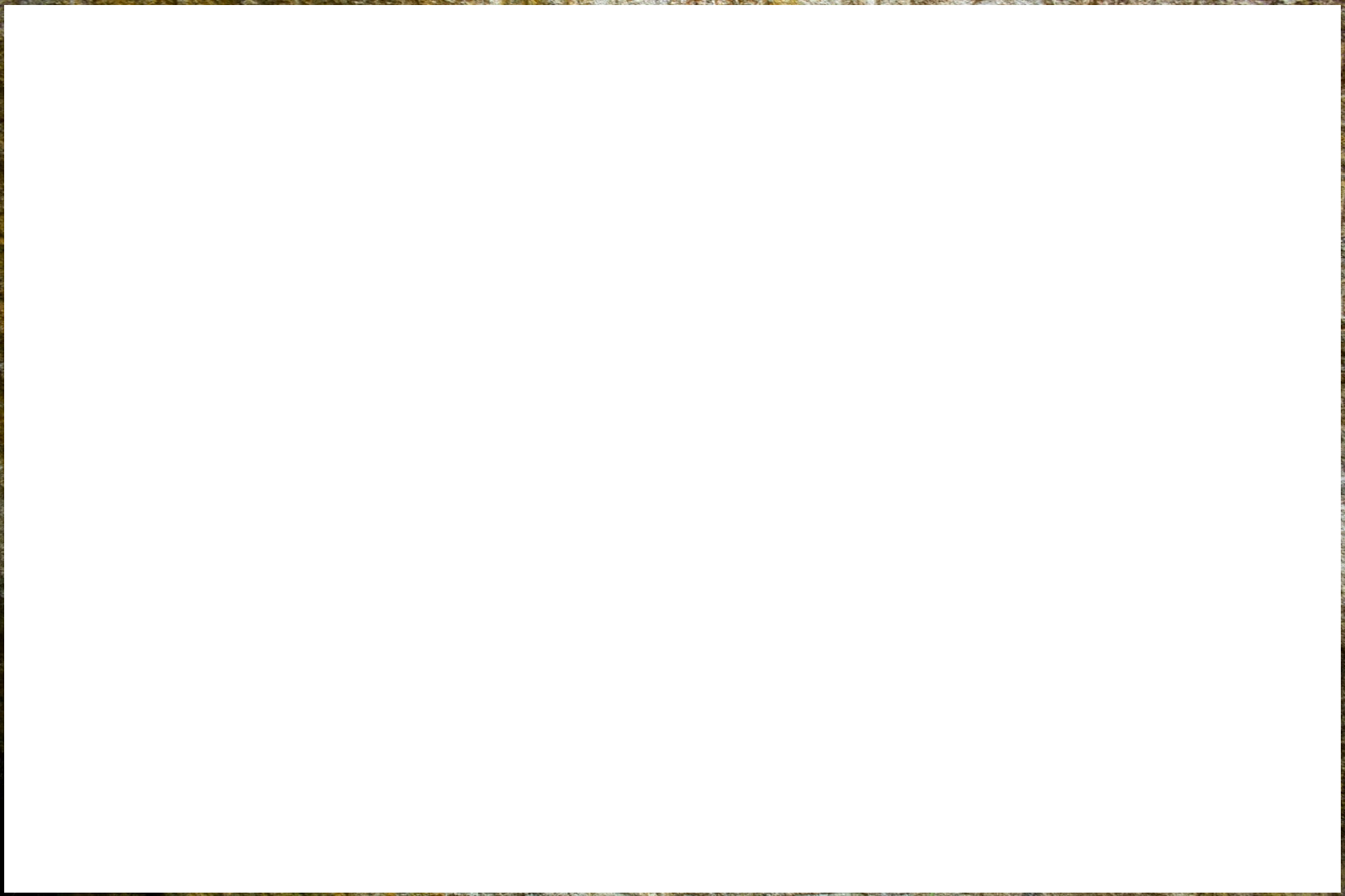} 
\caption[\emph{Sakwala Chakraya} or The Star Gate of Sri Lanka.]{\emph{Sakwala Chakraya} or The Star Gate of Sri Lanka. Situated in the ancient kingdom of Anuradhapura, it is believed to represent either a cosmic map or more of a philosophical concept about the continuity of life in the universe.
Time exact period of origin of this ruin or the garden it is situated in is yet unknown.  
{\bf Top:} A reproduction of the drawings by \citet{Bell1901}. 
{\bf Bottom:} Actual image of the ruin. Photo credit: Volodymyr Kovalov. 
{\bf These figures are removed from the online version due to copyright.}
}
\label{fig:ruins_sri_lanka}
\end{figure}

The grand beauty of the night sky has driven the human curiosity to know and learn more about the vast infinite universe even before the birth of civilization. The drawings of our pre-historic ancestors across the globe show the multitude of imaginative explanations given to explore and understand the curious phenomenons of the day and night sky \citep[eg.,][]{Norris2016}. 
Figure \ref{fig:ruins_sri_lanka} is one such example, where the imagination of the ancient civilizations in Sri Lanka has voyaged through the cosmos to comprehend the unknown.
In the era of global astronomy with multi-billion dollar telescopes and supercomputers we still pursue the same thirst for knowledge, wanting to know more, to demystify the complex universe, and understand our place in it.

The thesis presented herein, attempts to shed light on one small aspect of galaxies with hopes to address a timely concept of galaxy evolution in the distant universe. I hope this thesis will emphasise the beauty of science, that is the ability to change our view on our long held views of the universe in light of new evidence.

\section{Stars and Galaxies}

This thesis revolves around two well known ``building blocks'', \emph{stars} and \emph{galaxies}. Even though these two account for most of the what we see in the universe, they only comprise of traces of baryonic matter, which is only $\lesssim4\%$ of the composition of the universe \citep{PlanckCollaboration2016}. Most of our understanding of the large scale universe comes from the study of this $4\%$ and therefore, it is vital to expand our knowledge on stellar and galaxy formation and evolution in order to comprehend the evolution of the universe.

\emph{A star can be defined as a self-gravitating celestial object in which there is, or there once was (in the case of dead stars), sustained thermonuclear fusion of hydrogen in their core} \citep{LeBlanc2010}. 
Collections of individual stars may constitute a galaxy. However, galaxies are extremely complex, diverse class of objects \citep{Dalcanton2009,Mo2010,Bell2012}. Until the aftermath of the \emph{The Great Debate}  in 1920 \citep{Shapley1921}, the common consensus was that Milky Way comprised of the entire universe. However, in less than 100 years since then, our understanding of galaxies has grown from studying our closest neighbour to discovering the first generation of galaxies formed within $\sim400$ Myr of the birth of the universe \citep[eg.,][]{McLeod2015,Oesch2016}.

\subsection{Modern cosmology and galaxies}

The cosmic web describes the mass density field of the universe, which is believed to have evolved from the perturbations in the aftermath of the big bang in timescales of $\sim10^{-43}$ s. Cosmological simulations show the vast  network of filaments, clusters, and voids that connect galaxies together.  Observationally, the CMB demonstrates the early beginning of these structures, when the universe was only $\sim380,000$ years old. 
The expansion of the universe led to the cooling of the plasma observed by the CMB, allowing the recombination of electrons and protons made during the big bang to form Hydrogen and Helium. The cooling led the opaque plasma to become neutral gas and transparent to observations. 

The fluctuations of the temperatures in the CMB are driven by the quantum fluctuations that were dominant in the aftermath of the big bang.  In this era, the universe was dominated by quantum physics and/or perturbations in the gravitational field \citep[eg.,][]{Cirigliano2005}. These provide astronomers with the first clues of the anisotropies in the large scale structure in the universe, which grew with time to lead the formation of galaxies \citep{Mo2010}. 
Due to effects of gravity, regions even with extremely small relative over-densities started to attract matter, which was a mixture of dark matter and baryonic matter, increasing their density and making the surrounding regions less dense. 
Once the over-dense regions grew and become significantly denser than the surrounding regions, the matter collapsed under gravity and virialized via violent relaxation \citep{Eggen1962}. Smaller regions  collapsed first then merged together to form larger regions, resulting in high densities that drove the gas to decouple from the dark matter.  
The size of the collapsing regions were a function of matter density and the expansion rate of the universe \citep{Silk1968}. Photon diffusion\footnote{The travel of photons from hot regions to cold regions in space.} damped the small scale oscillations of the CMB up to masses $10^{12}-10^{14}$ \msol. However, the Jeans mass \citep{Jeans1902} after the decoupling of the radiation/matter fields was $\sim10^6$ \msol.  
Due to the absence of such small scale perturbations caused by photon diffusion, fragmentation happened in much larger scales of $\sim10^{12}-10^{14}$ \msol. However, due to the constant spatial variation in baryons and photons, the pressure stays uniform, thus perturbations are not affected by the dampening as suggested by \citet{Silk1968}.  Therefore, the presence of small scale structure in initial density perturbations can survive the recombination effect and collapse after the photon pressure can no longer support the baryon pressure. \citet{White1978} suggested a two stage galaxy formation scenario, where dark matter haloes form via hierarchical clustering while the visible content of galaxies form from the cooling and condensation of gas within these haloes. 
This formation scenario accounted for the fact that the dark matter haloes the galaxies were embedded in were more extended compared to the galaxies. This idea was later extended by \citet{Blumenthal1984}, who suggested that in CDM scenarios, typical peaks of the density field results in disk galaxies, while higher density peaks results in ellipticals. 
Either of these processes mark the beginning of the formation of galaxies.

With time, gas in galaxies cooled via mechanisms dependent on their temperature. At very high temperatures ($>10^7$K), gas is fully ionized, and therefore, the dominant mechanism of cooling is via Bremsstrahlung radiation. At moderate temperatures ($10^4-10^6$K), the gas looses energy by converting kinetic energy into photons that escape the gas via various physical processes such as collisional ionization/excitation and recombination \citep{Gnat2007}. When the gas reaches much cooler temperatures ($<10^4$K), collisions with neutral atoms drives the cooling \citep{Dalgarno1972}.  
Cooling of gas lowers thermal pressure resulting in the gas to collapse to the centre of the halo. Angular momentum conservation of the halo during the collapse results in a thin rotating disk with the cool gas rotating faster as it contracts.

\subsection{Stars}
\label{sec:stars}

The cool gas in the galactic disks contains molecular hydrogen and stars are formed from the collapse of molecular clouds into protostars. The conditions that lead the collapse of molecular clouds to form stars is complex and may have many physical origins, with fragmentation, turbulence, magnetic and radiation fields all contributing to the formation and regulation of stars \citep[eg.,][]{Larson1973,Krumholz2014,Federrath2016,Rosen2016}.

A galaxy's mass is commonly measured via its stellar mass and therefore, the rate that stars form within a galaxy plays a vital role in determining the growth of the galaxy. There are many physical processes that regulate the star formation within a galaxy, thus allowing star-formation to maintain for prolonged periods (in Gyr time-scales). UV radiation, prominent in massive stars drives the self regulation of star formation by heating/pushing the excess gas in the molecular clouds prohibiting them from making new stars. Furthermore, supernovae, which occurs at the end of the lives of massive stars generate strong energetic winds that could drive gas in galaxies out of star forming regions \citep[eg.,][ and references therein]{Shattow2015}.
Furthermore, active galactic nuclei (AGN) are also  strong drivers in regulating star formation \citep{Tabor1993}. Radiative energy from AGN can be converted to thermal energy within its host galaxy, heating up the cool gas, thus suppressing star formation \citep{Silk1998}. AGN with radio jets push the gas away from galaxy disks and heat up gas further suppressing star formation \citep{Croton2006}. AGN in low mass halos dominate the production  of UV photons at $z<3$ \citep{Madau1999}, that could ionize the gas in halos suppressing star formation \citep{Somerville2002}.

The temperature, luminosity, lifetime, and the evolution of a star is primarily governed by its mass, where stars with larger masses such as O and B stars, live for shorter lifetimes with bluer colours and hotter temperatures and evolve off the main sequence to ultimately end their life as neutron stars or black holes via undergoing supernovae explosions \citep[eg.,][]{Gennady2010}. 
The less massive A, F, and G type stars classified using their strong Hydrogen absorption lines, have redder colours and cooler temperatures and stay on the main sequence for a prolonged period of time \citep{LeBlanc2010}. 
Less massive stars generally end their life as white dwarves and planetary nebulae. Both supernovae and planetary nebulae are cosmic recyclers; i.e. regenerates vital material for future stellar and planet formation, thus making sure of continuity in the evolution of the universe.

\begin{figure}
\centering
\includegraphics[width=1.0\textwidth]{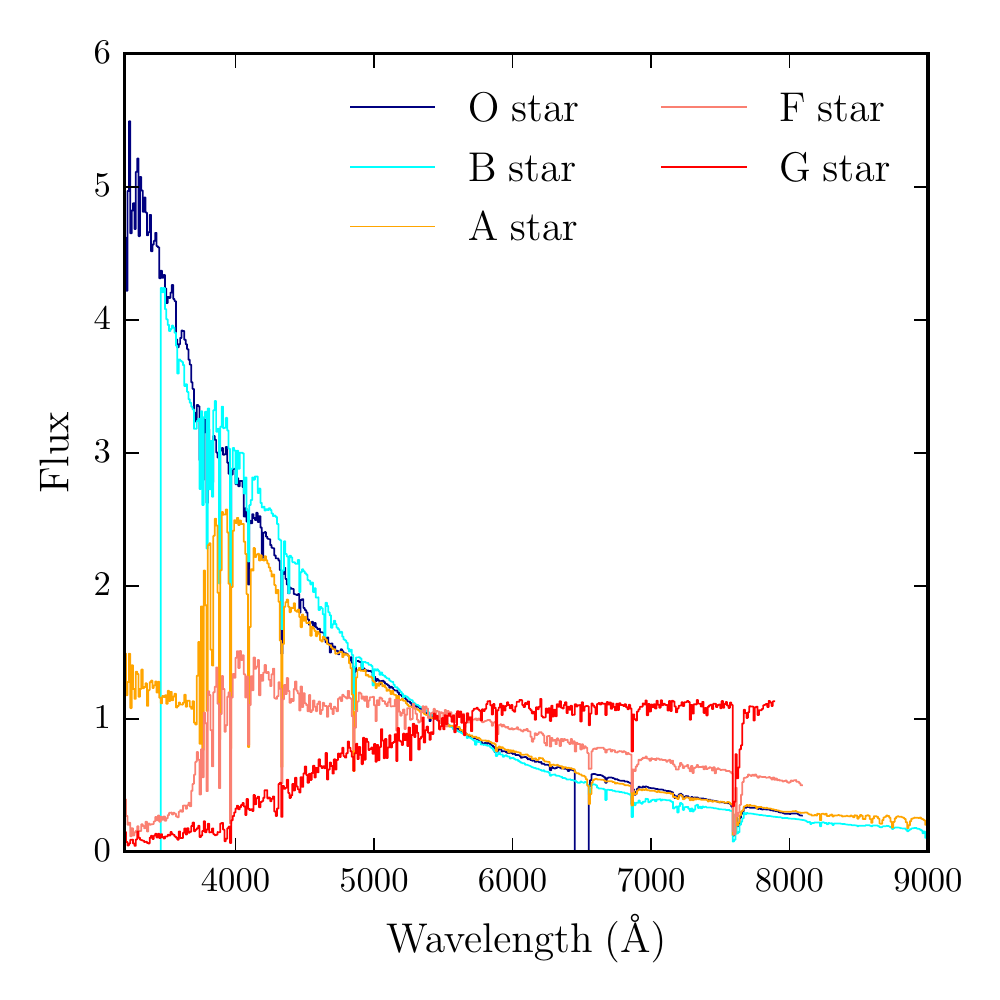} 
\caption[Example spectra of main types of stars discussed in this thesis.]{Example spectra of O (HD47839), B (HD034816), A (HD77350), F(HD36673), and G (HD42454) type stars. All spectra are from STELIB \citep{LeBorgne2003} and have been normalized to 1 at 5450 \AA.}
\label{fig:stel_spec}
\end{figure}

An example set of spectra for O, B, A, F, and G type stars are shown in Figure \ref{fig:stel_spec}. O stars have high UV flux compared to the other stars. Following Wien's law, with decreasing temperatures of the stars, the peak flux becomes redder, thus decreasing the UV flux. 
Furthermore, following the Stefan-–Boltzmann law, the energy emitted by a black-body is $\propto T^4$, where T is the temperature in Kelvin. Under the assumption that stars are blackbodies obeying hydrostatic equilibrium, the luminosity of a star is $\propto M^3$.
Therefore, in dusty molecular clouds, the UV flux is dominated by massive O and B type of stars. Since these stars have shorter lifetimes ($\sim20$ Myr), the UV flux can be a direct proxy for the star formation rate. 
However, UV photons undergo multiple scattering events with the dust particles adding extra complexity to measuring the SFR. 
Less massive F and G type stars show prominent absorption features at redder wavelengths. Dwarf stars have prominent Fe Wing-Ford band and NaI absorption features, which can be used as a proxy for the presence of dwarf stars in a stellar population \citep[eg.,][]{vanDokkum2012}. 

By studying the spectra of integrated stellar populations, such as in galaxies, astronomers can infer relative abundances of different type of stars present in these systems. However, attenuation by dust, uncertainties in stellar atmospheres, stellar rotation, binary star evolution, and errors in construction of empirical models to synthetic spectra may significantly influence the observed spectra of stars \citep[eg.,][]{Eldridge2008,Conroy2012,Leitherer2014}, thus introducing added complexity when relating stellar features directly to abundances of stars within stellar populations.

\subsection{Emission lines}
\label{sec:emission_line_background}

Emission lines that originate from ionized gas within nebula \HII\ regions in galaxies allows astronomers to probe the chemical evolution and star-formation history of galaxies \citep{Osterbrock1989}. These diffuse nebulae contain infant stars. 
Hot, massive O and B type stars formed within these regions ionize and illuminate these regions. 
The hot ionized gas that originate from these stellar regions expands into the cooler neutral gas regions decreasing the gas density and increasing the ionized volume. 
The ionization balance within the \HII\ regions is regulated by the photo-ionization from the UV photons of O, B stars and the recombination of electrons with these ions.  Since a finite number of UV photons can only ionize a specific region, in large gas clouds there is an outer edge for the ionized material. The region within the ionized material is  referred to as the \emph{Str\"{o}mgren sphere}. However, this is an idealized scenario since \HII\ regions are rarely spherical. 
Apart from O and B stars, hot white dwarfs, planetary nebula stars, and supernovae shocks can also produce ionizing photons within \HII\ regions to a lesser degree but are considered to have a much less significant contribution to luminosity in this thesis. Furthermore, AGN are also a dominant source of ionising photons in galaxies, however, we remove galaxies with AGN from our stellar population analysis.

The radiation emitted by \HII\ regions depends on the abundances of  elements determined by the previous evolutionary history of the gas, the local ionization, density, and temperature \citep[eg.,][]{Kewley2013a}. 
Nebula spectra contains prominent hydrogen recombination lines and collisionally excited lines such as \NII, \OII, \OIII\ etc., which are generated by the cooling of the gaseous regions. 
\citet{Zanstra1927} showed that the number of photons emitted by the Balmer series is equal to the energy quanta absorbed in the UV from the ionizing stars. This result, commonly know as the \emph{Zanstra Principle} can be used to determine the temperatures of the stars that produce the UV radiation.
All line photons emitted within the \HII\ regions can either undergo Case A or Case B recombination \citep{Peebles1968,Zeldovich1968}.
By definition, in Case A recombination scenarios, all line photons escapes the clouds without undergoing further scattering events. However, this scenario is only valid for optically thin \HII\ regions. 
In most nebulae due to high optical depths, all photons from $n^2p\rightarrow1^2S$ transitions are immediately reabsorbed by the atoms, which is called Case B recombination.
Therefore, all Lyman photons undergo multiple scattering events, thus loosing its energy to become lower energy photons such as \Halpha. 
\Halpha\ photons have lower optical depth and can escape the dense \HII\ regions. 
Consequently, the \Halpha\ flux can be used as a proxy for the amount of ionizing photons within a \HII\ region, which is a measure of young O, B stars. As aforementioned, O, B stars have very short life spans of $\sim20$ Myr, and therefore, \Halpha\ flux is linked to the star-formation rate of the \HII\ regions. 
Depending on densities and geometries of the \HII\ regions, ionizing photons can escape the star-forming nebula \citep[eg.,][]{Sharma2016} before getting converted to photons of lower energies and thus, SFRs derived via \Halpha\ flux should be considered as a lower limit to the true value. 

Galaxies with active star-formation may contain a high abundance of \HII\ regions. By observing the integrated light of galaxies, astronomers can infer galaxy properties such as star-formation rates, dust, ISM conditions, stellar mass etc.

\subsection{Galaxies across cosmic time}

\begin{landscape}
\begin{figure}
\centering
\includegraphics[height=0.45\textheight]{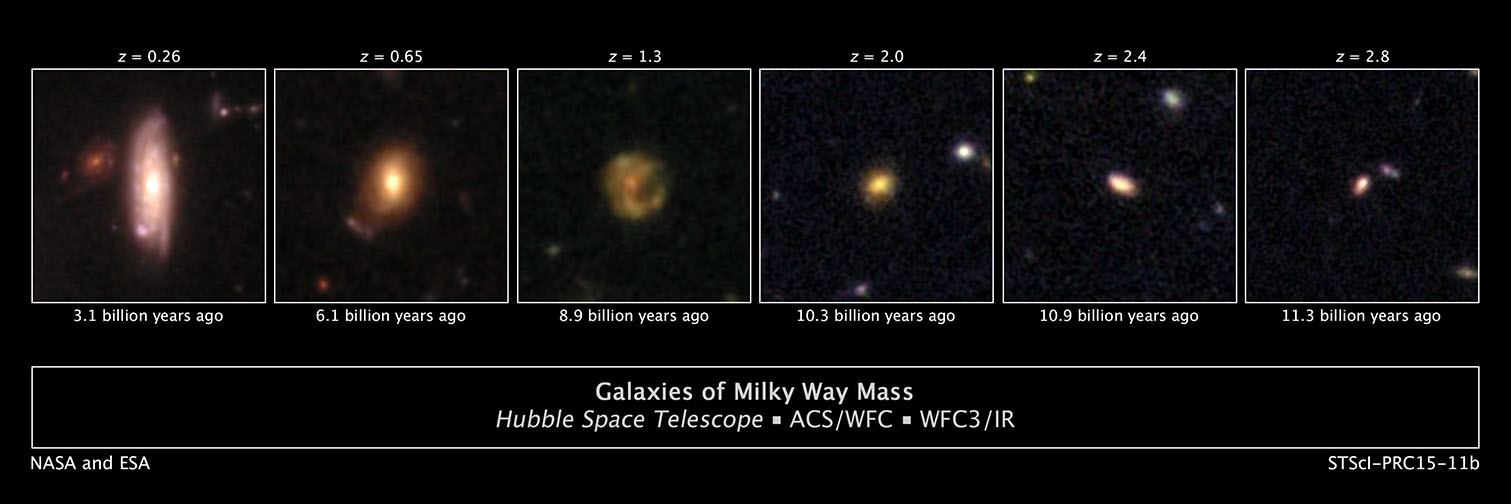} 
\caption[Cosmic evolution of progenitors of Milk-Way mass galaxy from ZFOURGE/CANDLES using a number desnity selection of progenitors.]{Cosmic evolution of a Milk-Way mass galaxy from ZFOURGE/CANDLES \citep{Papovich2015}. From right to left, the panels show the evolution of the galaxies from earlier times to present day. At earlier times, the Milky-Way like galaxies are small with bluer colours and show visual morphologies similar to Lyman-Break galaxies. Galaxies tend to be more diffuse at $z\sim2$ and are also more clumpier. At much later times, galaxies seem to form their spiral and spheroidal disk structures. 
Image credit: NASA/ESA/\citep{Papovich2015}}
\label{fig:papovich_MW_evolution}
\end{figure}
\end{landscape}

Large scale surveys of galaxies have contributed immensely to constrain galaxy evolution models by granting astronomers the opportunity to understand chemical, environmental, and physical factors that drive the evolution of the disk-like star-forming spiral galaxies to passive galaxies in our local neighbourhood \citep[eg.,][]{Patel2013a,Patel2013b,vanDokkum2013b,Papovich2015}. Figure \ref{fig:papovich_MW_evolution} is an example of such a scenario, which shows the evolution of a Milky-Way like galaxy in cosmic time using abundance-matching techniques. 
By observing mass/magnitude complete samples of galaxies deep into the universe, astronomers can use photometric and spectroscopic diagnostic tools to select similar type galaxies.

The evolution of the cosmic star-formation rate density is vital to understand the galaxy mass build up of the universe \citep[eg.,][]{Madau1998,Hopkins2006,Madau2014}. The evolution of the cosmic star-formation rate density to higher values from later times, and gradual decline to present day since $z\sim2$ suggest that massive galaxies in the universe were formed within short time periods at higher redshifts \citep[eg.,][]{Juneau2005,Fontana2006,Newman2010,vanderWel2014}, however the exact physics governing the rapid mass and size growth leading to quiescence in star-formation is not well understood. Different scenarios have been presented to explain the progenitors and subsequent size-growth of these populations such as, sub-millimetre galaxies \citep[eg.,][]{Tacconi2006,Toft2014}, rotating disk galaxies \citep[eg.,][]{Belli2014,Belli2016}, and compact star-forming galaxies \citep[eg.][]{Barro2013}.

Simply, star-formation should be regulated by the amount of gas available for a galaxy, where galaxies with high amounts of HI gas results in higher SFRs \citep[eg.,][]{Genzel2015,Saintonge2016}. However, as mentioned in Section \ref{sec:stars}, feedback mechanisms play a vital role in regulating the SFR. Galaxies further accrete cold gas from the inter galactic medium via filaments and minor mergers, which is expected to be the main driver for sustained star-formation in galaxies within observed time-scales \citep{Hopkins2008,Sancisi2008,Krumholz2012,Lilly2013b,Molla2016}. 
Even though minor mergers of galaxies may also result in the increase of star-formation rate density at $z\sim2$ \citep{Mo1996,Baugh1998}, simulations show that such mergers are not common and that the merger rates are approximately constant throughout cosmic time \citep{Hopkins2010}.
However, the definition of minor and major mergers, the definition of merger rate, and sample variances introduces large uncertainty into merger rate calculations \citep[eg.,][]{Bundy2009,Lotz2011}. 
Additionally, differences in computational algorithms used in simulations for physical processes such as supernovae feedback and galaxy interaction mechanisms adds further complexity in comparing minor merger rates between different studies \citep{Bertone2009}. 
Semi-analytic models further lack the capability to accurately predict the merger fraction and rate history of galaxies when baryonic physics is included alongside dark matter \citep{Conselice2014}.
Furthermore, studies at $z\sim1$ has shown that quiescent galaxies undergo more minor merger events compared to star-forming galaxies \citep{Lopez-Sanjuan2011}, thus, the exact contribution of minor mergers to the cosmic star-formation rate density is uncertain. 
The cosmic SFR declines at lower redshifts ($z<2$) due to the decline in hot and cold gas accretion rates, where the galaxy infall rate slows the cold accretion, while mass growth in hot accretion scenarios is slowed down by longer cooling times \citep{Kere2005}. \citet{Bouch2010} further showed that the decline in the gas accretion rates is a direct consequence of the expansion of the universe.

Accreted gas in galaxies leads to changes in metallicities, SFRs, and masses introducing added complexity to the evolution of galaxies \citep{Kacprzak2016}. 
For galaxies with smaller mass haloes, gas pressure is not sufficient to shock heat the gas, which causes the gas to flow directly into the galaxy by penetrating the hot halo \citep{Kere2005}. 
The injection of gas increases turbulence in the galaxies which is one explanation of the high rates of star-formation and initiation of new star-formation activity \citep{Dekel2009} prominent at high-$z$.

The  correlation between star-formation rate and stellar mass of galaxies, which is observed to higher redshifts \citep[eg.,][]{Whitaker2012b,Speagle2014,Stark2016,Tomczak2016} provides strong evidence that non-quiescent galaxies continuously form stars at all masses. The evolution of the so-called ``main sequence'' of star formation with redshift further implies that galaxies build up their mass smoothly rather than primarily being driven by stochastic major merger events. 
The growth of galaxies with cosmic time can be predicted by the SFR-mass relation by accounting for the growth of the mass function due to star-formation along with galaxy-galaxy mergers and applying continuity equations \citep[eg.,][]{Leja2015}. 
However, studies have shown that extrapolation of star-forming sequence to lower masses leads to disagreement with the observed evolution of the mass function by constantly over-producing the number of lower mass galaxies \citep{Tomczak2016}.  \citet{Tomczak2016} suggested that effects from galaxy mergers were required to account for this discrepancy between observed and expected values, but current galaxy merger rates are too low to account for such variations \citep[eg.,][]{Leja2015,Williams2011}.  
The build up of mass in the universe can be further divided into actively star-forming and quiescent galaxies, where low mass quiescent galaxy populations show a rapid build up of stellar mass between $0<z<1.5$ \citep{Tomczak2014}. The main driver for quenching low mass galaxies is unclear, but studies expect environment to play a larger role in the quenching process by accreting lower mass galaxies into larger haloes \citep[eg.,][]{Brown2008,Quadri2012,Kawinwanichakij2016}. 
Recent studies also show massive quiescent galaxy populations that exist in the very early universe \citep{Muzzin2013,Stefanon2013,Straatman2014,Straatman2015,Glazebrook2017}, however, the exact physics that resulted in such a quick mass build-up and quiescence is a mystery.

Galaxies further undergo morphological evolution during cosmic time. 
Since the morphological classification of galaxies by \citet{Hubble1926}, galaxies have been divided to two main types, elliptical and spiral with a transitional lenticular type of galaxies. The presence of a stellar bar further divided the spiral and lenticular galaxies to barred and normal galaxies. Galaxies with no elliptical or spiral features were classified as irregular galaxies. Further studies, introduced additional layers of divisions between these types \citep[eg.,][]{deVaucouleurs1959}, however, the division between elliptical and spiral remains the main morphological transformation of galaxies.

The evolution of galaxies between different morphological types is an important aspect of galaxy evolution. Elliptical and lenticular galaxies have older stellar populations with low or zero star formation compared to spiral and irregular galaxies, suggesting that galaxies are born as spirals at early times and undergo morphological transformation to become lenticular and elliptical galaxies. The first generation of stellar population synthesis models attempted to explore transformations between morphological types \citep{Tinsley1968}, by attempting to model the colour evolution of stellar populations in galaxies. 
The growth of the galaxy bulge in spirals and subsequent depletion of gas reservoirs, tidal interactions with galaxy neighbours, removal of interstellar gas via ram pressure stripping or strangulation may disrupt the galaxy spiral arms \citep{Somerville2008,Bekki2011}. 
Due to the removal of interstellar gas from the disk, the random motions of the stars will no longer be suppressed, resulting spiral galaxies to be transformed to lenticular galaxies \citep{Binney1998}. 
However, these scenarios require galaxies to be quenched before the morphological transformation \citep{Kovac2010} requiring the star-formation quenching time-scale to be shorter than the morphological transformation time-scale \citep{Bekki2011,Taranu2014,Bahe2015}.

At high-$z$, Hubble's morphological classification breaks down. This is driven by the higher fraction of irregular galaxies observed at these redshifts \citep[eg.,][]{Mortlock2013,Conselice2012}.
A large fraction of irregular luminous star-forming galaxies show clumpy structures \citep{Genzel2011,Wisnioski2013,Guo2015}. Clumpy galaxies show dynamical structure consistent with rotating disks with turbulent gas leading to dynamical instabilities resulting in high SFR clumps \citep{Wuyts2012}.  Studies of local analogues have shown the observations to be affected by resolution effects at high-$z$ \citep[eg.,][]{Fisher2017}, thus complicating stellar population analysis of individual clumps at high-$z$.

The growth of technology and larger partnerships between nations have led the advancement of our understanding of galaxies by building grand telescopes and sensitive detectors, thus allowing the faintest signals in the universe to be detected. The inherent collaborative nature of astronomers has granted opportunities to conduct large scale surveys of the universe allowing large quantities of galaxies to be observed to answer the most profound questions of the universe. 
This thesis is one such example, where a collaboration between astronomers in Australia, Europe, and North America have carried out a survey of galaxies in the distant past to enhance our understanding of galaxy evolution. Henceforth, this thesis will explain the motivation for such large-scale surveys and address one core aspect of galaxy evolution probed by this survey, the stellar initial mass function of galaxies.  


\section{The role of galaxy surveys in our understanding of the universe}

The rapid development of very deep multi-wavelength imaging surveys from the ground and space in the past decade has greatly enhanced our understanding of  important questions in galaxy evolution particularly through the provision of `photometric redshift' estimates (and hence the evolutionary sequencing of galaxies)  from multi-band SED fitting \citep[eg.,][]{Whitaker2011,McCracken2012,Skelton2014}. Studies using data from these surveys have led to a more detailed understanding of topics such as the evolution of the galaxy mass function \cite[eg.,][]{Marchesini2010,Muzzin2013,Tomczak2014,Grazian2015}, stellar population properties \cite[eg.,][]{Maseda2014,Spitler2014,Pacifici2015}, evolution of galaxy morphology \cite[eg.,][]{Huertas-Company2015,Papovich2015}, and the growth of the large-scale structure in the universe \citep{Adelberger2005b,Wake2011}.

\subsection{The local universe}

Galaxies at $z\sim0$ are diverse. The 13.7 billion years of evolutionary time has allowed galaxies to undergo complex star-formation and merger scenarios resulting in a large diversified local neighbourhood. 
Diversity of these galaxies has been explored extensively by the large-scale photometric surveys such as the Sloan Digital Sky Survey  \citep[SDDS.,][]{York2000}, 2 degree Field Galaxy Redshift Survey \citep{Colless2001}, and the Galaxy and Mass Assembly Survey  \citep[GAMA.,][]{Driver2009} revolutionizing our understanding of galaxy evolution.

Galaxies at $z\sim0$ are comparatively brighter in the sky, thus observations do not require large 8-10m class telescopes in order to study many of the basic galaxy properties. Furthermore, galaxy properties can be investigated by studying the rest-frame optical features, whose wavelength coverage is ideal for ground based observations.  
Therefore, using the 2.5 meter Sloan Foundation telescope \citep{Gunn2006}, surveys such as SDSS has been able to explore the low-redshift universe in great detail and so far has identified over 100 million galaxies with photometry and obtained over one million galaxy spectra in a range of physical and environmental parameters. 
These surveys have enhanced our understanding of the universe and have led to multiple key results in many areas of astrophysics including but not limited to the discovery  of the bimodal distribution of galaxies in the local universe and its implications to galaxy evolution \citep[eg.][]{Baldry2004,Menci2005,Martinez2006}, constraints on star-formation quenching mechanisms \citep[eg.][]{,Baldry2006,Cortese2009,Davies2016} and IMF and its universality \citep{Baldry2003,Hoversten2008,Gunawardhana2011}, the discovery of low surface brightness and faint dwarf galaxies in the local universe and their population properties \citep[eg.,][]{Zucker2004,Barazza2006,Williams2016}, the discovery of methane dwarfs \citep{Strauss1999,Tsvetanov2000}, the discovery of populations of high redshift quasars \citep{Fan2001,Fan2003,Fan2004,Fan2006}, merger histories of galaxies \citep[eg.,][]{Bell2006,Ferreras2017}, and constraints on the cosmological parameters from the local universe \citep[eg.,][]{Budavari2003,Tegmark2004, Eisenstein2005,Gheller2016}.  
However, to further constrain galaxy evolution and to understand the origin of the diverse local neighbourhood it is vital to study galaxies as a function of cosmic time.


\subsection{Advances with Deep Near-IR Imaging Surveys}

Stellar mass is especially useful for tracking galaxy evolution as it increases monotonically with time, however, as shown in Section \ref{sec:specz_of_z2_galaxies} at higher redshifts NIR data (wavelengths between $\sim7000-25000$ \AA) becomes vital to trace the stellar masses \citep{Brinchmann2000,Muzzin2009} and for the provision photometric redshift estimation \citep{Dahlen2013,Rafelski2015}.

New surveys have been made possible by the recent development of relatively wide-field sensitive NIR imagers in 4-8m telescopes such as FourStar \citep{Persson2013} , HAWK-I \citep{Pirard2004}, NEWFIRM \citep{NEWFIRM} and VIRCAM \citep{Dalton2006}. 
Surveys such as ZFOURGE \citep{Straatman2016}, the NMBS \citep{Whitaker2011}, ULTRAVISTA \citep{McCracken2012}, and IBIS \citep{Gonzalez2010} have obtained deep imaging over relatively large  sky areas (up to 1.5 deg$^2$) including UV, optical, and FIR wavelength regimes. The introduction of near-infrared medium-band filters ($\Delta\lambda\sim 1000$\AA) has resulted in photometric redshifts with accuracies of \around1--2\% \citep{Whitaker2011} and enabled galaxy properties to be accurately derived by SED fitting techniques such as EAZY \citep{Brammer2008} and FAST \citep{Kriek2009}.

The NMBS survey pioneered the use of NIR medium band filters \citep{vanDokkum2009} to observe $\sim13,000$ galaxies at $z>1.5$ to study galaxy growth, star formation, and environment. However, since the survey was carried out using the 4 meter Mayall telescope, observations were biased towards massive brighter galaxies, and thus did not provide a complete view of the universe at $z>1.5$.

The ZFOURGE survey implemented the same technique to study galaxy evolution using the purpose built FourStar imager \citep{Persson2013} on the 6.5 meter Magellan Telescope at the Las Campanas Observatory.  
ZFOURGE performed a 45 night observation program covering 121 arcmin$\mathrm{^2}$ each in COSMOS, CDFS, and UDS cosmic fields. Using CANDLES survey data \citep{Grogin2011,Koekemoer2011}  and the wealth of multi-wavelength legacy data sets \citep{Giacconi2002,Capak2007,Lawrence2007}, ZFOURGE was able to generate galaxy catalogues spanning from UV to FIR (0.3--350$\mu$m in the observed frame), to produce photometric redshifts accurate to $1-2\%$ \citep{Straatman2016}. The high quality SEDs of galaxies were used to derive rest-frame colours of galaxies via EAZY, and FAST was used to derive galaxy ages, SFRs, stellar masses, metallicities, and extinction values by minimum $\chi^2$ fits to galaxy SEDs.

\begin{figure}
\centering
\includegraphics[width=1.0\textwidth]{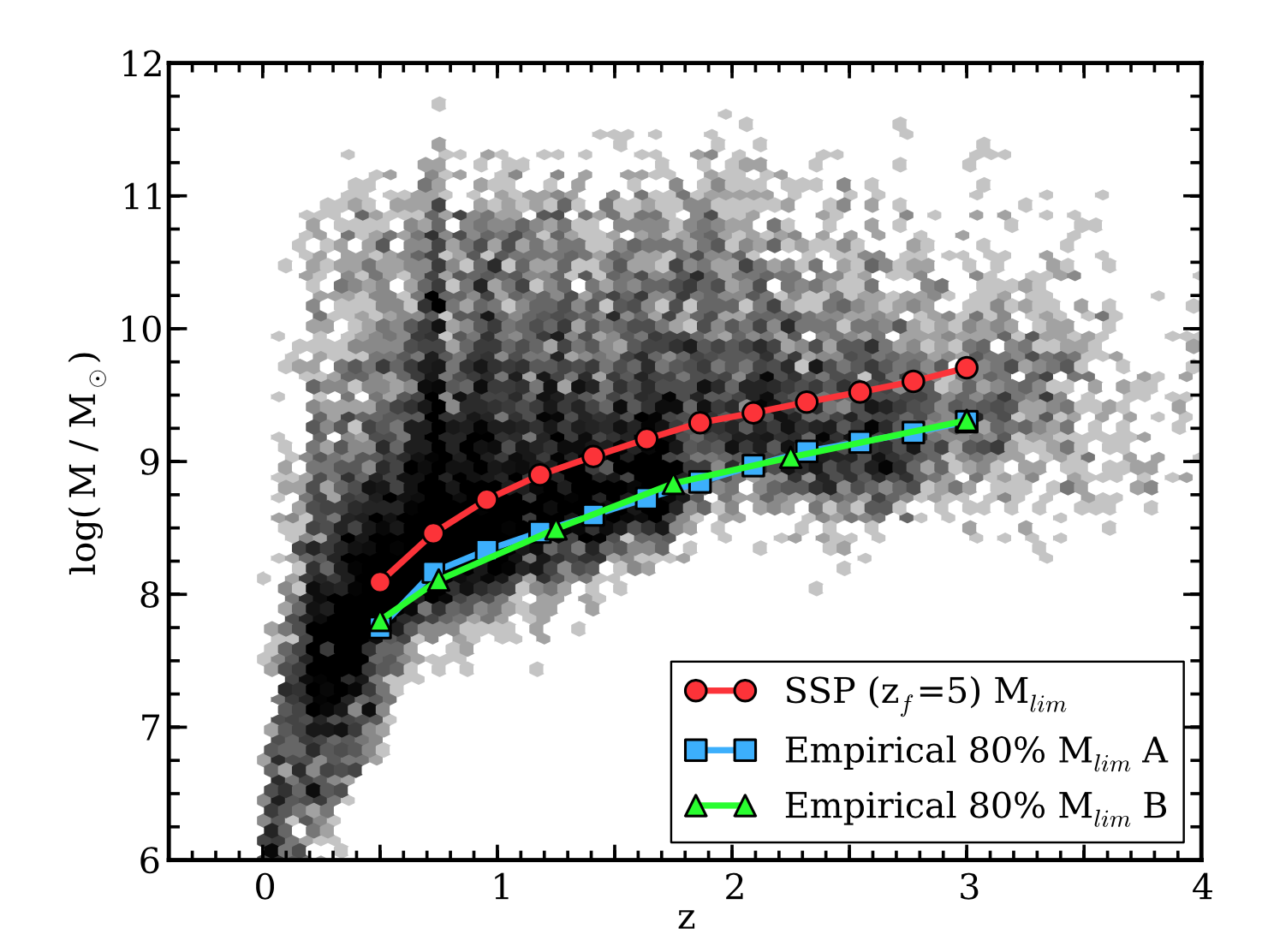} 
\caption[Reproduction of the \citet{Tomczak2014} Figure 2, which shows the distribution of galaxy masses of the ZFOURGE survey as a function of redshift.]{Reproduction of the \citet{Tomczak2014} Figure 2, which shows the distribution of galaxy masses of the ZFOURGE survey as a function of redshift. Blue squares and green triangles are empirically derived 80\% mass completeness limits, which were derived following methods outlined by \citet{Quadri2012} and \citet{Chang2013}, respectively. The red circle tracks show the mass completeness derived for quiescent galaxies using a SSP model that formed at $z=5$.}
\label{fig:tomczak_redshift_mass_evolution}
\end{figure}

The deep observations of ZFOURGE allowed us to discover extremely dusty \citep{Spitler2014} and massive quiescent galaxies \citep{Straatman2014} in the early universe. The deep Ks band imaging along with accurate photometric redshifts led to the discovery of a massive cluster at $z\sim2$ \citep{Spitler2012}. The evolution of Milky-way like progenitors were studied to investigate the  evolution of our own galaxy \citep{Papovich2015}. 
Furthermore,  multi-wavelength data was used to study the role of AGN in galaxy evolution \citep{Cowley2016,Rees2016}. As shown by Figure \ref{fig:tomczak_redshift_mass_evolution}, ZFOURGE was able to probe galaxies to very high mass completeness limits to higher redshifts. Therefore, \citet{Tomczak2014}  was able to study the 
the evolution of the mass functions to $z\sim3$ using high quality SED fitting via FAST.

Therefore, it is evident that photometric redshift surveys such as ZFOURGE have greatly enhanced our understanding of the universe at $z\sim2$, which is a critical epoch in the evolution of the universe. At this redshift, the universe was only 3 billion years old and was at the peak of cosmic star formation rate activity \citep{Hopkins2006,Lee2015}. We see the presence of massive, often dusty, star-forming galaxies \citep{Spitler2014,Reddy2015} which were undergoing rapid evolution and the development of a significant
population of massive, quiescent galaxies \citep{vanDokkum2008,Damjanov2009}.
Galaxy clusters have also now been identified at $z\sim2$, and results
indicate that this may be the epoch when environment starts to influence galaxy evolution \citep{Gobat2011,Spitler2012,Yuan2014,Allen2015,Casey2015}.


\subsection{Need for Spectroscopy}

Even though immense progress on understanding galaxy evolution has been made possible by deep imaging surveys, the spectroscopy of galaxies remains critically important. Spectroscopy provides the basic, precision redshift information that can be used to investigate the accuracy of photometric redshifts derived via SED fitting techniques. The galaxy properties derived via photometry have a strong dependence on the redshifts, and quantifying any systematic biases will help constrain the derived galaxy properties and understand associated errors. 
Spectral emission and absorption lines also provide a wealth of information on physical processes  and kinematics within galaxies \citep{Shapley2009}. Spectroscopy also provides accurate environmental information (for example, the velocity dispersions of proto-clusters; e.g. \cite{Yuan2014})  beyond the resolution of photometric redshifts.

Rest-frame UV spectroscopy of galaxies provides information on the properties of massive stars in galaxies and the composition and kinematics of the galaxies' ISM \citep[][]{Dessauges2010,Quider2010}. 
Rest-frame optical absorption lines are vital to determine the older stellar population properties of the galaxies \citep[eg.,][]{vandeSande2011,Belli2014}. Rest-frame optical emission lines provide information on the state of the ionized gas in galaxies, its density, ionization degree, and metallicity \citep{Pettini2004,Steidel2014,Kacprzak2015,Shimakawa2015,Kewley2016,Kacprzak2016}.


\subsection{Spectroscopy of $z\lesssim1$ Galaxies}

Large-scale spectroscopy is now routine at the low redshift universe. 
Surveys such as the SDSS, the 2-Degree Field Galaxy Redshift Survey \citep[][]{Colless2001}, the 6-Degree Field Galaxy Survey \citep{Jones2004}, and the GAMA survey \citep[][]{Driver2009} extensively explored the $z\la 0.2$ universe ($10^5$--$10^6$ galaxies). 
Galaxies at $z\sim1$ are \around100 times more faint than local counterparts observed by SDSS and thus require much larger telescopes and more observing time to build up representative samples of galaxies. 
Using large optical telescopes, surveys such as DEEP2 Galaxy Redshift Survey \citep{Newman2013}, the VVDS  \citep[]{LeFevre2005}, the VIMOS Public Extragalactic Survey \citep{Garilli2014},  zCOSMOS \citep{Lilly2007}, and AGES \citep{Kochanek2012}  have produced large spectroscopic samples ($10^4$--$10^5$ galaxies). 
The large number of galaxies sampled in various environmental and physical conditions by these surveys has placed strong constraints on galaxy models at $z<1$ while revealing rare phases and mechanisms of galaxy evolution \cite[e.g.,][]{Cooper2007,Coil2008,Cheung2012,Newman2013}.


\subsection{Spectroscopy of $z\sim2$ Galaxies}
\label{sec:specz_of_z2_galaxies}

At a $z\gtrsim1.5$ rest-frame optical features are redshifted to the NIR regime and therefore accessing these diagnostics becomes more challenging. Historically, the spectroscopy of galaxies in these redshifts focussed on the follow up of Lyman break galaxies, which are rest-frame UV selected  using the distribution of the objects in $\cal{U}$, $\cal{G}$, and $\cal{R}$ colour space \citep{Steidel1992}. This technique takes advantage of the discontinuity of the SEDs near the Lyman limit. \citet{Steidel2003} used this technique to target these candidates with multi-object optical spectrographs to obtain rest frame UV spectra for \around1000 galaxies at $z\sim3$. 
Furthermore, $\cal{U}$, $\cal{G}$, and $\cal{R}$ selections can be modified to select similar star-forming galaxies between $1.5<z<2.5$ via their U-band excess flux \citep{Steidel2004}. 
Such sample selections are biased toward UV bright sources and do not yield homogeneous mass complete samples.

\begin{landscape}
\begin{figure}
\centering
\includegraphics[height=0.40\textheight]{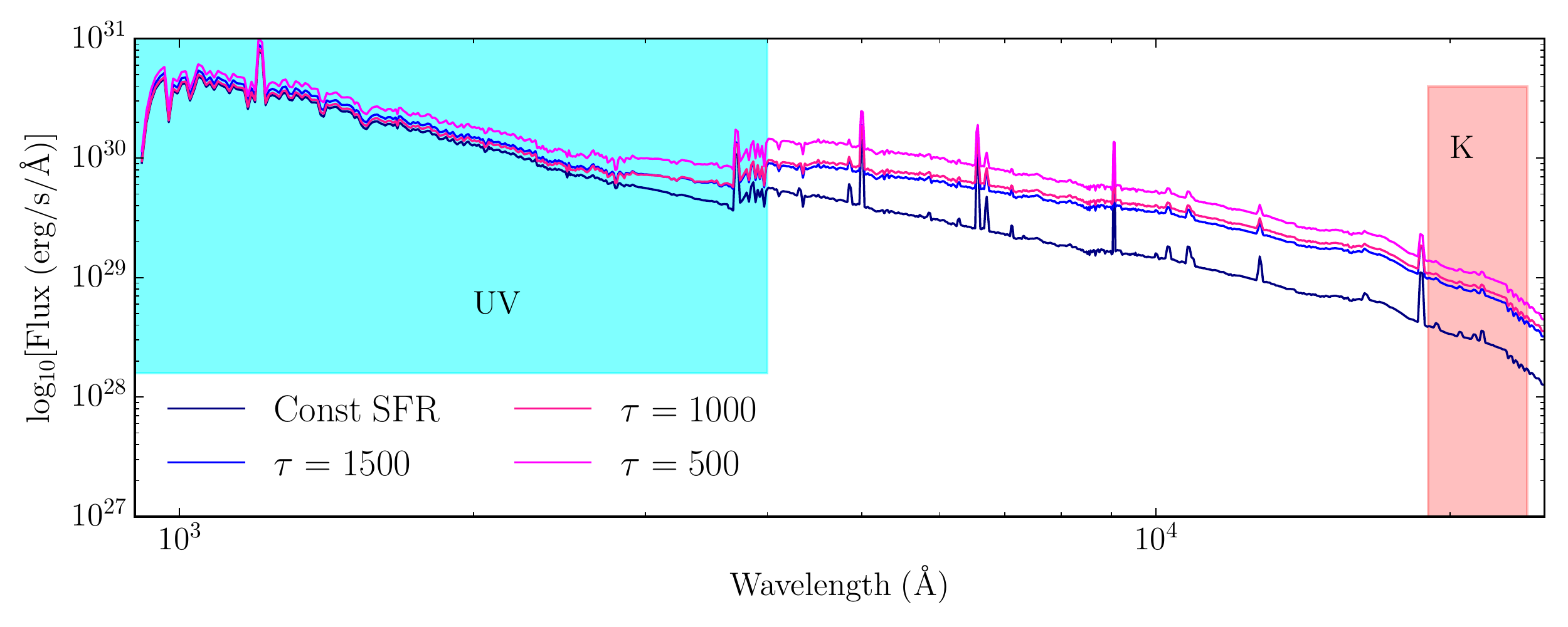} 
\includegraphics[height=0.40\textheight]{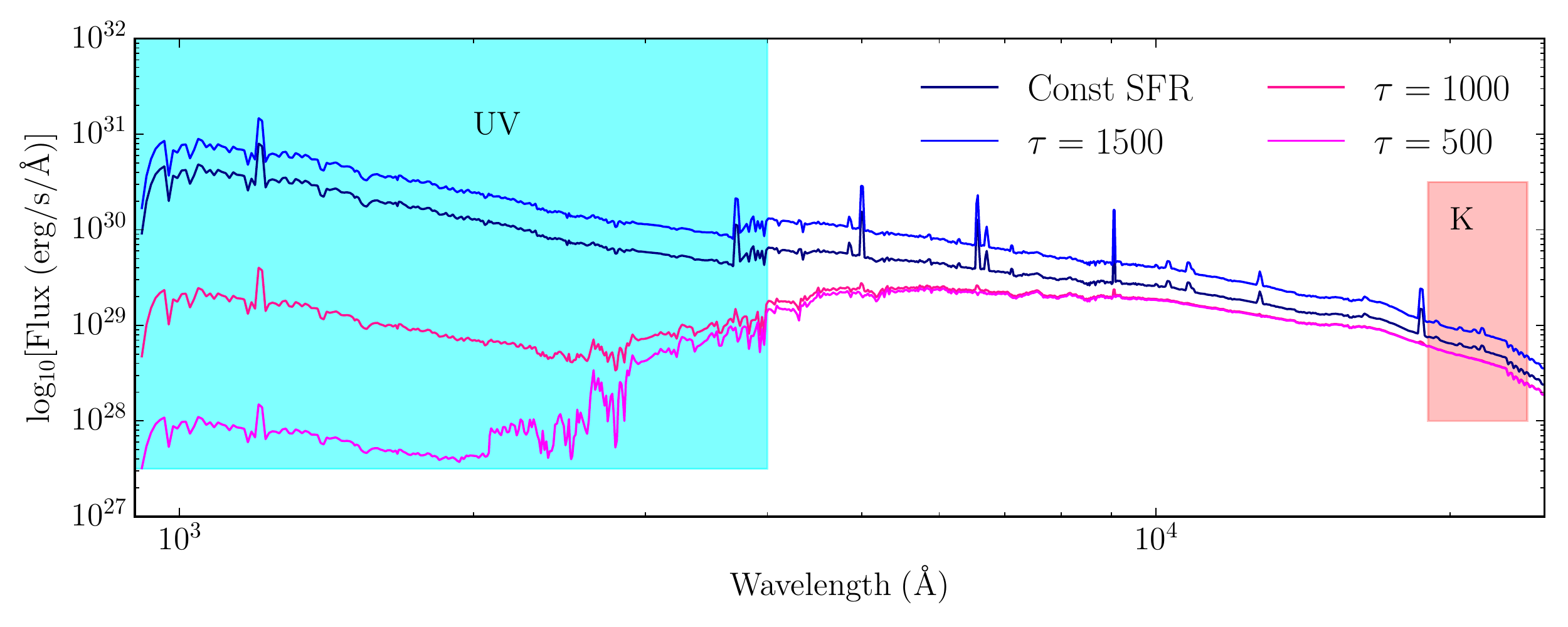} 
\caption[PEGASE model spectra of galaxies with different SFHs at similar SFRs and stellar masses.]{
PEGASE model spectra of galaxies with different SFHs at similar SFRs and stellar masses. Model galaxies are generated following a \citet{Salpeter1955} IMF and constant and exponentially declining ($\mathrm{SFR \propto exp(-t/\tau)}$, where t is the time in Myr and $\tau$ is the decay time-scale in Myr) SFHs. UV and NIR K band spectral regions are highlighted by cyan and pink regions, respectively. 
{\bf Top:} Model spectra at times when they have same SFRs. 
{\bf Bottom:} Model spectra at times when they have same stellar masses. 
 }
\label{fig:PEG_spectra_at_const_MassSFR}
\end{figure}
\end{landscape}

To illustrate mass vs SFR relation with spectral luminosity, Figure \ref{fig:PEG_spectra_at_const_MassSFR} shows PEGASE model spectra with different SFHs plot at similar SFRs and stellar masses.  
UV spectra traces the young O and B stars and therefore, are a direct probe of the SFR. At similar SFRs, all model galaxies with different SFHs show approximately similar UV flux.  
However, at similar stellar masses the UV spectra show deviations of \around3 dex, thus it is not a good tracer for stellar mass. 
The NIR region is ideal for such selections, since it traces the mass-to-light ratios of galaxies. This is evident from the model spectra, where galaxies with similar stellar masses show deviations of $\lesssim0.3$ dex in the K band.

Surveys such as the GDDS \citep[][]{Abraham2004} and the Galaxy Mass Assembly ultra-deep Spectroscopic Survey \citep[][]{Kurk2013} have attempted to address the mass incompleteness in UV by using the IR selection of galaxies (hence much closer to mass-complete samples) before obtaining optical spectroscopy.  
The K20 survey \citep{Cimatti2002} used a selection based on Ks magnitude (Ks$<20$ Vega) to obtain optical spectroscopy of extremely dusty galaxies at $z\sim1$. 
However, selection of galaxies purely based on flux does not preferentially yield higher redshift galaxies.
The GDDS sample is comprised of 309 K$<$20.6 Vega galaxies between $0.8<z<2$, but only $\sim20\%$ the sample had redshifts $z>1.3$.
In order to select spectra from higher redshift galaxies, VVDS selected faint galaxies with I$<$24.75.  Though their survey goal was to study galaxies at redshifts between $1<z<5$, only 6\% of the targeted galaxies had redshifts $z>1.4$.
Therefore, it is evident that these optical surveys are limited by low numbers at $z\sim2$.
DEEP2 survey used a combination of colour and surface brightness of galaxies to spectroscopically probe galaxies to $z\sim1.5$, but there was a significant dearth of numbers beyond $z>1.15$. 
Furthermore, these surveys have provided redshift information, but only rest-frame UV spectral diagnostics, and many red galaxies are extremely faint in the rest-UV requiring very long exposure times.

However, with the use of NIR medium band filters and sophisticated SED fitting techniques, surveys such as ZFOURGE have provided us with a unique opportunity to select large samples of mass/magnitude complete samples at $z\sim2$ with extremely high accuracy for spectroscopic follow-up.

The development of near-IR spectrographs has given us access to rest-frame optical spectroscopy of galaxies at $z\gtrsim1.5$, but the ability to perform spectroscopy of a large number of galaxies has been hindered due to low sensitivity and/or unavailability of multiplexed capabilities. 
For example the MOIRCS Deep Survey \citep{Kajisawa2006} had to compromise between area, sensitivity, number of targets, and resolution due to instrumental limits with MOIRCS in Subaru \citep{Ichikawa2006}. 
The Subaru FMOS galaxy redshift survey \cite{Tonegawa2015}, yielded mostly bright line emitters due to limitations in sensitivity of FMOS \citep{Kimura2010}. 
Furthermore, FMOS does not cover the longer K-band regime which places an upper limit for \Halpha\ detections at $z\sim1.7$. Sensitive long slit spectrographs such as GNIRS  \citep{Elias2006} and XShooter \citep{Vernet2011} have been utilised to observe limited samples of massive galaxies at $z\sim2$.
NIR-grism surveys from the \emph{HST} have yielded large samples such as in the 3DHST survey \citep{Momcheva2015,Treu2015} but have low spectral resolution ($R\sim70-300$) and do not probe wavelengths $>2$ \micron.

With the introduction of the MOSFIRE, a cryogenic configurable multislit system on the 10m Keck telescope \citep{McLean2012}, we are now able to obtain high-quality near-infrared spectra of galaxies in large quantities \citep{Kulas2013,Steidel2014,Kriek2015,Wirth2015}. 
The Team Keck Redshift Survey 2 observed a sample of 97 galaxies at $z\sim2$ to test the performance of the new instrument \citep{Wirth2015} and  investigate the ionization parameters of galaxies at $z\sim2$.
The Keck Baryonic Structure Survey is an ongoing survey of galaxies currently with 179 galaxy spectra, which is primarily aimed to investigate the physical processes between baryons in the galaxies and the intergalactic medium \citep{Steidel2014}. 
The MOSDEF survey is near-infrared selected and aims to observe \around1500 galaxies $1.5<z<3.5$ to study stellar populations, Active Galactic Nuclei, dust, metallicity, and gas physics  using  nebular emission lines and stellar absorption lines \citep{Kriek2015}. 

In Chapter \ref{chap:zfire_survey}, I describe the spectroscopic follow up of ZFOURGE, which was carried out using the MOSFIRE spectrograph. I explain sample selections, observations, and the data reduction procedures, which the ZFIRE survey used to obtain $\sim300$ galaxy redshifts at $z<1.5$.


\section{The initial mass function of galaxies}

The combination of high quality deep photometric data from ZFOURGE, and mass complete spectroscopic samples by ZFIRE, provided us with the unique opportunity to study commonly overlooked fundamental galaxy properties at $z\sim2$. In this thesis, I focus on the study of the stellar IMF of galaxies at $z\sim2$.

The IMF or the mass distribution of stars formed in a volume of space at a given time is one of the most fundamental empirically derived relations in astrophysics \citep{Salpeter1955,Miller1979,Kennicutt1983,Scalo1986b,Kroupa2001,Chabrier2003,Baldry2003}.  
Since the mass of a star is the primary factor that governs its evolutionary path, the collective evolution of a galaxy is driven strongly by its distribution of stellar masses \citep{Bastian2010}.
Therefore, understanding and quantifying the IMF is of paramount importance since it affects  galactic SFRs, galactic chemical evolution, formation and evolution of stellar clusters, stellar remnant populations,  galactic supernova rates, the energetics and phase balance of the ISM, mass-to-light ratios, galactic dark matter content, and how astronomers model galaxy formation and evolution \citep{Kennicutt1998b}. 
Historically, the IMF has been assumed to be a universal function due to lack of observational evidence for systematic variations and limitations in theoretical models to derive an IMF from first principles. 


\subsection{The origin and the functional form of the IMF}

The IMF is the product of fundamental properties of the giant molecular clouds that collapse due to gravity to become the birth place of stars. The physical and environmental conditions of these molecular clouds govern the mass distributions of the stars formed.
The IMF was first introduced by Edwin Salpeter for stars with masses between $\mathrm{0.4M_\odot\leq M\leq10.0M_\odot}$ as a power law in the form of,
\begin{equation}
\phi(log\ m) = dN/d(log\ m) \propto m^{-\Gamma} ,
\label{eq:imf_def_salp}
\end{equation}
where $m$ is the mass of a star, $N$ is the number of stars a logarithmic mass bin, and $\Gamma =1.35$ is the logarithmic slope \citep{Salpeter1955}.

\begin{figure}
\centering
\includegraphics[width=1.0\textwidth]{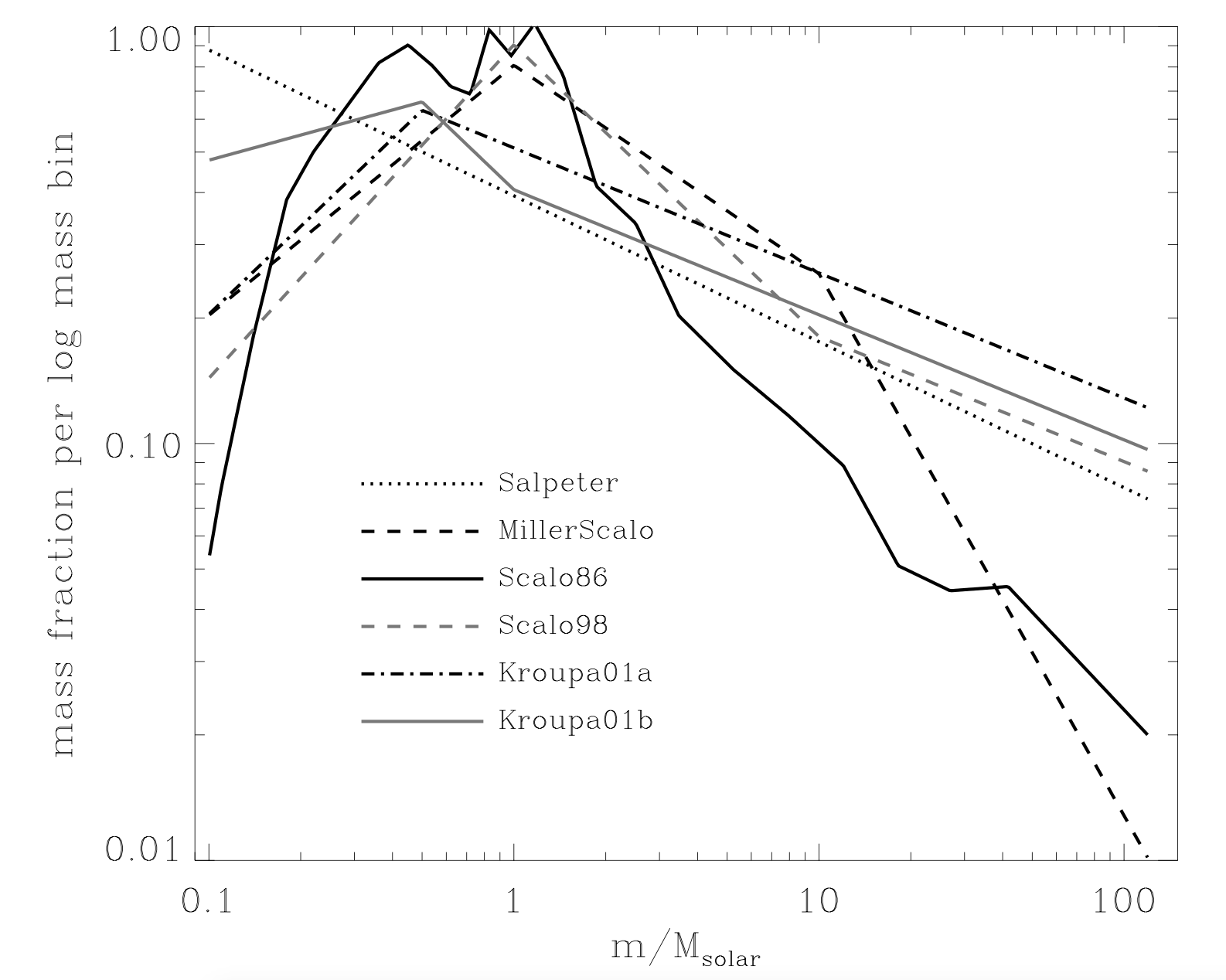} 
\caption[Reproduction of the \citet{Baldry2003} Figure 1, which shows the IMFs derived by different studies.]{
Reproduction of the \citet{Baldry2003} Figure 1, which shows the IMFs derived by different studies. The IMFs shown here are \citet{Salpeter1955}, \citet{Miller1979}, \citet{Scalo1986b}, \citet{Scalo1998}, and \citet{Kroupa2001}.  
 }
\label{fig:baldry2003_fig1}
\end{figure}

Further studies of the IMF placed stronger constraints on the functional form of the IMF and segmented power laws were introduced to describe the mass distributions \citep[eg.,][]{Kroupa2002}. This was driven by the mass distribution favouring a typical peak mass $\sim0.2-0.3$ \msol\ with a decline in the number of stars in either side of the peak. 
Star forming conditions of stellar populations are complex and therefore, we can assume it to depend on many independent variables. From the central limit theorem, the IMF can also therefore, be explained using a log normal distribution (with some modifications) as shown by \citet{Miller1979}. Some example IMFs derived by different studies are shown by Figure \ref{fig:baldry2003_fig1}.

\begin{figure}
\centering
\includegraphics[trim= 100 10 10 10 , clip, width=1.10\textwidth]{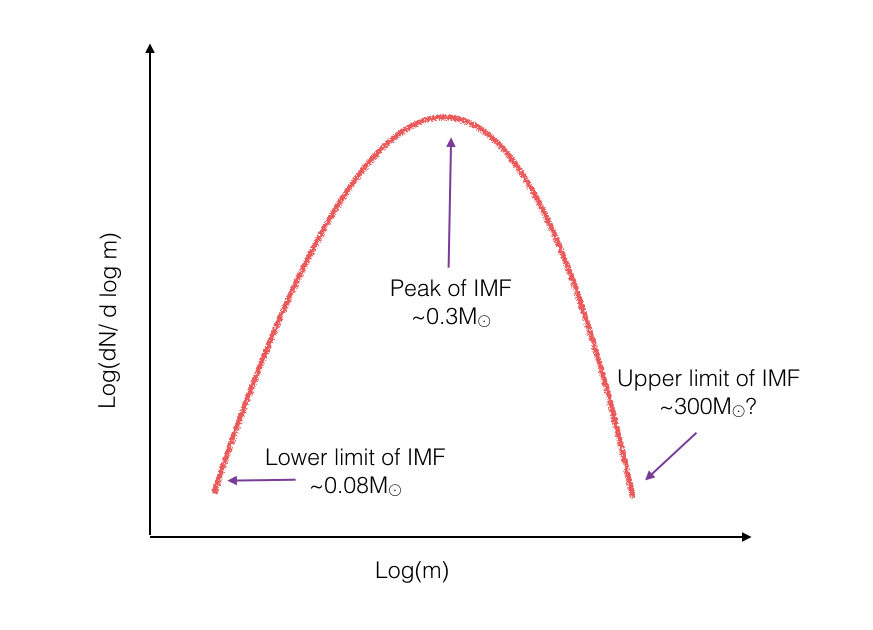} 
\caption[Illustration of the upper and lower limits and the peak of the IMF.]{Illustration of the upper and lower limits and the peak of the IMF. The x axis is in units of $\mathrm{log_{10}}$ solar mass.}
\label{fig:imf_mass_limits}
\end{figure}

Since the IMF is parametrized by a power law, it cannot reach zero, and therefore, is truncated between a physically motivated minimum and maximum masses.
Here, we divide the IMF into three sections: The low mass slope for $M \leq 0.3$ \msol, the peak of the IMF at $\sim0.3$ \msol, and the high mass end of the IMF for $M\geq0.3$ \msol, which is shown by Figure \ref{fig:imf_mass_limits}. 
The low mass end of the IMF is limited by the stellar mass limit of brown dwarfs, where stars with $M\leq0.08$\msol\ due to lack of self gravity are unable to ignite hydrogen to reach the main sequence. 
The high mass end of the stellar masses are not well constrained due to limitations in our numerical models. 
With increase in mass, stars reach the Eddington limit \citep{Eddington1917}, where radiation pressure competes with  gravitational energy for the stability of the star \citep[eg.,][]{Klassen2016}. Furthermore, radial pulsations make stellar atmospheres unstable. 
However, recent work by \citet{Rosen2016} suggest that Rayleigh Taylor instabilities in radiation pressure dominated bubble shells to overcome radiation pressure barriers and regulate massive star formation. 
More sophisticated modelling of feedback from stellar winds, magnetic fields, and collimated outflows are required in numerical models to understand realistic scenarios of massive star formation. 
The highest mass star observed to date is $\sim300$ \msol\ \citep{Crowther2016} and most synthetic stellar population models have upper mass limits of $\sim100$ \msol.  
The peak of the IMF is determined by complex physics \citep[eg.,][]{Whitworth1998,Larson2005,Krumholz2016}. However, \citet{Krumholz2016} showed that magnetic flux coupled with high temperatures favours the formation stars of $\sim0.3$ \msol, which was found to be independent of the time-evolution or initial conditions of the star-forming regions.

Apart from producing an IMF consistent with observations, simulations also require the SFR to be regulated to realistic terms. \citet{Federrath2015} showed that star cluster formation that includes turbulence, magnetic fields, and outflow feedback from multiple sources are required to regulate the SFR. Furthermore, the results indicated that turbulence and magnetic fields in molecular clouds play a larger role in the star-formation process. Hence, the determination of the IMF requires complex ultra-high resolution magnetohydrodynamical simulations to be performed accounting for such effects.


\subsection{Methods of investigating the IMF}

The IMF of stellar populations can be investigated either via direct \citep[studies involving individual counts of the stars, eg.,][]{Kraus2007,Bruzzese2015} or indirect methods \citep[modelling of integrated light from stellar populations, eg.,][]{Baldry2003,Hoversten2008,Meurer2009}. 
Each method has its advantages and drawbacks and the choice of IMF investigation is primarily driven by observational constraints.


\subsubsection{Direct methods}

In this method, studies derive a stellar luminosity function by calculating the number of stars that fall within each luminosity bin. The apparent magnitude of stars can be converted to luminosity using distance measurements, which is generally estimated via trigonometric parallax measurements. Stellar luminosity can be used to estimate the stellar masses, however, there are large uncertainties that can influence the calculations. 
Stellar multiplicity, SFH, stars' evolution off the main sequence (stellar tracks), and evolutionary time-scales play a vital role in understanding the true IMF of stellar populations probed via direct measures \citep{Salpeter1955,Scalo1986,Bastian2010}. 

\newpage

\clearpage

From an observational point of view measurements should be made within a volume limited sample under the assumption that the IMF is universal. Special consideration should be given to include sufficient higher mass stars in selected samples; i.e. in the Galactic disk there are a large number of low mass stars compared to high mass stars and therefore larger volumes should be considered that includes sufficient high mass stars \citep{Bastian2010}. 
Observations are also effected by Malmquist bias and Lutz-Kelker bias \citep{Lutz1973}, which can bias the luminosity function to high luminous stars and systematically overestimate the distance measurements. 
Furthermore, effects of stellar multiplicity should be considered. Many stars have gravitationally bound lower mass companions, which are hard to be resolved \citep[eg.,][]{Kroupa2001_conf}. 
These introduce subtle mass ratio corrections, which are hard to correct \citep{Bastian2010}. 
Stellar luminosity is also a function of metallicity, rotation, and age \citep{Kroupa2001_conf} and therefore, will introduce biases in the mass calculations.

The evolution of stars off the main sequence add complexities in estimating the initial masses of the stars due to mass loss. Estimating a present day mass function can remove the need of post main sequence stars to be included in the IMF. However, identifying post main sequence stars to be removed from the sample adds a further layer of complexity \citep{Salpeter1955,Miller1979,Hoversten2007}, and requires a complete understanding of the SFH and robust stellar evolutionary tracks.


\subsubsection{Indirect methods}

Apart from the galaxies in the local group, it is impossible to resolve individual stars in extragalactic stellar populations. Furthermore, with increasing distance lower mass faintest stars become undetectable. Due to these observational constraints, the number of direct extragalactic IMF measurements are limited \citep{Leitherer1998}.
In order to constrain the IMF, studies have implemented methods to investigative the IMF by probing the integrated light properties of galaxies.

Integrated light techniques allow the IMF to be probed in a large variety of galaxy populations throughout cosmic time. However, there are numerous systematic uncertainties and limitations in such methods. 
Indirect IMF measures can be  insensitive to low mass stellar populations since bright O, B and red supergiant stars may outshine low mass stars. 
In contrast at the highest mass end, there can be an insufficient number massive stars at low SFR to make a significant contribution to the detected light \citep[eg.,][]{Leitherer1998,Hoversten2008}. 
In addition, degeneracies in stellar population models play a significant role in the uncertainties of the derived IMFs, especially since stellar age, stellar metallicity, galactic dust, galactic SFH, and stellar IMF cannot be easily disentangled  \citep{Hoversten2007}.
Furthermore, indirect IMF results may depend strongly on stellar population models (mainly stellar rotation, binary evolution of O and B stars, and the treatment of W-R stars  \citep{Wolf2000}) and dark matter profiles of galaxies. 
Therefore, results may lead to be inconsistent between different IMF measurement techniques used on the same galaxy population \citep{McDermid2014, Smith2014}.

Indirect measures of IMF can roughly be divided into studies that investigate IMF of ETGs and that of star-forming galaxies. In the former, stellar kinematic and absorption line features provide an estimate of a \emph{relic IMF} that is limited to the lower mass end of the IMF. This is driven by the absence of massive stars in quiescent galaxies due to the shorter main sequence lifetimes and low current SFRs or old ages/early formation times. 

The higher-mass end of the IMF can be probed by studying star forming galaxies and can be probed via several methods. Due to the high amount of dust present around the birth place of the newly formed OB type stars, significant fractions of ionizing photons will be absorbed by the dust to be re-emitted in the FIR ($\sim40-350\mu$m). 
The FIR flux is sensitive to $\sim10$ \msol\ and therefore, can be used to probe the IMF of stars $\sim10$ \msol\ \citep{Leitherer1998}. 
Due to the fast evolution of massive stars of $\sim40$ \msol, galaxies undergoing a star-burst will have an excess of red super-giants that will dominate the continuum $\sim5500$\AA\ in the NIR \citep{Moorwood1996}, which can be probed to study the IMF.
UV stellar wind features are prominent on massive stars and therefore, from the mass loss rates of stellar winds, the number of massive stars in a stellar population can be estimated \citep{Leitherer1998}. 
However, nebular emission lines are the most popular method to study the IMF of star-forming galaxies \citep[eg.,][]{Kennicutt1983,Rigby2004,Hoversten2008,Gunawardhana2011,Nanayakkara2017}. 
In a star forming region, ionized gas is heated thermally by UV photons primarily emitted from O and B stars. 
Nebular emission lines, which span a wide range in wavelength from UV and IR, under basic assumptions, can be used to trace the number of ionizing photons and therefore the number of high mass stars. A combination of nebular emissions and continuum levels can be used to constrain the IMF slope for a given SFH.



\subsection{Universality of the IMF}

Since \citet{Salpeter1955} introduced the concept of an IMF, there has been many studies that probed the functional form of the IMF in local stellar and galaxy populations. Here, we briefly highlight some of the those studies. We note that studies that derived the mass distribution of stellar populations (the functional form of the IMF) should not be confused with studies that investigate systematic variances in the IMF. We discuss the latter in detail in Section \ref{sec:systematic_changes_in_imf}. 

Immediate studies of the IMF following Salpeter's work showed good agreement with the values derived by \citet{Salpeter1955} \citep{Sandage1957,vanDenBergh1957}. However, subsequent studies started showing deviations in the IMF in the lower mass end, with motivations from  probabilistic hierarchical fragmentation models and empirical models \citep{Limber1960,Larson1973,Taff1974}.

Assuming a time independent IMF and various SFRs, \citet{Miller1979} investigated the IMF in the solar neighbourhood to suggest an IMF that flattens at M $\lesssim1$ \msol. 
However, due to severe difficulties in probing the field star IMF, the mass distribution of the lower mass end was incomplete. 
Uncertain, inhomogeneous, and incomplete luminosity functions and mass-to-light relations along with sensitivity to SFH dominated the uncertainties of these stellar populations. Subsequent studies by \citet{Scalo1986b} showed the number distribution of stars per logarithmic mass to peak at $\sim0.3$ \msol\ with a decrease in either side.  
\citet{Reid1987} combined multiple studies of stellar luminosity function and showed that even though secondary stars in binary stellar systems provide a significant contribution to the luminosity function, the increase of the stellar mass function at the Hydrogen burning limit cannot solely be attributed to unresolved binaries.

\citet{Kroupa2001} used local star count data to show that the IMF should be expressed with broken power laws. The higher mass end of the IMF (M $>0.5$\msol) agreed with the \citet{Salpeter1955} slope and the lower mass end showed a three part broken power law. In combination with other data, the flattening of the IMF slope $\sim0.5$ \msol\ was confirmed \citep{Kroupa2001_conf}.

\citet{Chabrier2003} performed an extensive review of the IMF in various components of the Galaxy and found no statistically significant difference in the IMF as a function of environment. The IMF was expressed as a power-law form for high masses ($M>1$\msol) with a similar slope to Salpeter.  The lower-mass slope of the IMF was expressed as a log-normal distribution and was further divided into individual and binary systems due to corrections required to luminosity functions to account for unresolved binaries.

There have been numerous studies that used stellar clusters to study the IMF. 
\citet{Phelps1993} investigated 8 stellar clusters to derive their respective IMF slopes. The majority of the analysed clusters showed similar slopes to \citet{Salpeter1955}, however, two clusters showed statistically significant deviations from the $\Gamma=-1.35$. 
\citet{Sanner2001} implemented a different approach compared to \citet{Phelps1993} to measure the IMF for a sample of open star clusters. Their results showed variations in the slope between the star clusters that can be attributed to poor statistics of stars due to shallower depth in observations. 
\citet{Figer2005} measured the IMF for the Arches cluster, which is a young star cluster with massive O stars. This allowed the upper end of the IMF to be probed, and a best-fitting $\Gamma=-0.90$ was found to stars in the mass range $10<$\mass$<120$, with conservative upper limits to $\sim150$ \msol. 
However, the bulk of the work on the IMF mentioned above has been done in the Milky Way, and it is possible that there is not enough variance in properties (such as SFR surface density) to show systematic changes to the IMF.

Extensive studies performed on the Large and Small Magellanic Clouds (LMC, SMC) have shown contradicting results for the high mass end of the IMF, thus highlighting the underlying difficulties in probing the IMF. The central R136 star cluster in LMC was shown to have $\Gamma=-1.6$ for \mass$>5$ by \citet{Brandl1996} with evidence for radial variation in the IMF slope attributed to dynamical mass segregation. \emph{HST} imaging by \citet{Massey1998} showed the IMF slope to be $\Gamma=-1.4$ for stellar masses between $2.8<$\mass$<120$. 
The lower mass slope of the R136 cluster was reported to show a flattening at $\sim2$\msol\ by \citet{Sirianni2000}, however, subsequent studies showed that such flattening does not exist up to $\sim0.5$\msol\ \citep{Andersen2009}. 
The 30 Doradus region of the LMC was shown to have an IMF slope consistent with \citet{Salpeter1955} by \citet{Selman2005}, where deviations from a universal IMF for \mass$>40$ was attributed to complex SFHs, Be stars, and selective incompleteness. However, their study did not include the R136 stellar cluster in 30 Doradus. 
Multiple stellar cluster analysis of the LMC performed by \citet{Kerber2006} found that IMF slope on average agrees well with $\Gamma=-1.35$ value of \citet{Salpeter1955} for the mass range $0.9<$\mass$<2.5$. 
A combined study of OB star associations in the LMC and SMC by \cite{Massey2003} has shown the high mass end of the IMF to be consistent with \cite{Salpeter1955}. However, environment is also expected to play a role in determining the IMF of these stellar clusters, with multiple studies showing a steepening of the high mass IMF slope in field regions \citep{Parker1998,Massey2002,Gouliermis2006}, which can also be attributed to the modelling of the SFHs \citep{Elmegreen2006}.

\citet{Baldry2003} used the SDSS survey to probe the IMF by comparing cosmological luminosity densities from UV to NIR. By estimating galaxy luminosity densities at each bandpass for local galaxy populations at $z\sim0.1$, PEGASE \citep{Fioc1997} stellar population models were used to measure the slope of the IMF between 0.5 \msol\ to 120 \msol. Making the intrinsic  assumption that a universal IMF is applicable to the whole cosmic spectrum, they broke the IMF/SFH degeneracy by using the same SFH indicator to investigate a relative SFH at all redshifts and found the IMF slope to agree with \citet{Salpeter1955} within statistical uncertainties. However, due to associated degeneracies with mass, metallicity, dust, SFH, chemical evolution, and IMF in synthesising synthetic galaxy spectral models to compare with observed data, \citet{Baldry2003} do not rule out systematic variations in the IMF.


\subsection{The changing view of the IMF}
\label{sec:systematic_changes_in_imf}

Following from theoretical predictions, observational studies have started showing increasing evidence for a non-universal IMF \citep{Schombert1990,Lee2004,Rigby2004,Hoversten2008,Treu2010,vanDokkum2010,Meurer2011,Gunawardhana2011,Cappellari2012,Cappellari2013,LaBarbera2013,Ferreras2013,Conroy2013b,Spiniello2014,Navarro2015b,Navarro2015c,Navarro2015d,Navarro2015a,Zieleniewski2006}. These studies investigate both early and late type galaxies in different physical and environmental conditions using different techniques to probe the IMF at the lower and upper mass end. A selected few studies are discussed here.

In the low redshift universe, studies of IMF variations can be divided into two groups: the \emph{in situ IMF} or the IMF of galaxies when they are actively star forming and the \emph{relic IMF} or the IMF of galaxies that have ceased their star-formation.

\subsubsection{Evidence suggesting IMF variation in ETGs}
\label{sec:imf_variation_studies_etg}

For \emph{relic IMF} studies, absorption line spectroscopy and spectral synthesis modelling of massive ETGs in the local neighbourhood have shown a high abundance of low mass stars, which suggest a steeper IMF  than the Salpeter value \citep[eg.,][]{vanDokkum2010,Conroy2013b,Ferreras2013,LaBarbera2013}.

\citet{vanDokkum2010} reported a study of the IMF properties of eight bright, massive ellipticals using absorption line spectroscopy of \NaI doublet and FeH Wing-Ford band, which are prominent in dwarf galaxies. 
Their analysis with stellar population modelling suggested a steeper IMF for low mass stars between $0.1-1$ \msol\ compared to the Salpeter IMF. 
\citet{vanDokkum2012} further expanded the sample with higher signal-to-noise resolution spectra from LRIS on Keck adding the \CaII triplet to the analysis. While the \CaII feature is weak in dwarf stars, it is prominent in giant stars and therefore, is a proxy for the amount of old massive stars in stellar populations. 
These absorption features showed correlation with velocity dispersion of the galaxies, with high velocity dispersion galaxies showing stronger dwarf sensitive features and weaker giant sensitive features. 
However, the strength of IMF-sensitive features are also show variation with stellar age and element abundance patterns \citep{ConroyVanDokkum2012} and direct IMF comparisons require detailed stellar population analysis.

\citet{Conroy2012} used state-of-the-art stellar populations models with flexible abundance patterns and IMFs to higher wavelengths, which allowed stronger constraints to be made on the IMF and stellar mass-to-light ratios of galaxies. The study confirmed the observed trend of IMF variation with velocity dispersion by \citet{vanDokkum2012}, showing that massive galaxies tend to have larger fractions of low mass stars.
Systematic uncertainties of the models were comparatively larger compared to the statistical uncertainties of the sample and uncertainties in stellar atmospheres of giants and/or construction of empirical models introduced discrepancies between observed and modelled \CaII features. 
However, the overall IMF trend was not affected and results were consistent with scenarios where galaxies form larger fractions of low mass stars in short star-formation time-scales with high SFR surface densities and/or high ISM pressures.

It is important to note that \citet{vanDokkum2012} showed dwarf sensitive absorption features of galaxies can show considerable variation within a sample, with the FeH Wing-Ford band and \NaI doublet showing variations up to a factor of $\sim2$. Such variations are expected for dwarf-sensitive spectral features, since dwarfs only contribute $\sim5-10\%$ to the total integrated light in galaxies. Therefore, such analysis requires line measurements with extremely high accuracy and a thorough understanding of atmospheric and telescope detector features.

\citet{Conroy2013b} further improved the absorption line analysis by combining the stacked absorption line features of 1100 galaxies with dynamical masses to investigate systematic effects of the IMF. By comparing mass-to-light ratios derived via photometric, spectroscopic, and dynamical methods, the study found the systematic trend of IMF variation with velocity dispersion to hold regardless of the method used. Furthermore, results between the methods showed promising agreement between each other.

\citet{Ferreras2013} analysed a sample of $\sim40,000$ galaxies to generate 18 stacked spectra with high quality absorption features of TiO and Na. Using a completely independent set of stellar libraries to \citet{Conroy2012}, they confirmed the same systematic trend of IMF variation as a function of galaxy mass. 
\citet{LaBarbera2013} extended the analysis by considering a wide variety of absorption lines that are sensitive to various properties of stellar populations such as IMF, age, metallicity, and stellar abundances. 
They found that all IMF sensitive features show the same trend of IMF variation, with larger galaxies having a higher fraction of low mass stars.  
\citet{Spiniello2014} arrived at the same conclusion by conducting a thorough analysis of a large set of ETG optical absorption line features that are sensitive to the low mass slope of the IMF.

ATLAS$^\mathrm{3D}$ survey found a strong systematic variation of the mass-to-light ratios as a function of velocity dispersions in early-type galaxies \citep{Cappellari2012}. 
Their results showed that the fraction of low mass stars increase as a function of mass-to-light ratios derived from kinematic modelling, suggesting the IMF to most likely depend on the physical conditions of the galaxy when it formed bulk of its stars. However, their study couldn't distinguish between an excess of low mass stars or stellar remnants. 
If most of the massive early-type galaxies formed most of their stars in intense starbursts with higher fractions of high mass stars at higher redshifts than the spiral galaxies, this observed difference in mass-to-light ratios could be explained. 
Therefore the trend of the IMF should be followed by other population indicators such as optical colours and  \Hbeta\ absorption. This has been confirmed by follow up studies \citep[eg.,][]{Cappellari2013}.

\citet{Treu2010} used gravitational lensing and stellar dynamical models to investigate the IMF of a sample of 56 galaxies at $z\lesssim0.2$. By calculating the mass using three independent methods, they found the absolute IMF normalisation to be similar to Salpeter IMF. The comparison between the mass derived via lensing + dynamical method with a mass derived via stellar population models with \citet{Salpeter1955} and \citet{Chabrier2003} IMFs, demonstrated that IMF may vary as a function of metallicity, age, and abundance ratios of the stellar populations. This suggested that the abundance of low mass stars should increase with increasing mass. Alternatively, systematic changes in the dark matter haloes of the galaxies could explain the trends without the need to invoke changes to the IMF. Therefore, studying the IMF at higher redshifts where stellar population properties are different would allow stronger constraints on systematic variations of the IMF.

Possibilities for radial variations in the IMF in massive ETGs were reported by \citet{Navarro2015a}. Using the radial variation in the strength of the IMF sensitive TiO$_2$ absorption feature, the researchers showed that the two massive galaxies in the sample contained an enhanced fraction of high mass stars in the centre of the galaxies. However, a similar analysis performed on the low mass ETGs in their sample showed no such radial variation. Therefore, processes that drive the formation of galaxy cores are expected to form a high fraction of low mass stars during the early stages of star-formation. Even though, simple physically motivated models could explain such scenarios \citep{Hopkins2013,Weidner2013a}, the exact physics that govern such changes are yet to be understood.

\citet{Davis2017} extended the study of IMF in the ATLAS$^\mathrm{3D}$ survey by conducting a spatially resolved study of the IMF using molecular gas dynamics obtained by sub-millimetre interferometric data. Results suggested that no single IMF could describe the observed data, with some galaxies in Salpeter-IMF scenarios requiring stellar masses greater than the dynamical masses. They also find that galaxies with high specific star-formation rates show stronger mass-to-light ratio gradients. 
Furthermore, comparisons between the IMF normalizations derived using stellar kinematics show good agreement with that of molecular gas. However, the results of this study challenges observed correlations in IMF normalization in local and global scales by showing no evidence in correlation between IMF and stellar populations.

Recent results from \citet{Zieleniewski2006} adds further complexity to radial IMF variation studies. Analysis conducted by measuring absorption line strengths of IMF sensitive \NaI, \MgI, CaT, and Fe Wing-Ford band showed varying results for the 4 local ETGs. Furthermore, the IMF slopes inferred for the velocity dispersion of the galaxies showed an opposite trend to previous work. Therefore, \citet{Zieleniewski2006} concluded that interpreting the IMF in dwarf dominated stellar populations requires a galaxy-by-galaxy treatment.

Furthermore, results from ATLAS$^\mathrm{3D}$ survey have also shown a systematic variation of the IMF as a function of stellar mass to light ratios and other stellar population indicators such as optical colours and \Hbeta\ absorption \citep{Cappellari2012,Cappellari2013}. These results are independent of the dark matter profiles used and suggest that the IMF of ETGs are most likely to depend on the physical conditions of the galaxy when it formed bulk of its stars. 
\citet{Navarro2015c} investigated the IMF sensitive absorption line features of local ETGs along with the stellar metallicity to find that the IMF slope to be strongly correlated with the metallicity. This discovery was consistent with a scenario where, molecular clouds with lower metallicities preferentially forming higher mass stars due to lower cooling efficiencies, which is further discussed in Section \ref{sec:imf_variation_drivers}.

Above studies show strong evidence for variations in the IMF in ETGs as a function of multiple parameters. However, modelling of absorption features and uncertainties in the dark matter halo profiles may introduce systematic errors in understanding the IMF. 
The lack of high mass stars in ETGs make it difficult to constrain the high mass end of the IMF. 
Furthermore, the IMF probed is of stars that were formed in the past, thus is a  \emph{relic IMF}. 
To better constrain the IMF, it is necessary to probe galaxies while they are actively forming stars.

\subsubsection{Evidence suggesting IMF variation in star-forming galaxies}
\label{sec:imf_variation_studies_sfg}

Local star-burst galaxies have been extensively explored to constrain the IMF.
\citet{Rigby2004} used the \NeIII/\NeII emission line ratios of local start-burst galaxies measured by \citet{Thornley2000} to study the relative abundances of high mass stars. \NeIII 15.6$\mu$m and \NeII 12.8$\mu$m emission line ratios are weakly dependent on dust extinction and chemical abundances. However, stellar libraries show large discrepancies with each other when determining these line ratios under the same conditions. By using infra-red line ratios to constrain the ISM conditions, \citet{Rigby2004} showed that star-burst galaxies with high metallicity are deficient in stars with M $>$ 40 \msol\ compared to Salpeter IMF. Alternatively, if the massive O, B type stars lived most of their lifetimes in dense, highly excited \HII\ regions, the optical and infra-red emission lines would be highly attenuated up to \Av$\sim50$, thus making them invisible.

\citet{Schombert1990} and \citet{Lee2004} investigated stellar population properties of low surface brightness (LSB) galaxies. 
Radio measurements by \citet{Schombert1990}  showed no CO detections for a sample of six LSBs. CO is a proxy for molecular Hydrogen (H$_2$). Low metal abundances could lead to non-standard CO/H$_2$ ratios, thus explaining the non detection of CO even in H$_2$-rich regions. However, if these galaxies lack cool molecular gas resulting in stars to form in warm, low density \HII\ regions, high mass star-formation will be suppressed eventually leading to lower metallicities and high mass-to-light ratios. Therefore, a truncated IMF with constant star-formation of lower mass A, F type of stars can also result in the lack of CO detection.  
By using optical colours and stellar population models, \citet{Lee2004} demonstrated that steeper IMFs between $0.1-60 $\msol\ could best explain the distribution of B--V and B--I colours of LSBs.

\citet{Meurer2009} studied the \Halpha\ and FUV flux of \HI\ selected LSB and high surface brightness galaxies to investigate the upper mass end of the IMF. \Halpha\ flux of a stellar population was considered a direct tracer of massive O stars with \mass$>20$, while the FUV flux tracing  both O and B stars with \mass$>$3. \citet{Meurer2009} analysis showed the \Halpha\ to FUV ratio to be closely related with stellar mass density and optical surface brightness in both \Halpha\  and R band flux. By removing effects from dust, metallicity, star-gasps (sudden lack of star-formation), stochastic effects, and the escape of ionizing photons, they demonstrated a systematic variation of the IMF at the upper mass limit and/or the slope of the IMF \citep{Meurer2009,Meurer2011}. This suggested that galaxies with high R band surface brightnesses form stars in bound clusters with a high fraction of O stars compared to galaxies that preferentially form stars in loose or unbound clusters.

The IMF dependence of the galaxies in the \Halpha\ EW vs rest-frame optical colour parameter space (see Section \ref{sec:imf_probing_method}) was used by \citet{Hoversten2008} and \citet{Gunawardhana2011} to investigate systematic effects in the IMF of local star-forming galaxies. 
\citet{Hoversten2008} used data from SDSS to show that high luminosity galaxies follow a Salpeter-like IMF. However, low luminosity galaxies were found to have a higher fraction of low mass stars. 
Using data from the GAMA survey, \citet{Gunawardhana2011} showed a systematic variance of the IMF as a function of SFR.

\subsubsection{Complications in interpreting systematic variations in the IMF}

Multiple studies have investigated possibilities for IMF variation in galaxies using Balmer line flux in the context of probing SFHs \citep{Meurer2009,Weisz2012,Zeimann2014,Guo2016,Smit2016}. 
Modelling effects of IMF variation using \Halpha\ or \Hbeta\ to UV flux ratios have strong dependence on the assumed SFH and dust extinction of the galaxies and is only sensitive to the upper end of the high mass IMF. Apart from IMF variation \citep{Boselli2009,Meurer2009,Pflamm-Altenburg2009}, stochasticity in SFH \citep{Boselli2009,Fumagalli2011,Guo2016}, non-constant SFHs \citep{Weisz2012}, and Lyman leakage \citep{Relano2012} can provide viable explanations to describe offsets between expected Balmer line to UV flux ratios and observed values. 

\citet{Kauffmann2014} used SFRs derived via multiple nebular emission line analysis with the 4000\AA\ break and H$\delta_A$ absorption to probe the recent SFHs of SDSS galaxies with $\mathrm{log_{10}(M_*/M_\odot)<10}$ and infer possibilities for IMF variation. 
They did not find conclusive evidence for IMF variation, with contradictions in the 4000\AA\ features with \citet{Bruzual2003} stellar templates being attributed to errors in the spectro-photometric calibration. 
However, using absorption line analysis to probe possible IMF variations in actively star-forming galaxies suffers from strong Balmer line emissions that dominate and fill the absorption features. Furthermore, absorption lines probe older stellar populations, and linking them with current star-formation requires further assumptions about the SFH. 

\citet{Smit2016} used SED fitting techniques to probe discrepancies between \Halpha\ to UV SFRs ratios of $z\sim4-5$ galaxies and local galaxies.
They inferred an excess of ionizing photons in the $z\sim4-5$ galaxies but the origin couldn't be distinguished between a shallow high-mass IMF scenario or a metallicity dependent ionizing spectrum. 
Using broad band imaging and SED fitting techniques to infer \Halpha\ flux has underlying uncertainties from the contamination of other emission lines that fall within the same observed filter (eg., \NII, \SII) and assumptions of IMF, SFH, metallicity, and dust law of the SED templates.

On tracing the IMF of ETGs, \citet{vanDokkum2012} and \citet{Cappellari2013} implemented different approaches to probe any systematic changes in the IMF. \citet{Cappellari2013} used stellar dynamics, which is a gravitational tracer of the IMF that measures the mass of a system, while \citet{vanDokkum2012} used a spectroscopic tracer by probing absorption lines strengths of galaxies. In ETGs, absorption line strengths are sensitive to the low mass stars, while the gravitational tracers may be dominated by either low mass stars or remnants of massive stars. 
\citet{Smith2014} compared galaxies in the \citet{vanDokkum2012} and \citet{Cappellari2013} in a one-to-one basis to investigate whether any systematic effects could influence the observed variations in both the studies. Even though both gravitational and spectroscopic measures suggested on average a shallower IMF (excess of low mass stars compared to the Milky Way,  \citet{Smith2014} analysis showed that there was no correlation between mass-excess factors derived by the two measures on a galaxy-by-galaxy level.

These one-to-one variations can be driven by various factors. It is possible that kinematic studies have yet been unable to separate dark matter contributions from the IMF to be a completely independent probe of the IMF \citep{Smith2014}. 
Additionally, systematic errors in calibration of the observed filters and Gaussian errors in kinematic modelling have also been shown to influence IMF determinations of stellar dynamic studies \citep{Clauwens2015}. 
Further studies comparing stellar dynamics to absorption line studies have shown weaker correlations to stellar population properties for stellar dynamics studies \citep{McDermid2014,McDermid2015}.

Inferring IMF purely via absorption line properties is also problematic. In general spectroscopic features are sensitive to multiple parameters and therefore, IMF sensitive features may also show dependencies with age and abundance variations \citep{ConroyVanDokkum2012}.
\citet{Navarro2016} showed that a joint consideration of the most commonly used galaxy formation measure of [Mg/Fe] with IMF slope leads to unrealistically short formation time-scales and high SFRs for ETGs. However, recent studies of $z\sim4$ quiescent galaxies show evidence for extremely high SFRs and very short formation time scales \citep{Glazebrook2017}, thus introducing added complexity to our understanding of galaxy formation. 
Due to such reasons, extreme caution is warranted when interpreting evidence for systematic IMF change using indirect IMF measures.

\subsubsection{Summary of techniques used to probe the IMF}

Techniques used to probe the IMF have different mass sensitivities. Stellar kinematics in general should probe low and high mass end of the mass function. However, because absorption lines required to derive stellar kinematics are more prominent in ETGs, the short life-time of high mass stars and massive star stellar remnants adds complexity to distinguishing the excess of high/low mass stars \citep[eg.,][]{Cappellari2012N,Smith2014}. 
Similarly, gravity sensitive absorption features prominent in ETGs only allow the low mass end of the IMF ($\lesssim0.5$ \msol) to be probed \citep[eg.,][]{vanDokkum2012,Navarro2016}. 
Furthermore, combining gravitational lensing models with stellar absorption features can be used to probe the lower end of the mass function, however, added complexity arises due to the assumptions of the dark mater halo models required in the modelling process \citep[eg.,][]{Treu2010}.

Emission line features of galaxies are more sensitive to the high mass stars, and are widely used to constrain the high mass end of the IMF. The lack of newly formed stars in ETGs makes it impossible to use such features to probe the high mass IMF of ETGs. Emission lines such as \NeIII are sensitive to higher masses and have been used to show excesses of high mass stars ($>40$ \msol), however, discrepancies between stellar libraries in producing such features adds further complexity to the probe of the IMF \citep[eg.,][]{Rigby2004}. Optical colours of galaxies along with star-formation indicators also probe the mass function $>1$ \msol\ \citep[eg.,][]{Schombert1990,Lee2004,Nanayakkara2017}. \Halpha\ EWs vs rest-frame optical colours are effectively sensitive to stellar masses $>10$ \msol, however, our limited understanding of stellar evolution at the high mass end ($\gtrsim100$ \msol), along with the presence of exotic stellar features such as extreme low-metallicity stars, high stellar rotation, and binaries adds further complexity (see Chapter \ref{chap:imf_analysis}). 
Furthermore, \Halpha\ flux along with FUV fluxes of galaxies can effectively probe the IMF at higher masses ($\gtrsim20$ \msol), however, effects of dust, metallicity, and stochasticity of SFH leads to extra caveats \citep[eg.,][]{Meurer2011}.

\subsubsection{What physically should drive IMF variations}
\label{sec:imf_variation_drivers}

Theoretically, we expect the IMF to vary since a galaxy’s metallicity, SFRs, and environment can change dramatically with time \citep[eg.,][]{Schwarzschild1953,Larson1998,Larson2005, Weidner2013a, Chattopadhyay2015,Ferreras2015,Lacey2016}.  
Typically lower metallicities and higher SFRs at high redshift leads to lower cooling and higher heating efficiencies \citep{Schwarzschild1953,Larson1998}, which  may increase the mass of the fragmented molecular clouds leading to a redshift dependant IMF slope \citep{Chattopadhyay2015}. However, we do not know how this affects the IMF since the physics governing fragmentation is complex and poorly understood \citep[eg.,][]{Larson1973,Larson2006}. 

Environmental variations to the IMF can occur due to interactions between gas clumps in galaxies can lead to extra UV flux, strong winds, and powerful outflows, which suppresses the formation of low mass stars. Theoretical simulations show that with increasing cloud surface densities the fragmentation favours the formation of higher mass stars \citep{Krumholz2010}.

Radial variations of the IMF can occur due to dynamical mass segregation that will lead to massive stars to occupy the centre regions of a galaxy \citep[eg.,][]{Binney1987}.

\citet{Narayanan2012} used hydrodynamical simulations to investigate how the Jeans mass of fragmenting molecular clouds vary as a function of SFR. The Jeans mass showed a power law increase with SFR driven by the warming up of dust that thermally couples with gas in high radiation fields prominent in high SFR regions. Therefore, the actual SFRs of galaxies should be less than observed values, with higher fraction of high mass stars being formed in high SFR scenarios. 
With cosmic evolution, the SFR of galaxies drop, which allows the ISM to cool due to the lack of cosmic rays resulting in the preferential formation of low mass stars \citep{Narayanan2013}.

Physically motivated models of ETGs suggest similar scenarios to \citet{Narayanan2013}, where star formation occurs in different periods giving rise to variability in the mass of the stars formed \citep{Vazdekis1996,Weidner2013a,Ferreras2015}. 
Such models assume that galaxies start their SFR with high efficiency, reaching a high SFR in a short period of time producing a high fraction of high mass stars that lasts $\lesssim300$ Myr. 
These episodes build $\sim10\%$ of the total galaxy stellar population by mass and results in a metal-rich ISM. Another star-formation episode immediately follows, where a bulk of the mass is formed with a high fraction of low mass stars. 
The change in mass is physically motivated by the changes in the ISM, which is expected to be highly turbulent after a sustained period of high SFR resulting in enhanced fragmentation of molecular clouds due to the high velocity dispersions \citep{Hopkins2013}. However, the exact physics that drive these variations are arguable.

A somewhat similar model was presented by \citet{Bekki2013}.  
Following the physically motivated IMF model by \citet{Marks2012}, \citet{Bekki2013} used a chemodynamical evolution code  with an IMF that varies with local density and metallicity of the ISM, to show that starburst galaxies triggered by galaxy interactions form stars with preference towards higher masses. 
Supernovae feedback were expected to play a dominate role in setting the IMF slope by suppressing the formation of high mass stars. 
Simulations further suggested the IMF slope to vary as a function of star-formation rate surface density of the galaxies, with higher values forming higher fractions of high mass stars.
IMF also show radial variations within galaxies, with inner regions of galaxies having a higher fraction of low mass stars.
Radial variations were evident in the pre, during, and post burst phases. 
Local ETGs have been shown to have a higher fraction of low mass stars in the centre of the galaxy \citep{Navarro2015a}.  
However, radial IMF studies of star-forming galaxies and detailed modelling of \mass$<1$ stars, whose formation is dependent upon physical properties of the ISM are warranted to further constrain IMF variation models. 
Driven by observational results and physically motivated models, recent semi-analytical models allow users to vary the IMFs in different star-forming episodes \citep{Lacey2016}.

\subsubsection{IMF at $\mathbf{z\sim2}$}

In spite of IMF being fundamental to galaxy evolution, our understanding of it at higher redshifts ($z\gtrsim2$) is extremely limited. IMF studies of strong gravitational lenses at $z>1$ have shown no deviation from Salpeter IMF \citep{Pettini2000,Steidel2004,Quider2009}. 
However, \citet{Navarro2015b} investigated the IMF of $\sim50$ ETGs between $0.9<z<1.5$ using the IMF sensitive TiO$_2$ absorption lines obtained via stacking of galaxy spectra. The high mass sample of the stacked galaxy spectra showed an enhanced amount of low mass stars compared to the lower mass stacked bin. Therefore, these results at high-$z$ were in agreement with the local ETG studies showing a dwarf enhanced stellar populations as a function of galaxy mass suggesting a direct link between ETGs at $z\sim1$ to those at $z\sim0$. 

Using a local analogue to $z\sim2$ galaxies (NGC 1277), \citet{Navarro2015d} found evidence for an abundance of low mass stars in the early universe. Radial studies of the IMF showed that the \emph{relic} galaxy shows weaker variation compared to \citet{Navarro2015a}. Given the galaxy formed most of its stars within a short starburst at $z\sim2$ with centrally concentrated star-formation, the dry merging of such galaxies would explain the abundance of centrally concentrated low mass stars in locals ETGs.

Understanding the \emph{relic} IMF at high redshift requires populations of quiescent galaxies (which are relatively rare at high redshift), extremely long integration times to obtain absorption line/kinematic features, and complicated modelling of stellar absorption line features. 
Since IMF defines the mass distribution of formed stars at a given time, in the context of understanding the role of IMF in galaxy evolution, it should be investigated \emph{in situ} in an era when most galaxies are in their star forming phase and evolving rapidly to produce large elliptical galaxies found locally. 
Furthermore, as discussed in Section \ref{sec:imf_variation_studies_etg}, high mass stars are rare in ETGs and therefore star-forming galaxies are imperative to study the high mass end of the IMF. 
Simulations have shown that $z\sim2$ universe is ideal for such studies \citep{Hopkins2006}.
Rest-frame optical spectra of high redshift galaxies are dominated by strong emission lines produced by nebulae associated with high mass stars ($>$15 \msol) and therefore provide a direct tracer of the high mass end of the IMF \citep{Bastian2010}. 
Due to the recent development of sensitive near NIR imagers and multiplexed spectrographs that take advantage of the Y, J, H, and K atmospheric windows, the $z\sim2$ universe is ideal to study rest-frame optical features of galaxies.


\subsection{Probing the IMF of star-forming galaxies at $z\sim2$}
\label{sec:imf_probing_method}

\begin{figure}
\centering
\includegraphics[width=1.0\textwidth]{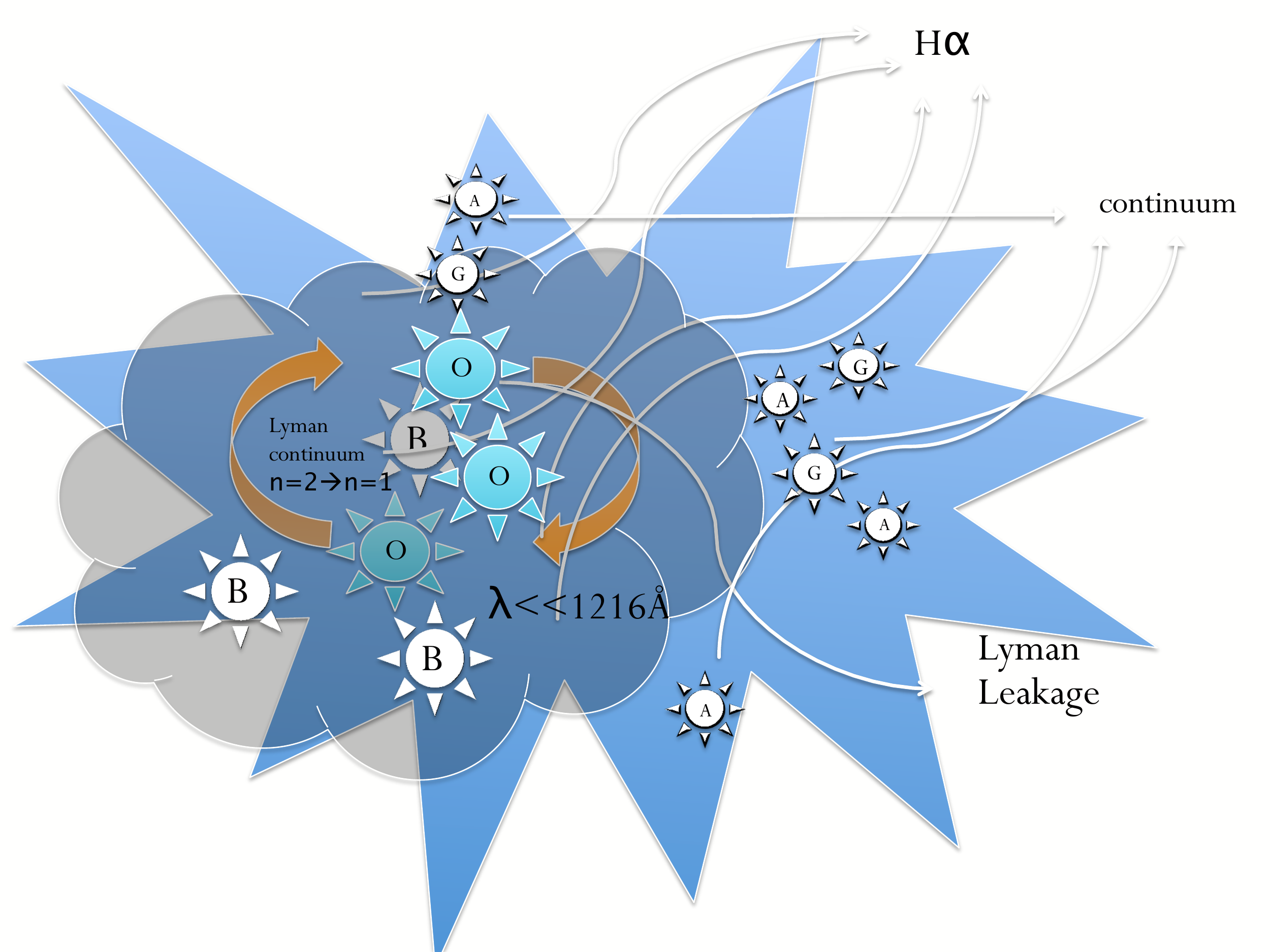} 
\caption[The physical origin of \Halpha\ EW.]{Illustration of a molecular cloud. The \Halpha\ flux and the continuum level at 6564.61\AA\ have different physical origins. 
In Case B recombination, Lyman photons generated by the hot massive O and B stars will get scattered to lower energy \Halpha\ photons due to high optical depths. \Halpha\ flux has lower optical depths, and hence escapes the molecular clouds and therefore, can be used as a direct tracer for the young and massive O and B stars in molecular clouds. 
Some Lyman photons could escape the clouds without undergoing scattering to lower energy photons, which is called ``Lyman Leakage''. 
The continuum contribution at 6564.61\AA\ is largely driven by the older and smaller A and G stars, which have sufficiently longer lifetimes to move from their dusty birth regions.}
\label{fig:mol_cloud}
\end{figure}

The method implemented in this thesis to study the IMF at $z\sim2$ is an improvement of the recipe which was first implemented by \citet{Kennicutt1983} and subsequently used by \citet{Hoversten2008} and \citet{Gunawardhana2011} to study the IMF at $z\sim0$. This method takes advantage of the IMF dependence of the \Halpha\ EW and rest-frame optical colour parameter space of star-forming galaxies to constrain the IMF, and can be explained as follows.

The total flux of a galaxy at \Halpha\ emission wavelength is the sum of the \Halpha\ emission flux, the continuum level at the same wavelength minus the \Halpha\ absorption. \Halpha\ absorption for galaxies at $z\sim2$ is $\leq$ 3\% of its flux level \citep{Reddy2015} and therefore can be ignored. 
In case B recombination \citep{Brocklehurst1971}, following the Zanstra principle \citep{Zanstra1927} explained in Section \ref{sec:emission_line_background},  the \Halpha\ flux of a galaxy is directly related to the number of Lyman continuum photons emitted by massive young O and B stars with masses $>$10 \msol (see Section \ref{sec:emission_line_background}). The continuum flux at the same wavelength is dominated by red giant stars with masses between $0.7-3.0$ \msol. Therefore, \Halpha\ EW, which is the ratio of the strength of the emission line to the continuum level, can be considered as the ratio of massive O and B stars to $\sim1$ \msol\ stars present in a galaxy. Figure \ref{fig:mol_cloud} illustrates the different physical origins of \Halpha\ flux and its associated continuum level.

The evolution of rest-frame optical colours of a galaxy with a monotonic SFH shows a similar trend to the evolution of \Halpha\ EW \citep{Kennicutt1983,Hoversten2008}. The \Halpha\ EW probes the specific-SFR (sSFR) of the shorter lived massive stars, while the optical colours probe the sSFR of the longer lived less massive stars. Therefore, in a smooth exponentially declining SFH, the optical colour of a galaxy will transit from bluer to redder colours with time due to the increased abundance of older less massive red stars. Similarly, with declining SFR the \Halpha\ flux will decrease and the relative continuum contribution of the older redder stars will increase, which will act to decrease the \Halpha\ EW in a similar SFH.

\begin{figure}
\centering
\includegraphics[trim=30 10 10 10, clip, width=0.8\textwidth]{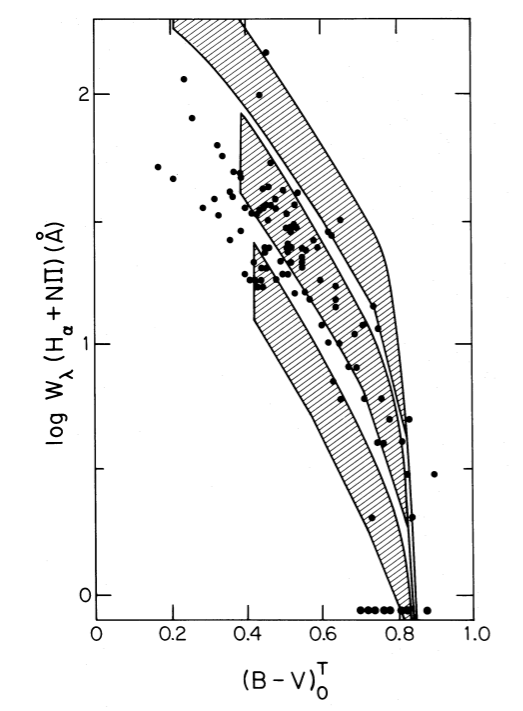} 
\caption[Reproduction of the \citet{Kennicutt1983} main analysis figure (Figure 4) used to investigate the IMF a population of local disk galaxies.]{Reproduction of the \citet{Kennicutt1983} main analysis figure (Figure 4), which was used to investigate the IMF in a population of local disk galaxies. The EW of \Halpha\ + \NII\ is plot against the optical (B--V) colours of the galaxies. The shaded areas represent galaxy evolutionary tracks for various IMFs with associated distributions in dust corrections. From top the to bottom the regions represent, $\Gamma=-1,-1.35$, and \citet{Miller1979} IMF for the solar neighbourhood. }
\label{fig:kennicutt_plot}
\end{figure}

The \Halpha\ EW and optical colours parameter space is degenerated in such a way that the slope is equivalent to lowering the fraction of highest mass stars that are formed and/or an increasing the fraction of intermediate-mass stars. Furthermore, the change in galaxy evolutionary tracks due to IMF is largely orthogonal to the changes in tracks due to effects of dust extinction.  
Figure \ref{fig:kennicutt_plot} shows the distribution of galaxies used by \citet{Kennicutt1983} in his analysis. 
In monotonic SFH scenarios, galaxies evolve along these tracks from the top right with high EWs and bluer colours to become low EW red galaxies to eventually be quiescent galaxies. 
In constant SFH scenarios, the decrease in \Halpha\ EW is driven by the rise of the continuum level due to the longer lifetimes of the lower mass stars that contribute to the continuum at 6565 \AA. 
Along with change in the IMF, sudden star-bursts or cease of star-formation (gasps), and changes to dust sight-lines may influence the evolution of galaxies in these tracks (see Section \ref{sec:dust} and \ref{sec:star_bursts}). 
If all galaxies follow a universal Salpeter-like IMF, galaxies should be distributed around the middle shaded area in Figure \ref{fig:kennicutt_plot}, which is seen in the \citet{Kennicutt1983} sample. 
However, effects of AGN, underlying \Halpha\ absorption, narrow band imaging that blends \Halpha\ with \NII, and dust extinction were not fully considered in that analysis, which could have affected the conclusions.
Studies by \citet{Hoversten2008} and \citet{Gunawardhana2011} at $z\sim0$ had multiple emission line diagnostics and higher spectral resolutions which enabled them to improve on the classical \citet{Kennicutt1983} method, and discovered evidence hinting at non-universal IMFs.

\begin{figure}
\centering
\includegraphics[trim=30 10 10 10, clip, width=1.0\textwidth]{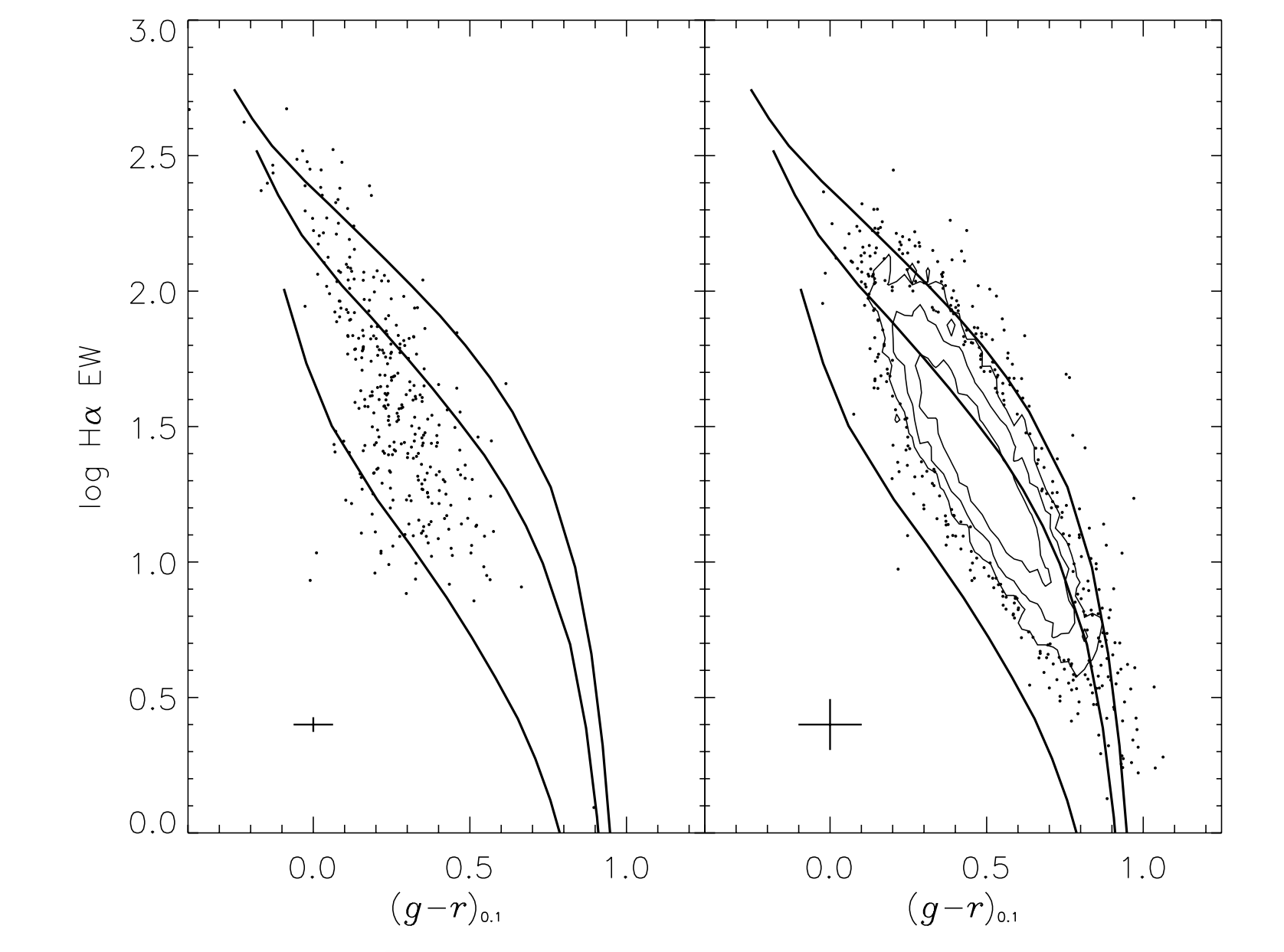} 
\caption[Reproduction of the \citet{Hoversten2008} Figure 16, which shows the distribution of SDSS galaxies in \Halpha\ EW and optical colour parameter space in different R band luminosity bins.]{Reproduction of the \citet{Hoversten2008} Figure 16, which shows the distribution of SDSS galaxies in \Halpha\ EW and optical colour parameter space in different R band luminosity bins. From top to bottom, the tracks plot are from PEGASE with IMF slopes of $\Gamma=-1.0, -1.35,$ and $-2.0$. All model tracks have exponentially declining SFHs with $\tau=1.1$ Gyr. The crosses indicate  median error bars.
{\bf Left:} The low luminosity sample.
{\bf Right:} The high luminosity sample.}
\label{fig:hoversten_plot}
\end{figure}

Figure \ref{fig:hoversten_plot} shows the main result of \citet{Hoversten2008}. Even though the SDSS galaxy sample as a whole showed good correlation with a Salpeter-IMF, when the sample was divided into R band high and low luminous galaxies, the low luminous sample showed a large scatter with a large vertical distribution.
Due to non-uniform SFHs in galaxies through cosmic time, individual measurements of the IMF slope could not be performed. However, due to the large number of galaxies in the \citet{Hoversten2008} sample, this suggested that the low luminous galaxies were not consistent with a universal IMF, with a possibility that they preferentially form stars with a high fraction of low mass stars.

\begin{landscape}
\begin{figure}
\centering
\includegraphics[trim=30 10 10 10, clip, width=1.4\textwidth]{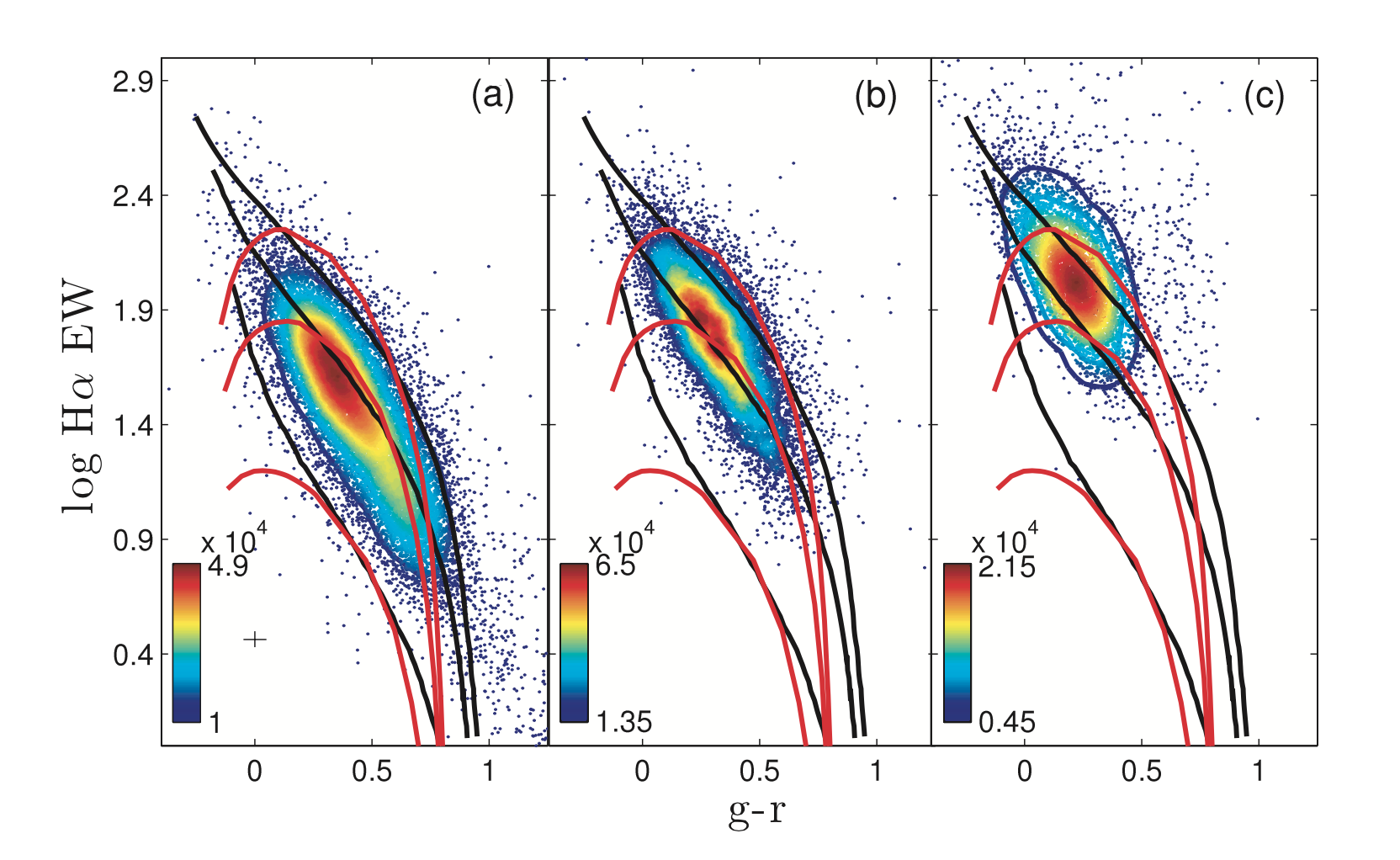} 
\caption[Reproduction of the \citet{Gunawardhana2011} Figure 5, which shows the distribution of GAMA galaxies in \Halpha\ EW and optical colour parameter space in different SFR bins.]{Reproduction of the \citet{Gunawardhana2011} Figure 5, which shows the distribution of GAMA galaxies in \Halpha\ EW and optical colour parameter space in different SFR bins. The black model tracks are from PEGASE and are similar to Figure \ref{fig:hoversten_plot}. The red tracks are from \citet{Maraston2005}. 
{\bf Left:} Low SFR sample (0\msol/yr$<$SFR$<$3\msol/yr).
{\bf Centre:} Mid SFR sample (3\msol/yr$<$SFR/yr$<$13\msol/yr).
{\bf Right:} High SFR sample (13\msol/yr$<$SFR).}
\label{fig:gunawardhana_plot}
\end{figure}
\end{landscape}

The GAMA sample used by \citet{Gunawardhana2011} to study the IMF is shown by Figure \ref{fig:gunawardhana_plot}. The sample is divided into three SFR bins to show that galaxies with high SFR show a higher locus in the \Halpha\ EW vs optical colour parameter space than the low SFR sample. However, it is important to note that both the SFR and the IMF slope is determined using \Halpha\ nebular emission.  Given that galaxies at high-$z$ have higher SFRs, the authors argue that the IMF slope of galaxies at high-$z$ should be flatter.

\newpage 

\section{Plan of thesis}
\label{sec:thesis_plan}

Next, in Chapter \ref{chap:zfire_survey}, I introduce the ZFIRE survey, a spectroscopic survey of star-forming galaxies in rich environments at $z\sim2$ carried out using the MOSFIRE instrument in Keck I telescope. Chapter \ref{chap:spec_analysis} shows an analysis of spectroscopic/photometric properties of ZFIRE galaxies and demonstrates the high quality of photometric data and .

\noindent In Chapters \ref{chap:imf_observations} and \ref{chap:imf_analysis}, I use the NIR spectra of galaxies obtained as a part of the ZFIRE survey \citep{Nanayakkara2016} along with multi-band photometric data from the ZFOURGE survey \citep{Straatman2016} to study the IMF of a mass complete sample of star-forming galaxies at $z\sim2$.  Galaxies follow nearly the same locus in \Halpha\ EW, optical colour as long as their SFHs are smoothly increasing or decreasing \citep{Kennicutt1983}. Furthermore, the change in galaxy evolutionary tracks due to IMF is largely orthogonal to the changes in tracks due to effects of dust extinction.  This method is an improvement of the recipe which was first implemented by \citet{Kennicutt1983} and subsequently used by \citet{Hoversten2008} and \citet{Gunawardhana2011} to study the IMF at $z\sim0$. 
Observational data is compared to galaxy models with various physical and chemical properties created via synthetic stellar populations such as PEGASE, Starburst99, and BPASS.

\noindent The final conclusions of this thesis are provided in Chapter \ref{chap:conclusions}. I further discuss my plans for future direction of research in Section \ref{chap:conclusions}.

\chapter{ZFIRE Survey I: A KECK/MOSFIRE Spectroscopic Survey of Galaxies in Rich
Environments at $z\sim2$}
\label{chap:zfire_survey}

In this chapter, I present the ZFIRE survey, which utilizes MOSFIRE to  observe galaxies in rich environments at $z>1.5$ with a complementary sample of field galaxies. A mass/magnitude complete study of rich galaxy environments is essential to overcome selection-bias.
Galaxy clusters are the densest galaxy environments in the universe and are formed via multiple physical processes \citep[eg.,][]{Kravtsov2012}. 
They are a proxy for the original matter density fields of the universe and can be used to constrain fundamental cosmological parameters. Focusing on these rich environments at high-redshift provides access to numerous galaxies with various physical conditions that are rapidly evolving and interacting with their environments. 
These galaxies can be used to study the formation mechanisms of local galaxy clusters in a period where they are undergoing extreme evolutionary processes. Such environments are rare at $z\sim 2$ \citep[eg.,][]{Gobat2011,Newman2014,Yuan2014}: for example, we target the \cite{Spitler2012} cluster at $z=2.1$, which was the only such massive structure found in the 0.1 deg$^2$  ZFOURGE survey (and that at only  4\% chance, \citep{Yuan2014}). Hence, a pointed survey on such clusters and their environs is highly complementary to  other field surveys being performed with MOSFIRE.


\section{ZFIRE\ Observations and Data Reduction}
\label{sec:survey}

The MOSFIRE spectrograph \citep{McLean2008,McLean2010,McLean2012} operates from 0.97--2.41 microns (i.e. corresponding to atmospheric $YJHK$ bands, one band at a
time) and provides a 6.1$'\times 6.1'$ field of view with a resolving power of $R$\around3500.  It is equipped with a cryogenic configurable slit unit that can include up to 46 slits and be configured in
\around6 minutes.  MOSFIRE has a Teledyne H2RG HgCdTe detector with
2048 $\times$ 2048 pixels ($0''.1798$/pix) and can be used as a multi-object spectrograph and a wide-field imager by removing the masking bars from the field of view.  ZFIRE\ utilizes the multi-object spectrograph capabilities of MOSFIRE.

The galaxies presented in this paper consist of observations of two cluster fields from the  COSMOS field \citep{Scoville2007} and the Hubble UDS Field \citep{Beckwith2006}.
These clusters are the \cite{Yuan2014} cluster at \zspec=2.095 and IRC 0218 cluster \citep{Papovich2010,Tanaka2010,Tran2015} at \zspec=1.62.
\cite{Yuan2014} spectroscopically confirmed the cluster, which was identified by \citet{Spitler2012} using photometric redshifts and deep Ks band imaging from ZFOURGE.
The IRC 0218 cluster was confirmed independently by \citet{Papovich2010} and \citet{Tanaka2010}.
Field galaxies neighbouring on the sky, or in redshift shells, are also observed and provide a built-in comparison sample.

\subsection{ZFIRE\ Survey Goals and Current Status}

The primary science questions addressed by the ZFIRE\ survey are as follows:

\begin{enumerate}
\item  What are the ISM physical conditions of the galaxies? 
We test the Mappings IV models by using \Halpha, \NII, \Hbeta, \OII, \OIII, and \SII\ nebular emission lines to study the evolution of chemical enrichment and the ISM as a function of redshift \citep{Kewley2016}. 

\item  What is the IMF of galaxies?
We use the \Halpha\ equivalent width as a proxy for the IMF of star-forming galaxies at $z\sim2$ \citep{Nanayakkara2017}.  
		
\item  What are the stellar and gas kinematics of galaxies?
Using \Halpha\ rotation curves we derive accurate kinematic parameters of the galaxies. Using the Tully-Fisher relation \citep{Tully1977} we track how stellar mass builds up inside dark matter halos to provide a key observational constraint on galaxy formation models \citep[][C. Straatman et al., in preparation]{Alcorn2016}. 

\item  How do fundamental properties of galaxies evolve to $z\sim2$ ?
Cluster galaxies at z\around2 include massive star-forming members that are absent in lower redshift clusters. 
We measure their physical properties and determine how these members must evolve to match the galaxy populations in clusters at $z<$1 \citep{Tran2015, Kacprzak2015}. 


\end{enumerate}

Previous results from ZFIRE\ have already been published. 
\citet{Yuan2014} showed that the galaxy cluster identified by ZFOURGE \citep{Spitler2012} at $z=2.095$ is a progenitor for a Virgo like cluster. \citet{Kacprzak2015} found no significant environmental effect on the stellar MZR for galaxies at $z\sim2$.  \citet{Tran2015} investigated \Halpha\ SFRs and gas phase metallicities at a lower redshift of $z\sim1.6$ and found no environmental imprint on gas metallicity but detected quenching of star formation in cluster members. 
\citet{Kewley2016} investigated the ISM and ionization parameters of galaxies at $z\sim2$ to show significant differences of galaxies at $z\sim2$ with their local counterparts.
Here the data used to address the above questions in past and future papers is presented.

\subsection{Photometric Catalogues}

Galaxies in the COSMOS field are selected from the ZFOURGE  survey (Straatman et al. in press) which is 
a 45 night deep Ks band selected photometric legacy survey carried out using the 6.5 meter Magellan Telescopes located at Las Campanas observatory in Chile. 
The survey covers 121 arcmin$\mathrm{^2}$ each in COSMOS, CDFS, and UDS cosmic fields using the near-IR medium-band filters of the FourStar imager \citep{Persson2013}.  
All fields have \emph{HST} coverage from the CANDELS survey \citep{Grogin2011,Koekemoer2011} and a wealth of multi-wavelength legacy data sets \citep{Giacconi2002,Capak2007,Lawrence2007}. 
For the ZFIRE\ survey, galaxy selections were made from the v2.1 of the internal ZFOURGE catalogues. A catalogue comparison between v2.1 and the the updated ZFOURGE public data release 3.1 is provided in the Appendix \ref{sec:ZFOURGE comparison}. 
The v2.1 data release reaches a $5\sigma$ limiting depth of $Ks=25.3$ in FourStar imaging of the COSMOS field \citep{Spitler2012} which is used to select the ZFIRE K-band galaxy sample. 
\emph{HST} WFC3 imaging was used to select the ZFIRE H-band galaxy sample.

EAZY \citep{Brammer2008} was used to derive photometric redshifts by fitting linear combinations of nine SED templates to the observed SEDs\footnote{An updated version of EAZY is used in this analysis compared to what is published by \citet{Brammer2008}. Refer \citet{Skelton2014} Section 5.2 for further information on the changes. The updated version is available at \url{https://github.com/gbrammer/eazy-photoz}.}. 
With the use of medium-band imaging and the availability of multi-wavelength data spanning from UV to Far-IR (0.3-8$\mu$m in the observed frame), ZFOURGE produces photometric redshifts accurate to $1-2\%$ \cite[Straatman et al., in press;][]{Kawinwanichakij2014,Tomczak2014}.

Galaxy properties for the ZFOURGE catalogue objects are derived using FAST \citep{Kriek2009} with synthetic stellar populations from \citet{Bruzual2003}  using a $\chi^2$ fitting algorithm to derive ages, star-formation time-scales, and dust content of the galaxies. 
Full information on the ZFOURGE imaging survey can be found in  Straatman et al. (in press).

The IRC 0218 cluster is not covered by the ZFOURGE survey. Therefore  publicly available UKIDSS imaging \citep{Lawrence2007} of the UDS field is used for sample selection.
The imaging covers 0.77 deg$^2$ of the UDS field and reaches a $5\sigma$ limiting depth of $\rm K_{AB}= 25$  \citep[DR10;][]{UDS_DR10}. 
Similar to ZFOURGE, public K-band selected catalogues of UKIDSS were used with EAZY and FAST to derive photometric redshifts and galaxy properties \citep{Quadri2012}.

\subsection{Spectroscopic Target Selection}
\label{sec:sample_def}

In the first ZFIRE\ observing run, the COSMOS field between redshifts $2.0<$\zphoto$<2.2$ was surveyed to spectroscopically confirm the overdensity of galaxies detected by \cite{Spitler2012}. 
The main selection criteria were that the \Halpha\ emission line falls within the NIR atmospheric windows and within the coverage of the MOSFIRE filter set. 
For each galaxy, H and K filters were used to obtain multiple emission lines to constrain the parameters of  interest.

Nebular emission lines such as \Halpha\ are strong in star-forming galaxies and hence it is much quicker to detect them than underlying continuum features of the galaxies. 
Therefore, rest frame UVJ colour selections \citep{Williams2009} were used to select primarily star-forming galaxies in the cluster field for spectroscopic follow up.
While local clusters are dominated by passive populations, it is known that high-$z$ clusters contain a higher fraction of star-forming galaxies \citep[eg.,][]{Wen2011,Tran2010,Saintonge2008}.
This justifies our use of K band to probe strong emission lines of star-forming galaxies, but due to the absence of prominent absorption features, which fall in the K band at $z\sim2$, we note that our survey could be incomplete due to missing weak star-forming and/or quiescent cluster galaxies.

\begin{figure}
\centering
\includegraphics[width=0.55\textwidth]{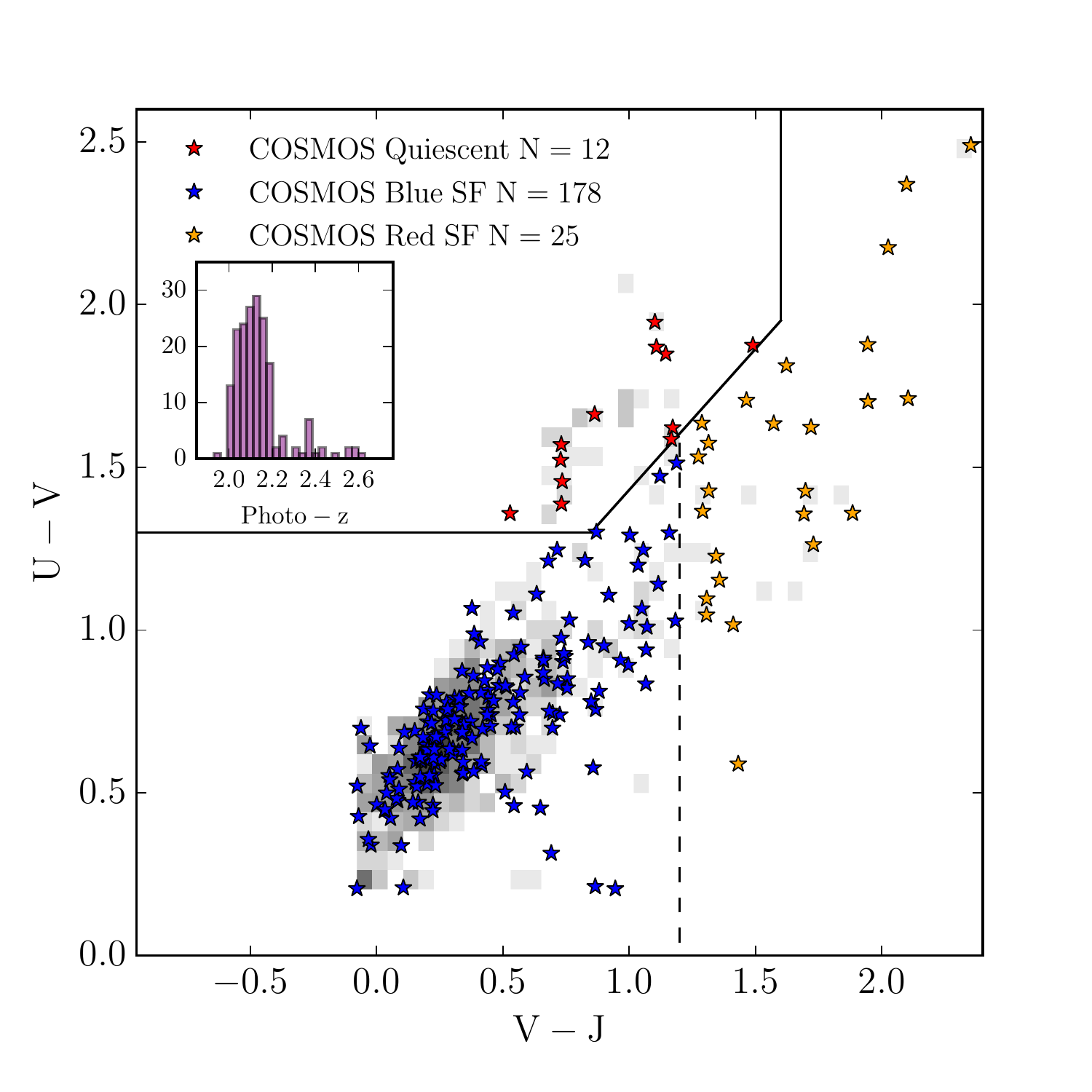}
\includegraphics[width=0.55\textwidth]{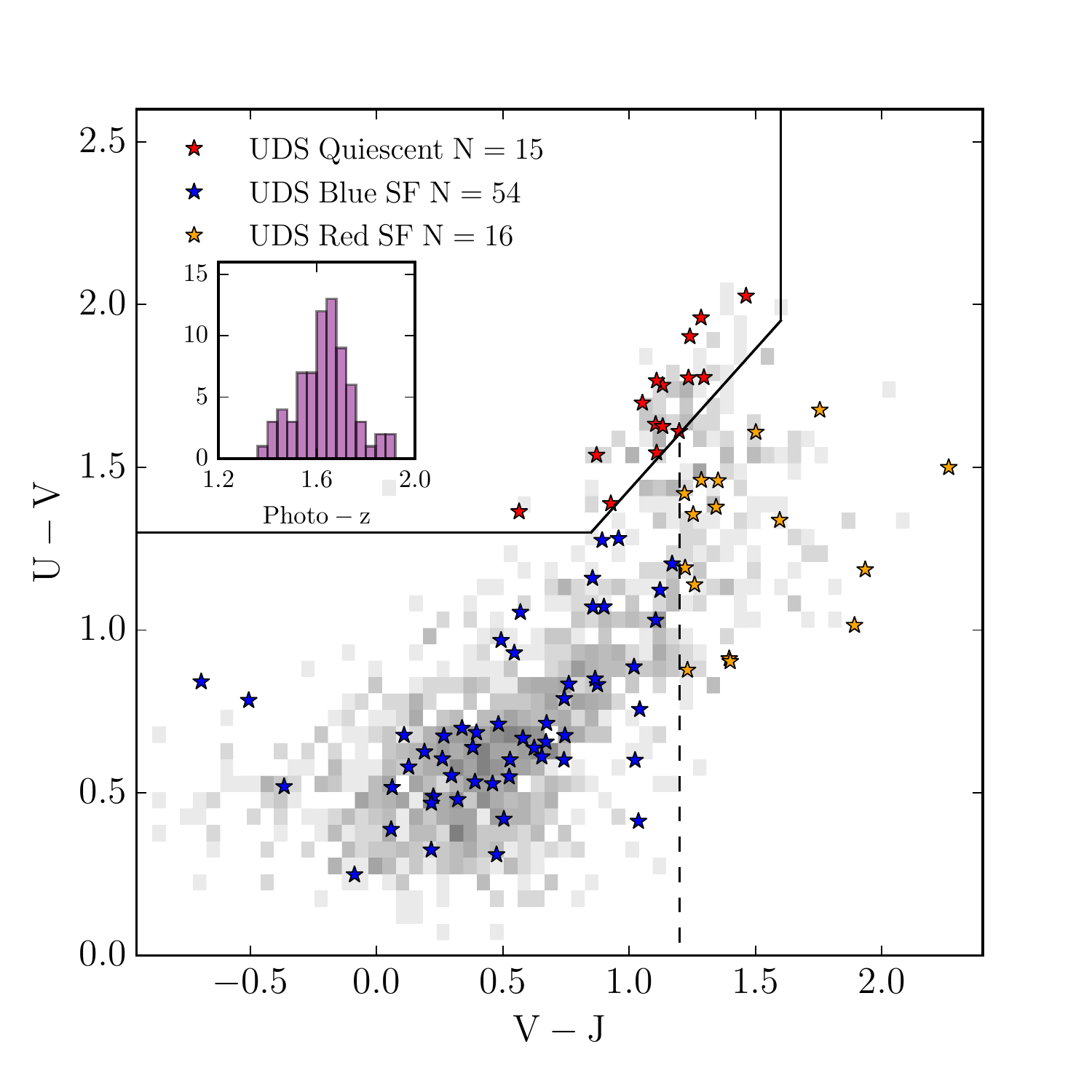}
\caption[Rest frame UVJ diagram of the galaxy sample selected from ZFOURGE and UKIDSS surveys to be observed by ZFIRE.]{Rest frame UVJ diagram of the galaxy sample selected from ZFOURGE and UKIDSS surveys to be observed.
Quiescent, blue star-forming, and red (dusty) star-forming galaxies are selected using \citet{Spitler2014} criteria which are shown as red, blue, and orange stars, respectively. 
Galaxies above the outlined section are considered to be quiescent. The remaining galaxies are divided to blue and red star-forming galaxies by the dashed vertical line. 
Photometric redshifts are used to derive the rest-frame colours using EAZY. The photometric redshift distribution of the selected sample is shown by the histogram in the inset.
{\bf Top:} the ZFOURGE sample in the COSMOS field selected to be observed by ZFIRE. 
The logarithmic (2D density) greyscale histogram shows the total UVJ distribution of the ZFOURGE galaxies between 1.90$<$\zphoto$<$2.66. 
In the sample selection, priority is given for the star-forming galaxies that lie below the outlined section in the diagram.
{\bf Bottom:} similar, but now for the UKIDSS sample in the UDS field with galaxies within $10'$ radii from the cluster BCG and at redshifts  $1.57<$\zphoto$<1.67$ shown as the greyscale. 
}
\label{fig:UVJ_selection}
\end{figure}

The primary goal was to build a large sample of redshifts to identify the underlying structure of the galaxy overdensity, therefore, explicitly choosing  star-forming galaxies increased the efficiency of the observing run. 
Quiescent galaxies were selected either as fillers for the masks or because they were considered to be the brightest cluster galaxies (BCG). 
Rest-frame U$-$V and V$-$J colours of galaxies are useful to distinguish star-forming galaxies from quenched galaxies \citep{Williams2009}.
The rest-frame UVJ diagram and the photometric redshift distribution of the selected sample is shown in the left panel of Figure \ref{fig:UVJ_selection}. 
All rest-frame colours have been derived using photometric redshifts using EAZY with special dustier templates as per \citet{Spitler2014}.  
Out of the galaxies selected to be observed by ZFIRE, \around83\% are (blue) star-forming. The rest of the population comprises  \around11\% dusty (red) star-formers and \around6\% quiescent galaxies. 
For all future analysis in this paper, the \citet{Spitler2014} EAZY templates are replaced with the default EAZY templates in order to allow direct comparison with other surveys. 
More information on UVJ selection criteria is explained in Section \ref{sec:UVJ}.

The COSMOS sample at $z\sim2$ requires K-band observations from MOSFIRE to detect \Halpha\ emission lines. 
A subset of the K-band selected galaxies are then followed up in H-band to retrieve \Hbeta\ and \OIII\ emission lines. 
During the first observing run, object priorities for the galaxies in the COSMOS field were assigned as follows. 
\begin{enumerate}
\item K-band observations for rest frame UVJ selected star-forming K$<$24 galaxies with 2.0$<$\zphoto$<$2.2. 
\item K-band observations for rest frame UVJ selected star-forming K$>$24 galaxies with 2.0$<$\zphoto$<$2.2. 
\item K-band observations for rest frame UVJ selected non-star-forming galaxies with 2.0$<$\zphoto$<$2.2.  
\item Galaxies outside the redshift range to be used as fillers. 
\end{enumerate}
In subsequent observing runs, the following criteria were used to assign priorities. 
\begin{enumerate}
\item H-band observations for galaxies with \Halpha\ and  \NII\ detections from K-band. 
\item H-band observations for galaxies with only \Halpha\ detection for follow up spectroscopic redshift verification with \Hbeta\ and/or  \OIII\ emission lines. 
\item K-band observations for galaxies with only \Halpha\ emission lines for deeper spectroscopic redshift verification and gas phase metallicity study with deeper \NII\ emission lines. 
\end{enumerate}

The UDS sample was selected from  the XMM-LSS J02182-05102  cluster \citep{Papovich2010,Tanaka2010} in order to obtain \OIII, \Halpha\ and \NII\ emission lines. At $z=1.62$, these nebular emission lines are redshifted to J and H-bands. 
Cluster galaxies were specifically targeted to complement with the Keck Low Resolution Imaging Spectrometer (LRIS) observations \citep{Tran2015}. 
Y-band spectra were obtained for a subset of galaxies in the cluster in order to detect \MgII\ absorption features and the D4000 break. 
The UVJ diagram and the photometric redshift distribution of the selected sample is shown by the right panel of Figure \ref{fig:UVJ_selection}. In the selected sample, \around65\% of galaxies are star-forming while dusty star-forming and quiescent galaxies are each \around17\%. 
The highest object priorities for the UDS sample were assigned as follows. 
\begin{enumerate}
\item BCGs of the \citet{Papovich2010} cluster. 
\item LRIS detections with \zspec$\sim$1.6 by \cite{Tran2015}. 
\item Grism spectra detections with $z_{\mathrm{grism}}\sim1.6$ \citep[3DHST][]{Momcheva2015}
\item Cluster galaxy candidates within R$<1$ Mpc and \zphoto$\sim1.6$ \citep{Papovich2010}. 
\end{enumerate} 
For further information on target selection, refer to \citet{Tran2015}.

\subsection{Slit Configurations with MAGMA}
\label{sec:mask_design}

MOSFIRE slit configurations are made through the publicly available MAGMA\footnote{http://www2.keck.hawaii.edu/inst/mosfire/magma.html} slit configuration design tool. 
The primary purpose of MAGMA is to design slit configurations to be observed with MOSFIRE and to execute the designed slit configurations in real time at the telescope. 
Once the user specifies a target list and priorities for each of the objects, the software will dither the pointing over the input parameters (which can be defined by the user) to determine the most optimized slit configuration. 

The slit configurations can then be executed during  MOSFIRE observing. With MAGMA, the physical execution of the slit configurations can be done within $<$15 minutes. 
For the objects in the COSMOS field \around10,000 iterations were used to select objects from a target list compromising of \around2000 objects. 
\citet{vanderWel2012} used \emph{HST} imaging to derive position angles of galaxies in the CANDELS sample using GALFIT \citep{Peng2010b}. 
The number of slits within $\pm30^{\circ}$ of the galaxy major axis were maximized using position angles of \citet{vanderWel2012} catalogue by cross-matching it with ZFOURGE.

Due to the object prioritization, a subset of galaxies was observed in multiple observing runs. These galaxies were included in different masks and hence have different position angles. When possible, position angles of these slits were deliberately varied to allow coverage of a different orientation of the galaxy.

\subsection{MOSFIRE Observations}

\begin{figure}
\centering
\includegraphics[width=1.00\textwidth]{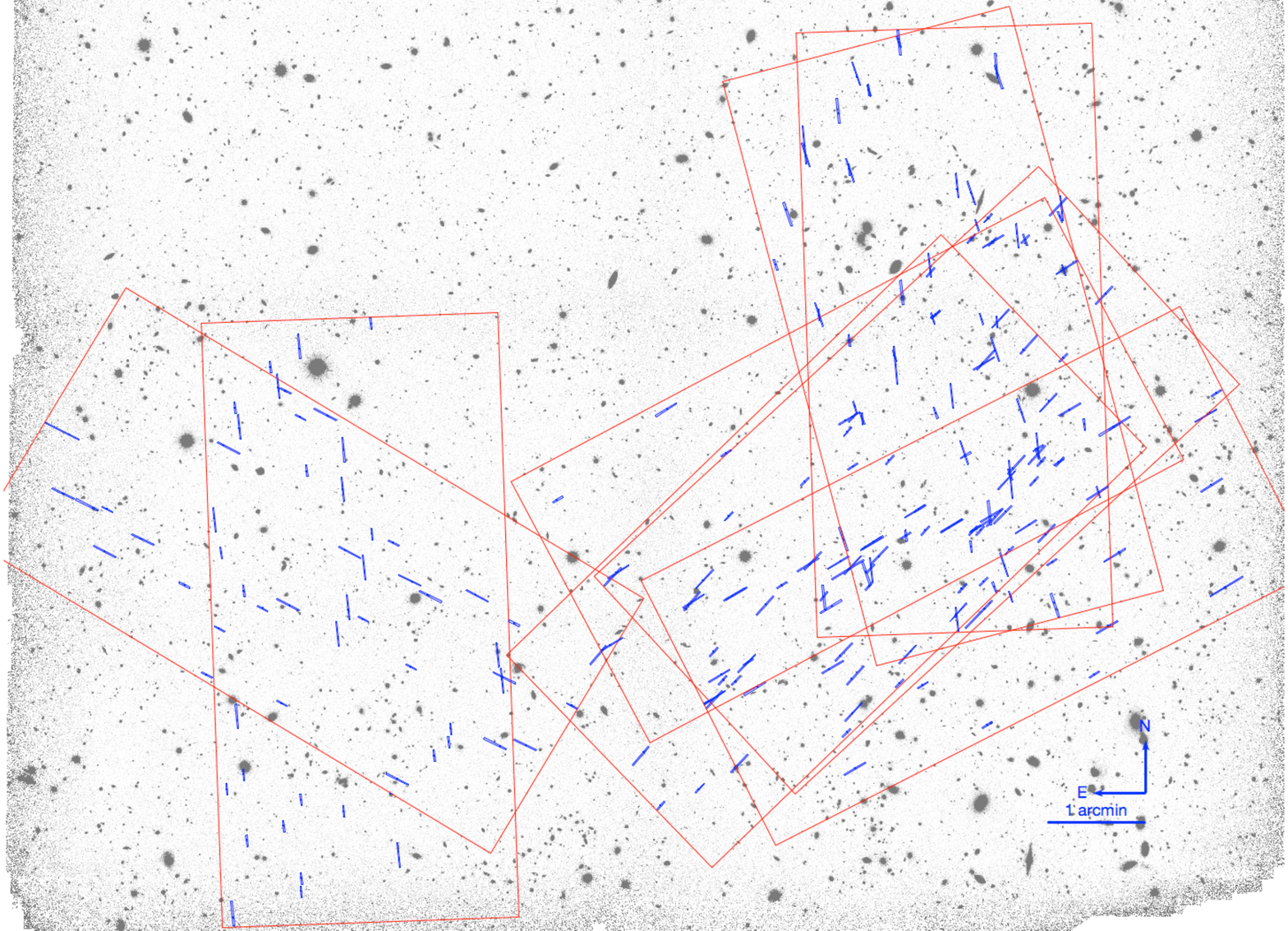}
\caption[MOSFIRE slit configurations for the 6 K-band masks in the COSMOS field.]{MOSFIRE slit configurations for the 6 K-band masks in the COSMOS field. 
The blue lines show each individual slit.  
Each slit in a mask is expected to target a single galaxy. However, some galaxies are targeted in multiple masks. 
The red boxes are the individual masks. 
The inverse greyscale image is from the Ks imaging from FourStar obtained as a part of the ZFOURGE survey.}
\label{fig:masks}
\end{figure}

Between 2013 and 2016 15 MOSFIRE nights were awarded to the ZFIRE program  by a 
combination of Swinburne University (Program IDs- 2013A\_W163M, 2013B\_W160M, 2014A\_W168M, 2015A\_W193M, 2015B\_W180M), Australian National University (Program IDs- 2013B\_Z295M, 2014A\_Z225M, 2015A\_Z236M, 2015B\_Z236M), and NASA (Program IDs- 2013A\_N105M, 2014A\_N121M) telescope time allocation committees. 
Data for 13 nights observed between 2013 and 2015 are released with this paper, where six nights resulted in useful data collection. 
Observations during 2013 December resulted in two nights of data in excellent conditions, while four nights in 2014 February were observed in varying conditions. Exposure times and observing conditions are presented in Table \ref{tab:observing_details}. 
With this paper, data for 10 masks observed in the COSMOS field and four masks observed in the UDS field are released. An example of on-sky orientations of slit mask designs used for K-band observations in the COSMOS field is shown in Figure \ref{fig:masks}. 
Standard stars were observed at the beginning, middle, and end of each observing night. 

The line spread functions were calculated using Ne arc lamps in the K-band, and were found to be \around2.5 pixels. The partial first derivative for the wavelength (CD1\_1) in Y, J, H, and K-bands are respectively 1.09 \AA/pixel, 1.30 \AA/pixel, 1.63 \AA/pixel, and 2.17 \AA/pixel. 

$0.7''$ width slits were used for objects in science masks and the telluric standard, while, for the flux standard star a slit of width $3''$ was used to minimize slit loss. On average, $\sim$ 30 galaxies were included per mask. A flux monitor star was included in all of the science frames to monitor the variation of the seeing and atmospheric transparency. In most cases only frames that had a FWHM of $\lesssim0''.8$ was used for the flux monitor stars. A standard 2 position dither pattern of ABBA was used.\footnote{For more information, see: \url{http://www2.keck.hawaii.edu/inst/mosfire/dither\_patterns.html\#patterns}} 

\begin{landscape}
\begin{deluxetable}{llllrrr}
\centering
\tabletypesize{\small}
\tablecaption{ ZFIRE Data Release 1: Observing details}  
\tablecomments{ This table presents information on all the masks observed by ZFIRE between 2013 and 2015 with the integration times and observing conditions listed.  
\label{tab:observing_details}}
\tablecolumns{6}
\tablewidth{0pt} 
\startdata
\hline \hline \\ [+1ex]
Field  & Observing & Mask & Filter & Exposure & Total integra-     & Average \\
	   & Run       & Name &        & Time (s) & -tion Time (h) 		& Seeing ($''$) \\ [+1ex]  \hline \\ [+1ex]
COSMOS & Dec2013 & Shallowmask1 (SK1)    & K & 180 & 2.0  & 0$''$.70\\ 
COSMOS & Dec2013 & Shallowmask2 (SK2)   & K & 180 & 2.0  & 0$''$.68\\
COSMOS & Dec2013 & Shallowmask3 (SK3)   & K & 180 & 2.0  & 0$''$.70\\ 
COSMOS & Dec2013 & Shallowmask4 (SK4)   & K & 180 & 2.0  & 0$''$.67\\ 
COSMOS & Feb2014 & KbandLargeArea3 (KL3) & K & 180 & 2.0  & 1$''$.10\\ 
COSMOS & Feb2014 & KbandLargeArea4 (KL4) & K & 180 & 2.0  & 0$''$.66\\ 
COSMOS & Feb2014 & DeepKband1      (DK1) & K & 180 & 2.0   & 1$''$.27\\ 
COSMOS & Feb2014 & DeepKband2      (DK2) & K & 180 & 2.0   & 0$''$.70\\ 
COSMOS & Feb2014 & Hbandmask1      (H1) & H & 120 & 5.3 & 0$''$.90\\ 
COSMOS & Feb2014 & Hbandmask2      (H2) & H & 120 & 3.2 & 0$''$.79\\ 
UDS    & Dec2013 & UDS1            (U1H) & H & 120 & 1.6 & 0$''$.73\\ 
UDS    & Dec2013 & UDS2            (U2H) & H & 120 & 1.6 & 0$''$.87\\ 
UDS    & Dec2013 & UDS3            (U3H) & H & 120 & 0.8 & 0$''$.55\\ 
UDS    & Dec2013 & UDS1            (U1J) & J & 120 & 0.8 & 0$''$.72\\ 
UDS    & Dec2013 & UDS2            (U2J) & J & 120 & 0.8 & 0$''$.90\\ 
UDS    & Dec2013 & UDS3            (U3J) & J & 120 & 0.8 & 0$''$.63\\ 
UDS    & Feb2014 & uds-y1          (UY) & Y & 180 & 4.4 & 0$''$.80\\
\enddata
\end{deluxetable}
\end{landscape}

\subsection{MOSFIRE Spectroscopic Reduction}
\label{data_reduction}

The data were reduced in two steps. 
Firstly, a slightly modified version of the publicly available 2015A MOSFIRE DRP release \footnote{A few bug fixes were applied along with an extra function to implement barycentric corrections to the spectra. This version is available at \url{https://github.com/themiyan/MosfireDRP\_Themiyan}.} was used to reduce the raw data from the telescope.
Secondly, a custom made IDL package was used to apply telluric corrections and flux calibrations to the data and extract 1D spectra. Both are described below. 

Extensive tests were performed to the MOSFIRE DRP while it was in a beta stage, and multiple versions of the DRP were used to test the quality of the outputs. 
The accuracy of the error spectrum generated by the DRP was investigated by comparing the noise we expect from the scatter of the sky values with the DRP noise. 
The following steps are currently performed by the modified MOSFIRE DRP. 
\begin{enumerate}
\item Produce a pixel flat image and identify the slit edges. 
\item For K-band: remove the thermal background produced by the telescope dome. 
\item Wavelength calibrate the spectra. This is performed using the sky lines. For K-band: due to the lack of strong sky lines at the red end of the spectra, a combination of night sky lines along with Neon and/or Argon\footnote{As of version 2015A, using both Ar and Ne lamps together with sky line wavelength calibration is not recommended. See the MOSFIRE DRP github issues page for more details.} arc lamp spectra are used to produce per pixel wavelength calibration.
\item Apply barycentric corrections to the wavelength solution. 
\item Remove the sky background from the spectra. This is done in two steps. Firstly, the different nod positions of the telescope are used to subtract most of the background. 
Secondly, any residual sky features are removed following the prescription by \citet{Kelson2003}.
\item Rectify the spectra.
\end{enumerate}
All the spectra from the DRP were calibrated to vacuum wavelengths with a typical residual error of $<$ 0.1 \AA. 

The customized IDL package was used to continue the data reduction process using outputs of the public DRP. The same observed standard star was used to derive telluric sensitivity and flux calibration curves  to be applied to the science frames as follows.
\begin{enumerate}
\item The 1D standard star spectrum was extracted from the wavelength calibrated 2D spectra.
\item Intrinsic hydrogen absorption lines in the stellar atmosphere  were removed from the telluric A0 standard by fitting Gaussian profiles and then  interpolating over the filled region. 
\item The observed spectrum was ratioed to a theoretical black body function corresponding to the temperature of the star.
\item The resulting spectrum was then normalised and smoothed to be used as the sensitivity curve, i.e., the wavelength-dependent sensitivity that is caused by the atmosphere and telescope-instrument response. 
\item The sensitivity curve was used on the flux standard star to derive the flux conversion factor by comparing it to its 2MASS magnitude \citep{Skrutskie2006}.
\end{enumerate}
These corrections are applied to the 2D science frames to produce telluric corrected, flux calibrated spectra. 
Further information is provided in Appendix \ref{sec:MOSFIRE cals}.  
The derived response curves that were applied to all data include corrections for the MOSFIRE response function, the telescope sensitivity, and atmospheric absorption. 
If the mask was observed in multiple nights, the calibrated 2D spectra were co-added by weighting by the variance spectrum.  Extensive visual inspections were performed to the 2D spectra to identify possible emission line-only detections and to flag false detections due to, e.g. sky line residuals.

To extract 1D spectra, Gaussian extractions were used to determine the FWHM of the spatial profile. If the objects were too faint compared to the sky background, the profile from the flux monitor star of the respective mask was used to perform the extraction. 
The same extraction procedure was performed for any secondary or tertiary objects that fall within any given slit.
Depending on how object priorities were handled, some objects were observed during multiple observing runs in different masks. There were 37 such galaxies.  
Due to variations in the position angles between different masks, these objects were co-added in 1D after applying the spectrophotometric calibration explained in Section \ref{sec:sp calibration}.

\subsection{Spectrophotometric Flux Calibration}
\label{sec:sp calibration}

\subsubsection{COSMOS Legacy Field}

Next zero-point adjustments were derived for each mask to account for any atmospheric transmission change between mask and standard observations. Synthetic slit aperture magnitudes were computed from the ZFOURGE survey to calibrate the total magnitudes of the spectra, which also allowed us to account for any slit-losses due to the $0''.7$ slit-width used during the observing.  
The filter response functions for FourStar \citep{Persson2013} were used to integrate the total flux in each of the 1D calibrated spectra.  

For each of the masks in a respective filter, first,  all objects with a photometric error $>0.1$ mag were removed. 
Then, a background subtracted Ks and F160W (H-band) images from ZFOURGE were used with the seeing convolved from $0''.4$ to $0''.7$ to match the average Keck seeing. 
Rectangular apertures, which resemble the slits with various heights were overlaid in the images to integrate the total counts within each aperture. 
Any apertures that contain multiple objects or had bright sources close to the slit edges were removed. 
Integrated counts were used to calculate the photometric magnitude to compare with the spectroscopy. 
A slit-box aligned with similar PA to the respective mask with a size of $0''.7 \times 2''.8$ was found to give the best balance between the spectrophotometric comparison and the number of available slits with good photometry per mask.

Next, the median offset between the magnitudes from photometry and spectroscopy were calculated by selecting objects with a photometric magnitude less than 24 in the respective filters.  
This offset was used as the scaling factor and was applied to all spectra in the mask. Typical offsets for K and H bands were $\sim \pm0.1$ mag. 
We then performed 1000 iterations of bootstrap re-sampling of the objects in each mask to calculate the scatter of the median values. We parametrized the scatter using normalized absolute median deviation (\NMAD) which is defined as 1.48$\times$ median absolute deviation. 
The median \NMAD\ scatter in K and H-bands for these offsets are \around0.1 and \around0.04 mag, respectively.

\begin{figure}
\centering
\includegraphics[width=0.9\textwidth]{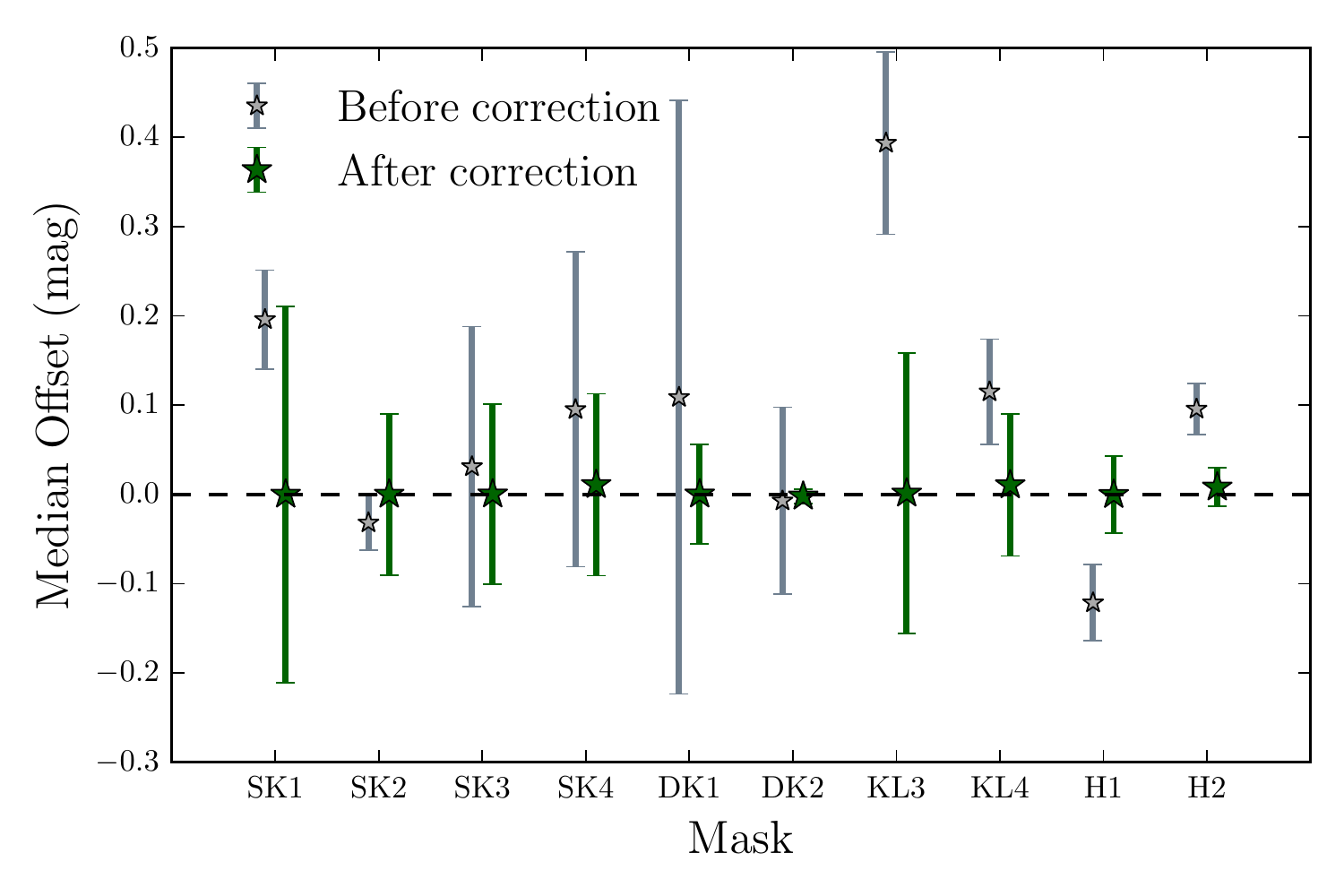}
\includegraphics[width=0.9\textwidth]{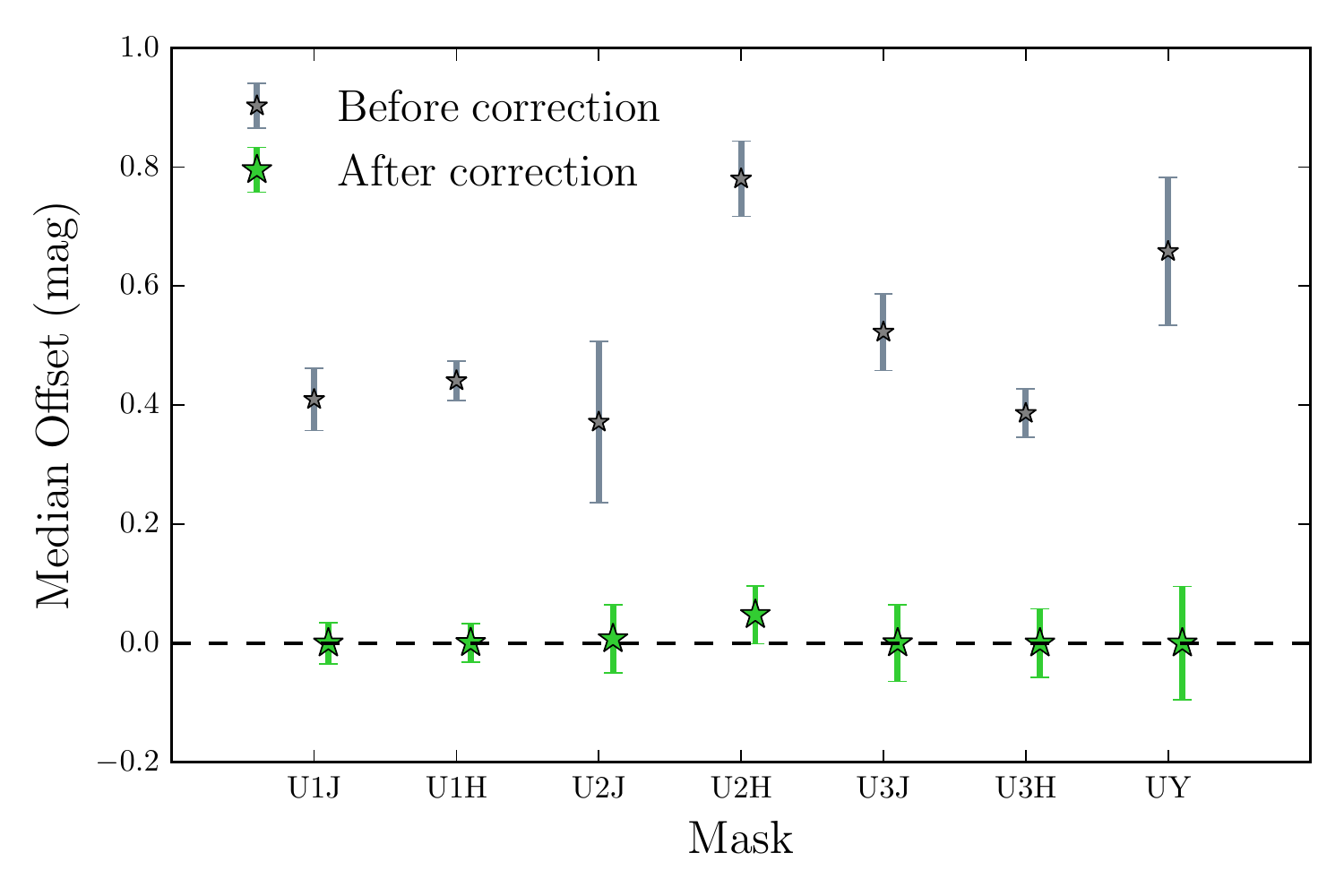}
\caption[Spectrophotometric calibration of the ZFIRE masks.]{Spectrophotometric calibration of the ZFIRE masks. The median offsets between spectroscopic flux and the photometric flux before and after the scaling process is shown in the figure. Filter names correspond to the names in Table \ref{tab:observing_details}.
The grey stars denote the median offsets for the standard star flux calibrated data before any additional scaling is applied. 
The median mask sensitivity factors are applied to all objects in the respective masks to account for slit loss. The green stars show the median offsets after the flux corrections are applied. The errors are the \NMAD\ scatter of the median offsets calculated via bootstrap re-sampling of individual galaxies. 
{\bf Top:} all COSMOS masks. Photometric data are from a slit-box aligned with similar PA to the respective mask with a size of $0''.7 \times 2''.8$.
{\bf Bottom:} all UDS masks. Photometric data are total fluxes from UKIDSS.}
\label{fig:scaling_values}
\end{figure}

The median offset values per mask before and after the scaling process with its associated error is shown in the top panel of Figure \ref{fig:scaling_values}. Typical offsets are of the order of $\lesssim0.1$ mag which are consistent with expected values of slit loss
and the small amount of cloud variation seen during the observations. 
The offset value after the scaling process is shown as green stars with its bootstrap error. 

The scaling factor was applied as a multiple for the flux values for the 2D spectra following Equation \ref{eq:scale_single_masks},  
\begin{subequations}
\label{eq:scale_single_masks}
\begin{equation}
F_i = f_i \times \hbox{scale}_{\mathrm{mask}} 
\end{equation}
\begin{equation}
\Sigma_i = \sigma_i \times \hbox{scale}_{\mathrm{mask}}
\end{equation}
\end{subequations}
where $\mathrm{f_i}$ and $\mathrm{\sigma_i}$ are, respectively, the flux and error per pixel before scaling and scale$\mathrm{_{mask}}$ is the scaling factor calculated.

1D spectra are extracted using the same extraction aperture as before. 
The bootstrap errors after the scaling process is \around0.08 mag (median) for the COSMOS field,  which is considered to be the final uncertainty of the spectrophotometric calibration process.  Once a uniform scaling was applied to all the objects in a given mask, the agreement between the photometric slit-box magnitude and the spectroscopic magnitude increased.

As aforementioned, if an object was observed in multiple masks in the same filter, first the corresponding mask scaling factor was applied and then co-added optimally in 1D such that a higher weight was given to the objects, which came from a mask with a lower scaling value (i.e. better transmission). The procedure is shown in equation \ref{eq:1D_scale_and_coadd}, 
\begin{subequations}
\label{eq:1D_scale_and_coadd}
\begin{equation}
 F_i = \frac{\sum\limits_{j=1}^n (P_j/\sigma_{ji})^2 (F_{ji}/P_j)}{\sum\limits_{j=1}^n(P_j/\sigma_{ji})^2 }
\end{equation}
\begin{equation}
\sigma_i^2 = \frac{\sum \limits_{j=1}^n \big\{(P_j/\sigma_{ji})^2 (F_{ji}/P_j)\big\}^2} {\big\{\sum \limits_{j=1}^n(P_j/\sigma_{ji})^2\big\}^2 }
\end{equation}
\end{subequations}
where $P$ is the 1/scale value, $i$ is the pixel number, and $j$ is the observing run. Further examples for the spectrophotometric calibration process are shown in Appendix \ref{sec:MOSFIRE cals}.

\subsubsection{UDS Legacy Field}

The filter response functions for WFCAM \citep{Casali2007} was used to integrate the total flux in each of the 1D calibrated spectra in the UDS field. 
The {\it total} photometric fluxes from the UKIDSS catalogue were used to compare with the integrated flux from the spectra since images were not 
available to simulate slit apertures. 
To calculate the median offset a magnitude limit of 23 was used.  
This magnitude limit was brighter than the limit used for COSMOS data since the median photometric magnitude of the UDS data are \around0.5 mag brighter than COSMOS.

Typical median offsets between photometric and spectroscopic magnitudes were \around0.4 magnitude.
The lower panel of Figure \ref{fig:scaling_values} shows the median offset values per mask before and after the scaling process with its associated error. 
The median of the bootstrap errors for the UDS masks after scaling is \around0.06 mag. 

Comparing with the COSMOS offsets, the UDS values are heavily biased toward a positive offset. 
This behaviour is expected for UDS data because the broadband total fluxes from the UKIDSS data are used, and therefore the flux expected from the finite MOSFIRE slit should be less than the total flux detected from UKIDSS. 
Since UDS objects are not observed in multiple masks in the same filter, only Equation \ref{eq:scale_single_masks} is applied to scale the spectra. \\

\subsection{Measuring Emission Line Fluxes}
\label{sec:line_fits}

A custom made IDL routine was used to fit nebular emission lines on the scaled 1D spectra. This was done by fitting Gaussian profiles to user defined emission lines. 
The code identifies the location of the emission line in wavelength space and calculates the redshift.  

In emission line fitting, if there were multiple emission lines detected for the same galaxy in a given band, the line centre and velocity width were kept to be the same. Emission lines with velocity structure were visually identified and were fit with multiple component Gaussian fits.  
If the line was narrower than the instrumental resolution, the line width was set to match the instrument resolution.  
The code calculated the emission line fluxes (f) by integrating the Gaussian fits to the emission lines. The corresponding error for the line fluxes ($\sigma$(f)) were calculated by integrating the error spectrum using the same Gaussian profile. The code further fits a 1$\sigma$ upper level for the flux values (f$_{limit}$). The signal-to-noise ratio (SNR) of the line fluxes was defined as the line flux divided by the corresponding error for the line flux.

Next, in Chapter \ref{chap:spec_analysis} I provide a thorough analysis of the ZFIRE spectroscopic sample and investigate the role of spectroscopic redshift in determining galaxy properties via SED fitting techniques. Redshift accuracies of other photometric and grism redshift surveys are also investigated using the ZFIRE spectroscopic data. 

\newpage


\section{Summary of ZFIRE publications}
\label{sec:other_zfire_work}

Apart from the work presented in this thesis, ZFIRE data catalogue produced as a result of my work has been used in numerous publications by the ZFIRE team and their collaborators. 
A complete list of publications are provided in the declaration. Here, I present a brief  summary of ZFIRE publications that I have directly contributed to.

\begin{figure}
\centering
\includegraphics[trim=30 10 10 10, clip, width=0.80\textwidth]{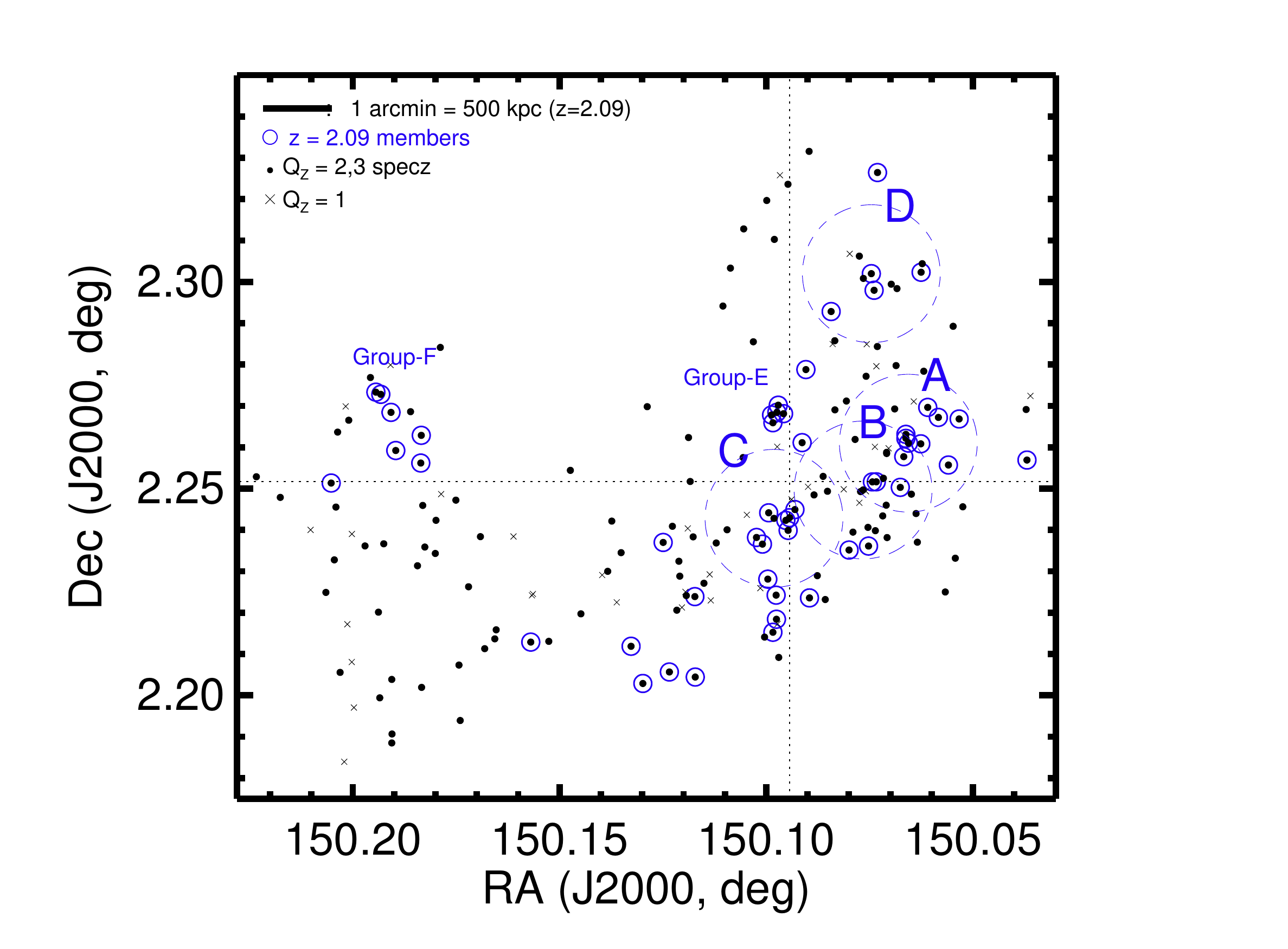} 
\includegraphics[trim=30 10 10 10, clip, width=0.80\textwidth]{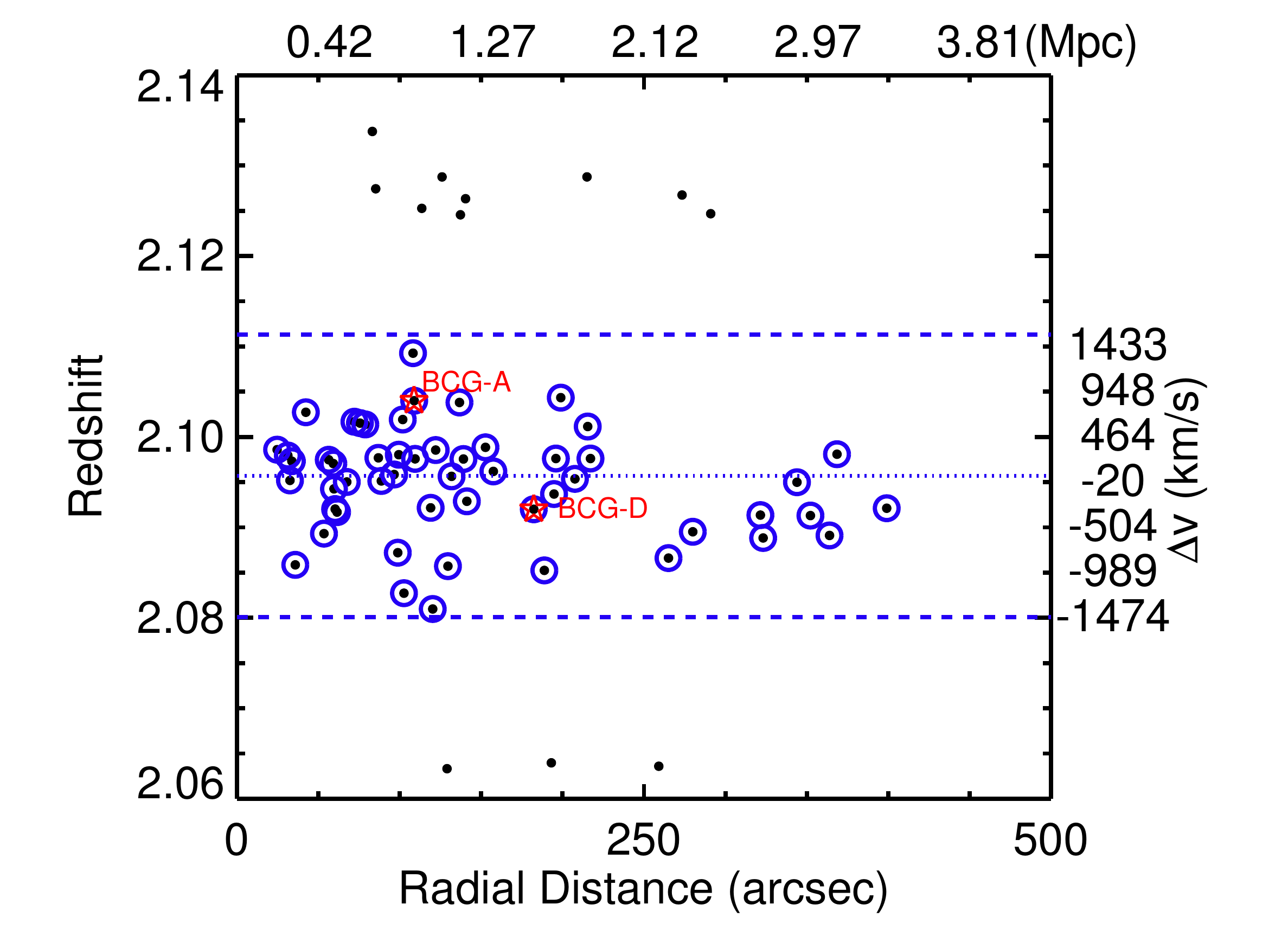} 
\caption[Reproduction of the \citet{Yuan2014} Figures 4 and 5, which shows the spatial and velocity distribution of the confirmed cluster at $z=2.095$ in the COSMOS field.]{Reproduction of the \citet{Yuan2014} Figures 4 and 5, which shows the spatial and velocity distribution of the confirmed cluster at $z=2.095$ in the COSMOS field.
{\bf Top:} The spatial distribution of the ZFIRE surveyed region in the COSMOS field. Cluster members are marked as blue circles. A, B, C, and D regions marked by dashed circles correspond to peaks in the seventh nearest neighbour density maps identified by \citet{Spitler2012}. 
{\bf Bottom:} The spectroscopic redshifts of the cluster members along with the radial distances and velocities with respect to the centre of the cluster are shown here. The brightest cluster galaxies are highlighted as red stars.
}
\label{fig:yuan_plot}
\end{figure}

\citet{Yuan2014} used the ZFIRE redshifts to spectroscopically confirm the $z=2.1$ overdensity discovered by \citet{Spitler2012}. ZFIRE masks which were designed to spatially cover this region to high completeness resulted in over 50 confirmed cluster members with a cluster velocity dispersion of $552\pm52$ kms$^{-1}$.  The spatial coverage and the velocity dispersions of the cluster members are shown by Figure \ref{fig:yuan_plot}. $\Lambda$CDM hydrodynamical simulations further showed that this structure would evolve into a Virgo like cluster at $z\sim0$. Considering the ZFOURGE survey area of 0.1 deg$^2$ the probability of finding such a structure at $z\sim2$ was determined to be $\sim4\%$.

\begin{figure}
\centering
\includegraphics[trim=10 10 10 10, clip, width=0.60\textwidth]{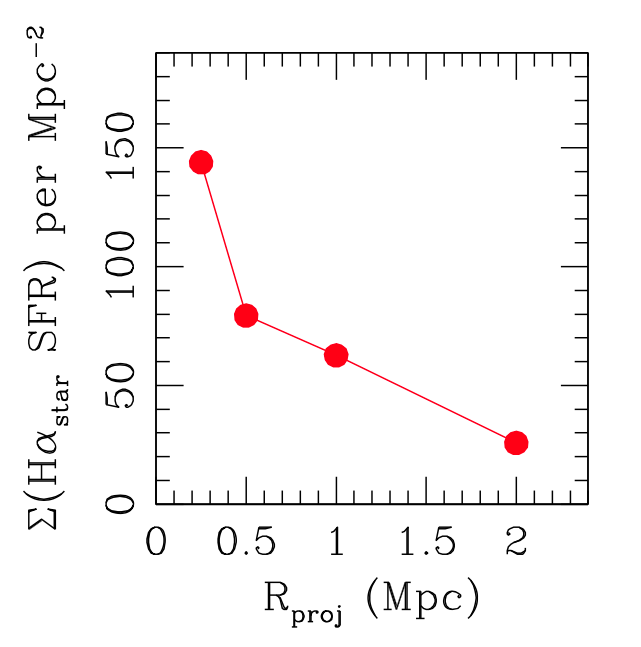} 
\includegraphics[trim=30 10 10 10, clip, width=0.60\textwidth]{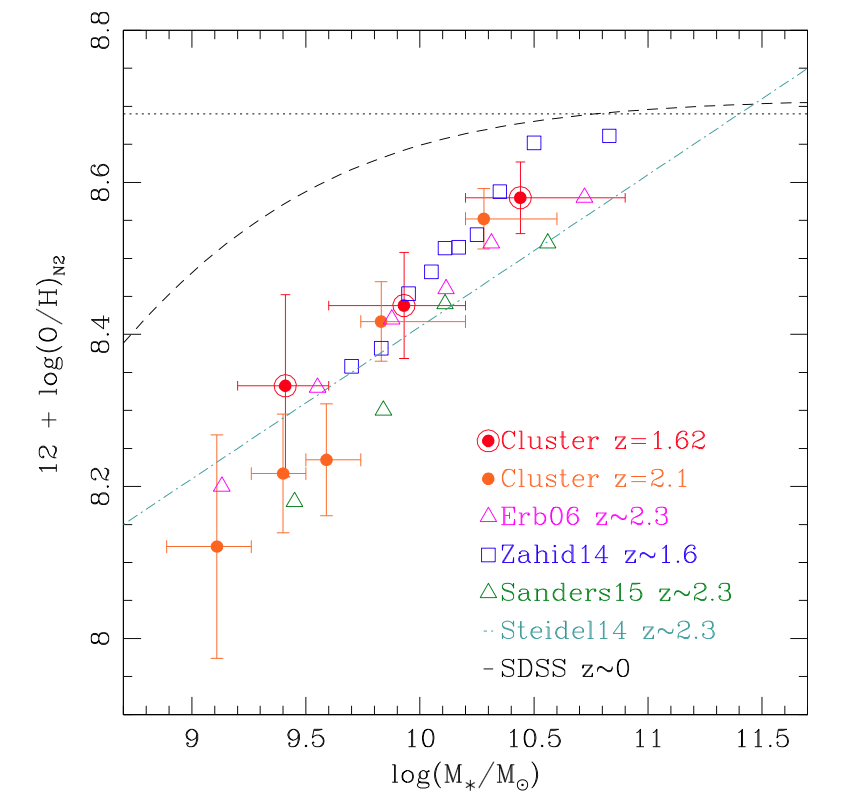} 
\caption[Reproduction of the \citet{Tran2015} Figures 6 (left) and 9, which shows the \Halpha\ SFR per unit area from the core of the XMM-LSS J02182-05102 cluster and its MZR.]{Reproduction of the \citet{Tran2015} Figures 6 (left) and 9, which shows the \Halpha\ SFR per unit area from the core of the XMM-LSS J02182-05102 cluster and its MZR.
{\bf Top:} The integrated \Halpha\ SFR per unit area for the XMM-LSS J02182-05102 cluster. The distances are marked from the cluster core. 
{\bf Bottom:} The MZR for the stacked ZFIRE sample of the XMM-LSS J02182-05102 cluster is shown by the red filled circles. 
Best fits from other studies of the MZR at $z\sim2$ are shown for comparison and are as follows:  \citet{Kacprzak2015} shown by orange circles, \citet{Erb2006} shown by pink triangles, \citet{Zahid2014} shown by the blue squares, \citet{Sanders2015} shown by the green triangles, \citet{Steidel2014} fit shown by the green dash-doted line, and the \citet{Moustakas2011} SDSS relation shown by the black dashed line. 
Solar abundance is shown by the dotted line \citep{Asplund2009}. 
}
\label{fig:tran15_plot}
\end{figure}

\citet{Tran2015} used the \Halpha\ emission lines from the XMM-LSS J02182-05102 cluster observed at $z=1.62$ by ZFIRE in the UDS field to study the galaxy kinematics, star-formation rates, and metallicities. The velocity dispersion of the galaxy cluster was found to be one of the lowest measured for a $z\sim1.6$ cluster with an elevated SFR towards the core, which is shown by Figure \ref{fig:tran15_plot} (top panel). Extinction values derived via SED fits were inconsistent with values derived via Balmer decrements and showed a large scatter consistent with other work at $z\sim2$ \citep[eg.,][]{Reddy2015}. Furthermore, the MZR showed no environmental dependence for this cluster, which is shown by Figure \ref{fig:tran15_plot} (bottom panel).

\citet{Tran2017} probed the galaxy scaling relations of the $z\sim2$ \citet{Yuan2014} cluster to find that  environment had no statistically significant effect on galaxy SFRs, stellar masses, sizes,  and SFR densities. However, galaxies with high IR luminosity showed $\sim\times5$ more stellar mass and 60\% larger sizes compared to low IR luminous galaxies. Analysis of the SFR main sequence showed that galaxies above the main sequence have smaller radii, higher Sersic indices, younger ages, and higher star-forming surface densities suggesting that galaxies grow from their stellar cores.

\begin{figure}
\centering
\includegraphics[ width=1.0\textwidth]{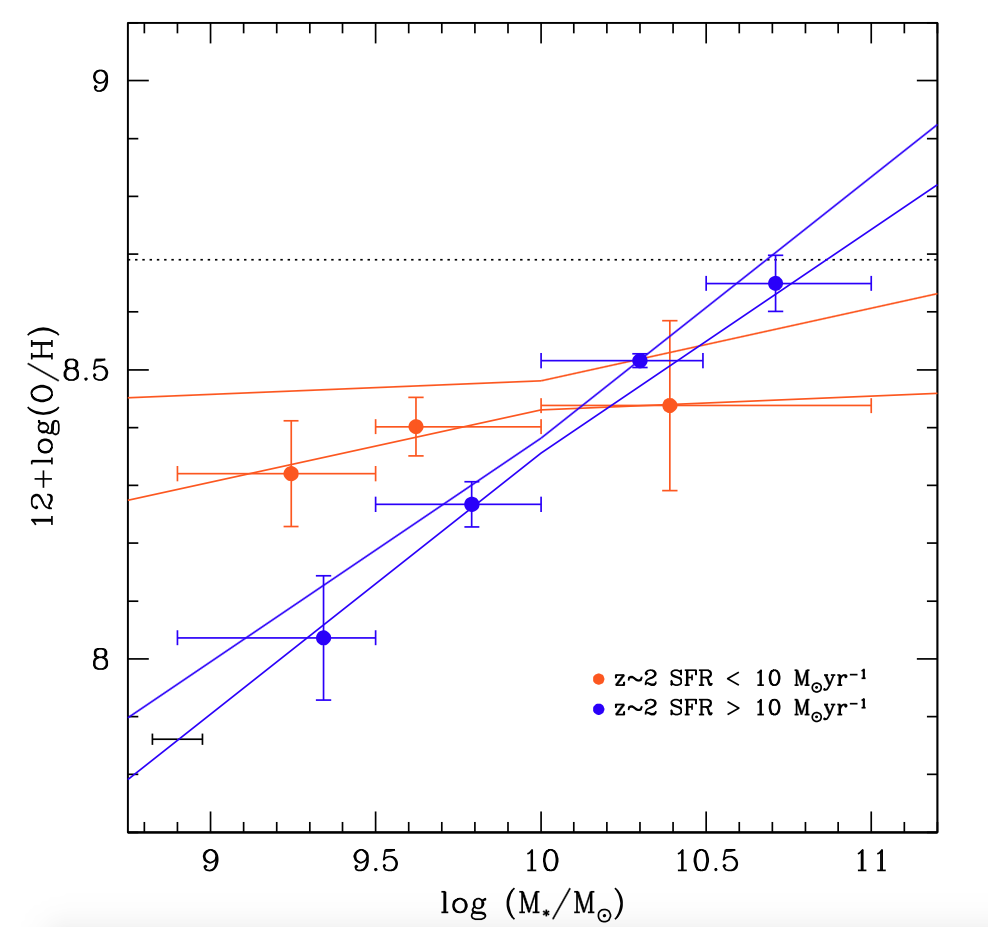} 
\caption[Reproduction of the \citet{Kacprzak2016} Figure 1, which shows the MZR for the ZFIRE COSMOS sample at $z\sim2$.]{Reproduction of the \citet{Kacprzak2016} Figure 1, which shows the MZR for the stacked ZFIRE COSMOS sample at $z\sim2$. Galaxies are binned into low and high SFRs as shown by the legend and bootstrap fits, with $1\sigma$ limits of the two SFR samples are shown by the solid lines. Solar abundance is shown by the dotted line \citep{Asplund2009}. }
\label{fig:kacprzak16_plot}
\end{figure}

\citet{Kacprzak2015} investigated the MZR of the \citet{Yuan2014} cluster and ruled out environmental variances at the $1\sigma$ level between cluster and field galaxies. Hydrodynamical simulations showed that cluster galaxies should be marginally metal rich compared to field galaxies, which could be attributed to the removal of gas from galaxy outskirts or shorter gas recycling time-scales in rich environments. 
\citet{Kacprzak2016} further analysed the MZR relation to find that at a fixed mass, low SFR galaxies show higher metallicities compared to their higher SFR counterparts. 
The stacked galaxy sample with associated errors to the fit is shown by Figure \ref{fig:kacprzak16_plot}. 
The difference in the MZR was attributed to the cold-mode gas accretion, where galaxies with high SFRs are actively accreting pristine gas from the inter-galactic medium, thus driving their gas phase metallicities to lower values compared to the low SFR counterparts.

\begin{figure}
\centering
\includegraphics[trim=10 10 10 10, clip, width=0.90\textwidth]{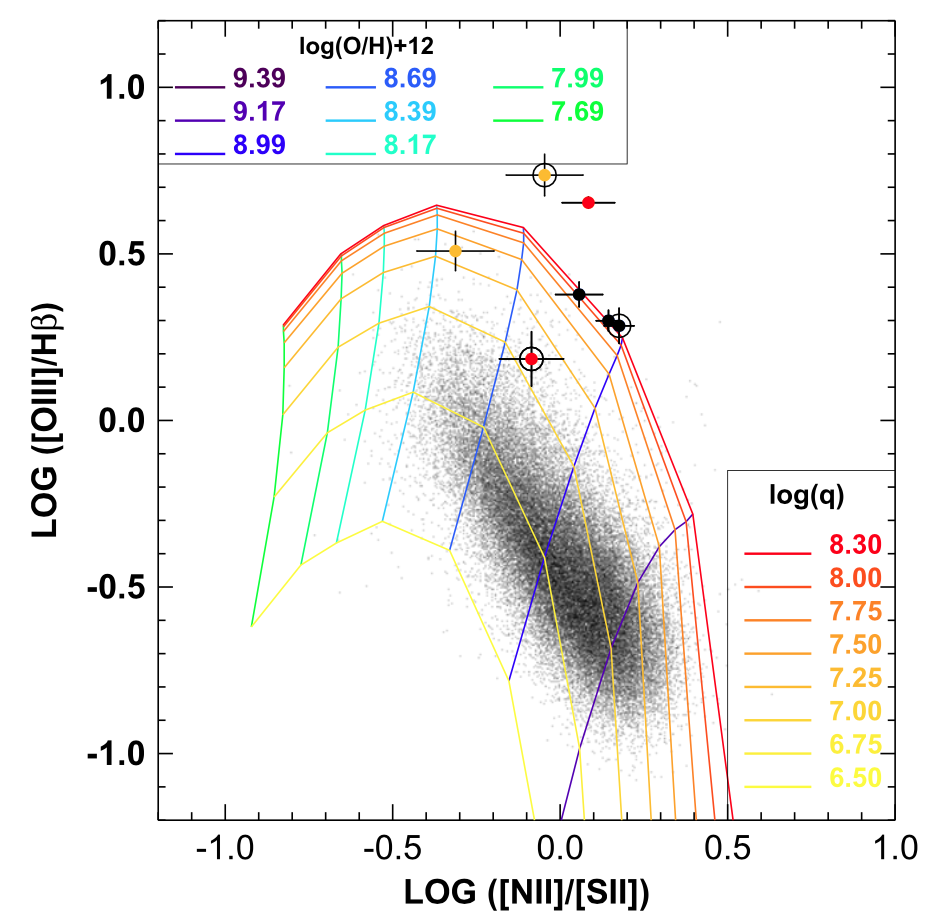} 
\caption[Reproduction of the \citet{Kewley2016} Figure 5, which show the \NII/\SII\ vs \OIII/\Hbeta\ ratios for the ZFIRE COSMOS sample at $z\sim2$. ]{ Reproduction of the \citet{Kewley2016} Figure 5, which show the \NII/\SII\ vs \OIII/\Hbeta\ ratios for the ZFIRE COSMOS sample at $z\sim2$. $3\sigma$ \SII\ detections are shown as orange circles, while  $2\sigma$ detections are shown as black filled circles. SDSS galaxies are shown by the black dots and the stellar masses have been matched to be similar to the ZFIRE sample. The coloured curves are theoretical photo-ionization models for varying metallicities (blue-green lines) and ionization parameters (yellow-red lines).  
}
\label{fig:kewley2016_plot}
\end{figure}

\citet{Kewley2016} studied the ISM properties of the $z\sim2$ ZFIRE COSMOS sample and found no statistically significant differences between the \NII/\Halpha\ and \OIII/\Hbeta\ emission line ratios in the cluster and field galaxies, thus ruling out environmental variations in the ISM pressure and ionization parameter at $z\sim2$.  However, as shown by Figure \ref{fig:kewley2016_plot}, \citet{Kewley2016} found that these emission line ratios were significantly higher than local star-forming galaxies due to the ionization parameter being up to $\sim1$dex higher than local galaxies.

Straatman et al., (submitted) used the \Halpha\ emission lines to probe the \citet{Tully1977} relation at $z\gtrsim2$ and found a gradual evolution of the Tully-Fisher stellar mass zero-point from $z=0$ to $z\gtrsim2$. The results of this study indicated that a higher fraction of galaxies are pressure supported at these redshifts and that the evolution of the stellar mass zero-point was partly due to the conversion of gas to stars during cosmic time.  

\begin{figure}
\centering
\includegraphics[trim=10 10 10 10, clip, width=0.90\textwidth]{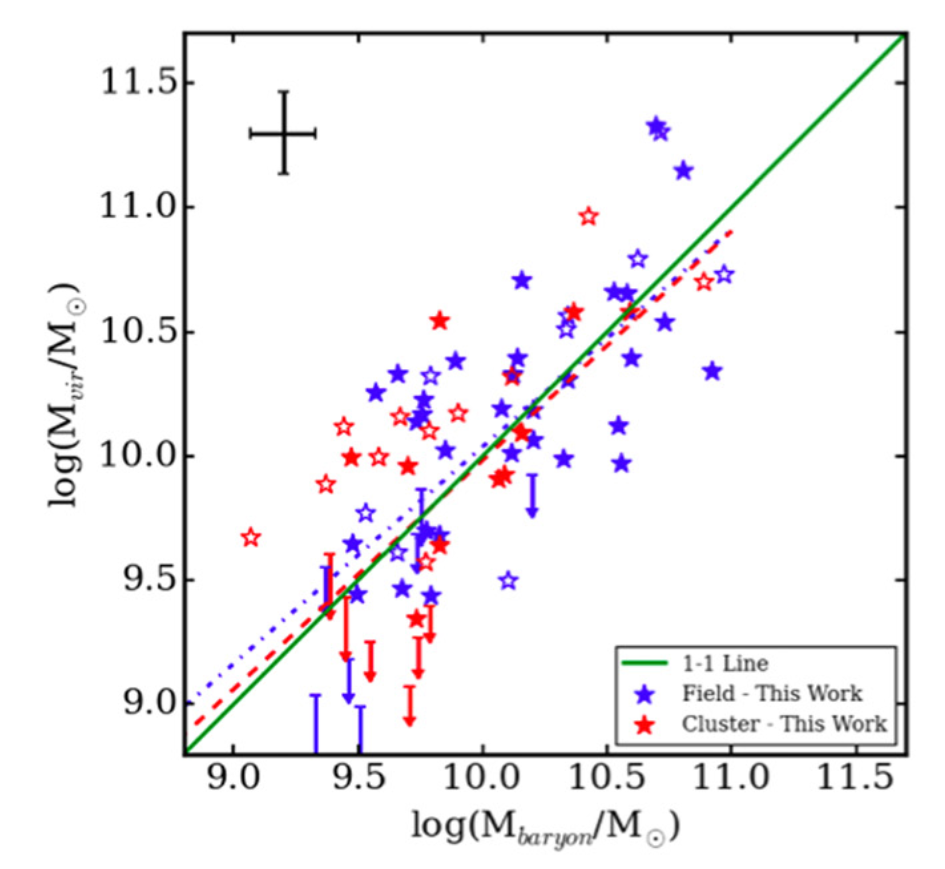} 
\caption[Reproduction of the \citet{Alcorn2016} Figure 4 (left panel), which shows the comparison between virial masses of the ZFIRE COSMOS $z\sim2$ galaxies with their baryonic masses. ]{ Reproduction of the \citet{Alcorn2016} Figure 4 (left panel), which shows the comparison between virial masses of the ZFIRE COSMOS $z\sim2$ sample with their baryonic masses. The sample is divided into cluster and field galaxies with their best-fitting lines shown by the dashed (red) and dashed dotted (blue) lines, respectively. Galaxies with confirmed line widths are shown by the filled points while the outlined points are faint or sky-line interfered line-widths. The one-to-one line is shown by green. Characteristic error bars are shown by black in the upper left corner.  
}
\label{fig:alcorn2016_plot}
\end{figure}

\citet{Alcorn2016} performed a kinematic analysis of the $z\sim2$ COSMOS galaxies using \Halpha\ emission line widths. Emission line kinematic scaling relations showed no environmental dependence and derived stellar mass to velocity dispersion relationships were found to be similar to other IFU studies at $z\sim2$. The derived baryonic and virial mass estimates for the galaxies were similar to each other suggesting that baryonic matter dominated the ZFIRE sample within one effective radius, which is shown by Figure \ref{fig:alcorn2016_plot}.

\chapter{ZFIRE Survey II: Analysis of spectroscopic/photometric properties of ZFIRE galaxies}
\label{chap:spec_analysis}

In this chapter, I present observational properties of the first ZFIRE data release, which includes a discussion of the observing completeness and a comparison of the accuracy of photometric redshifts of various photometric surveys and investigate the role of spectroscopic redshifts in determining fundamental galaxy properties via SED fitting techniques.

\section{Properties of ZFIRE Galaxies}
\label{sec:results_zfire}

\subsection{Spectroscopic Redshift Distribution}
\label{sec:Q flags}

Using nebular emission lines, 170 galaxy redshifts were identified for the COSMOS sample and 62 redshifts were identified for the UDS field. 
A combination of visual identifications in the 2D spectra and emission line fitting procedures explained in Section \ref{sec:line_fits} were used to identify these redshifts. 
The redshift quality is defined using three specific flags: 
\begin{itemize}
\item Q$_z$ Flag $=$ 1: These are objects with no line detection with SNR $<5$. These objects are not included in our final spectroscopic sample. 
\item Q$_z$ Flag $=$ 2: These are objects with one emission line with SNR $>$ 5 and a $|$\zspec $-$ \zphoto $|$ $> 0.2$.
\item Q$_z$ Flag $=$ 3: These are objects with more than one emission line identified with SNR $>$ 5 or one  emission line identified with SNR $>$ 5 with a $|$\zspec $-$ \zphoto $|$ $< 0.2$.
\end{itemize}

\begin{figure}
\centering
\includegraphics[width=1.0\textwidth]{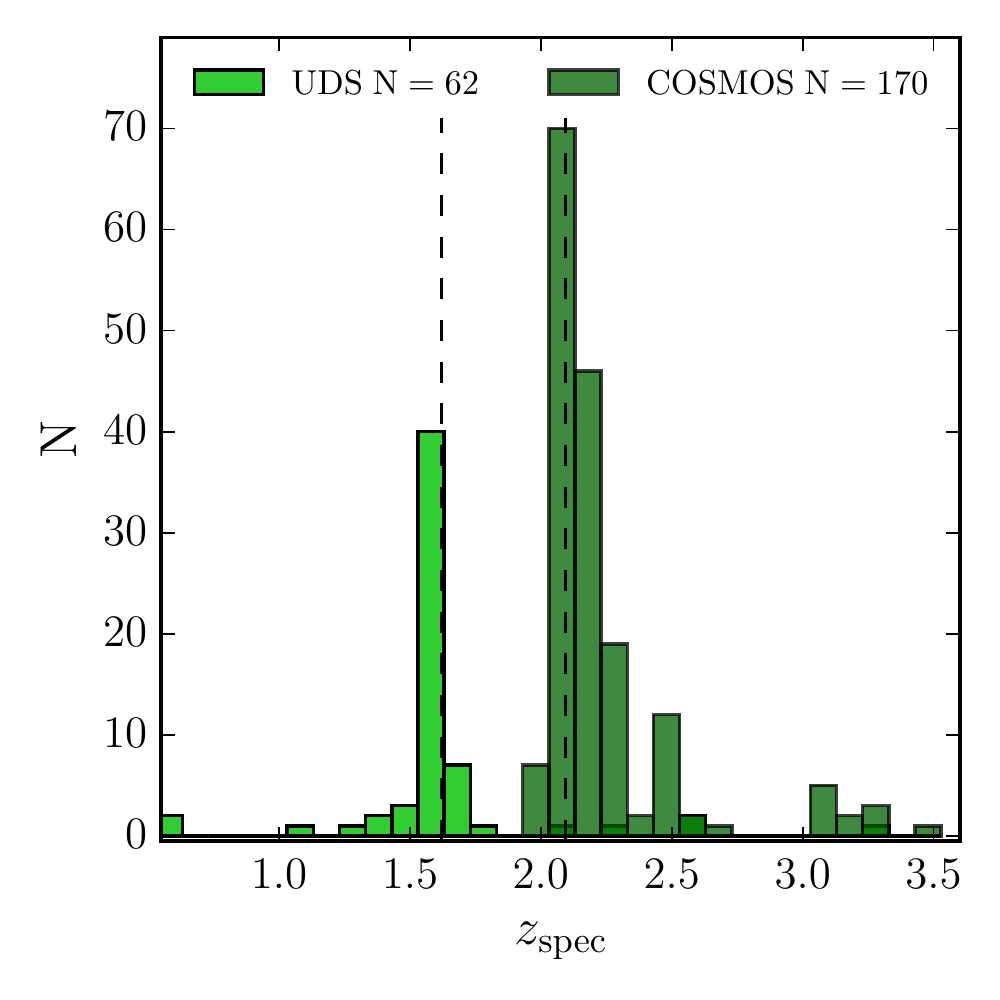}
\caption[Spectroscopic redshift distribution of the ZFIRE data release.]{Redshift distribution of the ZFIRE data release. All detected galaxies with Q$_z$=2 and Q$_z$=3 from UDS (light green) and COSMOS (dark green) are shown in the figure. The two dashed vertical lines at x=1.620 and x=2.095 shows the location of the IRC 0218 cluster \citep{Tran2015} and the COSMOS cluster \citep{Yuan2014}, respectively.}
\label{fig:zspec}
\end{figure}

The redshift distribution of all ZFIRE\ Q$_z$=2 and Q$_z$=3 detections are shown in Figure \ref{fig:zspec}. 62 galaxy redshifts were detected in the UDS field, out of which 60 have a Q$_z$ of 3 and 2 have a Q$_z$ of 2. Similarly, for the COSMOS field, there are 161  Q$_z$=3 objects and 9 Q$_z$=2 objects. 

The systematic error of the redshift measurement was estimated by comparing Q$_z$=3 objects with a SNR $>$ 10 in both H and K-bands in the COSMOS field. 
\citet{Yuan2014} showed that the agreement between the redshifts in the two bands is $\Delta z$(median) = 0.00005 with a rms of $\Delta z$(rms) = 0.00078. Therefore, the error in redshift measurement is quoted as $\Delta z$(rms) = 0.00078/$\sqrt 2$= 0.00055, which corresponds to  $\sim\mathrm{53km~s^{-1}}$ at $z=2.1$.
This is $\sim$2 times the spectral resolution of MOSFIRE, which is $\sim\mathrm{26km~s^{-1}}$ \citep{Yuan2014}. However, for the \citet{Yuan2014} analysis barycentric corrections were not applied to the redshifts and H and K masks were observed on different runs. Once individual mask redshifts were corrected for barycentric velocity, the rest-frame velocity uncertainty  decreased to $\sim\mathrm{15km~s^{-1}}$.

\begin{figure}
\centering
\includegraphics[width=1.0\textwidth]{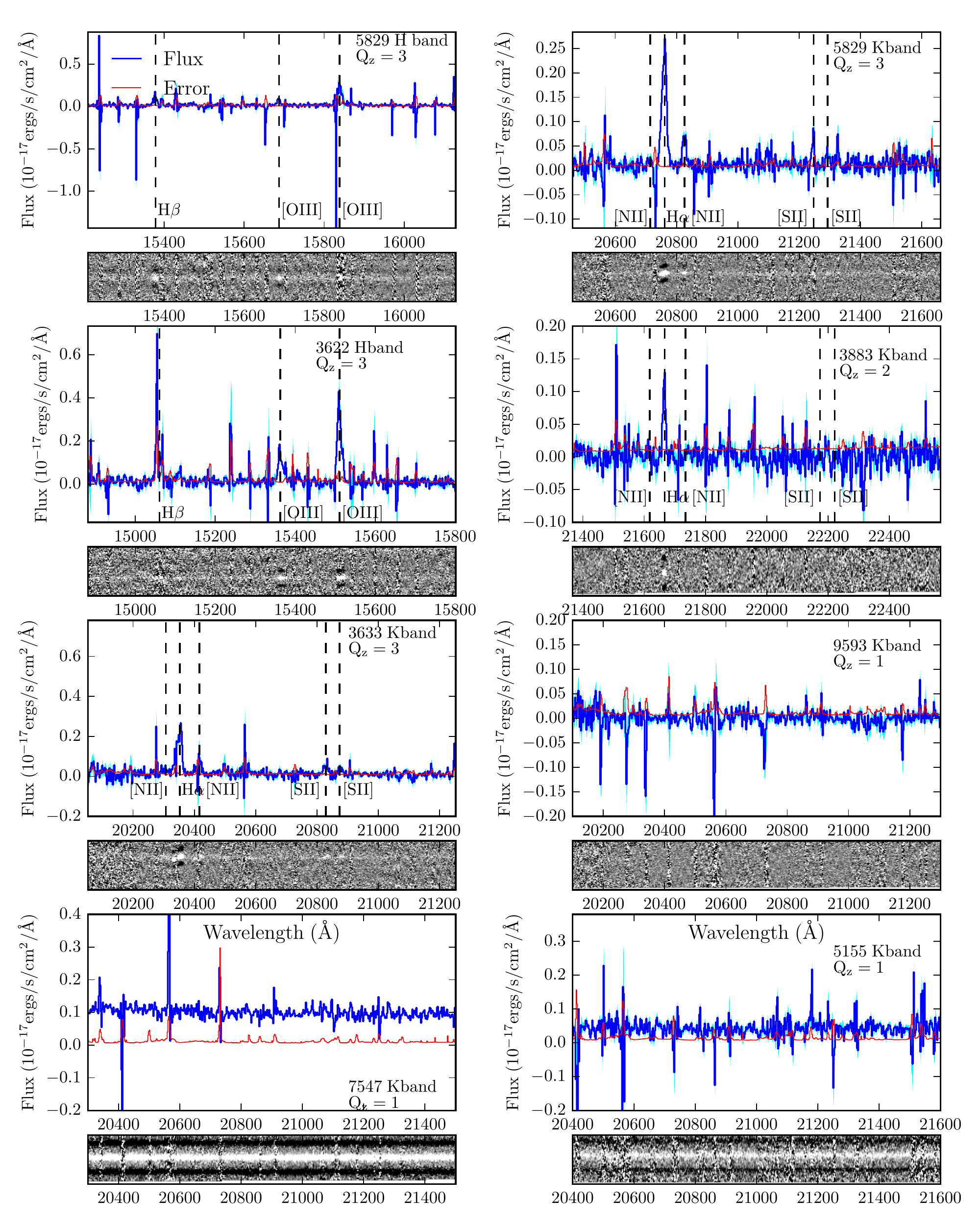}
\caption[Example MOSFIRE H and K-band spectra from the COSMOS field.]{Example MOSFIRE H and K-band spectra from the COSMOS field. 
In the 1D spectra, the flux is shown in blue and the corresponding error in red. The 1$\sigma$ scatter of the flux value parametrized by the error level is highlighted around the flux value in cyan.
Each 1D spectra are accompanied by the corresponding 2D spectra covering the same wavelength range. 
Each panel shows the name of the object, the wavelength it was observed in, and the redshift quality of the object. Vertical dashed lines show where strong optical emission lines ought to lie given the spectroscopic redshift.}
\label{fig:spectra}
\end{figure}

A few example spectra are shown in Figure \ref{fig:spectra}. Object 5829 is observed in both H and K-bands with strong emission lines detected in both instances. Object 3622 has strong H-band detections, while 3883 has only one emission line detection. Therefore, 3883 is assigned a Q$_z$ of 2. The 2D spectrum of object 3633 shows two emission line detections around \Halpha\ at different y pixel positions, which occur due to multiple objects falling within the slit. Object 9593 shows no emission line or continuum detection.
Objects 7547 and 5155 have strong continuum detections with no nebular emission lines. These galaxies were selected to be the BCGs of the D and A substructures by \citet{Yuan2014} and \citet{Spitler2012}, respectively, and have absorption line redshifts from \citet{Belli2014}.

The ZFIRE data release catalogue format is given in Table \ref{tab:catalogue}.
An overview of the data presented is provided in the table, along with the 1D spectra, which is available online at \url{zfire.swinburne.edu.au}.
Galaxy stellar mass and dust extinction values are from ZFOURGE, but for Q$_z>1$ galaxies these values are rederived using the spectroscopic redshifts with FAST.
The ZFIRE-COSMOS galaxy sample comprises  both field and cluster galaxies selected in the Ks band with an 80\% mass completeness down to $\log_{10}($\mass$)>9.30$ (Figure \ref{fig:detection_limits}). 

The survey selection for this data release was done using the ZFOURGE internal catalogues, and therefore the results presented here onwards could vary slightly from the ZFOURGE public data release.  
For the 2016 ZFOURGE public data release, the catalogue was upgraded by including pre-existing public K-band imaging for the source detection image.
This increased the amount of galaxies in the COSMOS field by \around 50\%, which was driven by the increase of fainter smaller mass galaxies. In Appendix \ref{sec:ZFOURGE comparison}, a comparison between the internal ZFOURGE catalogue and the public data release version is shown.

\begin{deluxetable}{ || l | l || }
\tabletypesize{\footnotesize}
\tablecaption{ The ZFIRE\ v1.0 data release }
\tablecomments{ This table presents an overview of the data available online.  
All galaxy properties and nebular emission line values of the galaxies targeted by ZFIRE between 2013 to 2015 are released with this paper.
\label{tab:catalogue}}
\tablecolumns{2}
\tablewidth{0pt} 
\tablewidth{0pt}
\startdata
\hline 
&   \\ 
ID												&	Unique ZFIRE\ identifier.								\\ [+1ex]
RA 												&	Right ascension (J2000)		\\ [+1ex]
DEC												&	Declination (J2000)  		\\ [+1ex]
Field											&	COSMOS or UDS			\\ [+1ex]
\Ks \tablenotemark{a}							&  \Ks\ magnitude from ZFOURGE \\ [+1ex]
$\mathrm{\sigma}$\Ks			    			&	Error in \Ks\ magnitude.    								\\ [+1ex]
\zspec					    					&	ZFIRE\ spectroscopic redshift. 							\\ [+1ex]
$\sigma$(\zspec)								&	Error in spectroscopic redshift.					\\ [+1ex]
Q$_z$											& 	ZFIRE\ redshift quality flag (see Section \ref{sec:Q flags})      \\ [+1ex]
Cluster\tablenotemark{b}						&   Cluster membership flag   		\\ [+1ex]
Mass\tablenotemark{c}							&	Stellar mass  from FAST.    		\\ [+1ex]
Av              								&	Dust extinction from  FAST.     	\\ [+1ex]
AGN\tablenotemark{d}							&	AGN flag. \\[+1ex]
\Halpha	\tablenotemark{e}						&   Emission line \Halpha\ flux from ZFIRE\ spectrum   \\ [+1ex]
$\sigma$(\Halpha)\tablenotemark{f}				&	Error in \Halpha\ flux.      						\\ [+1ex]
 \Halpha$_{\mathrm{limit}}$\tablenotemark{g}	&	1$\sigma$ upper limit for the \Halpha\ flux detection   \\ [+1ex]
 \NII \tablenotemark{e}							&	Emission line \NII\ flux (6585\AA) from ZFIRE\ spectrum   \\ [+1ex]
$\sigma$(\NII) \tablenotemark{f}				&	Error in \NII\ flux     							\\ [+1ex]
\NII$_{\mathrm{limit}}$\tablenotemark{g}		&	1$\sigma$ upper limit for the \NII\ flux detection   	\\ [+1ex]
\Hbeta \tablenotemark{e}						&	Emission line \Hbeta\ flux from ZFIRE\ spectrum 	\\ [+1ex]
$\sigma$(\Hbeta) \tablenotemark{f}				&	Error in \Hbeta\ flux 								\\ [+1ex]
\Hbeta$_{\rm limit}$\tablenotemark{g}			&	1$\sigma$ upper limit for the \Hbeta\ flux detection	\\ [+1ex]
\OIII \tablenotemark{e}							&	Emission line \OIII\ flux (5008\AA) from ZFIRE\ spectrum 	\\ [+1ex]
$\sigma$(\OIII)	\tablenotemark{f}				&	Error in \OIII\ flux 								\\ [+1ex]
\OIII$_{\rm limit}$\tablenotemark{g}			&	1$\sigma$ upper limit for the \OIII\ flux detection.  \\ 
\enddata
\tablenotetext{a}{Magnitudes are given in the AB system.}	
\tablenotetext{b}{Cluster$=$True objects are spectroscopically confirmed cluster members in either the COSMOS \citep{Yuan2014} or UDS \citep{Tran2015} fields.}
\tablenotetext{c}{Stellar mass (M$_*$) is in units of $\mathrm{log_{10}}$\msol\ as measured by FAST.}
\tablenotetext{d}{AGNs are flagged following \citet{Cowley2016} and/or \citet{Coil2015} selection criteria.}
\tablenotetext{e}{The nebular emission line fluxes (along with errors and limits) are given in units of $10^{-17}ergs/s/cm^2$.}
\tablenotetext{f}{The error of the line fluxes  are from the integration of the error spectrum within the same limits used for the emission line extraction.}
\tablenotetext{g}{Limits are $1\sigma$ upper limits from the Gaussian fits to the emission lines.}
\end{deluxetable}

\subsection{Spectroscopic Completeness}
\label{sec:completeness_zfire}

The main sample of galaxies in the COSMOS field were selected in order to include \Halpha\ emission in the MOSFIRE K-band, which corresponds to a redshift range of $1.90<$\zphoto$<2.66$. Due to multiple objects in the slits and object priorities explained in Section \ref{sec:mask_design}, there were nine galaxies  outside this redshift range. 

\begin{figure}
\centering
\includegraphics[width=1.0\textwidth]{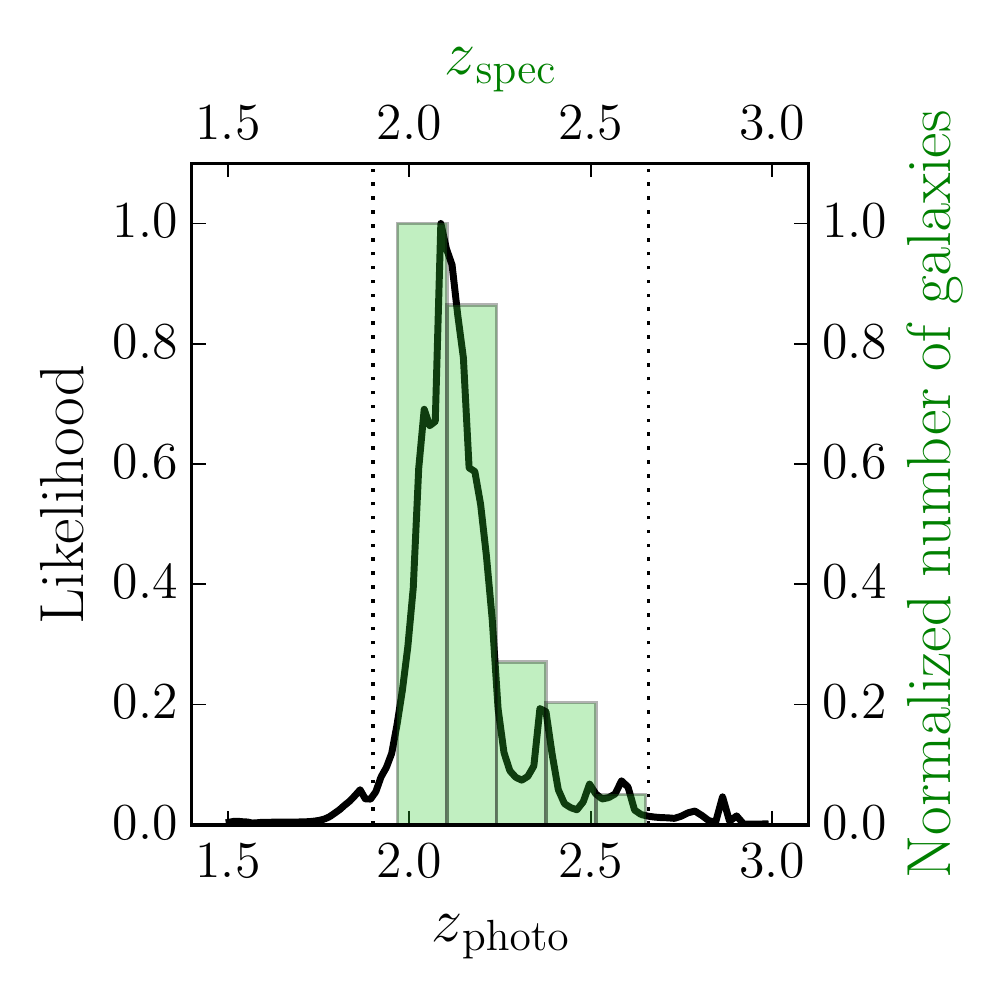}
\caption[Stacked probability distribution functions of the photometric redshifts for galaxies targeted in the ZFIRE-COSMOS field.]{Stacked probability distribution functions of the photometric redshifts for galaxies targeted in the ZFIRE-COSMOS field (shown by the black solid line). 
The black dotted lines show the redshift limits for \Halpha\ detection in the K-band. 
The wavelength coverage is corrected by the slit positions for each of the galaxies and the total probability that falls within the detectable range is calculated to be \around73\%. 
The actual \Halpha\ detection in the COSMOS field is \around73\%.
The bias toward z=2.1 is due to the object priorities weighting heavily towards the cluster galaxies. 
The green histogram shows the distribution of \zspec\ values for galaxies with \Halpha\ detections in K-band in the COSMOS field. 
}
\label{fig:completeness}
\end{figure}

We assess completeness against an expectation computed using the photometric redshift likelihood functions ($P(z)$) from EAZY, i.e. the  expected number of galaxies with \Halpha\ within the bandpass in the ZFIRE-COSMOS sample, taking account of the slightly different wavelength coverage of each slit. 
There were 203 galaxies targeted in the K-band. Of the galaxies, 10 had spectroscopic redshifts that were outside the redshift range of interest ($1.90<$\zspec$<2.66$).
The remaining 193 $P(z)$s of the detected and non-detected galaxies were stacked. 
Figure \ref{fig:completeness} shows the average $P(z)$ of the stacked 193 galaxies. 
If the \Halpha\ emission line falls on a sky line, the emission line may not be detected. Therefore, in the $P(z)$ of each of the galaxies' sky line regions parametrized by the MOSFIRE K-band spectral resolution was masked out ($\pm$5.5\AA).  
We then calculate the area of the $P(z)$ that falls within detectable limits in K-band of the object depending on the exact wavelength range of each slit.  
Since each $P(z)$ is normalized to 1, this area gives the probability of an \Halpha\ detection in K-band for a given galaxy. 
The probability to detect all 193 galaxies is calculated to be \around73\%. 
141 galaxies are detected with \Halpha\ SNR $>5$ which is a \around73\% detection rate. 
As seen by the overlaid histogram in Figure \ref{fig:completeness}, the detected redshift distribution of the ZFIRE-COSMOS sample is similar to the expected redshift distribution from $P(z)$.

\begin{figure}
\centering
\includegraphics[width=0.49\textwidth]{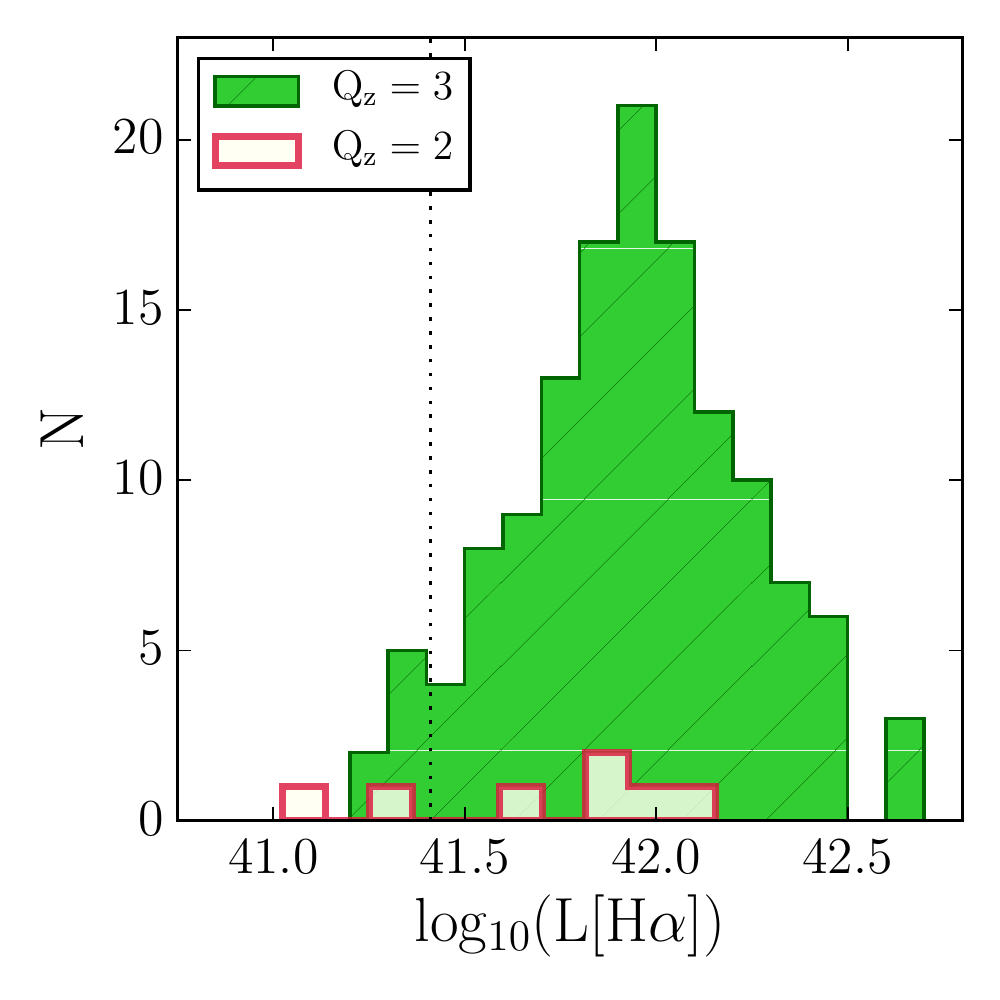}
\includegraphics[width=0.49\textwidth]{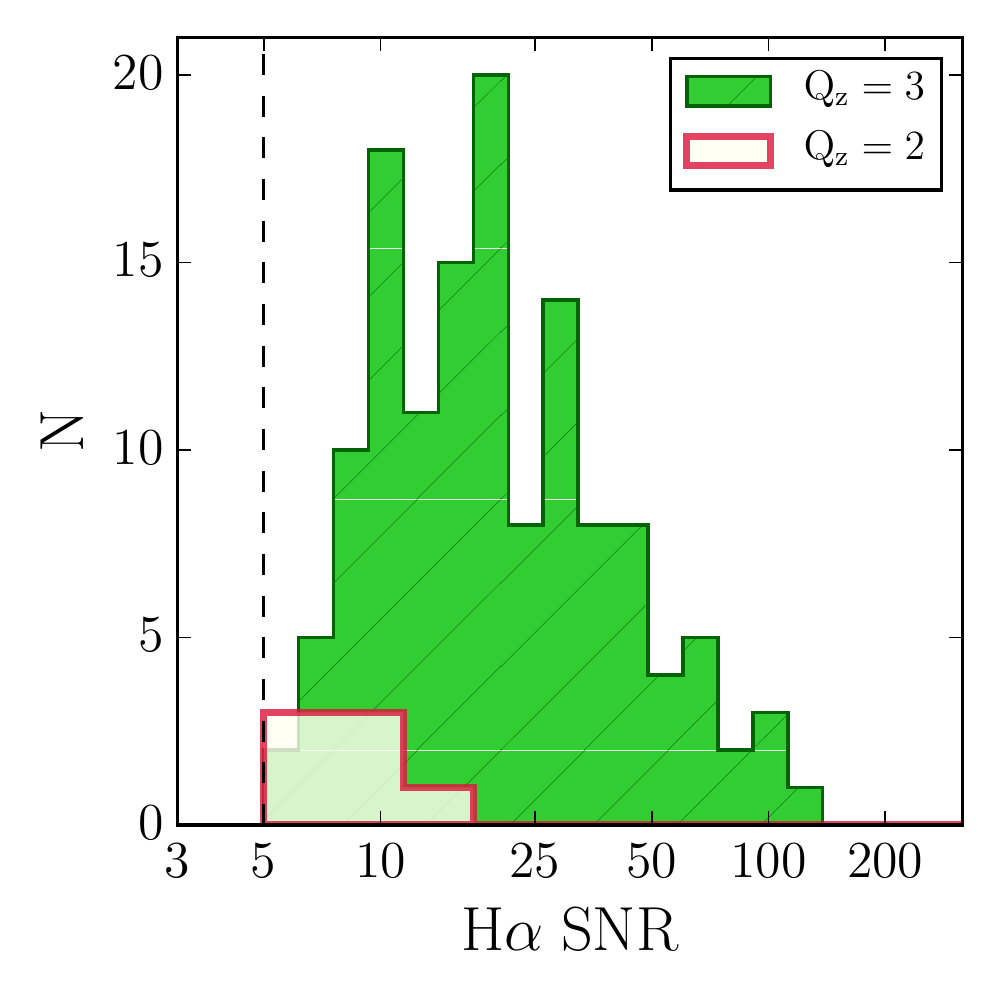}
\includegraphics[width=0.49\textwidth]{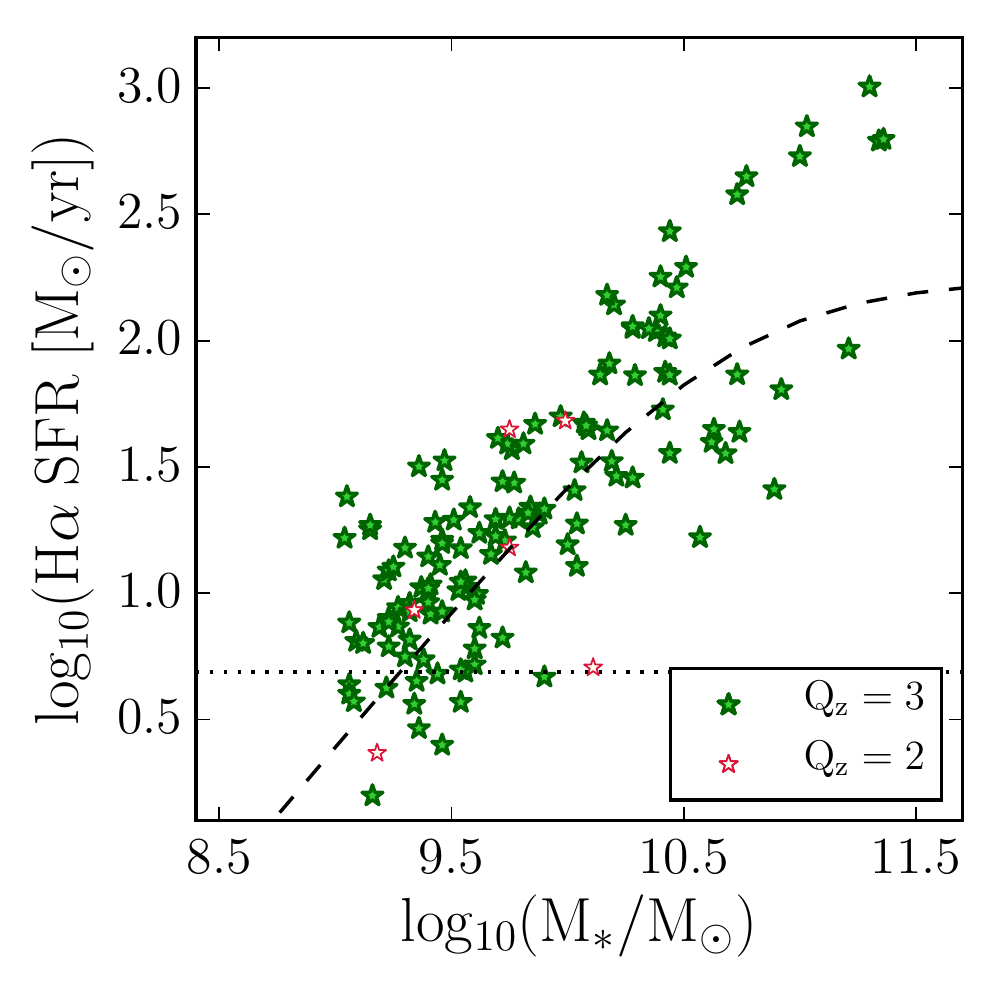}
\caption[ZFIRE K band \Halpha\ detection properties.]{{\bf Top Left:} the distribution of \Halpha\ luminosity of all ZFIRE-COSMOS galaxies in log space. The green histogram (with diagonal lines) is for galaxies with a quality flag of 3, while the ivory histogram is for galaxies with a quality flag of 2.
The vertical dotted line is the \Halpha\ SFR for a typical \Halpha\ SNR of \around 5 at z=2.1. 
{\bf Top Right:} similar to the left figure, but the distribution of \Halpha\ SNR of all ZFIRE-COSMOS detections are shown. The dashed vertical line is SNR = 5, which is the \Halpha\ detection threshold for ZFIRE. 
{\bf Bottom:} the \Halpha\ SFR vs. stellar mass distributions for the objects shown in the left histograms. The stellar masses and dust extinction values are derived from FAST. The  dashed line is the star-forming main sequence from \citet{Tomczak2014}. The horizontal dotted line is the \Halpha\ SFR for a typical \Halpha\ SNR of \around 5 at z=2.1.
}
\label{fig:Halpha}
\end{figure}

Figure \ref{fig:Halpha} shows the \Halpha\ luminosity (left) and SNR distribution (middle) of the ZFIRE-COSMOS galaxies with \Halpha\ detections. 
The detection threshold is set to SNR $\geq$ 5 which is shown by the vertical dashed line in the centre panel. There are 134 galaxies in the Q$_{z}$=3 sample, 7 in the Q$_{z}$=2 sample. 

The \Halpha\ luminosity in Figure \ref{fig:Halpha} (left panel) is peaked $\sim10^{42}$ergs/s. From the SNR distribution it is evident that the majority of galaxies detected have a \Halpha\ SNR $>10$, with the histogram peaking \around SNR of 20.
Normally astronomical samples are dominated by low SNR detections near the limit. 
It is unlikely that objects with SNR$<$20 are missed. Our  interpretation of  this distribution is that because  the sample is mass-selected the drop off of low flux \Halpha\ objects is because the region below the stellar mass-SFR main sequence \citep{Tomczak2014} at $z\sim 2$ is probed. 
This is shown in Figure \ref{fig:Halpha} where we make a simple conversion of \Halpha\ to SFR assuming the \citet{Kennicutt1998}  conversion and stellar extinction values from FAST
which we convert to nebula extinction using the  \citet{Calzetti2000} prescription  with  $R_V=4.05$. 
It is indeed evident that the ZFIRE-COSMOS sample limits do probe the limits of the galaxies in the star-forming main sequence at $z\sim2$ with a $3\sigma$ \Halpha\ SFR detection threshold at $\sim4$ \msol/yr. A more detailed analysis of the \Halpha\ main sequence will be presented in a future paper (K. Tran et al. in preparation). 


\subsection{Magnitude and Stellar Mass Detection Limits}

\begin{figure}
\centering
\includegraphics[width=0.8\textwidth]{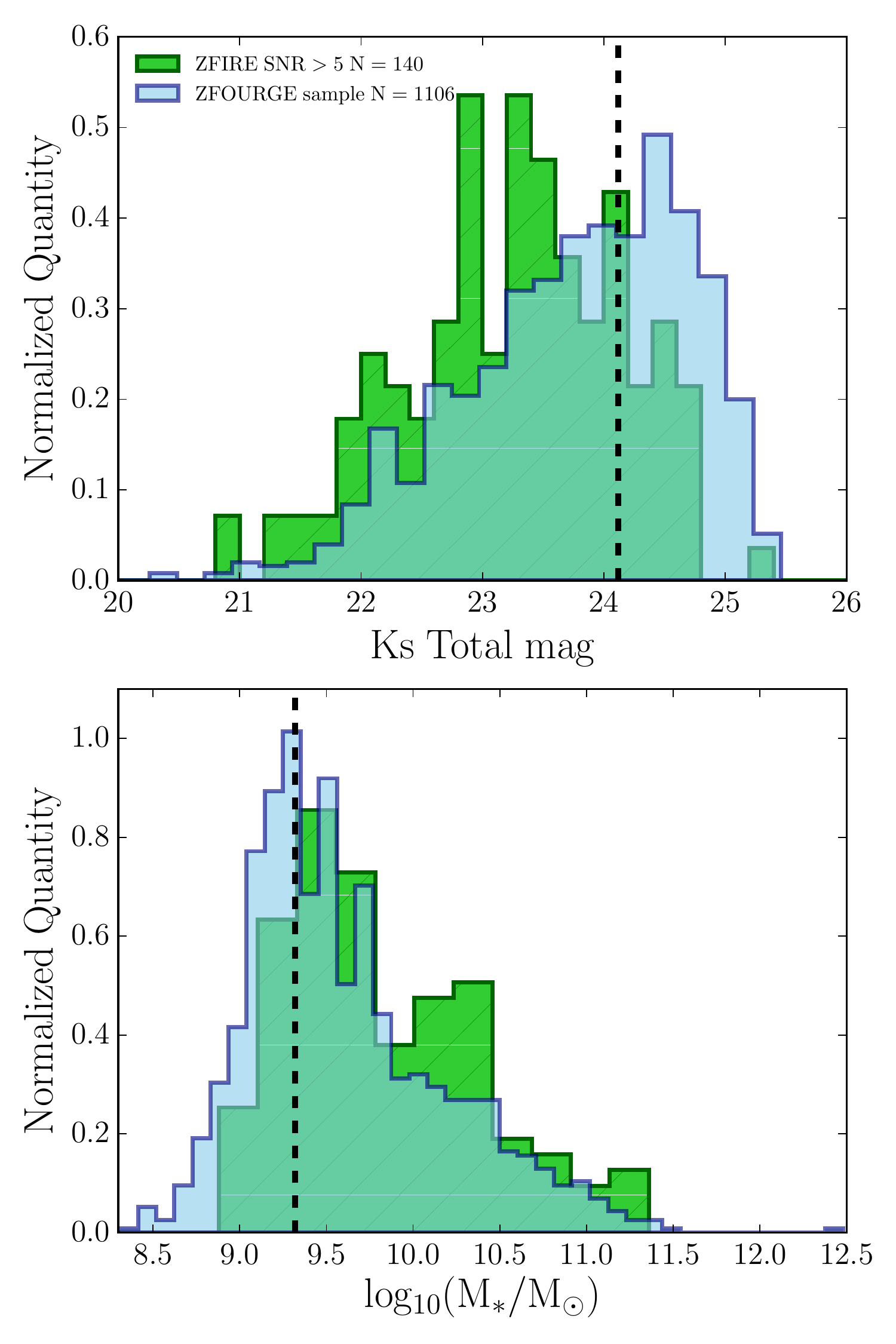}
\caption[The Ks magnitude and mass distribution of the ZFIRE K band sample compared with ZFOURGE.]{The Ks magnitude and mass distribution of the $1.90<z<2.66$ galaxies from ZFOURGE (cyan) overlaid with the ZFIRE (green) detected sample for the COSMOS field. The ZFOURGE distribution is derived using the photometric redshifts and spectroscopic redshifts (when available). The ZFIRE histogram uses the spectroscopic redshifts. 
The histograms are normalized for area. UVJ quiescent galaxies (only 1 in ZFIRE) are removed from both the samples. 
{\bf Top:} Ks magnitude distribution. The black dashed line (Ks=24.11) is the limit in which 80\% of the detected sample lies below. 
{\bf Bottom:} stellar mass distribution of the galaxies in log space as a fraction of solar mass.  Masses are calculated using FAST and spectroscopic redshifts are used where available. The black dashed line ($\mathrm{Log}_{10}($\mass$)=9.3$) is the limit down to where the detected sample is 80\% mass complete.  
}
\label{fig:detection_limits}
\end{figure}

The ZFIRE-COSMOS detection limits in Ks magnitude and stellar mass are estimated using ZFOURGE photometry. 
Out of 141 objects with \Halpha\ detections (Q$_{z}$=2 or Q$_{z}$=3) and $1.90<$\zspec$<2.66$, galaxies identified as UVJ quiescent are removed since the spectroscopic sample does not significantly sample these (see Section \ref{sec:UVJ}). The remaining sample comprises 140 UVJ blue (low dust attenuation) and red (high dust attenuation) star-forming galaxies. 
Similarly, galaxies from the ZFOURGE survey are selected with redshifts between $1.90<$\zspec$<2.66$ and all UVJ quiescent galaxies are removed. The Ks magnitude and the stellar mass distributions of the remaining 1106 ZFOURGE galaxies with the selected ZFIRE sample are compared in Figure \ref{fig:detection_limits}.  

The top panel of Figure \ref{fig:detection_limits} demonstrates that the \Halpha\ detected galaxies reach Ks$>$24. 
80\% of the detected ZFIRE-COSMOS galaxies have Ks$\leq$24.11. The ZFOURGE input sample reaches deeper to 
Ks$\leq$24.62 (80\%-ile). The photometric detection completeness limit of ZFOURGE is discussed in detail in Straatman
et al. (2014), but we note that at $K=24.62$, 97\% of objects are detected. It is important to understand if the distribution in Ks of the spectroscopic sample is biassed relative to the photometric sample. A two-sample K-S test for Ks$\leq$24.1 is performed to find a $p$ value of 0.03 suggesting that there is no significant bias between the samples. 

Similarly, the mass distribution of the \Halpha\ detected sample is investigated in the bottom panel of Figure  \ref{fig:detection_limits}. Galaxies are detected down to $\log_{10}($\mass$)\sim9$.
80\% of the \Halpha\ detected galaxies have a stellar masses $\log_{10}($\mass$)>9.3$. A K-S test on the two distributions for galaxies $\log_{10}($\mass$)>9.3$ gives a $p$ value of 0.30 and therefore, similar to the Ks magnitude distributions, the spectroscopic sample shows no bias in stellar mass compared to the ZFOURGE photometric sample. 

This shows that the ZFIRE-COSMOS detected sample of UVJ star-forming galaxies has a similar distribution in magnitude and stellar mass as the ZFOURGE distributions except at the very extreme ends. 
Removing UVJ dusty galaxies from the star-forming sample does not significantly change this conclusion. 

A final test is to evaluate the photometric magnitude at which continuum emission in the spectra can be typically detected. To estimate this, a constant continuum level is fit to blank sky regions across the whole $K$-band spectral range. This shows that  the $2\sigma$ spectroscopic continuum detection limit for the ZFIRE-COSMOS sample is Ks$\simeq 24.1$ ($0.05 \times 10^{-17} \mathrm{erg/s/cm^2/\AA}$). More detailed work on this is presented in Chapter \ref{chap:imf_observations}.

\subsection{Rest frame UVJ colours}
\label{sec:UVJ}

The rest-frame UVJ colours are used to assess the stellar populations of the detected galaxies. 
In rest frame U$-$V and V$-$J colour space, star-forming galaxies and quenched galaxies show strong bimodal dependence \citep{Williams2009}. Old quiescent stellar populations with strong 4000\AA\ and/or Balmer breaks show redder U$-$V colours and bluer V$-$J colours, while effects from dust contribute to redder V$-$J colours. 

\begin{figure}
\centering
\includegraphics[width=1.0\textwidth]{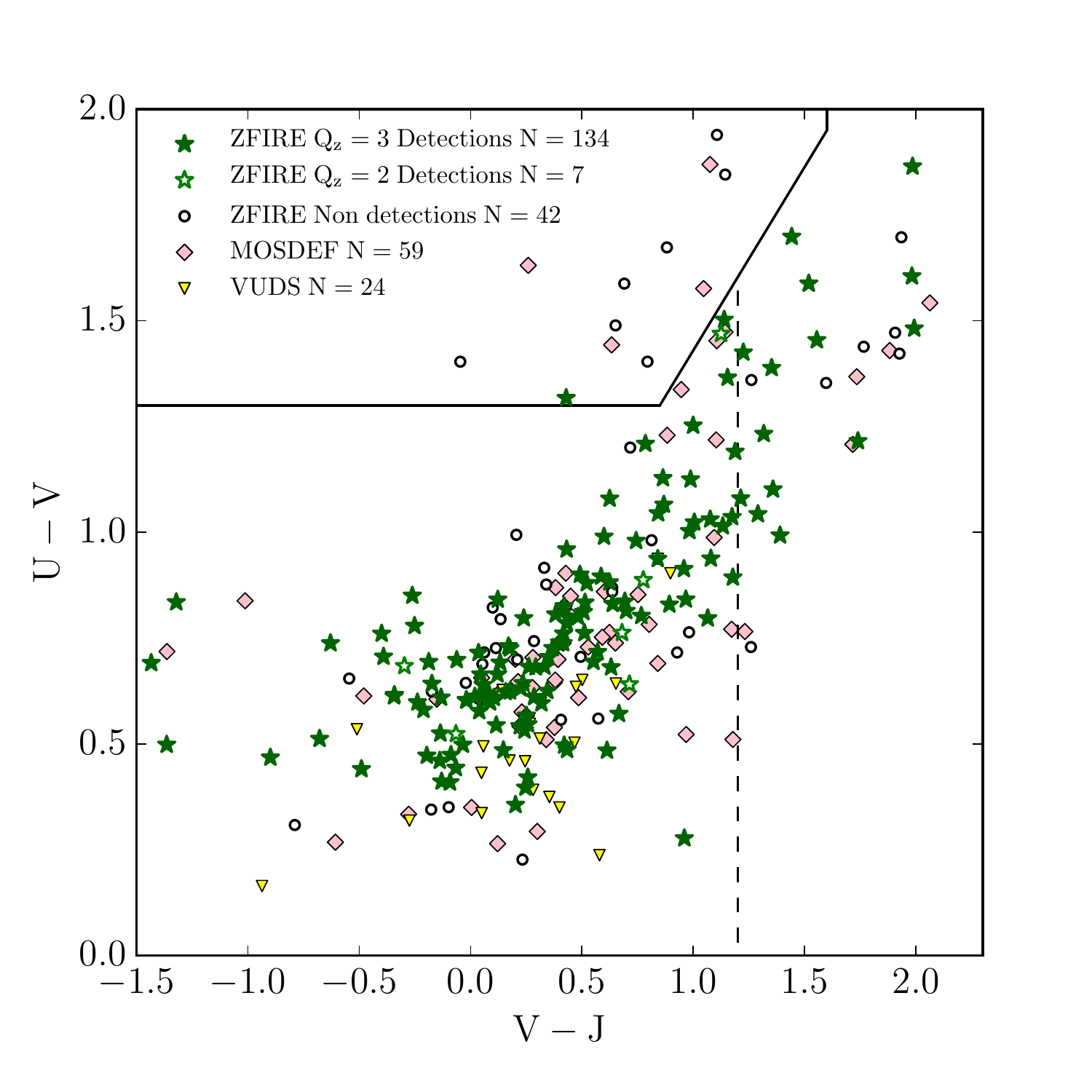}
\caption[The rest frame UVJ diagram of the ZFIRE-COSMOS K band detected sample.]{The rest frame UVJ diagram of the ZFIRE-COSMOS sample with redshifts $1.90<z<2.66$. 
Quiescent, star-forming, and dusty star-forming galaxies are selected using \citet{Spitler2014} criteria. 
The green stars are ZFIRE detections (filled$\rightarrow$Q$_z=3$, empty$\rightarrow$Q$_z=2$) and the black circles  are the non-detections. 
Pink diamonds and yellow triangles are MOSDEF and VUDS detected galaxies respectively, in the same redshift bin with matched ZFOURGE counterparts.  
Rest frame colours are derived using spectroscopic redshifts where available. 
}
\label{fig:UVJ}
\end{figure}

Figure \ref{fig:UVJ} shows the UVJ selection of the COSMOS sample, which lies in the redshift range between $1.99<$\zspec$<2.66$. 
The selection criteria are adopted from \citet{Spitler2014} and are as follows.
Quiescent galaxies are selected by (U$-$V)$>$1.3 , (V$-$J)$<$1.6, (U$-$V) $>$ 0.867$\times$(V$-$J)$+$0.563. 
Galaxies which lie below this limits are considered to be star-forming. 
These star-forming galaxies are further subdivided into two groups depending on their dust content. Red galaxies with (V$-$J)$>$1.2 are selected to be dusty star-forming galaxies, which correspond to A$_{v}\gtrsim$1.6. Blue galaxies with (V$-$J)$<$1.2 are considered to be relatively unobscured. MOSFIRE detected galaxies are shown as green stars while the non-detections (selected using \zphoto\ values) are shown as black filled circles. 

The total sampled non-detections are \around23\% for this redshift bin. 
\around82\% of the blue star-forming galaxies and \around70\% of the dusty star-forming galaxies were detected, but only 1 quiescent galaxy was detected out of the potential 12 candidates in this redshift bin. 
Galaxies in the red sequence are expected to be quenched with little or no star formation and hence without any strong \Halpha\ features; therefore the low detection rate of the quiescent population is expected. \citet{Belli2014} has shown that \around8 hours of exposure time is needed to get detections of continua of quiescent galaxies with J\around22 using MOSFIRE. 
The prominent absorption features occur in the H-band at $z\sim2$. ZFIRE currently does not reach such integration times per object in any of the observed bands and none of the quiescent galaxies show strong continuum detections.  We note that this is a bias of the ZFIRE survey, which may have implications on the identification of weak star-forming and quiescent cluster members by \citet{Yuan2014}.

For comparison MOSDEF and VUDS detections in the COSMOS field with matched ZFOURGE candidates are overlaid in Figure \ref{fig:UVJ}. All rest-frame UVJ colours for the spectroscopic samples are derived from photometry using the spectroscopic redshifts.  The MOSDEF sample, which is mainly H-band selected, 
primarily includes star-forming galaxies independently of the dust obscuration level. 
VUDS survey galaxies are biased toward blue star-forming galaxies, which is expected because it is an optical spectroscopic survey. This explains why their spectroscopic sample does not include any rest-frame UVJ selected dusty star-forming or quiescent galaxies.

\subsection{Spatial distribution}
\label{sec:spatial}

\begin{figure}
\includegraphics[width=1.15\textwidth]{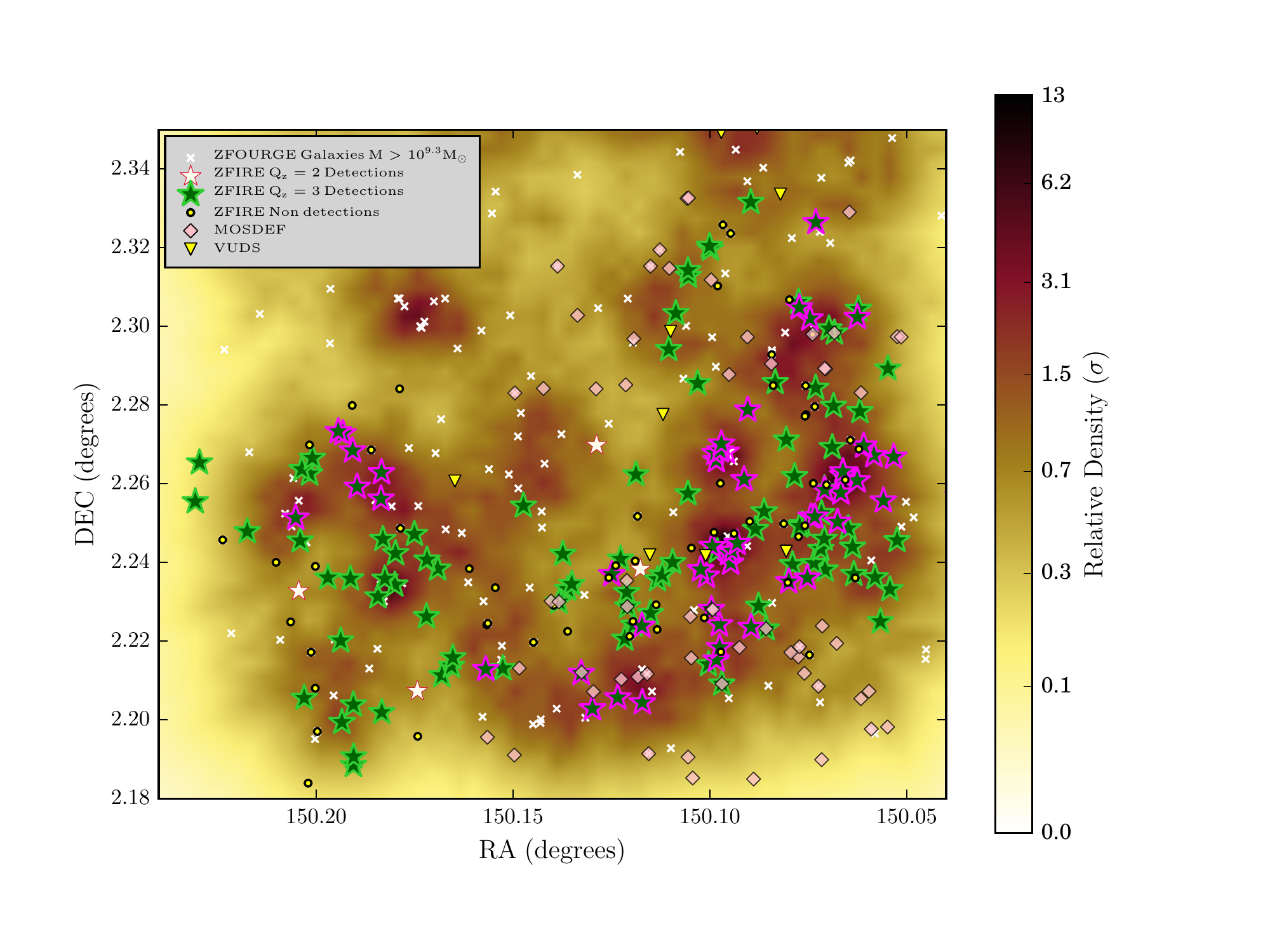}
\caption[Spatial distribution of the ZFIRE-COSMOS sample.]{Spatial distribution of the ZFIRE-COSMOS sample. 
Galaxies that fall within $2.0<z<2.2$ are used to produce the underlying seventh nearest neighbour density map. The units are in standard deviations above the mean of redshift bins (see Section~\ref{sec:spatial}).The white crosses are the ZFOURGE galaxies with M$>$10$^{9.34}$\msol, which is the 80\% mass completeness of the ZFIRE\ detections. 
Spectroscopically detected galaxies with redshifts between $1.90<z_\mathrm{spec}<2.66$ have been overlaid on this plot.
The stars are ZFIRE-COSMOS detections (green filled$\rightarrow$Q$_z=3$, white filled $\rightarrow$Q$_z=2$) and the black circles are the non-detections. Galaxies outlined in bright pink are the confirmed cluster members by \citet{Yuan2014}. 
The light pink filled diamonds are detections from the MOSDEF survey. Yellow triangles are from the VUDS survey.  
}
\label{fig:detection_map}
\end{figure}

The COSMOS sample is primarily selected from a cluster field. The spatial distribution of the  field is shown in Figure \ref{fig:detection_map}. 
(The ZFOURGE photometric redshifts are replaced with our spectroscopic values where available.)  A redshift cut between $2.0<z<2.2$ is used to select galaxies in the cluster redshift range. 
Using necessary ZFOURGE catalogue quality cuts there are  378 galaxies within this redshift window.
Following \citet{Spitler2012}, these galaxies are used to produce a seventh nearest neighbour density map.
Similar density distributions are calculated to the redshift window immediately above and below $2.0<z<2.2$. These neighbouring distributions are used to  calculate the mean and the standard deviation of the densities. The density map is plotted in units of standard deviations above the mean of the densities of the neighbouring bins similar to \citet{Spitler2012}. Similar density maps were also made by \citet{Allen2015}.

The figure shows that ZFIRE has achieved a thorough sampling of the underlying density
structure at $z\sim2$ in the COSMOS field.  Between
$1.90<z_\mathrm{spec}<2.66$, in the COSMOS field the sky density of
ZFIRE is 1.47 galaxies/arcmin$^2$. For MOSDEF and VUDS it is 1.06
galaxies/arcmin$^2$ and 0.26 galaxies/arcmin$^2$, respectively.  A
detailed spectroscopic analysis of the cluster from
ZFIRE redshifts has been published in  \citet{Yuan2014}.

\begin{figure}
\centering
\includegraphics[width=0.49\textwidth]{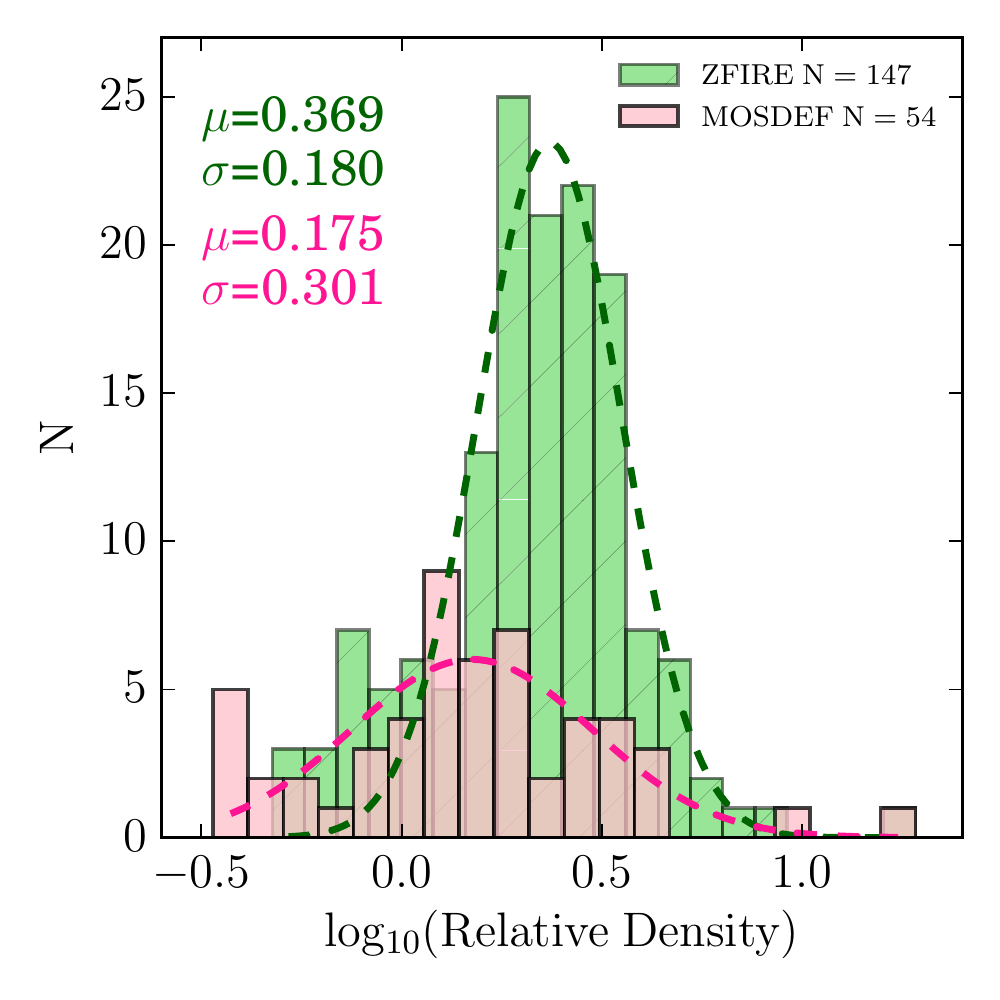}
\includegraphics[width=0.49\textwidth]{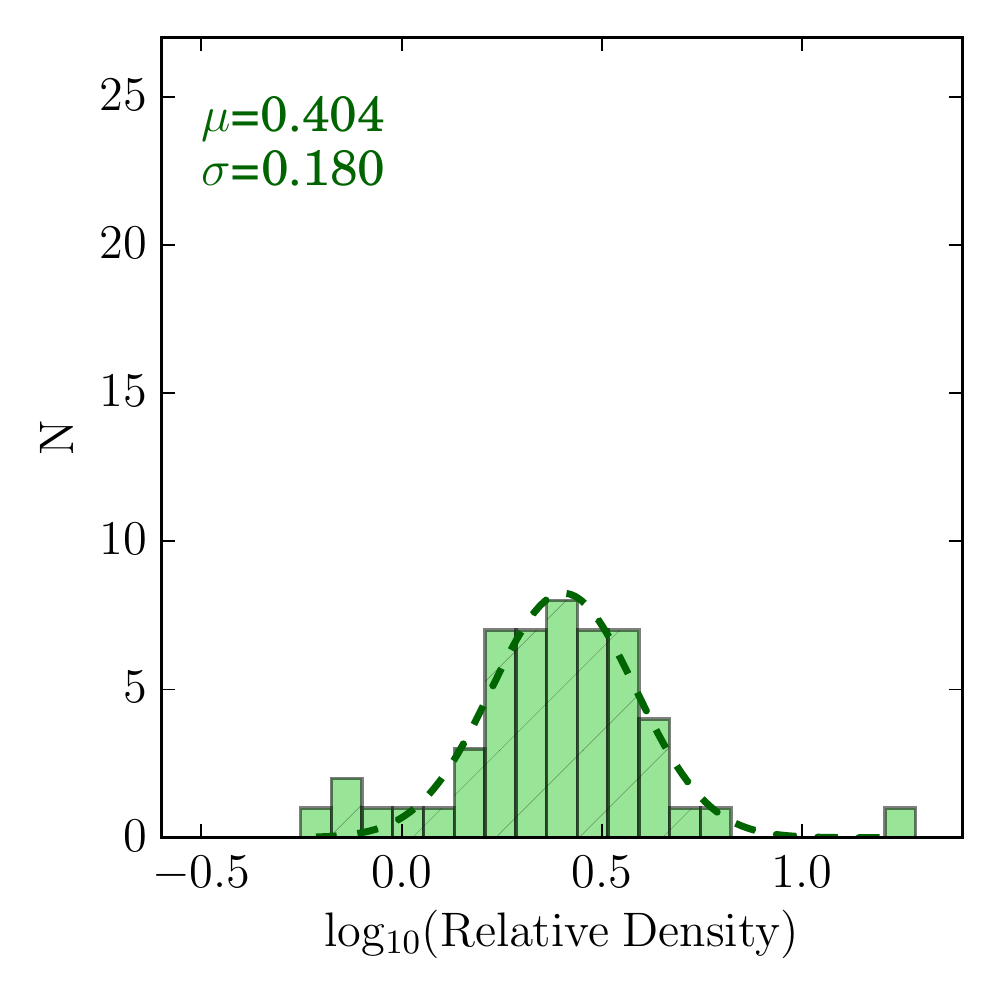}
\caption[The relative galaxy density distribution of the ZFIRE-COSMOS sample.]{{\bf Left:} the relative galaxy density distribution of the galaxies with confident redshift  detections in the COSMOS field. Galaxies with $1.90<z_\mathrm{spec}<2.66$  in ZFIRE (green) and MOSDEF (pink) surveys are shown in the histogram. The density calculated is similar to what is shown in Figure \ref{fig:detection_map}.
Gaussian fits have been performed to both the samples. The density of the ZFIRE sample is distributed in logarithmic space around $\mu=0.369$ and $\sigma=0.180$, which is shown by the green dashed line. Similarly, the fit for the MOSDEF sample shown by the pink dashed line has $\mu=0.175$ and $\sigma=0.301$. Compared to MOSDEF, ZFIRE probes galaxies in richer environments. 
{\bf Right:} similar to the left plot but only the confirmed cluster members by \citet{Yuan2014} are shown in the histogram. The normalisation is lower because  the cluster identification of \citet{Yuan2014} came from a smaller earlier sample. (MOSDEF has only detected two cluster members and hence only the ZFIRE sample is shown in the figure.) The Gaussian best-fitting parameters shown by the green dashed line has $\mu=0.404$ and $\sigma=0.180$. 
}
\label{fig:density_hist}
\end{figure}

Figure \ref{fig:density_hist} shows the relative density distribution of the $1.90<$\zspec$<2.66$ galaxies. The MOSDEF sample is overlaid on the left panel and a Gaussian best-fitting functions are fit for both ZFIRE (cluster and field) and MOSDEF samples. It is evident from the distributions, that in general ZFIRE galaxies are primarily observed in significantly higher density environments (as defined by the Spitler et al. metric) compared to MOSDEF. 
Because of the explicit targeting of `cluster candidate' fields, this is expected. 
In the right panel, the density distribution of the confirmed
cluster members of \citet{Yuan2014} is shown.

\section{Comparing ZFIRE Spectroscopic Redshifts to the Literature}
\label{sec:photometric_redshifts}

The new spectroscopic sample, which is in well-studied deep fields, is ideal to test the redshift accuracy of some of the most important photometric redshift surveys, including the ZFOURGE survey from which it is selected.

\subsection{Photometric Redshifts from ZFOURGE and UKIDSS}

\begin{figure}
\centering
\includegraphics[width=1.0\textwidth]{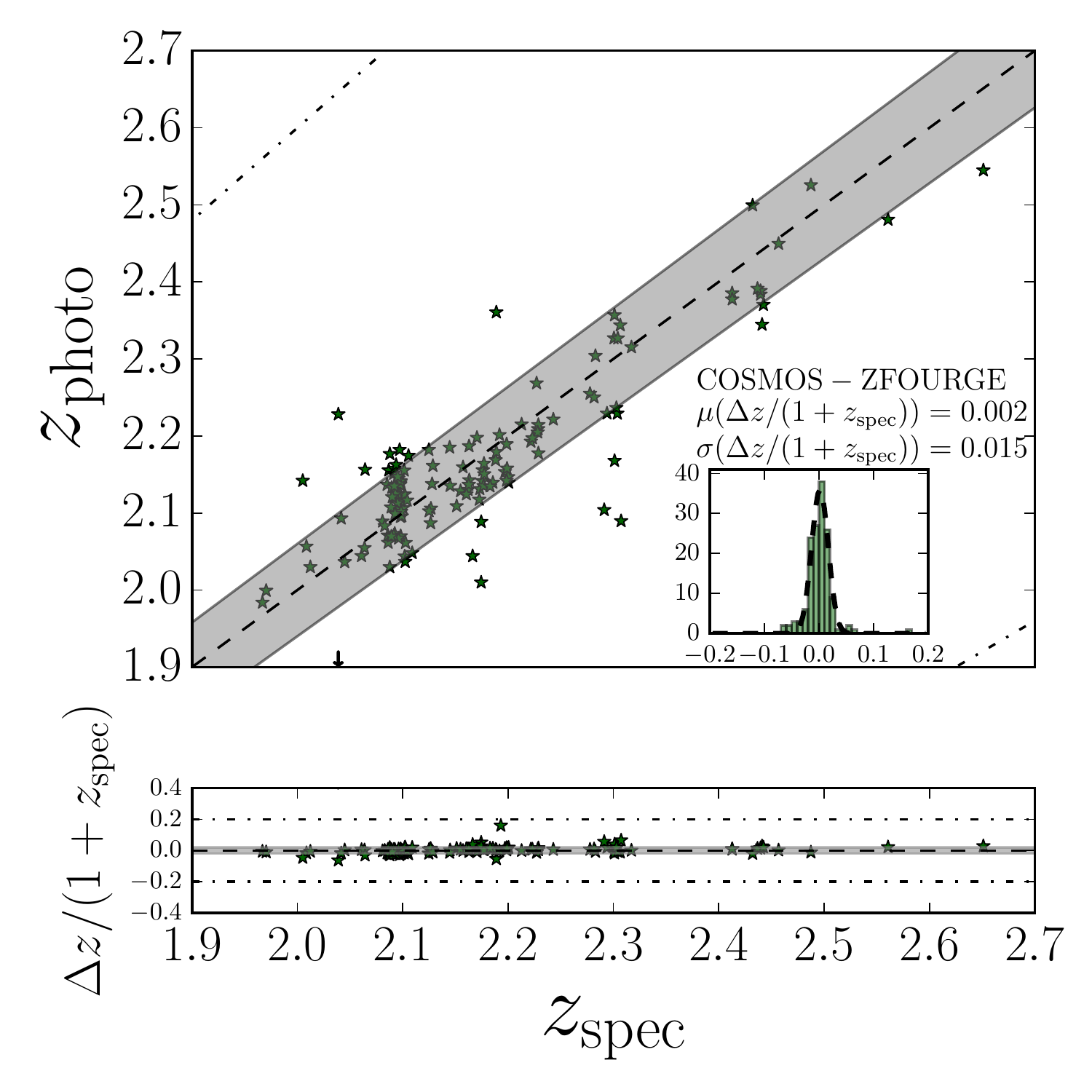}
\caption[Comparison between the photometrically derived redshifts from ZFOURGE with the ZFIRE spectroscopic redshifts.]{Comparison between the photometrically derived redshifts from ZFOURGE  with the ZFIRE Q$_{z}$=3 spectroscopic redshifts. 
{\bf Top:} $z_{\mathrm{photo}}$ vs. $z_{\mathrm{spec}}$ for the COSMOS field. $z_{\mathrm{photo}}$ values are from ZFOURGE v3.1 catalogue. 
The black dashed line is the one-to-one line. The grey shaded region represents the 2\% error level expected by the photometric redshifts (Straatman et al., in press). 
The dashed dotted line  shows the $\mid$$\Delta z$/(1+$z_{\mathrm{spec}}$)$\mid$ $>$ 0.2  drastic outlier cutoff.
The inset histogram shows the histogram of the  $\Delta z$/(1+$z_{\mathrm{spec}}$) values and Gaussian fits as described in the text.
Only galaxies with $1.90<z_{\mathrm{spec}}<2.70$ are shown in the figure.  
{\bf Bottom:} similarly for the residual  $\Delta z / (1+z_{\mathrm{spec}})$ between photometric and spectroscopic redshifts plotted against the spectroscopic redshift. 
}
\label{fig:specz_photoz_cosmos}
\end{figure}

The comparison of photometric redshifts and the spectroscopic redshifts for the ZFIRE-COSMOS sample is shown by the left panel of Figure \ref{fig:specz_photoz_cosmos}. The photometric redshifts of the v3.1 ZFOURGE catalogue are used for this purpose because they represent the best calibration and photometric-redshift performance of the imaging. 
For the 42 detected secondary objects in the slits, 25 galaxies are identified with Ks selected ZFOURGE candidates.
Deep HST F160W band selected catalogues from ZFOURGE show probable candidates for eight these galaxies. 
Five galaxies cannot be confidently identified. HST imaging shows unresolved blends for four of these galaxies, which are listed as single objects in ZFOURGE. 
Only galaxies uniquely identified in ZFOURGE are shown in the figure.

\citet{Straatman2016} has determined that photometric redshifts are accurate to $<$2\% based on previous spectroscopic redshifts.  
Results from ZFIRE\ agree within this estimate. 
This error level is shown as a grey shaded region in Figure \ref{fig:specz_photoz_cosmos}. Defining $\Delta z=\mathrm{z_{spec}-z_{photo}}$ (which will be used throughout this paper)
galaxies with $|\Delta z$/(1+\zspec)$|>$ 0.2 are considered to be ``drastic outliers''. There is one  drastic outlier in the Q$_{z}$=3 sample.
The advantage of medium-band NIR imaging relies on probing the D4000 spectral feature at $z>1.6$ by the J1, J2, and J3 filters, which span  \around 1--1.3\micron. 
Drastic outliers may arise due to blue star-forming galaxies having power-law-like SEDs, which do not have a D4000 breaks \citep{Bergh1963}, leading to uncertain photometric redshifts at $z\sim2$ and also from confusion between Balmer and Lyman breaks. Furthermore, blending of multiple sources in ground based imaging can also lead to drastic outliers. 

The inset in Figure \ref{fig:specz_photoz_cosmos} is a histogram that shows the residual for the Q$_{z}$=3 sample. 
A Gaussian best-fitting is performed for these galaxies (excluding  drastic outliers). The $\sigma$ of the Gaussian fit is considered to be the the accuracy of the photometric redshift estimates for a typical galaxy. The Q$_{z}$=3 sample is bootstrapped 100 times with replacement and the \NMAD\ scatter is calculated, which is defined as the error on  $\sigma$. 
The photometric redshift accuracy of the ZFOURGE-COSMOS sample is $1.5\pm0.2\%$ which is very high. 
The bright Ks $<23$ Q$_{z}$=3 galaxies show better redshift accuracy, but are within error limits of the redshift accuracy of the total sample. 
Furthermore, the Q$_{z}$=3 blue and red star-forming galaxies (as shown by Figure \ref{fig:UVJ}) also show similar redshift accuracy within error limits. 
The Q$_{z}$=2 ZFOURGE-COSMOS sample comprises 8 galaxies with a redshift accuracy of 14$\pm$12\%.

\begin{figure}
\centering
\includegraphics[width=1.0\textwidth]{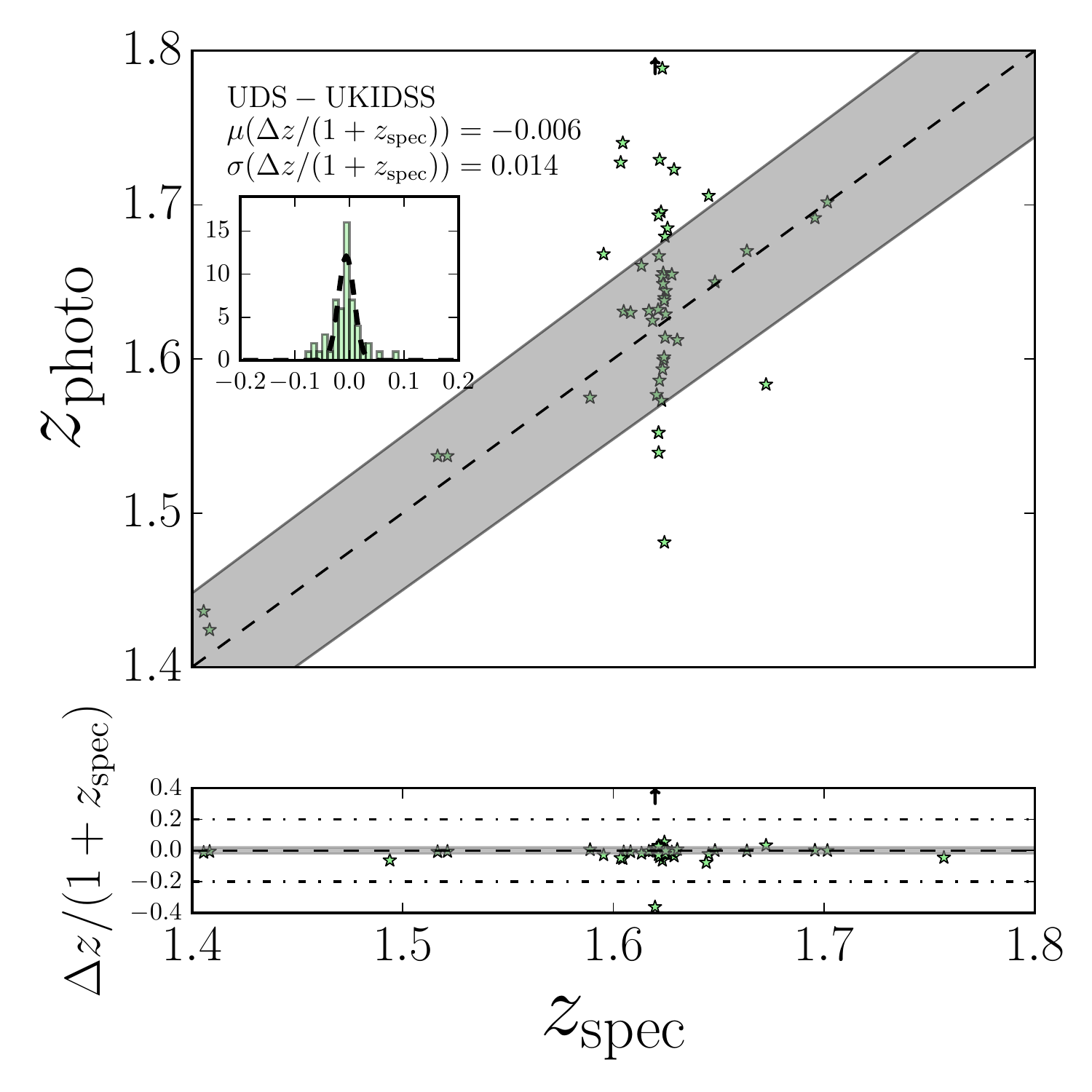}
\caption{ Similar to \ref{fig:specz_photoz_cosmos} but for the UDS field. $z_{\mathrm{photo}}$ values are from UKIDSS. 
}
\label{fig:specz_photoz_uds}
\end{figure}

In Figure \ref{fig:specz_photoz_uds}, a similar redshift analysis is performed to investigate the accuracy of the UKIDSS photometric redshift values with the ZFIRE-UDS spectroscopic sample. For the Q$_{z}$=3 objects, there are four drastic outliers (which give a rate of $\sim7\%$) and the accuracy is calculated to be 1.4$\pm$0.8\%. There are 12 Q$_{z}$=2 objects with one drastic outlier (which gives a rate of $\sim14\%$) and an accuracy of $3\pm12\%$. 
UKIDSS, which does not contain medium-band imaging has a comparable accuracy to the ZFOURGE medium-band survey. This is likely to arise from the lower redshifts probed by UKIDSS compared to ZFOURGE.

\subsection{Photometric Redshifts from NMBS and 3DHST}

\begin{figure}
\centering
\includegraphics[width=1.0\textwidth]{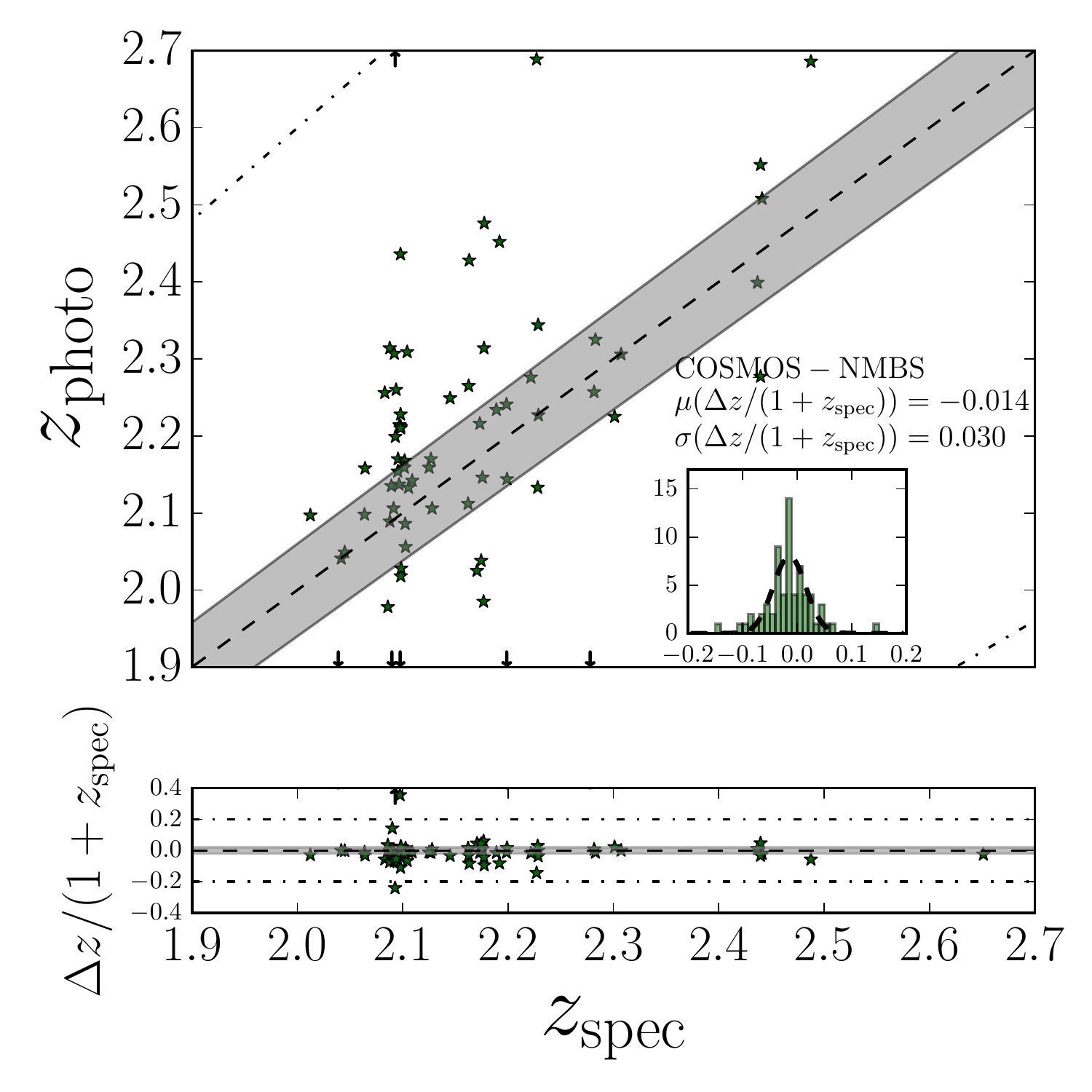}
\caption[Comparison between photometric redshifts derived by NMBS with the ZFIRE spectroscopic sample.]{Comparison between photometric redshifts derived by NMBS with the ZFIRE spectroscopic sample. 
Lines and inset figures are similar to Figure \ref{fig:specz_photoz_cosmos}. }
\label{fig:photo_z_comp_nmbs}
\end{figure}

\begin{figure}
\centering
\includegraphics[width=1.0\textwidth]{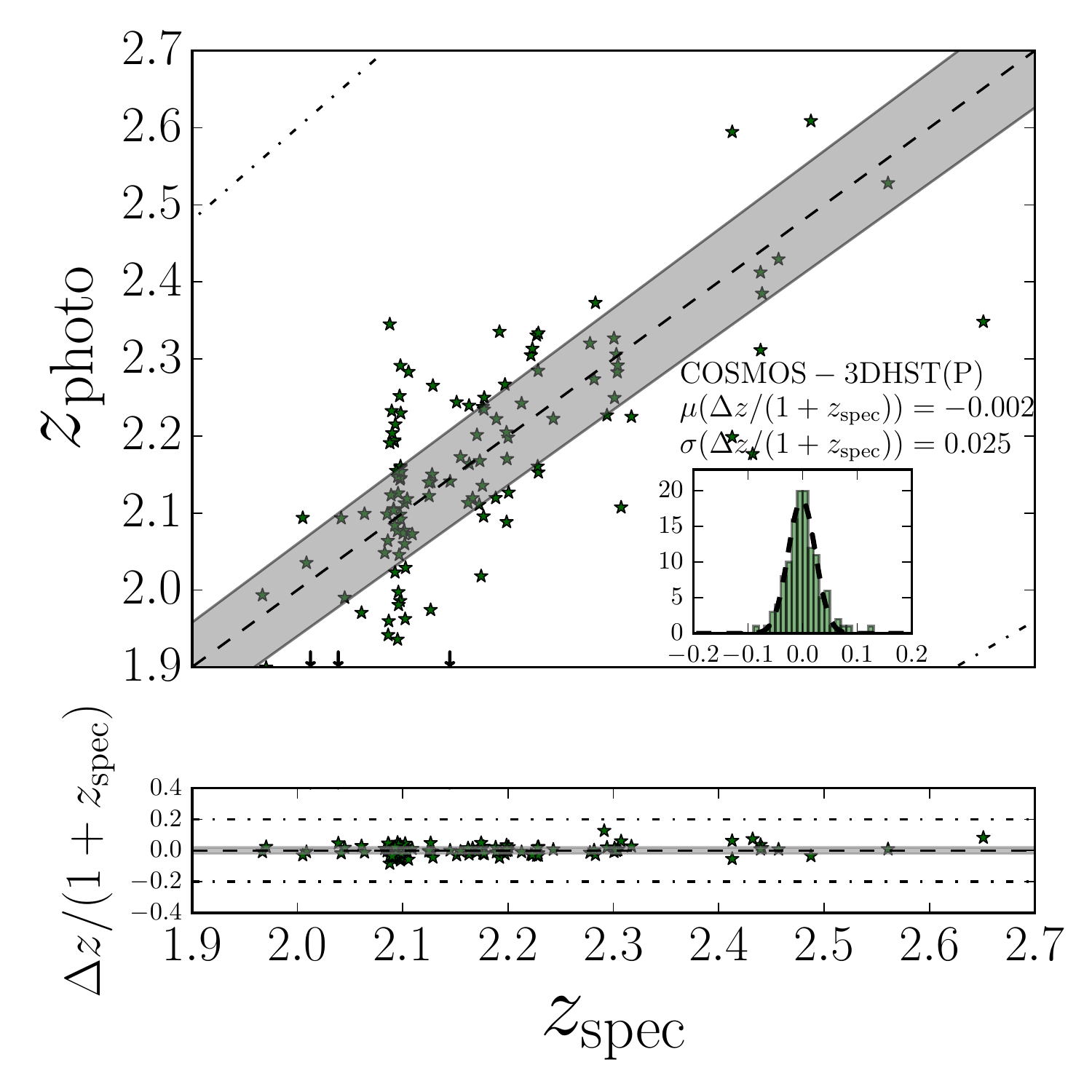}
\caption[Comparison between photometric redshifts derived by 3DHST photometric sample with the ZFIRE spectroscopic sample.]{Comparison between photometric redshifts derived by 3DHST photometric sample \citep{Skelton2014} with the ZFIRE spectroscopic sample. 
Lines and inset figures are similar to Figure \ref{fig:specz_photoz_cosmos}. }
\label{fig:photo_z_comp_3dhst}
\end{figure}

Figure \ref{fig:photo_z_comp_nmbs} and \ref{fig:photo_z_comp_3dhst} shows a redshift comparison for the 3DHST photometric redshift input sample NMBS \citep{Whitaker2011} and 3DHST \citep{Skelton2014} surveys with the ZFIRE Q$_{z}$=3 spectroscopic redshifts. 3DHST comes from the photometric data release  of \cite{Skelton2014}. 
The catalogues are compared to ZFOURGE by matching objects within a 0$''$.7 radius.
The ZFOURGE survey is much deeper than NMBS, so comparison to NMBS is only possible for a smaller number of brighter objects. 3DHST and ZFOURGE are similarly deep, with much better overlap.
The residuals between the photometric redshifts and spectroscopic redshifts are calculated using the same methods as for ZFOURGE.

Table \ref{tab:photo_z_comparision} shows the Gaussian best-fitting values, redshift accuracies, and the drastic outlier fractions of all comparisons.  
All surveys resulted in high accuracy for the photometric redshifts. In particular, at $z\sim2$ some comparisons can be made between the ZFOURGE, 3DHST, and NMBS surveys. NMBS has the worst performance, both in scatter, bias, and outlier fraction, presumably because of the shallower data set, which also includes fewer filters (no HST-CANDELS data).  
NMBS  samples brighter objects, and in ZFOURGE  such bright objects have better photometric redshift performance compared to the main sample (for galaxies with $K<23$ photometric redshift accuracies for ZFOURGE and NMBS are respectively, $1.3\pm0.2\%$ and $2\pm1$). 
3DHST fares  better in all categories.  ZFOURGE performs the best of the three in this comparison.
This is attributed to the much better seeing and depth of ZFOURGE NIR medium-band imaging, which is consistent with the findings of \citet{Straatman2016}. 

\begin{landscape}
\begin{deluxetable}{lrrcccccc}
\tabletypesize{\small}
\tablecaption{ 
Photometric (P)/Grism (G) redshift comparison results for ZFIRE Q$_{z}$=3 galaxies.
\label{tab:photo_z_comparision}}
\tablecolumns{8}
\tablewidth{0pt} 
\tablehead{
\colhead{Survey}&
\colhead{ N (Q$_z=3$)\tablenotemark{a}}&
\colhead{ $\mu$ ($\Delta z$/(1+$z_{\mathrm{spec}}$))} &
\colhead{ $\sigma$ ($\Delta z$/(1+$z_{\mathrm{spec}}$))} &
\colhead{ $z_{\mathrm{err}}$\tablenotemark{b}} &
\colhead{$\Delta z_{\mathrm{err}}$\tablenotemark{c}} &
\colhead{ Drastic Outliers \tablenotemark{d}} &
\colhead{ N$\mathrm{_{Q_z=3}\ Ks<23}$ \tablenotemark{e}}&
}
\startdata
ZFOURGE (P)-Total  			& 147		&   0.002 & 0.016  & 1.6\%   & $\pm$0.2\%  & $0.7\%$ & 53 \\ 
ZFOURGE (P)-Ks $<23$         & 53		&   0.004 & 0.013  & 1.3\%   & $\pm$0.2\%  & $2.0\%$ & -- \\ 
& & & & & & &\\ 
\hline
& & & & & & &\\ 
NMBS    (P)  & 67   	&  -0.014 & 0.030  & 3.0\%   & $\pm$0.8\%  & 10.0\% & 48 \\
3DHST   (P)  & 127  	&  -0.002 & 0.025  & 2.5\%   & $\pm$0.3\%  & 3.2\% & 49 \\
3DHST   (P+G)  & 64  	    &  -0.001 & 0.009  & 0.9\%   & $\pm$0.2\%  & 4.7\% & 43 \\
& & & & & & &\\ 
UKIDSS  (P)  & 58   	&  -0.006 & 0.014  & 1.4\%   & $\pm$0.8\%  & 7.0\% & 38 \\
\enddata
\tablenotetext{a}{The number of spectroscopic objects matched with each photometric/grism catalogue.}
\tablenotetext{b}{The accuracy of the photometric redshifts. }
\tablenotetext{c}{The corresponding bootstrap error for the redshift accuracy.}
\tablenotetext{d}{Drastic outliers defined as  $\Delta z$/(1+$z_{\mathrm{spec}}$) $>0.2$. They are given as a percentage of the total matched sample $(N)$ for each photometric/grism catalogue. Limits correspond to having $<1$ outlier.}
\tablenotetext{e}{The number of bright galaxies with Ks$<$23.}
\end{deluxetable}
\end{landscape}

\subsection{Grism Redshifts from 3DHST}

\begin{figure}
\centering
\includegraphics[width=1.0\textwidth]{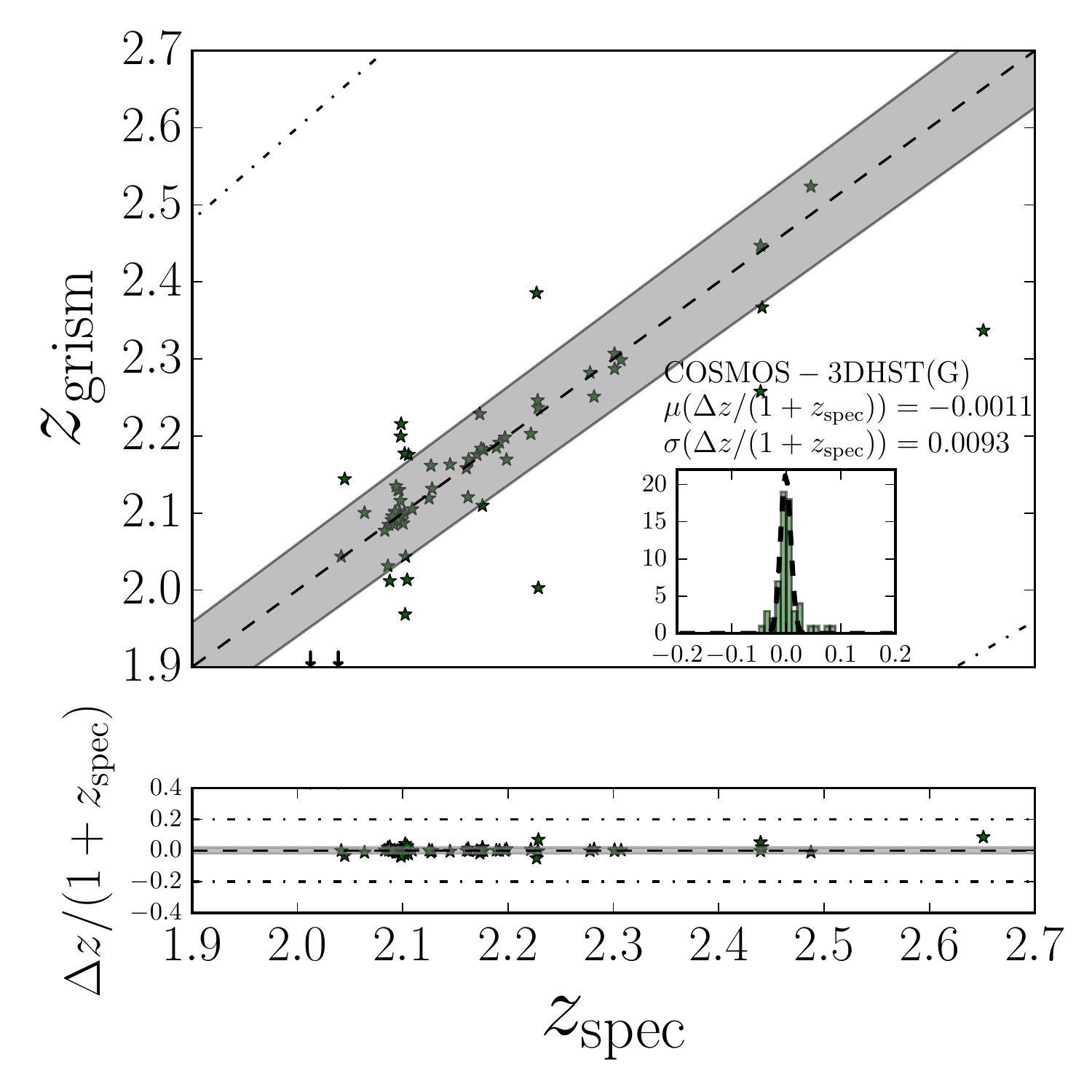}
\caption[Spectroscopic redshift comparison between ZFIRE and 3DHST grism + photometric redshifts.]{Spectroscopic redshift comparison between ZFIRE and 3DHST grism + photometric redshifts. 
This figure is similar to Figure \ref{fig:specz_photoz_cosmos} with the exception of all photometric redshifts being replaced with the 3DHST \citet{Momcheva2015} data. The Gaussian fit to $\Delta$z/(1+z) has a $\mu$=-0.0011 and $\sigma$=0.009$\pm$0.001. Only galaxies with 1.90$<z_{\mathrm{spec}}<$2.70 are shown in the figure. 
}
\label{fig:specz_3DHST_grism}
\end{figure}

3DHST grism data is used to investigate the improvement of redshift accuracy with the introduction of grism spectra to the SED fitting technique. \citet{Momcheva2015} uses a combination of grism spectra and multi-wavelength photometric data to constrain the redshifts of the galaxies. 
\citet{Momcheva2015} states that 3DHST grism data quality has been measured by two independent users. All objects, which are flagged to be of good quality by both of the users are selected to compare with the ZFIRE sample. 
This gives 175 common galaxies out of which 123 have Q$_{z}$=3 and 64 of them pass the 3DHST grism quality test.
The \zgrism\ vs. \zspec\ distributions of these 64 galaxies are shown in Figure \ref{fig:specz_3DHST_grism}.  
There are three drastic outliers, which have been identified as low-redshift galaxies by 3DHST grism data with \zgrism$<0.5$. ZFIRE \zspec\ of these outliers are $>$2.

Comparing with the 3DHST redshifts derived only via pure photometric data, it is evident that the introduction of grism data increases the accuracy of the redshifts by \around$\times$3 to an accuracy of $0.9\pm0.1$\%.  The \zgrism\ accuracy is lower than the \around0.4\% accuracy computed by \citet{Bezanson2016} for grism redshifts. We note that \citet{Bezanson2016} is performed for galaxies with $H_{F160W}<24$ and that the ZFIRE-COSMOS sample probes much fainter magnitudes.

\subsection{Spectroscopic Redshifts from MOSDEF and VUDS}
\label{sec:specz_comparisions}

The final comparison is with other public spectroscopic redshifts in these fields. Galaxies from MOSDEF \citep{Kriek2015} and VUDS \citep{Cassata2015} surveys are matched with the ZFIRE sample within a 0$''$.7 aperture. 

The MOSDEF overlap comprises  84 galaxies in the COSMOS field with high confidence redshift detections, out of which 74 galaxies are identified with matching partners from the ZFOURGE survey. In the ZFOURGE matched sample, 59 galaxies are at redshifts between 1.90$<$\zspec$<2.66$. 
7 galaxies are identified to be in common between ZFIRE and MOSDEF detections. 
The RMS of the scatter between the spectroscopically derived redshifts is \around0.0007.
This corresponds to a rest frame velocity uncertainty of  \around67 km s$^{-1}$, which is attributed to barycentric redshift corrections not being applied for the MOSDEF sample. 
We note that barycentric velocities should be corrected as a part of the wavelength solution by the DRP for each observing night, and therefore we are unable to apply such corrections to the MOSDEF data. Considering ZFIRE data, once the barycentric correction is applied we find, by analysing repeat observations in K band, that our redshifts are accurate to $\pm13$ km s$^{-1}$.

Similarly, the VUDS COSMOS sample comprises 144 galaxies with redshift detections $>3\sigma$ confidence, out of which 76 galaxies have ZFOURGE detections. In the ZFOURGE matched sample, 43 galaxies lie within $1.90<$\zspec$<2.66$. 
There are two galaxies in common between ZFIRE and VUDS detections and redshifts agree within 96 km s$^{-1}$ and 145 km s$^{-1}$. The redshift confidence for the matched two galaxies are $<\mathrm{2}\sigma$ in the VUDS survey, while the ZFIRE has multiple emission line detections for those galaxies.  Furthermore, the VUDS survey employs VIMOS in the low-resolution mode ($R\sim200$) in its spectroscopy leading to absolute redshift accuracies of $\sim200$ km s$^{-1}$. Therefore, we expect the ZFIRE redshifts of the matched galaxies to be more accurate than the VUDS redshifts.

\begin{figure}
\centering
\includegraphics[width=0.6\textwidth]{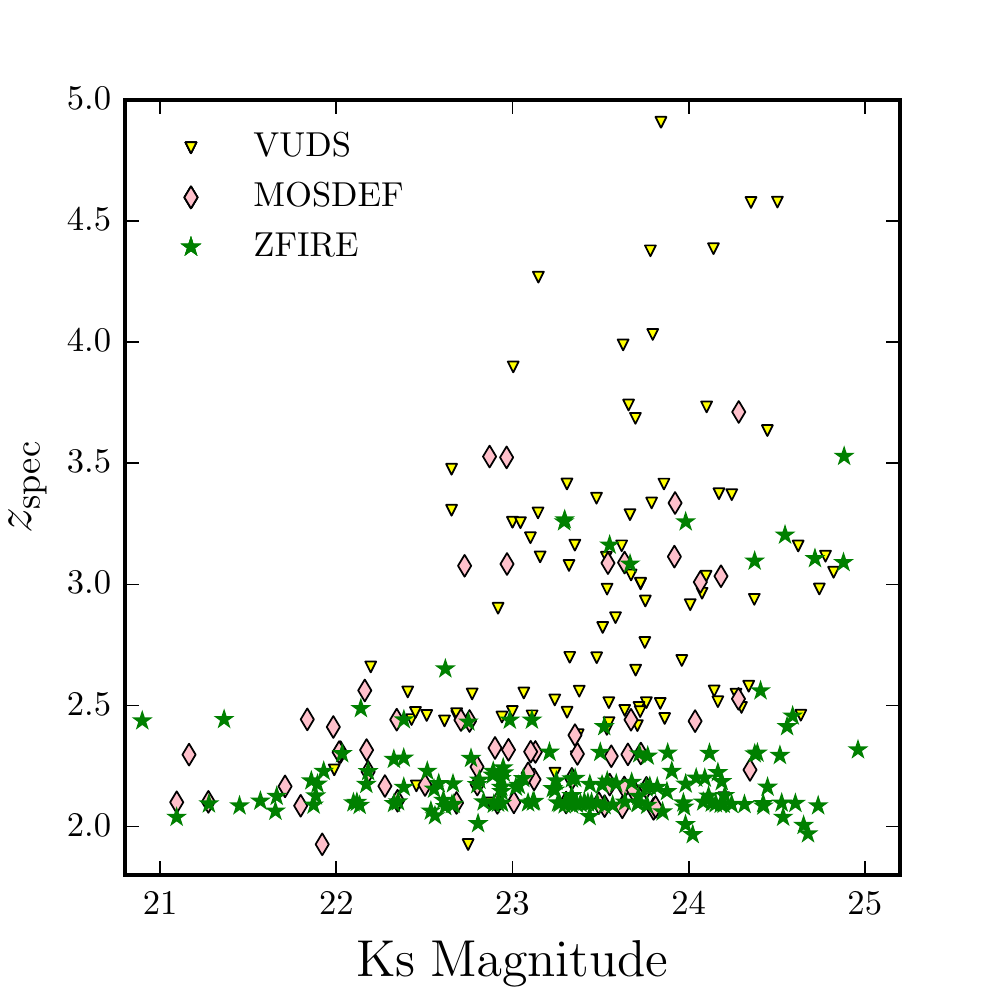}
\includegraphics[width=0.6\textwidth]{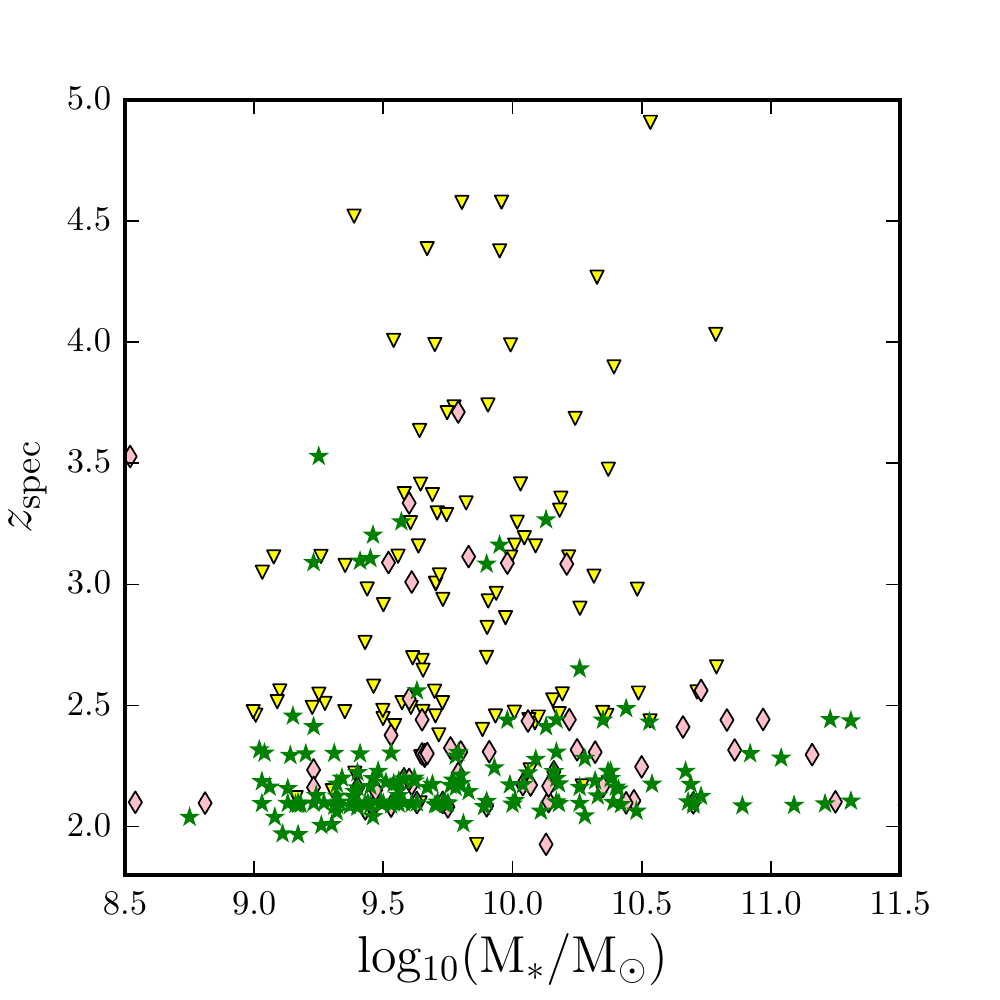}
\caption[Redshift comparison as a function of Ks magnitude and stellar mass.]{Redshift comparison as a function of Ks magnitude and stellar mass. Along with the ZFIRE\ Q$_{z}$=3 detections, the MOSDEF and VUDS samples are shown for comparison. 
For the MOSDEF sample, only galaxies with identified ZFOURGE detections are shown.  All VUDS galaxies with $\mathrm{z_{spec}}>1.8$ with $>3\sigma$ detections are shown.  
Note that VUDS observes galaxies in the optical regime, while ZFIRE\ and MOSDEF observes in the NIR. 
{\bf Top:} $\mathrm{z_{spec}}$ vs. Ks magnitude for the spectroscopically detected galaxies. The VUDS sample is plotted as a function of K magnitude. 
{\bf Bottom:} $\mathrm{z_{spec}}$ vs. stellar mass for the same samples of galaxies.  
}
\label{fig:survey_depth_comp}
\end{figure}

Figure \ref{fig:survey_depth_comp} shows the distribution of the redshifts of the ZFIRE sample as a function of Ks magnitude and stellar mass. ZFIRE detections span a wide range of Ks magnitudes and stellar masses at $z\sim2$. The subset of galaxies observed at $z\sim3$ are fainter and are of lower mass. MOSDEF and VUDS samples are also shown for comparison. VUDS provides all auxiliary stellar population parameters, which are extracted from the CANDELS survey and hence all data are included. However, MOSDEF only provides the spectroscopic data and thus, only galaxies with identified ZFOURGE counterparts are shown in the figure, which is $\sim$90\% of the MOSDEF COSMOS field galaxies  with confident redshift detections. 

In Figure \ref{fig:survey_depth_comp}, MOSDEF detections follow a similar distribution to ZFIRE. Since both the surveys utilize strong emission lines in narrow NIR atmospheric passbands, similar distributions are expected.  VUDS, however, samples a different range of redshifts as it uses  optical spectroscopy. We note the strong \zspec=2.095 overdensity due to the cluster in the ZFIRE sample, but not in the others.

\newpage

\section{Broader Implications}
\label{sec:implications}

The large spectroscopic sample presented can be used to assess the fundamental accuracy of galaxy physical parameters (such as stellar mass, SFR, and galaxy
SED classification) commonly derived from photometric redshift surveys. It can also be used to understand the performance of the previous
cluster selection that was done.

\subsection{Galaxy Cluster Membership}

The completeness and purity of galaxy cluster membership of the $z=2.1$ cluster based on photometric redshifts is next investigated and compared with spectroscopic results.
First,  photometric redshifts are used to compute a seventh nearest neighbour density map as shown in Figure \ref{fig:detection_map}. 
Any galaxy that lies in a region with density $>3\sigma$ is assumed to be a photometric cluster candidate.
From the ZFOURGE photometric redshifts in the COSMOS field (coverage of $\sim 11'\times11'$) for $2.0<$\zphoto$<2.2$, there are 66 such candidates. All of these galaxies have been targeted to obtain spectroscopic redshifts. 
\citet{Yuan2014} cluster galaxies are chosen to be within $3\sigma$ of the Gaussian fit to the galaxy peak at $z = 2.095$. 
Only 25 of the photometric candidates are identified to be a part of the \citet{Yuan2014} cluster, which converts to \around38\% success rate. 
The other 32 spectroscopically confirmed cluster galaxies at $z=2.095$ from Yuan et al. are not selected as cluster members using photometric redshifts, $i.e.$ membership identification based on photometric redshifts and seventh nearest
neighbour is \around56\% incomplete. 

\citet{Yuan2014} finds the velocity dispersion of the cluster structure to be $\sigma_{\mathrm{v1D}}= 552\pm52$ km s$^{-1}$, while the photometric redshift accuracy of ZFOURGE at $z=2.1$ is $\sim4500$ km s$^{-1}$. Therefore, even high-quality photometric redshifts such as from ZFOURGE, we are unable to precisely identify cluster galaxy members, which demonstrates that  spectroscopic redshifts are crucial for identifying and studying cluster galaxy populations at $z\sim2$.

\subsection{Luminosity, Stellar Mass, and Star Formation Rate}
\label{sec:M-SFR-dz}

\begin{figure}
\centering
\includegraphics[width=0.9\textwidth]{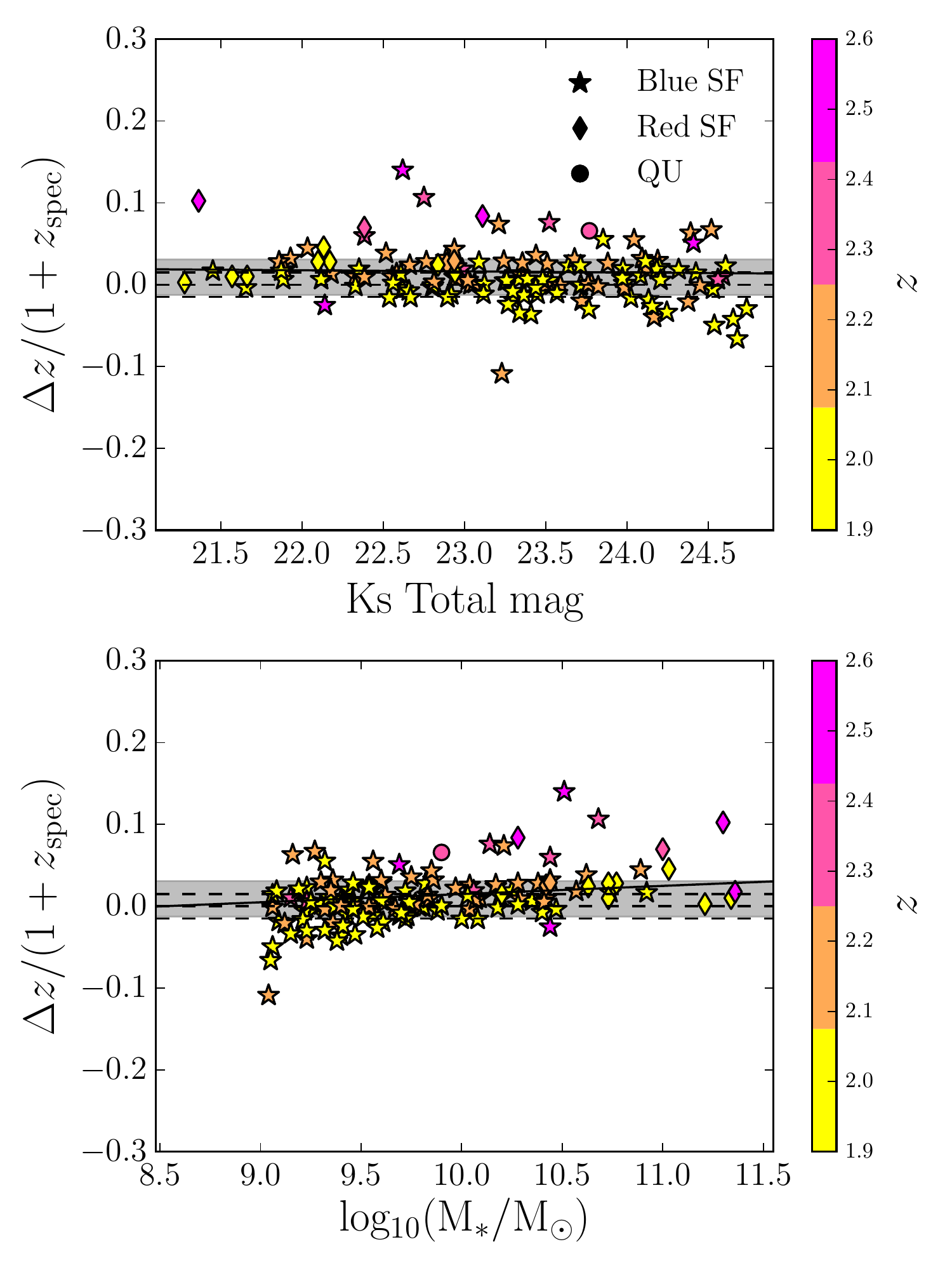}
\caption[Photometric redshift accuracies as a function of Ks magnitude and stellar mass.]{Photometric redshift accuracies as a function of Ks magnitude and stellar mass. 
All Q$_z=3$ ZFIRE-COSMOS galaxies with redshifts between $1.90<z<2.66$ have been selected. 
All galaxies are divided into blue star-forming, red (dusty) star-forming, and quiescent galaxies, which are shown with different symbols. Galaxies are further sub-divided into redshifts and are colour coded as shown. 
{\bf Top:} $\Delta z/(1+z_\mathrm{spec}$) vs. Ks total magnitude from ZFOURGE.  
{\bf Bottom:} similar to above but with stellar mass on the x-axis. 
The median  $\Delta z/(1+z_\mathrm{spec}$) is 0.009.
The grey shaded region in both the plots shows the \NMAD\ of the $\Delta z/(1+z_\mathrm{spec}$) scatter (0.022) around the median of the selected galaxies. The solid lines are the least squares best-fitting lines for the data. 
}
\label{fig:delta_z_vs_param}
\end{figure}

An important question in utilising photometric redshifts is whether their accuracy depends on key galaxy properties such as luminosity, stellar mass, and/or SFR. This could lead
to biases in galaxy evolution studies.
The Ks total magnitudes and stellar masses from ZFOURGE (v2.1 catalogue) are used for this comparison, which is shown in Figure \ref{fig:delta_z_vs_param}.
The redshift error is plotted as a function of Ks magnitude and stellar mass for all Q$_z$=3 ZFIRE galaxies. The sample is binned into redshift bins and further subdivided into star-forming, dusty star-forming, and quiescent galaxies depending on their rest-frame UVJ colour.

The least squares best-fitting lines for the Ks magnitude and stellar mass are \newline $y=-0.001(\pm0.003)x+0.05 (\pm0.06)$ and $y=0.010 (\pm0.005)x-0.08 (\pm0.05)$, respectively.
Therefore, it is evident that there is a slight trend in stellar mass in determining the accuracy of photometric redshifts with more massive galaxies showing positive offsets for $\Delta z/(1+$\zspec$)$.
However, the relationship of $\Delta z/(1+$\zspec$)$ with Ks magnitude is not statistically significant.
The typical \NMAD\ of $\Delta z/(1+$\zspec$)$ is 0.022 with a median of 0.009. 
Note that the $\Delta z/(1+$\zspec$)$ scatter parametrized here is different from the \zphoto\ vs. \zspec\ comparison in Figure \ref{fig:specz_photoz_cosmos} for the ZFOURGE sample. We use the ZFOURGE catalogue version 2.1 for the  $\Delta z/(1+$\zspec$)$ vs. mass, magnitude comparison while for the \zphoto\ vs. \zspec\ comparison,  we use v3.1. Furthermore, the scatter here is calculated using \NMAD\ , while in Figure \ref{fig:specz_photoz_cosmos} a Gaussian function is fit to the $\Delta z/(1+$\zspec$)$ after removing the drastic outliers. The changes in \zphoto\ between v2.1 and v3.1 is driven by the introduction of improved SED templates. This comparison is expanded on in Appendix \ref{sec:ZFOURGE comparison}.

\begin{figure}
\centering
\includegraphics[width=0.9\textwidth]{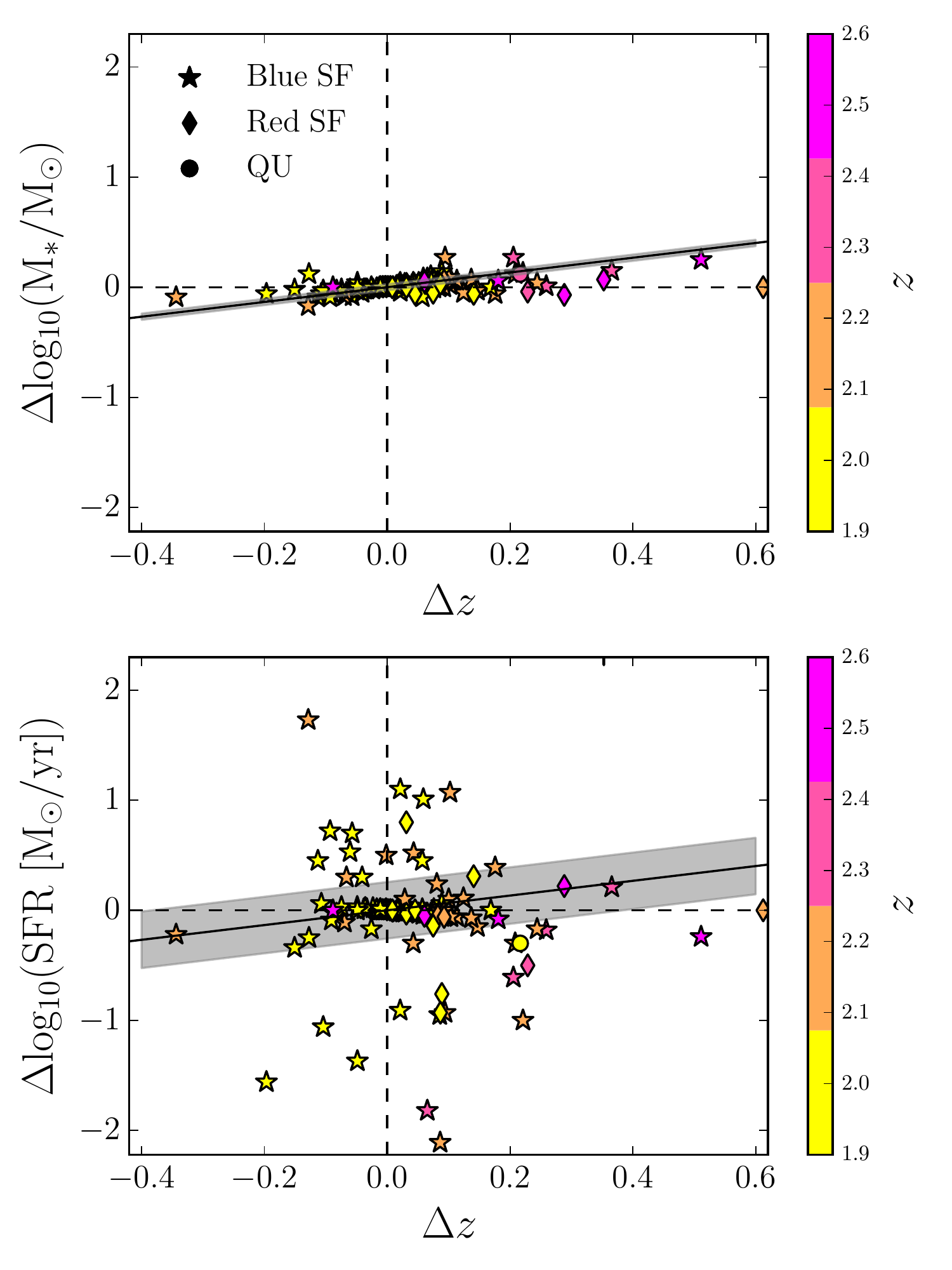}
\caption[Effect of $\Delta z$ on galaxy stellar mass and dust extinction derived by FAST.]{Effect of $\Delta z$ on galaxy stellar mass and dust extinction derived by FAST. 
All ZFIRE-COSMOS galaxies with redshifts between $1.90<z<2.66$ have been selected. All galaxies are divided into blue star-forming, red (dusty) star-forming and quiescent galaxies which are shown as different symbols. Galaxies are further sub-divided into redshifts and are colour coded as shown. 
The diagonal solid lines are Equation (\ref{eq:delta_m_z}), which is the simplified theoretical expectation for mass/SFR  correlation with redshift error.
The grey shaded regions corresponds to the $\sigma$ value of the best-fitting Gaussian functions that describes the deviation of the observed values from the theoretical expectation.
{\bf Top:} $\Delta\log_{10}$Mass vs. $z_\mathrm{spec}-z_\mathrm{photo}$. 
$\Delta\log_{10}$Mass is defined as the difference in stellar mass when computed by FAST using $z_\mathrm{spec}$ and  $z_\mathrm{photo}$ values. 
{\bf Bottom:} similar to top but with $\Delta\log_{10}$(SFR) on the y axis. 
$\Delta\log_{10}$SFR is defined as the difference in SFR when computed by FAST using $z_\mathrm{spec}$ and  $z_\mathrm{photo}$ values. 
}
\label{fig:delta_param_vs_delta_z}
\end{figure}

There should be a  dependency of galaxy properties derived via SED fitting techniques on $\Delta z$. Figure \ref{fig:delta_param_vs_delta_z} shows the change of stellar mass and SFR (both calculated using FAST using either photometric or spectroscopic redshifts) as a function of $\Delta z$. To first order, an analytic calculation of the expected residual can be made.
SED fitting techniques estimate galaxy stellar masses from luminosities and
mass-to-light ratios. The luminosity calculated from the flux will depend on the redshift used, and hence the mass and redshift change should correlate.
Ignoring changes in mass to light ratios and K-correction effects, from the luminosity distance change we expect
\begin{subequations}
\begin{equation}
\frac{d[\log_{10}(M)]}{dz} = \frac{2}{D_L} \left(\frac{dD_L}{dz}\right)_{z=2}
\end{equation}
where $M$ is the stellar mass of the galaxy  and $D_L$ is the luminosity distance. Evaluating for $z=2$, with $D_L=15.5$ Gpc:
\begin{equation}
\label{eq:delta_m_z}
\Delta \log_{10}(M) = 0.67 \Delta z
\end{equation} 
\end{subequations} 

Equation (\ref{eq:delta_m_z}) is plotted in Figure \ref{fig:delta_param_vs_delta_z}.  The top panel of the figure shows that the mass and redshift changes correlate approximately as expected with a \NMAD\ of 0.017 dex. SED SFRs are also calculated from luminosities, albeit with a much greater  
weight to the UV section of the SED, and thus should scale similarly to mass. 
The \NMAD\ scatter around this expectation is 0.086 dex, which is higher than the mass scatter  with a much greater number of outliers. To fully comprehend the role of outliers in the scatter, we fit a Gaussian function to the deviation of $\Delta\log_{10}$(Mass) and $\Delta\log_{10}$(SFR) for each galaxy from its  theoretical expectation. The $\Delta\log_{10}$(SFR) shows a larger scatter of $\sigma=0.2$ in the Gaussian fit compared to the $\sigma=0.03$ of $\Delta\log_{10}$(Mass). 
It is likely that the higher scatter in $\Delta\log_{10}$(SFR)  is because the  rest-frame UV luminosity is much more sensitive to the star formation history and dust extinction encoded in the best-fitting SED than the stellar mass.

It is evident that photometric-redshift derived stellar masses are robust against the typical redshift errors, however, caution is warranted when using SED based SFRs with
photometric redshifts because they are much more sensitive to small redshift changes 
 (in our sample \around26\% of galaxies have $|\Delta\log_{10}$SFR$|>0.3$ even though the photometric redshifts have good precision). 
Studies that investigate galaxy properties solely relying on photometric redshifts may result in inaccurate conclusions about inherent galaxy properties and therefore, it is imperative that  they are supported by spectroscopic studies. It should be noted that previous ZFOURGE papers have extensively used photometric redshift derived stellar masses (for example, the mass function evolution of \citet{Tomczak2014}), which we find to be reliable, but not SED-based SFRs. Most commonly, the best-fitting SEDs are used to derive the UV+IR fluxes in order to derive SFRs, since SFRs derived directly via FAST templates \cite[eg.,][]{Maraston2010} are degenerated with age, metallicity, and dust law. See \citet{Conroy2013} for a review on this topic.

\subsection{Rest-Frame UVJ Colours}

ZFOURGE rest frame UVJ colours are derived using photometric redshifts. 
UVJ colours from \zphoto\ are commonly used to identify the evolutionary stage of a galaxy \citep{Williams2009}. Here we investigate the effect of photometric redshift
accuracy on the UVJ colour derivation of galaxies. 

\begin{figure}
\centering
\includegraphics[width=1.0\textwidth]{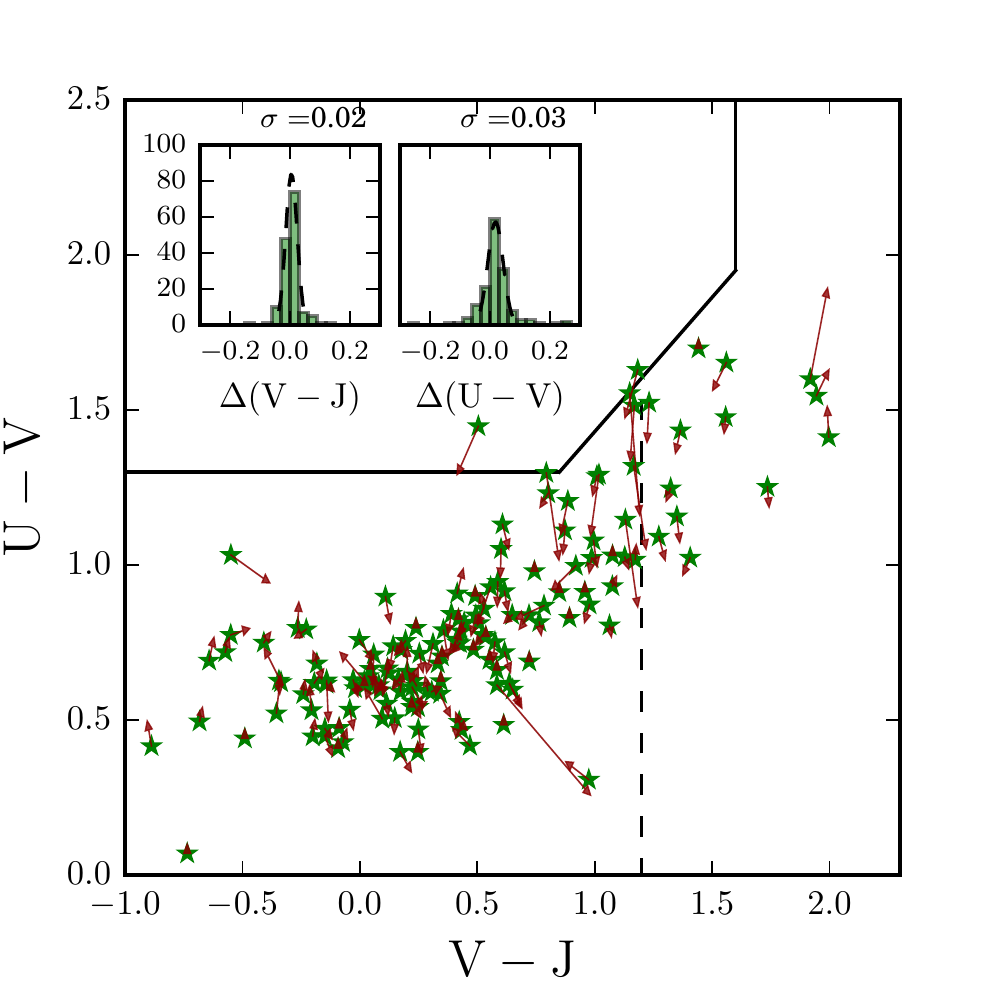}
\caption[Effect of $\Delta z$ on rest frame UVJ colours.]{Effect of $\Delta z$ on rest frame UVJ colours. 
All ZFIRE-COSMOS galaxies are shown in the redshift bin $1.90<z<2.66$. 
The green stars are rest frame UVJ colours derived using photometric redshifts from EAZY. The rest frame colours are re-derived using spectroscopic redshifts from ZFIRE. The brown arrows denote the change of the position of the galaxies in the rest frame UVJ colour space when $z_{\mathrm{spec}}$ is used. The large arrows (one of which moves outside the plot range) are driven by $\Delta z$ outliers. 
The two inset histograms show the change in (V$-$J) and (U$-$V) colours for these sample of galaxies. Gaussian fits with $\sigma$ of 0.02 and 0.03 are performed, respectively, for the (V$-$J) and (U$-$V) colour differences.  
}
\label{fig:delta_UVJ}
\end{figure}

Figure \ref{fig:UVJ} shows the rest frame UVJ colours of Q$\mathrm{_z}$=3 objects re-derived using spectroscopic redshifts from the same SED template library. 
Figure \ref{fig:delta_UVJ} shows the change of location of the galaxies in rest frame UVJ colour when ZFIRE redshifts are used to re-derive them (the lack of quiescent galaxies overall is a bias in the ZFIRE\ sample selection as noted earlier).
Only one to two galaxies change their classifications from the total sample of 149. 
The inset histograms show the change of (U$-$V) and (V$-$J) colours. Gaussian functions are fit to the histograms to find that the scatter in (U$-$V) colours ($\sigma$=0.03) to be higher than that of (V$-$J) colours ($\sigma$=0.02) and (U$-$V)  has a greater number of outliers. 
The conclusion is that the U$-$V rest-frame colours are more sensitive to redshift compared to V$-$J colours by \around50\%, which may contribute to a selection bias in high-redshift samples. This sensitivity of the UV part of the SED is in accordance with the results of Section~\ref{sec:M-SFR-dz}.

To further quantify the higher sensitivity of U magnitude on redshift, Gaussian fits are performed on the $\Delta$U, $\Delta$V, and $\Delta$J magnitudes of the ZFIRE galaxies, by calculating the difference of the magnitudes computed when using \zphoto\ and \zspec. $\Delta$U shows a larger scatter of $\sigma=0.04$, while $\Delta$V and $\Delta$J show a scatter of $\sigma=0.01$. This further validates our conclusion that the UV part of the SED has larger sensitivity to redshift.

\section{Summary}
\label{sec:summary_zfire}

Here we present the ZFIRE survey of galaxies in rich environments and our first public data release.  A detailed description of the data reduction used by ZFIRE is provided.  The use of a flux standard star along with photometric data from ZFOURGE and UKIDSS has made it possible to flux calibrate the spectra to $\lesssim10$\% accuracy.  The ZFIRE-COSMOS sample spans a wide range in Ks magnitude and stellar mass and secures redshifts for UVJ star-forming galaxies to Ks=24.1 and stellar masses of $\log_{10}($\mass$)>9.3$.  We show that selecting using rest-frame UVJ colours is an effective method for identifying \Halpha-emitting galaxies at $z\sim2$ in rich environments. Redshifts have been measured for 232 galaxies of which 87 are identified as members of the rich clusters we have targeted in COSMOS and UDS fields.

Photometric redshift probability density functions from EAZY are used to show that the expected \Halpha\ detections are similar to the ZFIRE detection rate in the COSMOS field. In the COSMOS field, the ZFIRE survey has detected \around80\% of the targeted star-forming galaxies. We also show that the density structure discovered by \citet{Spitler2012} has been thoroughly sampled by ZFIRE.

Using spectroscopic redshifts from ZFIRE with ZFOURGE and other public photometric survey data, we investigated the accuracies of photometric redshifts. The use of medium-band imaging in SED fitting techniques can result in photometric redshift accuracies of $\sim1.5\%$. ZFIRE calculations of photometric redshift accuracies are consistent with the expectations of the ZFOURGE survey (Straatman at al., in press) but are slightly less accurate  than the NMBS \citep{Whitaker2011} and 3DHST \citep{Skelton2014} survey results. The higher redshift errors can be attributed to sampling differences, which arises from the deeper NIR medium-band imaging in ZFOURGE compared to the other surveys (i.e. overlapping galaxies tend to be fainter than typical in the respective galaxies in NMBS). 
If we select a brighter subset of NMBS (Ks $<23$) we find that the redshift accuracy increases by 30\%.

Using UKIDSS, \citet{Quadri2012} shows that the photometric redshift accuracy is dependent on redshift and that at higher redshifts the photometric redshift error is higher. Between UKIDSS at $z\sim1.6$  and ZFOURGE at $z\sim2$ the photometric redshift accuracies are similar. Therefore, the use of medium-band imaging in ZFOURGE has resulted in more accurate redshifts at $z\sim2$, due to finer sampling of the D4000 spectral feature by the J1, J2, and J3 NIR medium-band filters. The introduction of medium-bands in the K band in future surveys may allow photometric redshifts to be determined to higher accuracies at $z\gtrsim4$. 

The importance of spectroscopic surveys to probe the large-scale structure of the universe is very clear. For the COSMOS \citet{Yuan2014} cluster, we compute a 38\% success rate (i.e., 38\% of galaxies in $3\sigma$ overdensity regions are identified spectroscopically as cluster galaxies) and a 56\% incompleteness (56\% of spectroscopic cluster galaxies are not identified from data based on purely photometry) using the best photometric redshifts (with seventh nearest neighbour algorithms) to identify clustered galaxies. 

We find a systematic trend in photometric redshift accuracy, where massive galaxies give higher positive offsets up to $\sim$0.05 for $\Delta z/(1+z_\mathrm{spec}$) values as a function of galaxy stellar mass. However, it is not evident that there is any statistically significant trend for a similar relationship with galaxy luminosity. 
Results also suggest that the stellar mass and SFR correlates with redshift error. This is driven by the change in the calculated galaxy luminosity as a function of the assigned redshift and we show that the values correlate approximately with the theoretical expectation. SFR shows larger scatter compared to stellar mass in this parameter space, which can be attributed to the stronger weight given to UV flux, which is very sensitive to the underlying model, in the derivation of the SFR.

This stronger correlation of the UV flux with redshift error is further evident when comparing the change in (U$-$V) and (V$-$J) colour with change in redshift. When rest-frame U,V, and J colours are re-derived using spectroscopic redshifts, our results show a stronger change in (U$-$V) colour compared to the (V$-$J) colour. Therefore, a redshift error may introduce an extra selection bias on rest-frame UVJ selected galaxies.  Further studies using larger samples of quiescent and dusty star-forming galaxies at $z\sim2$ are needed to quantify this bias.

Clearly the use of photometric redshifts can lead to biases even when using the same SED template set. However, it is important to acknowledge the underlying uncertainties that lie in deriving galaxy properties even with spectroscopic redshifts. 
Future work could consider the role of SED templates used in SED fitting techniques. Generally the templates used are empirically derived, which limits the capability to understand the inherent properties of the observed galaxies. With the use of physically motivated models such as MAGPHYS \citep{daCunha2008}, more statistically meaningful relationships between different physical parameters of the observed galaxies could be obtained. Improving such models to include photo-ionization of galaxies the in future will allow us to directly make comparisons of star-forming galaxies at $z\sim2$, which will be vital to study the inherent galaxy properties. 

Furthermore, the accuracy of underlying assumptions used in SED fitting techniques such as the IMF, dust properties, and star formation histories at $z\sim2$ should be investigated. These assumptions are largely driven by observed relationships at $z\sim0$, and if the galaxies at higher redshifts are proven to be inherently different from the local populations, results obtained via current SED fitting techniques may be inaccurate.  Future work should focus on the physical understanding of the galaxy properties at $z\gtrsim2$ with large spectroscopic surveys to better constrain the galaxy evolution models. The recent development of sensitive NIR integral field spectrographs with multiplexed capabilities will undoubtedly continue to add a wealth of more information on this topic over the next few years.

The ZFIRE survey will continue focusing on exploring the large spectroscopic sample of galaxies in rich environments at $1<z<3$ to investigate galaxy properties in rich environments.  Upcoming papers
include analyses of the IMF \citep{Nanayakkara2017}, kinematic scaling relations \citep{Alcorn2016,Straatman2017}, the mass--metallicity fundamental plane \citep{Kacprzak2016}, and galaxy growth in cluster and field samples \citep{Tran2017}.

\chapter{The Stellar Initial Mass Function\\ Observational Data \& Synthetic Stellar Population Models}
\label{chap:imf_observations}

In this chapter, I describe the procedures used to select a galaxy sample from ZFIRE survey to examine the IMF. This includes an analysis of sample completeness and a thorough description of the \Halpha\ flux calculations and continuum fitting techniques. Furthermore, the synthetic stellar population models used to analyse the observed data are also presented in this chapter.

\section{Observations \& Data}
\label{sec:EW_calc}

\subsection{Galaxy Sample Selection}
\label{sec:sample_selection}

The sample used in this study was selected from the ZFIRE \citep{Nanayakkara2016} spectroscopic survey, which also consists of photometric data from the ZFOURGE survey \citep{Straatman2016}.
In this section, we describe the sample selection process from the ZFIRE survey for our analysis.

ZFIRE is a spectroscopic redshift survey of star-forming galaxies at $1.5<z<2.5$, which utilized the MOSFIRE instrument \citep{McLean2012} on Keck-I to primarily study galaxy properties in rich environments. ZFIRE has observed $\sim300$ galaxy redshifts with typical absolute accuracy of $\mathrm{\sim15\ kms^{-1}}$ and derived basic galaxy properties using multiple emission line diagnostics. \citet{Nanayakkara2016} give full details on the ZFIRE survey. 
In this study we use the subset of ZFIRE galaxies observed in the COSMOS field \citep{Scoville2007} based on a stellar mass limited sample reaching up to 5$\sigma$ emission line flux limits of $\mathrm{\sim3\times10^{-18}erg/s/cm^2}$ selected from deep NIR data $\mathrm{K_{AB}<25}$ obtained by the ZFOURGE survey.

ZFOURGE\footnote{\url{http://zfourge.tamu.edu}} (PI I. Labb\'e) is a Ks selected deep 45 night photometric legacy survey carried out using the purpose built FourStar imager \citep{Persson2013} in the 6.5 meter Magellan Telescopes located at Las Campanas observatory in Chile. The survey covers 121 arcmin$^2$ in each of the COSMOS, UDS \citep{Beckwith2006}, and CDFS \citep{Giacconi2001} legacy fields. Deep FourStar medium band imaging (5$\sigma$ depth of Ks$\leq$25.3 AB ) and the wealth of public multi-wavelength photometric data (UV to Far infra-red) available in these fields were used to derive photometric redshifts with accuracies $\lesssim1.5\%$ using EAZY \citep{Brammer2008}. Galaxy masses, ages, SFRs, and dust properties were derived using FAST \citep{Kriek2009} with a \citet{Chabrier2003} IMF, exponentially declining SFHs, and \citet{Calzetti2000} dust law. At $z\sim2$ the public ZFOURGE catalogues are 80\% mass complete to $\sim10^9$\msol \citep{Nanayakkara2016}. Refer to \citet{Straatman2016} for further details on the ZFOURGE survey.

ZFIRE and ZFOURGE are ideal surveys to use in this study since both provide mass complete samples. The total ZFIRE sample in the COSMOS field contains 142 \Halpha\ detected ($>5\sigma$, redshift quality flag (Q$_z$)=3 ) star-forming galaxies that is mass complete down to \logmass$>9.30$  (at 80\% for $Ks=24.11$). Thus, our \Halpha\ selected sample  contains no significant systematic biases towards star formation history, stellar mass, and magnitude. Furthermore, ZFIRE contains a large cluster at $z=2$ containing 51 members with $5\sigma$ \Halpha\ detections \citep{Yuan2014} and therefore we are able to examine if the IMF is affected by the local environment of galaxies. 

For this study, we apply the following additional selection criteria to the 142 \Halpha\ detected galaxies.\\  

$\bullet$ We remove AGN using photometric \citep{Cowley2016} and emission line \\ ($\mathrm{log_{10}}$(f(\NII)/f(\Halpha))$>-0.5$; \citet{Coil2015}) criteria resulting in identifying 26 AGN with our revised sample containing $N=116$ galaxies.
We note that all galaxies selected as AGN from ZFOURGE photometry by \citet{Cowley2016} are flagged as AGN by the \citet{Coil2015} selection. 100\% of the BPT identified AGN from MOSDEF are flagged as AGN from this selection by \citet{Coil2015}. We further discuss contamination to \Halpha\ from sub-dominant AGN in Appendix \ref{sec:AGN}.\\

$\bullet$ Galaxies must have a matching ZFOURGE counterpart such that we can obtain galaxy properties, resulting in N=109 galaxies.\\

$\bullet$ We compute the total spectroscopic flux for these galaxies and remove 4 galaxies with negative fluxes resulting in N=105 galaxies. 
We perform stringent \Halpha\ emission quality cuts to the spectra for these 105 galaxies and remove 2 galaxies due to strong sky line subtraction issues. We further remove 1 galaxy due to an overlap of the galaxy spectra with a secondary object that falls within the same slit.\\

\begin{figure}
\centering
\includegraphics[scale=1.5]{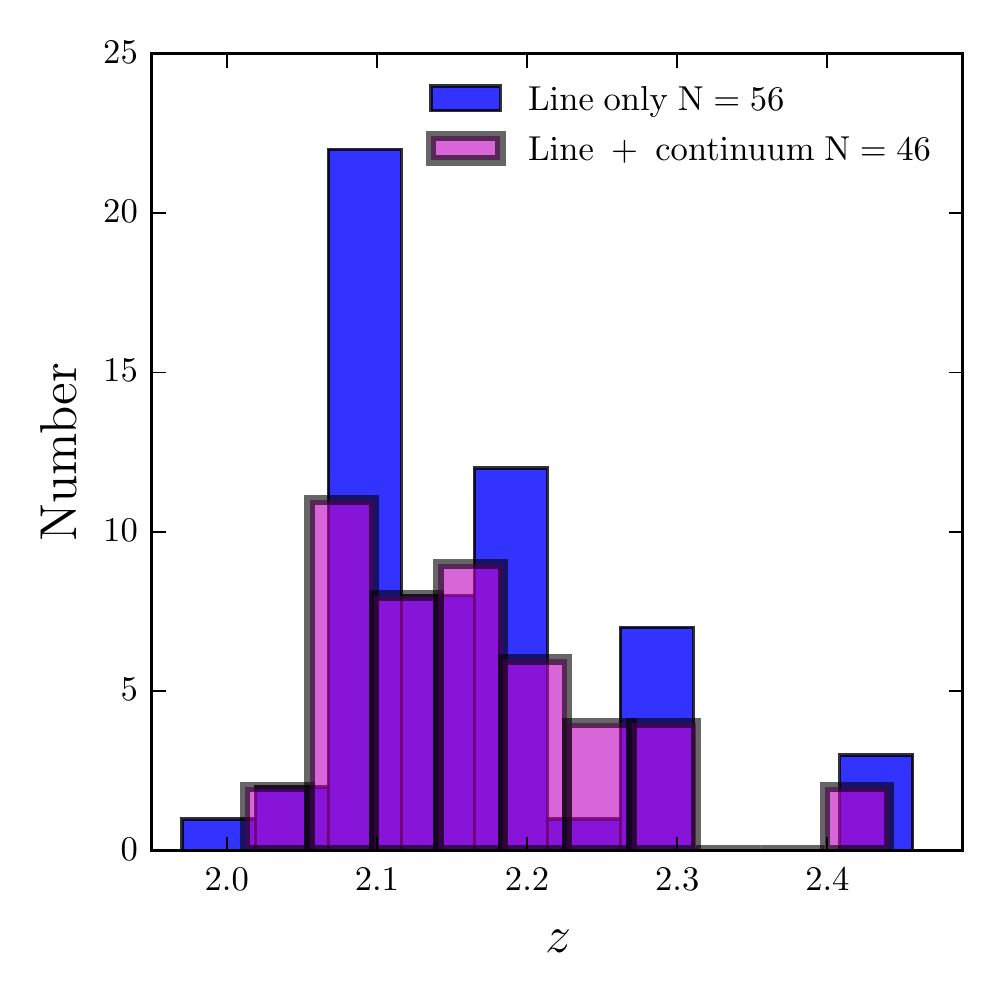}
\caption[The redshift distribution of the \sample.]{The redshift distribution of the \sample. Galaxies with line+continuum detection are shown by magenta and galaxies only with \Halpha\ line detection are shown by blue.}
\label{fig:specz_distribution}
\end{figure}

Our final sample of galaxies used for the IMF analysis in this paper comprise of 102 galaxies. 
Henceforth we refer to this sample of galaxies as the ZFIRE stellar population (SP) sample. \\
The redshift distribution for the \sample\ is shown by Figure \ref{fig:specz_distribution}. The \sample\ is divided into continuum detected and non-detected galaxies as described in section \ref{sec:cont_fit}. Galaxies in our sample lie within redshifts of $1.97<z<2.46$ corresponding to a $\Delta t \sim 650$ Myr.

\subsection{Completeness}
\label{sec:completeness}

In order to determine any significant detection biases in our \sample, we evaluate the completeness of the galaxies selected in this analysis. 
We define a redshift window for analysis between $1.90<z<2.66$ ($\sim8.6$ Gpc), which corresponds to the redshifts that \Halpha\ emission will fall within the MOSFIRE K band.  Note that here we discuss galaxies with \Halpha\ detections and Q$_z>1$, while in Section \ref{sec:sample_selection} we discussed the Q$_z=3$ \Halpha\ detected sample.

In the ZFOURGE catalogues used for the ZFIRE sample selection (see \citet{Nanayakkara2016} for details), there were 1159 galaxies (including star-forming and quiescent galaxies) in the COSMOS field with photometric redshifts ($z_{photo}$) within $1.90<z_{\mathrm{photo}}<2.66$.  
160 of these galaxies with $1.90<z_{\mathrm{photo}}<2.66$ were targeted in K band out of which 
128\footnote{Note that this is different from the 142 galaxies mentioned in Section \ref{sec:sample_selection} because the sample of 142 galaxies has a Q$_z=3$, includes galaxies with no ZFOURGE counterparts (see \citet{Nanayakkara2016} for further details) and galaxies with non-optimal ZFOURGE photometry (see \citet{Straatman2016} for further details).} 
were detected with at least one emission line with SNR $>5$. 
 None of the \Halpha\ detected galaxies had spectroscopic redshifts outside the considered redshift interval. 
However, 3 additional galaxies (1 object with Q$_z=2$, 2 objects with Q$_z=3$) fell within $1.90<z<2.66$ due to inaccurate photometric redshifts. 
There were 8 galaxies targeted in K band that did not have \Halpha\ detections but do have other emission line detections (i.e. No \Halpha\ but have \NII, \OIII, \Hbeta\ etc). Furthermore, there were no galaxies that were targeted in K band expecting \Halpha\ but resulted in other emission line detections.

There were 151 objects within $1.90<z<2.66$ with \Halpha\ detections (Q$_z>1$) and 26 of them were flagged as AGN following selection criteria from \citet{Coil2015} and \citet{Cowley2016}. In the remaining 125 galaxies, 8 galaxies did not have matching ZFOURGE counterparts and 8 galaxies had low confidence for redshift detection (Q$_z=2$) from \citet{Nanayakkara2016}. We removed those 16 galaxies from the sample. 
Out of the 109 remaining galaxies, seven are removed due to the following reasons: four galaxies due to negative spectroscopic flux, one galaxy due to multiple objects overlapping in the spectra, and two galaxies due to extreme sky line interference.

Our sample constitutes of the remaining 102 galaxies out of which, 46 have continuum detections (see Section \ref{sec:cont_fit}). Furthermore, 38 (out of which, 16 are continuum detected) galaxies are confirmed cluster members \citep{Yuan2014} and the remaining 64 (out of which, 30 are continuum detected) galaxies comprise of field galaxies. 
32 galaxies targeted with photometric redshifts between $1.90<z<2.66$ show no \Halpha\ emission detection. 
We divide our sample into 3 mass bins with masses between $\log_{10}(\mathrm{M}_\odot)<9.5$, $\mathrm{9.5\leq log_{10}(M_\odot)\leq 10.0}$, $\mathrm{10.0<log_{10}(M_\odot)}$ and show the corresponding data as described above in Table \ref{tab:sample_details}.

We define observing completeness as the percentage of detected galaxies (Q$_z>1$) with photometric redshifts between $1.90<z<2.66$ and calculate it to be  $\sim80\%$.
However, it is possible that the 32 null detections with photometric redshifts  within $1.90<z<2.66$ to have been detected if the ZFIRE survey was more sensitive. We stack the the photometric redshift likelihood functions ($P(z)$) of the ZFIRE targeted galaxies within this redshift range, to compute the expectation of detections based of photometric redshift accuracies (See \citet{Nanayakkara2016} Section 3.2 to further details on how $P(z)$ stacking is performed) . The calculated expectation for \Halpha\ to be detected within K band is $\sim80\%$, which is extremely similar to the observed completeness. Therefore, non-detections rate is consistent with uncertainties in the photometric redshifts. 
To further account for any detection bias, we employ a stacking technique of the non-detected spectra in order to calculate a lower-limit to the stacked EW values. This is further discussed in Section \ref{sec:EW_stacking}.

\begin{landscape}
\begin{deluxetable}{ccccccccccc}
\tabletypesize{\scriptsize}
\tablecaption{ Galaxies selected for the IMF study.
\label{tab:sample_details}}
\tablecolumns{10}
\tablewidth{0pt} 
\tablehead{
\colhead{\mass\tablenotemark{a}} &
\colhead{$\mathrm{N_{ZFOURGE}}$} &
\colhead{$\mathrm{N_{ZFIRE}}$} &
\colhead{$\mathrm{N_{detections}}$} &
\colhead{$\mathrm{N_{outliers}}$} &
\colhead{$\mathrm{N\tablenotemark{b}_{AGN}}$} &
\colhead{$\mathrm{N_{sky}}$} &
\colhead{$\mathrm{N_{selected}}$} &
\colhead{$\mathrm{N_{line\_only}}$} &
\colhead{$\mathrm{N_{null\_detection}}$} &
}
\startdata
$<9.5$     & 568 & 59 &  44 & 0 &  2 & 1 & 41 & 34 &  9 & \\ 
$9.5-10.0$ & 318 & 47 &  39 & 0 &  1 & 0 & 35 & 16 &  6 & \\ 
$10.0<$    & 273 & 54 &  45 & 0 & 20 & 1 & 26 &  6 &  5 & \\ 
Total      &1159 &160 & 128 & 0 & 23 & 2 &102 & 56 & 20 & \\
\enddata
\tablecomments{The columns keys are as follows:\\
\mass: The Mass bin of the galaxies in log$\mathrm{_{10}}$(\msol).\\ 
$\mathrm{N_{ZFOURGE}}$: Number of ZFOURGE galaxies with photometric redshifts within $1.90<z<2.66$ .\\
$\mathrm{N_{ZFIRE}}$: Number of ZFIRE targeted galaxies in K band with photometric redshifts within $1.90<z<2.66$ .\\
$\mathrm{N_{detections}}$: Number of ZFIRE detected galaxies in K band (Q$_z>1$) with  spectroscopic redshifts within $1.90<z<2.66$ .\\
$\mathrm{N_{outliers}}$: Number of ZFIRE detected galaxies with spectroscopic redshifts outside $1.90<z<2.66$ .\\
$\mathrm{N_{AGN}}$: Number of ZFIRE detected galaxies identified as AGN with  spectroscopic redshifts within $1.90<z<2.66$ .\\
$\mathrm{N_{sky}}$: Number of ZFIRE detected galaxies with spectroscopic redshifts within $1.90<z<2.66$  removed from \sample\ due to sky line interference.\\
$\mathrm{N_{selected}}$: Number of ZFIRE detected galaxies selected for the IMF study with  spectroscopic redshifts within $1.90<z<2.66$.\\
$\mathrm{N_{line\_only}}$: Number of galaxies selected for the IMF study which shows no continuum detection with  spectroscopic redshifts within $1.90<z<2.66$ .\\
$\mathrm{N_{null\_detection}}$: Number of ZFIRE K band targeted galaxies with photometric redshifts within $1.90<z<2.66$  and no \Halpha\ detection.}
\tablenotetext{a}{Where applicable spectroscopic redshifts have been used to calculate the stellar masses from FAST.}
\tablenotetext{b}{One galaxy flagged as an AGN doesn't have a matching ZFOURGE counterpart.}
\end{deluxetable}
\end{landscape}

\subsection{Continuum fitting and \Halpha\ EW calculation}
\label{sec:cont_fit}

In this section, we describe our continuum fitting method for our 102 \Halpha\ detected galaxies selected from the ZFIRE survey. 
Fitting a robust continuum level to a spectrum requires nebular emission lines and sky line residuals to be masked. Furthermore, the wavelength interval used for the continuum fit should be sufficient enough to perform an effective fit but should be smaller enough to not to be influenced by the intrinsic SED shape. After extensive testing of various measures used to fit a continuum level, we find the method outlined below to be the most effective to fit a continuum level for our sample.

By visual inspection and spectroscopic redshift of the galaxies in our sample, we mask out the \Halpha\ and \NII\ emission line regions and in the spectra. 
We further mask all known sky-lines by selecting a region $\times 2$ the spectral resolution ($\pm5.5$\AA) of MOSFIRE K band. 
We then use the \texttt{astropy} \citep{Astropy2013} sigma-clipping algorithm to mask out remaining strong features in the spectra.  
These spectra are then used to fit an inverse variance weighted constant level fit, which we consider as the continuum level of the galaxy. 
Three objects fail to give a positive continuum level using this method and for these we perform a $3\sigma$ clip with two iterations without masking nebular emission lines and sky lines.  
Using this method we are able to fit positive continuum levels to all galaxies in our sample.  
We further investigate the robustness of our measures continuum levels in Appendix \ref{sec:cont_Halpha_test} using ZFOURGE photometric data and conclude that our measured continuum level is consistent (or in agreement) with the photometry.

We use two approaches to calculate the \Halpha\ line flux: 1) direct flux measurement and 2) Gaussian fit to sky-line blended and kinematically unresolved emission-lines. Our two methods provide consistent results for emission-lines that are not blended with sky lines (see Appendix \ref{sec:cont_Halpha_test}).
By visual inspection, we selected kinematically resolved (due to galaxy rotation etc.) \Halpha\ emission-lines that were not blended with sky-lines and computed the EW by integrating the line flux.
Within the defined emission-line region, we calculated the  \Halpha\ flux by subtracting the flux at each pixel ($F_i$) by the corresponding continuum level of the pixel ($F_{cont_i}$). 
For the remaining sample, which comprises of galaxies with no strong velocity structure and galaxies with \Halpha\ emission with little velocity structure and/or \Halpha\ contaminated by sky lines, we perform Gaussian fits to the emission lines, to calculate the \Halpha\ flux values. We then subtract the continuum level from the computed \Halpha\ line flux.

Next, we use the calculated \Halpha\ flux along with the fitted continuum level to calculate the observed \Halpha\ EW ($H\alpha_{EW_{obs}}$) as follows:
\begin{subequations}
\begin{equation}
\label{eq:Ha_EW_obs}
H\alpha\ EW_{obs} =  \sum_{i} (1-\frac{F_i - F_{cont_i}}{F_{cont_i}}) \times \Delta \lambda_i
\end{equation} 
where $\Delta \lambda_i$ is the increment of wavelength per pixel.
Finally, using the spectroscopic redshift ($z$) we calculate the rest frame \Halpha\ EW ($H\alpha_{EW_{rest}}$), which we use throughout the paper:
\begin{equation}
\label{eq:Ha_EW_rest}
H\alpha\ EW = \frac{H\alpha\ EW_{obs}}{1+z}
\end{equation} 
\end{subequations}

We calculate EW errors by bootstrap re-sampling the flux of each spectra randomly within limits specified by its error spectrum. We re-calculate the EW iteratively 1000 times and use the 16$\mathrm{^{th}}$ and 84$\mathrm{^{th}}$ percentile of the distribution of the EW measurements as the lower and upper limits for the EW error, respectively. 
Since the main uncertainty arises from the continuum fitting, we do not consider the error of the \Halpha\ flux calculation in our bootstrap process.

The robustness of an EW measurement relies on the clear identification of the nebular emission line and the underlying continuum level. The latter becomes increasingly hard to quantify at high redshift for faint star forming sources due to the continuum not being detected. 
Therefore we derive continuum detection limits to identify robustly measured continua from non-detections.

In order to establish the limit to which our method can reliably measure the continuum, we select 14 2D slits with no continuum or nebular emission line detections to extract 1D spectra. We define extraction apertures using a representative profile of the flux monitor star and perform multiple extractions per slit depending on their spatial size. 
A total of 93 1D sky spectra are extracted and their continuum level is measured by masking out sky lines and performing a sigma-clipping algorithm. 
The error of the sky continuum fit is calculated by bootstrap re-sampling of the sky fluxes 1000 times. We consider the $1\sigma$ scatter of the bootstrapped continuum values to be the error of the sky continuum fit and $1\sigma$ scatter of the flux values used for the continuum fit as the RMS of the flux.

\begin{figure}
\centering
\includegraphics[scale=1.5]{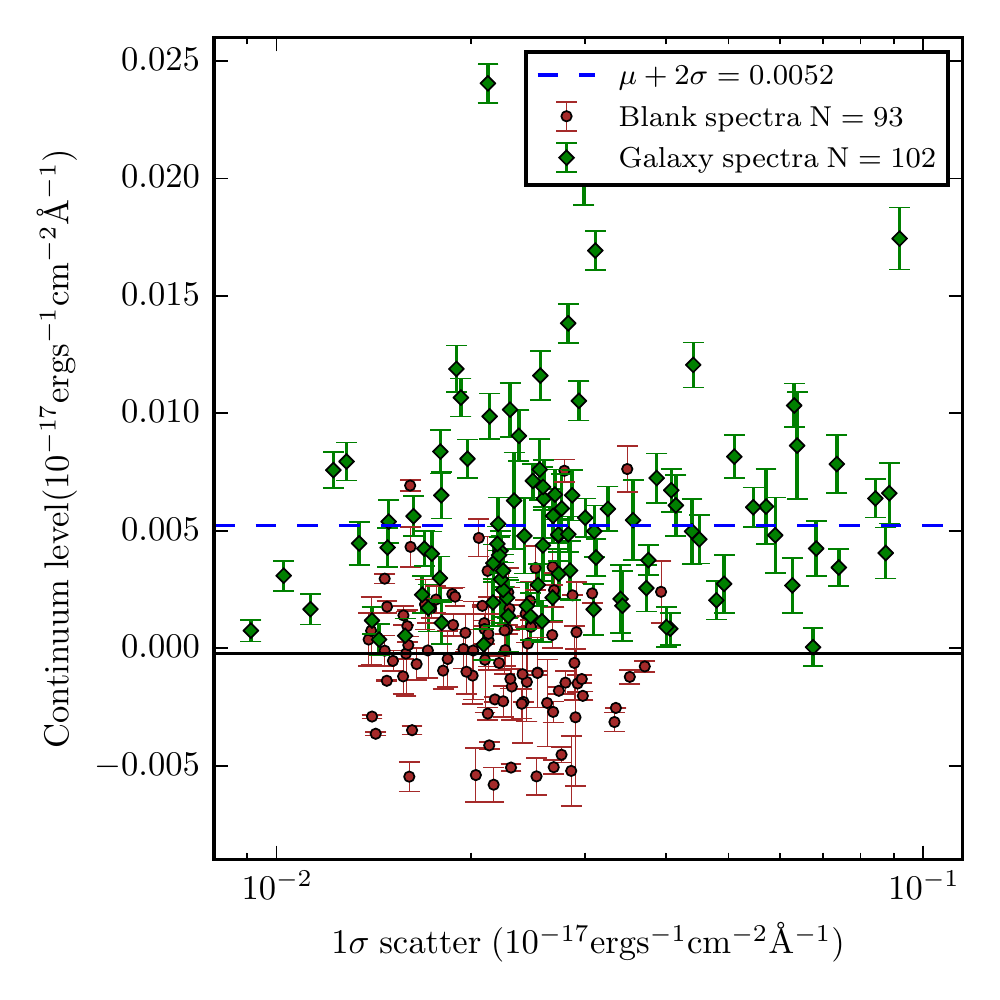}
\caption[The continuum detection levels for the \sample.]{The figure illustrates the continuum detection levels for the \sample. 
The measured continuum level is plot against the $1\sigma$ scatter of the flux values used to fit the continuum level. 
The brown circles represent the continuum levels calculated for the blank slits and the green diamonds represent the continuum level calculated for the IMF sample. 
The blue horizontal line is the $2\sigma$ scatter above the median ($\sim$ 0) for the blank sky regions. Any continua detected above this level of $5.2\mathrm{\times 10^{-20} ergs/s/cm^2/\AA}$ are considered as detected continuum levels. 
\label{fig:noise_levels}
}
\end{figure}

The comparison between the flux continuum level for the sky spectra with the \sample\ spectra are shown in Figure \ref{fig:noise_levels}. 
The median and the 2$\sigma$ standard deviation for the continuum levels of the sky spectra are $\mathrm{-2.3\times 10^{-21} ergs/s/cm^2/\AA}$ and $\mathrm{5.4 \times 10^{-20} ergs/s/cm^2/\AA}$ respectively. 
We consider the horizontal blue dashed line in the Figure \ref{fig:noise_levels}, which is $2\sigma$ above the median sky level, to be our lower limit for the continuum detections in our sample. The \ndetections\ galaxies in our \sample\ with continuum levels above this flux level detections are considered to have a robust continuum detection. For the remaining \nlimits\ galaxies we consider the continuum  measurement as a limit and use it to calculate a lower limit to the \Halpha\ EW values. The redshift distribution of these galaxies is shown by Figure \ref{fig:specz_distribution}.

\subsection{Calculating optical colours}
\label{sec:col_calculation}

Rest frame optical colours for the \sample\ are computed using an updated version of EAZY\footnote{Development version: \url{https://github.com/gbrammer/eazy-photoz/}} \citep{Brammer2008}, which derives the best-fitting SEDs for galaxies  using high quality ZFOURGE photometry to compute the colours. We investigate the robustness of the rest-frame colour calculation of EAZY in Appendix \ref{sec:EAZY colour comparision}.    
The main analysis of our sample is carried out using optical colours derived using two idealized, synthetic box car filters, which probes the bluer and redder regions of the rest-frame SEDs. We select these filters to avoid regions of strong nebular emission lines as explained in Section \ref{sec:PEGASE_models} and Appendix \ref{sec:box car filters}. 

In order to allow direct comparison between ZFIRE $z\sim2$ galaxies with $z=0.1$ SDSS galaxies from HG08, we further calculate optical colours for the \sample\ at $z=0.1$ using blue-shifted SDSS g and r filters.  
Blue shifting the filters simplifies the (g$-$r) colour calculation at $z=0.1$ (\gr) by avoiding additional uncertainties, which may arise due to K corrections if we are to redshift the galaxy spectra to $z=0.1$ from $z=0$.


\section{Galaxy Spectral Models}
\label{sec:PEGASE_models}

In this section, we describe the theoretical galaxy stellar spectral models employed to investigate the effect of IMF, SFHs, and other fundamental galaxy properties in \Halpha\ EW vs optical colour parameter space. We use PEGASE.2 detailed in \citet{Fioc1997} as our primary spectral synthesis code to perform our analysis and further employ Starburst99 \citep{Leitherer1999} and BPASSv2 \citep{Eldridge2016} models to investigate the effects of other exotic stellar features.

PEGASE is a publicly available spectral synthesis code developed by the Institut d'Astrophysique de Paris. 
Once the input parameters are provided, PEGASE produces galaxy spectra for varying time steps, which can be used to evaluate the evolution of fundamental galaxy properties over cosmic time.

\subsection{Model Parameters}

In this paper, we primarily focus on the effect of varying the IMF, SFH, and metallicity on \Halpha\ EW and  optical colour of galaxies. A thorough description of the behaviour of PEGASE models in this parameter space can be found in \cite{Hoversten2008}. 
The parameters we vary are as follows. 
\begin{itemize}
	\item {\bf The IMF :} We follow HG08 and use an IMF with a single power law  as shown by Equation \ref{eq:imf_def_salp}. Models were calculated with varying IMF slopes ($\Gamma$) ranging between -0.5 to -2.0 in logarithmic space. The lower and upper mass cutoffs were set to 0.5\msol\ and 120\msol, respectively. 
	The IMF indication method used in this analysis is dependant on the ability of a star with a specific mass to influence the \Halpha\ emission and the optical continuum level. Stars below 1\msol\ cannot strongly influence the optical continuum and hence this method is not sensitive to probe the IMF below 1\msol. 
	Furthermore stars below $\sim0.5$\msol\ gives no significant variation to the parameters investigated by this method \citep{Hoversten2008}. Therefore we leave the lower mass cutoff at 0.5\msol.  
	The higher mass cutoff of 120\msol\ used is the maximum mass allowed by PEGASE. 
	Though the high mass end has a stronger influence on the IMF identified using this method we justify the 120\msol\ cutoff due to the ambiguity of the stellar evolution models above this mass. 
	Varying the upper mass cutoff has a strong effect on \Halpha\ EW and optical colours. As HG08 showed this is strongly degenerated with changing $\Gamma$. In this work we focus on $\Gamma$ parametrization, noting that changing the cutoff could produce similar effects. 
	We further discuss the degeneracy between the high mass cutoff and the \Halpha\ EW vs optical colours slope in Section \ref{sec:mass_cutoff}. 
 
	\item {\bf The SFH :} Exponentially increasing/declining SFHs, constant SFHs, and starbursts are used. Exponentially declining SFHs are in the form of $\mathrm{SFR(t) = p_2\ exp(-t/p_1) / p_1 }$, with p$_1$ varying from 500 to 1500 Myr. 
	Star bursts are used on top of constant SFHs with varying burst strength and time scales. 
	Further details are provided in Section \ref{sec:simulations}. 

	\item {\bf Metallicity :} Models with consistent metallicity evolution (closed box models with recycling of gas) and models with fixed metallicity of 0.02 are used.  
\end{itemize}

The other parameters we use for the PEGASE models are as follows.
We use Supernova ejecta model B from \cite{Woosley1995} with the stellar wind option turned on.
The fraction of close binary systems are left at 0.05 and the initial metallicity of the ISM is set at 0.
We turn off the galactic in-fall function and the mass fractions of the substellar objects with star formation are kept at 0. Galactic winds are turned off, nebular emissions are turned on and we do not consider extinction as we extinction correct our data.

\begin{figure}
\begin{center}
 \includegraphics[scale=1.2]{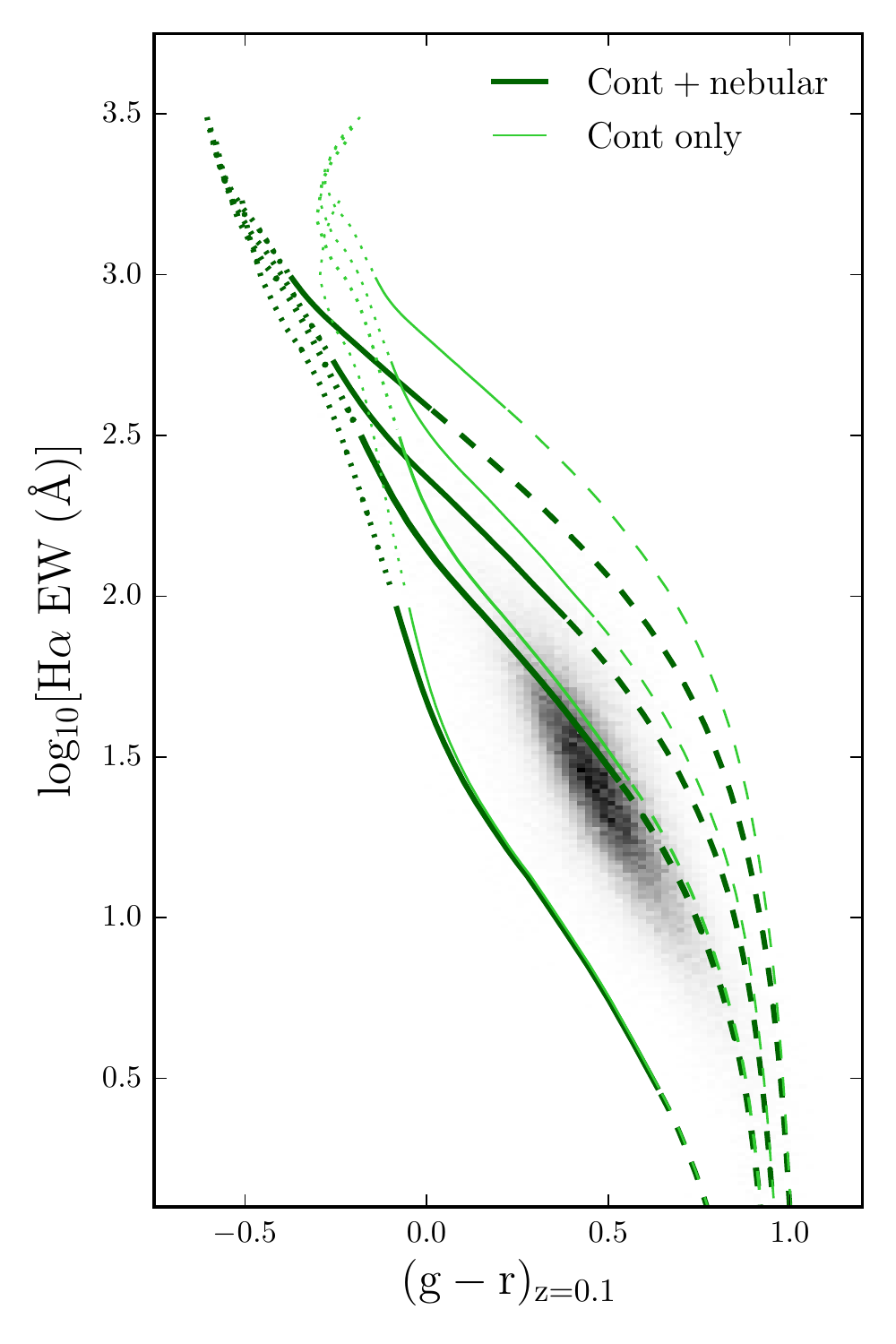} 
 \caption[The evolution of PEGASE SSP galaxies in the \Halpha\ EW vs \gr\ colour space.]{The evolution of PEGASE SSP galaxies in the \Halpha\ EW vs \gr\ colour space. We show four galaxy models with  exponentially declining SFHs computed using identical parameters but varying IMFs. The thick dark green tracks show from top to bottom  galaxies with $\Gamma$ values of -0.5,-1.0,-1.35, and -2.0, respectively. The thin light green tracks follow the same evolution as the thick ones, but the nebular line contribution is not considered for the \gr\ colour calculation. All tracks commence at the top left of the figure and are divided into three time bins. The dotted section of the track corresponds to the first 100 Myr of evolution of the galaxy. The solid section of the tracks show the evolution between 100 Myr - 3100 Myr ($z\sim2$) and the final dashed section shows evolution of the galaxy up to 13100 Myr ($z\sim0$). 
 The distribution of the galaxies from the SDSS HG08 sample are shown by 2D histogram.
 }
\label{fig:PEG_neb_comp_gr}
\end{center}
\end{figure}

As a comparison with HG08, in Figure \ref{fig:PEG_neb_comp_gr} we show the evolution of 4 model galaxies from PEGASE in the \Halpha\ EW vs \gr\ colour space. The models computed with exponentially declining SFHs with p$_{1}=1000$ Myr, varying IMFs, and nebular emission lines agrees well with the SDSS data. However, the evolution of the \gr\ colour shows strong dependence on the nebular emission contribution, specially for shallower IMFs. HG08 never considered the effect of emission lines in \gr\ colours and the significant effect at younger ages/bluer colours are likely to be important for $z\sim2$ galaxies.

\begin{figure}
\begin{center}
 \includegraphics[scale=1.5]{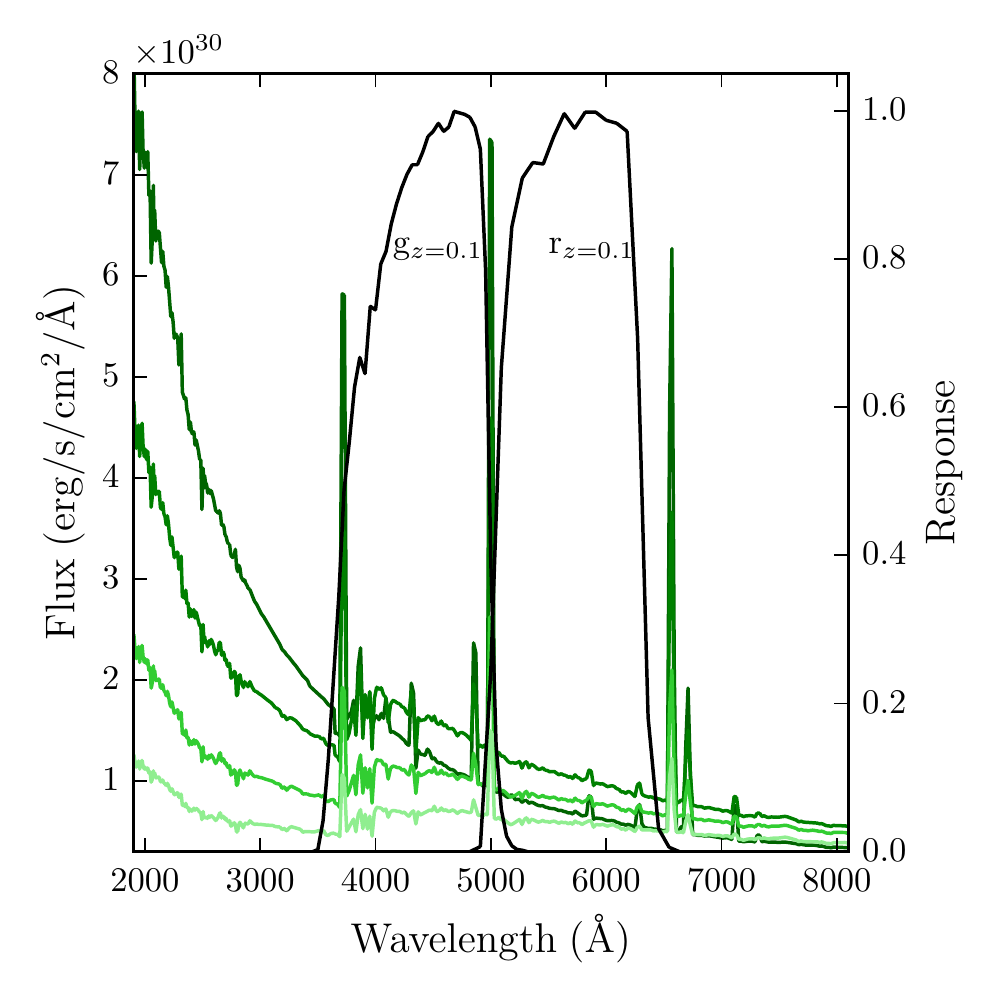} 
 \caption[An example of a model galaxy spectrum generated by PEGASE.]{An example of a model galaxy spectrum generated by PEGASE. Here we show the evolution of the optical wavelength of a galaxy spectra with an exponentially declining SFH and a \salpeter\ with no metallicity evolution. The time steps of the models from top to bottom are: 100 Myr (dark green), 1100 Myr (green), 2100 Myr (limegreen), and 3100 Myr (lightgreen).
 The $g_{z=0.1}$ and $r_{z=0.1}$ filter response functions are overlaid on the figure.  
 }
\label{fig:PEG_spectra}
\end{center}
\end{figure}

Figure \ref{fig:PEG_spectra} shows an example of a synthetic galaxy spectra generated by PEGASE. The galaxy is modelled to have an exponentially declining SFH with $p_1=1000$Myr and a \salpeter\  IMF. Due to the declining nature of the SFR, the stellar and nebular contribution of the galaxy spectra decreases with cosmic time. We overlay the filter response functions of the $g_{z=0.1}$ and $r_{z=0.1}$ filters used in the analysis by HG08. As evident from the spectra, this spectral region covered by the $g_{z=0.1}$ and $r_{z=0.1}$ filters includes strong emission lines such as \OIII\ and \Hbeta. Therefore, the computed \gr\ colours will have a strong dependence on photo-ionization properties of the galaxies.

To mitigate uncertainties in photo-ionization models in our analysis, we employ synthetic filters specifically designed to avoid regions with strong nebular emission lines. We design two box-car filters centred at 3400\AA\ and 5500\AA\ with a width of 450\AA. The rest-frame wavelength coverage of these filters corresponds to a similar region covered by the FourStar $\mathrm{J_1}$ and $\mathrm{H_{long}}$ filters in the observed frame for galaxies at $z=2.1$ and therefore requires negligible K corrections. 
Further details on this filter choice is provided in Appendix \ref{sec:filter choice 340}. 
Henceforth, we refer to the blue filter as [340], the redder filter as [550], and the colour of blue filter - red filter as \boxfil. The \boxfil\ colour evolution of a galaxy is independent of the nebular emission lines.

We also compare results using Starburst99 (S99) \citep{Leitherer1999} models in Appendix \ref{sec:SSP comparision}. We find that PEGASE and S99 models show similar evolution and find that our choice of SSP model (PEGASE or S99) to interpret the IMF of the \sample\ at $z\sim2$ to be largely independent to our conclusions. 
However, stellar libraries that introduce rotational and/or binary stars used in these models do have an influence of the \Halpha\ EWs and \boxfil\ colours, which we discuss in detail in Section \ref{sec:ssp_issues}.


\subsection{Comparison to \Halpha\ EW \& optical colours at $z\sim2$}
\label{sec:results}

\begin{figure}
\centering
\includegraphics[trim=5 10 10 0, clip, scale=0.75]{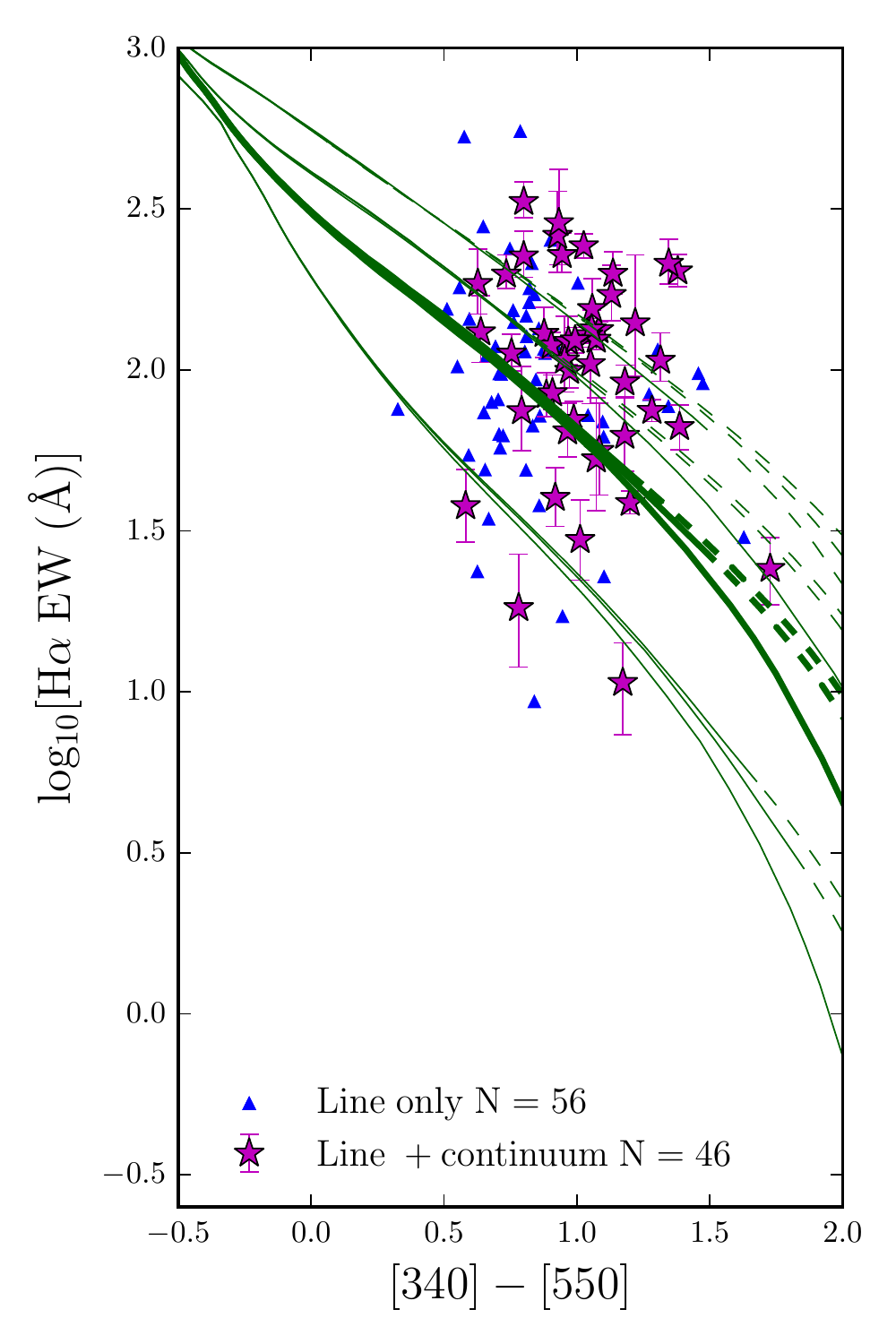}
\includegraphics[trim=5 10 10 0, clip, scale=0.75]{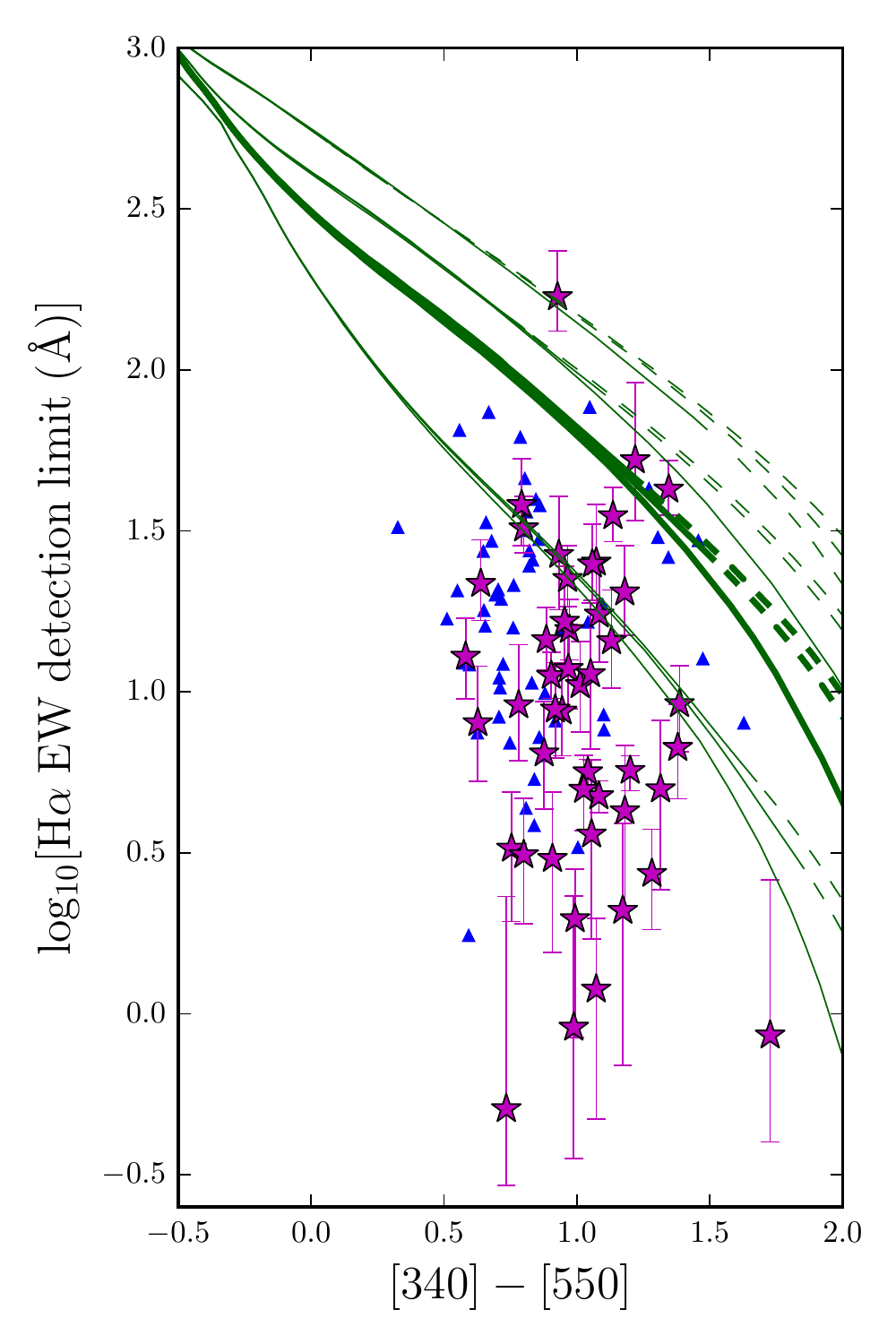}
\caption[The \Halpha\ EW vs \boxfil\ colour distribution of the \sample.]{The \Halpha\ EW vs \boxfil\ colour distribution of the \sample. No dust corrections have been applied to the observed data. Galaxies with \Halpha\ and continuum detections are shown by magenta stars while galaxies only with \Halpha\ detections (and continuum from $1\sigma$ upper limits) are shown as $1\sigma$ lower limits on EW by blue triangles. The errors for the continuum detected galaxies are from bootstrap re-sampling.
The solid (t $<3200$ Myr) and dashed lines (t $>3200$ Myr) are SSP models computed from PEGASE. Similar to Figure \ref{fig:PEG_neb_comp_gr}, we compute models for 4 IMFs with $\Gamma$ values of $-0.5,-1.0,-1.35$ (this is the thick set of tracks which is similar to the IMF slope inferred by Salpeter), and $-2.0$. Each set of tracks from top to bottom represent these IMF in order. 
For each IMF we compute three models with exponentially declining SFHs with varying p$_1$ values. From top to bottom for each IMF these tracks represent p$_1$ values of 1500 Myr, 1000 Myr, and 500 Myr.
{\bf Left:} The \Halpha\ EW vs \boxfil\ colours of the \sample.  
{\bf Right:} Similar to the left figure but the \Halpha\ EW has been calculated using $1\sigma$ detection limits of the \Halpha\ flux values to demonstrate the sensitivity limits of our EW measurements. }
\label{fig:EW_no_dust_corrections}
\end{figure}

We explore the IMF of $z\sim2$ star-forming galaxies using \Halpha\ EW values from ZFIRE spectra and rest-frame optical colours from ZFOURGE photometry. Our observed sample used in our analysis is shown in Figure \ref{fig:EW_no_dust_corrections}.
The left panel shows the distribution of \Halpha\ EW and \boxfil\ colours of the \sample\ before dust corrections are applied. 
We overlay model galaxy tracks generated by PEGASE for various IMFs. All models are computed using an exponentially declining SFH, but with varying time constants (p$_1$) as shown in the figure caption. For a given IMF, smoothly varying monotonic SFHs have very similar loci in this parameter space. 
The thick set of models (third from top) shows a slope with $\Gamma=-1.35$, which is similar to the Salpeter slope. 
Galaxies above these tracks are expected to contain a higher fraction of higher mass stars in comparison to the mass distribution expected following a Salpeter  IMF. Similarly galaxies below these tracks are expected to contain a lower fraction of high mass stars. 
Galaxies have a large spread in this parameter space but we expect this scatter to decrease when dust corrections are applied to the data as outlined in Section \ref{sec:dust_corrections}.

We note the large scatter of the \Halpha\ EW values with respect to the Salpeter IMF, especially the large number of high EW objects ($\gtrsim0.5$ dex above the Salpeter locus). Could this simply be due to the \sample\ only detecting \Halpha\ emissions in bright objects?\ i.e. a sample bias. First, we note our high completeness of $\sim80\%$ for \Halpha\ detections (Section \ref{sec:completeness}). Secondly, our \Halpha\ flux limits are actually quite faint. To show this explicitly, we define \Halpha\ flux detection limits for our sample using $1\sigma$ detection thresholds for each galaxy parametrised by the integration of the error spectrum within the same width as the emission line. 
Figure \ref{fig:EW_no_dust_corrections} (right panel) shows the \Halpha\ EW calculated using \Halpha\ flux detection limits, which illustrates the distribution of the \sample\ if the \Halpha\ flux was barely detected.
The \Halpha\ EW of the continuum detected galaxies decrease by $\sim1$ dex which suggest that our EW detection threshold is not biased towards higher \Halpha\ EW values. \\

\newpage

Similar to IMF, there are a number of effects that may account for the clear disagreement between the observed data and models. In subsequent sections we explore effects from\\
$\bullet$ dust (Section \ref{sec:dust}),\\
$\bullet$ observational bias (Section \ref{sec:observational_bias}),\\
$\bullet$ star bursts (Section \ref{sec:star_bursts}),\\
$\bullet$ stellar rotation (Section \ref{sec:stellar_rotation}),\\
$\bullet$ binary stellar systems (Section \ref{sec:binaries}),\\ 
$\bullet$ metallicity (Section \ref{sec:model_Z}), and\\
$\bullet$ high mass cutoff (Section \ref{sec:mass_cutoff})\\
in SSP models to explain the distribution of \Halpha\ EW vs optical colours of the \sample\ without invoking IMF change.

\chapter{The Stellar Initial Mass Function\\ Analysis}
\label{chap:imf_analysis}

This Chapter performs a thorough analysis of the factors that effect the distribution of galaxies in the \Halpha\ EW vs optical colour parameter space. By analysing the effects from dust, star bursts, stellar rotation, binaries, metallicity, and upper mass cutoff, strong constraints are made on the stellar IMF of the \sample. 
I further discuss implications of a non-universal IMF for high-$z$ star-forming galaxies in this chapter.

\section{Is Dust the Reason?}
\label{sec:dust}

As summarized by \citet{Kennicutt1983}, the dust vector is nearly orthogonal to the IMF change vector, and therefore, we expect the tracks in the \Halpha\ EW vs optical colour parameter space to be independent of galaxy dust properties. 
In this section, we describe galaxy dust properties. We explain how dust corrections were applied to the data and their IMF dependence and explore the difference in reddening between stellar and nebular emission line regions as quantified by \citet{Calzetti2000} for $z\sim0$ star-forming galaxies.

We use FAST \citep{Kriek2009} with  ZFIRE spectroscopic redshifts from \citet{Nanayakkara2016} and multi-wavelength photometric data from ZFOURGE \citep{Straatman2016} to generate estimates for  stellar attenuation (Av) and stellar mass for our galaxies. 
FAST uses SSP models from \citet{Bruzual2003} and a $\chi^2$ fitting algorithm to derive ages, star-formation time-scales, and dust content of the galaxies. All FAST SED templates  have been calculated assuming solar metallicity, \citet{Chabrier2003} IMF, and \citet{Calzetti2000} dust law.  We refer the reader to \citet{Straatman2016} for further information on the use of FAST to derive stellar population properties in the ZFOURGE survey.


\subsection{Applying SED derived dust corrections to data}
\label{sec:dust_corrections}

We use stellar attenuation values calculated by FAST to perform dust corrections to our data. 
First, we consider the dust corrections for rest frame \Halpha\ EWs and then we correct the \boxfil\ colours. 

By using \cite{Cardelli1989} and \citet{Calzetti2000} attenuation laws to correct nebular and continuum emission lines, respectively, we derive the following equation to obtain dust corrected \Halpha\ EW ($EW_i$) values:
\begin{equation}
\label{eg:EW_dust_corrected}
log_{10}(EW_i) = log_{10}(EW_{obs}) + 0.4A_c(V)(0.62f-0.82)
\end{equation}
where $EW_{obs}$ is the observed EW, $A_c$ is the SED derived continuum attenuation, and $f$ is the difference in reddening between continuum and nebular emission lines. 

\cite{Calzetti2000} found a $f\sim2$ for $z\sim0$ star-forming galaxies, which we use for our analysis under the assumption that the actively star forming galaxies at $z\sim0$ are analogues to star-forming galaxies at $z\sim2$. 
Henceforth, for convenience we refer to $f=1/0.44$ \cite{Calzetti2000} value as $f=2$.
We further show key plots in this analysis using a dust correction of $f=1$ to consider equal dust extinction between stellar and ionized gas regions. This is driven by the assumption that A and G stars that contribute to the continuum of $z\sim2$ star-forming galaxies are still associated within their original birthplaces similar to O and B stars due to insufficient time for the stars to move away from the parent birth clouds within the $<3$ Gyr time scale.

Similarly, using \citet{Calzetti2000} attenuation law we obtain dust corrected fluxes for the [340] and [550] filters as follows:
\begin{subequations}
\begin{equation}
\label{eq:BC340 dust corrected}
f([340]) = f([340]_{obs}) \times 10^{0.4 \times 1.56 A_c(V)}
\end{equation}
\begin{equation}
\label{eq:BC550 dust corrected}
f([550]) = f([550]_{obs}) \times 10^{0.4 \times 1.00 A_c(V)}
\end{equation}
\end{subequations}
A complete derivation of the dust corrections presented here are shown in Appendix \ref{sec:dust_derivation}.

Figure \ref{fig:EW_with_dust_corrections} shows the distribution of our sample before and after dust corrections are applied.  
In the left panels we show our sample before any dust corrections are applied, with arrows in cyan denoting dust vectors for varying $f$ values. 
It is evident from the figure that the galaxies in this parameter space are very dependent on the $f$ value used. For $f$ values of 1 and 2, the effect of dust is orthogonal to IMF change, while values above 2 may influence the interpretation of the IMF. 
We note that $f>2$ makes the problem of high \Halpha\ EW objects worse, so we do not consider such values further. 

Figure \ref{fig:EW_with_dust_corrections} right panels show the dust corrections applied to both \Halpha\ EW and the \boxfil\ colours for the \sample. Without the effect of dust, we expect the young star forming galaxies to show similar bluer colours and therefore, the narrower \boxfil\ colour space occupied by our dust corrected sample is expected. 
With a dust correction of $f=1$, majority of the galaxies lie below the $\Gamma=-1.35$ IMF track with only $\sim1/5$th of galaxies showing higher \Halpha\ EWs. 
However, with $f\sim2$ dust correction, there is a significant presence of galaxies with extremely high \Halpha\ EW values for a given \boxfil\ colour inferred from a $\Gamma=-1.35$ IMF and \around60\% of the galaxies lie above this IMF track. 

Even $\sim\times2$ larger errors for the individual \Halpha\ EW measurements cannot account for the galaxies with the largest deviations from the Salpeter tracks. 
The change of $f$ from $2\Rightarrow1$ decreases the median \Halpha\ EW value by $\sim0.2$dex. However, galaxies still show a large scatter in \Halpha\ EW vs \boxfil\ colour parameter space with points lying well above the Salpeter IMF track.

\begin{figure}
\includegraphics[scale=0.75]{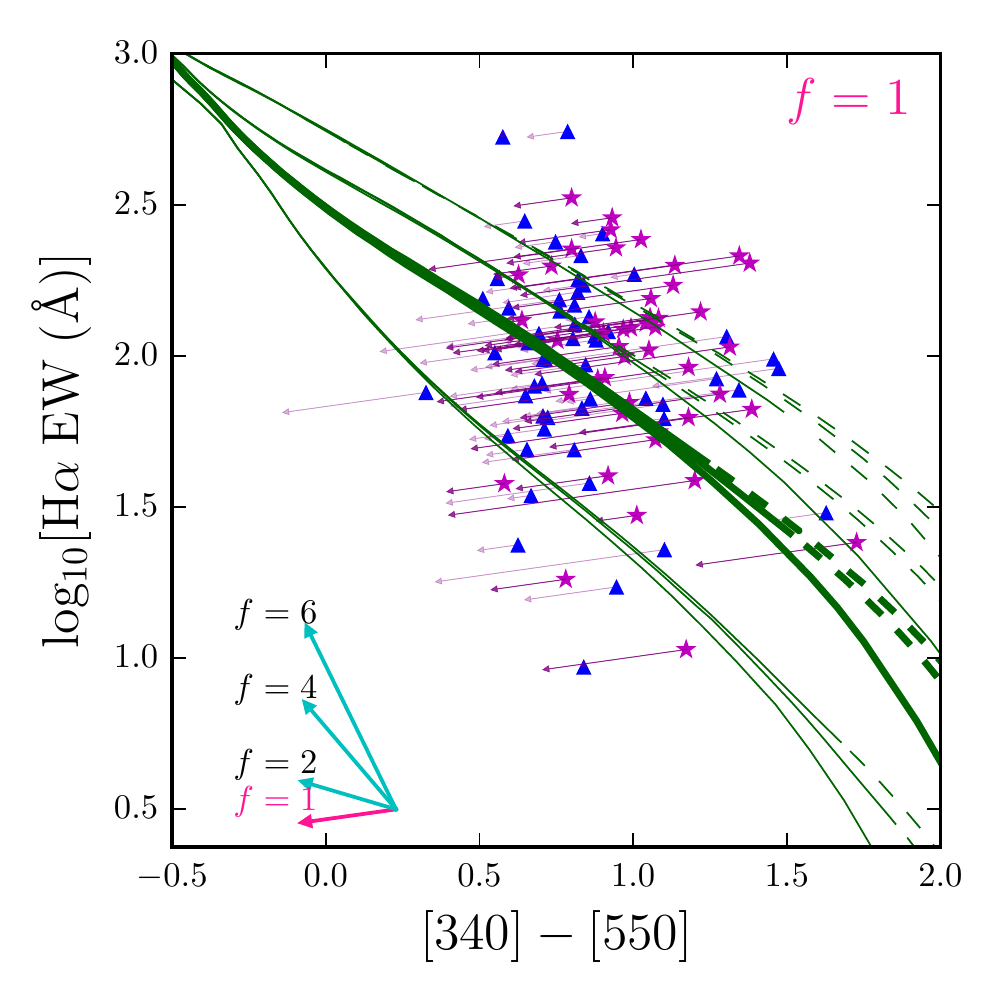}
\includegraphics[scale=0.75]{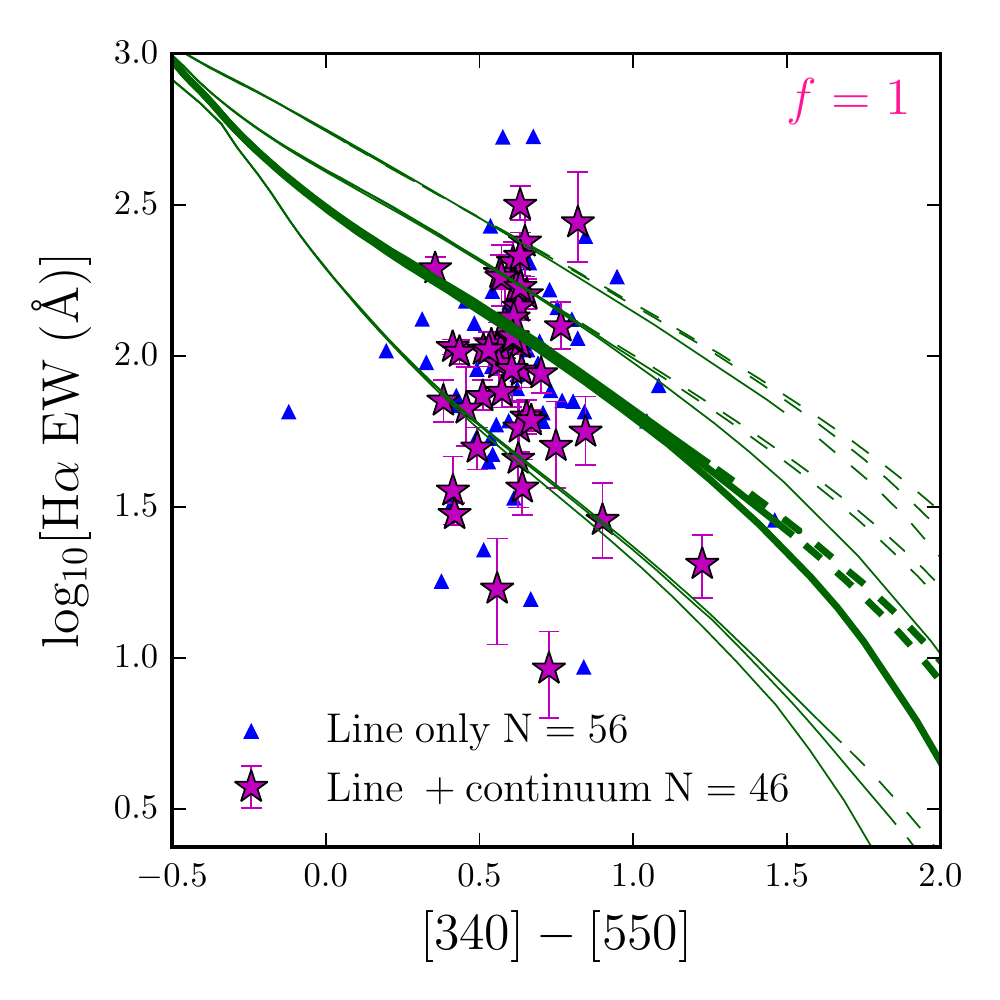}
\includegraphics[scale=0.75]{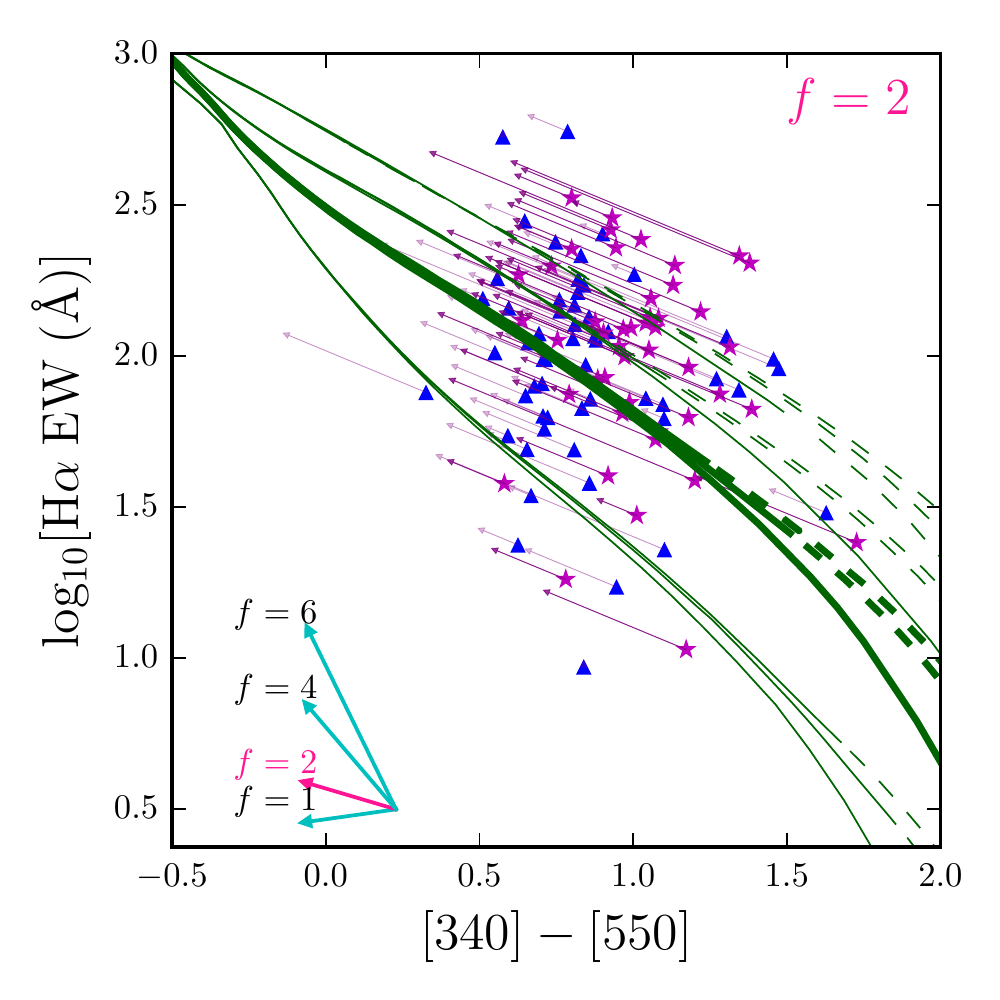}
\includegraphics[scale=0.75]{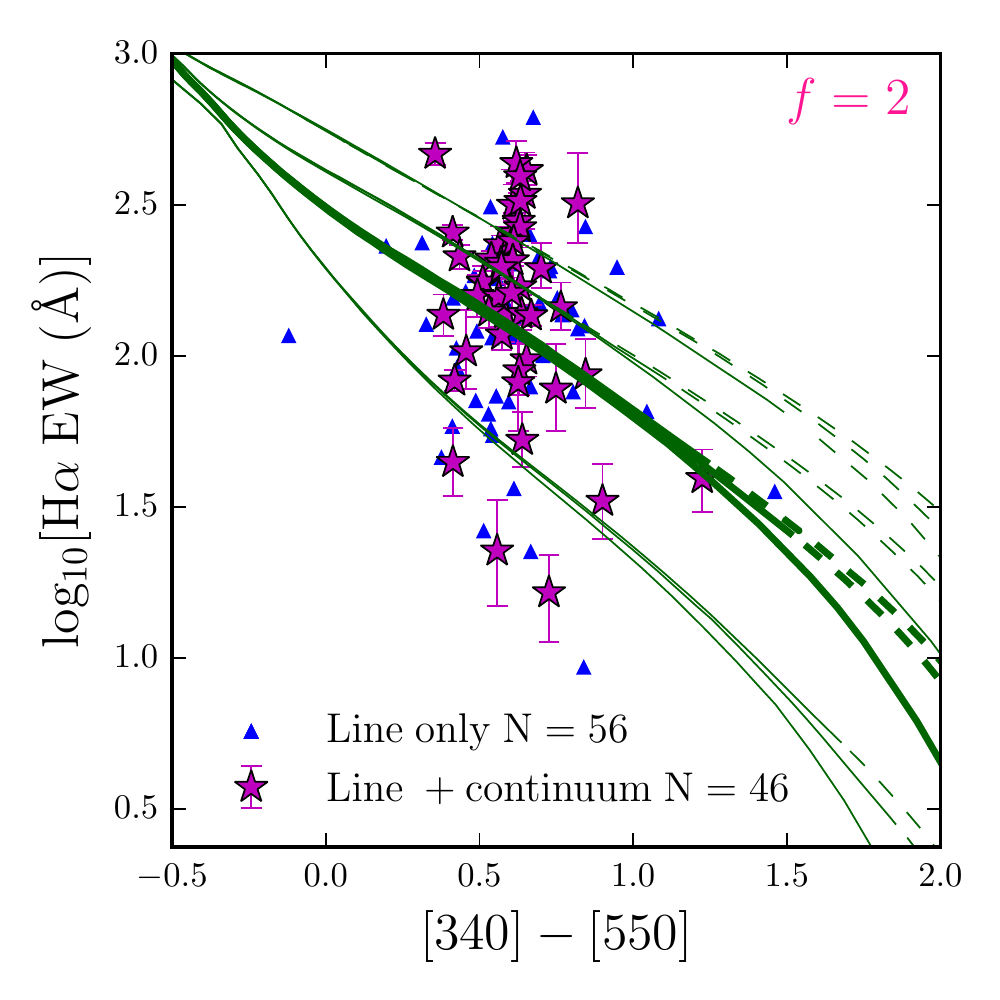}
\caption[The dust correction process of the \sample.]{The dust correction process of the \sample. This figure is similar to Figure \ref{fig:EW_no_dust_corrections} but shows the intermediate and final step of the dust correction process. 
{\bf Top Left:} Here we show the dust correction vector for each galaxy in our sample, computed following the prescriptions explained in Section \ref{sec:dust_corrections}. In summary we use \cite{Calzetti2000} attenuation law to correct the continuum levels and the optical \boxfil\ colours. We use \cite{Cardelli1989} attenuation law to dust correct the nebular emission lines. We use attenuation values calculated by FAST and apply equal amount of extinction to continuum and nebular emission line regions. The purple arrows denote the dust vector for the individual galaxies. Galaxies with no arrows have 0 extinction. 
The arrows in the bottom left corner show the dust vector for a galaxy with Av=0.5 but with varying \cite{Calzetti1994} factors, which is shown as \emph{f} next to each arrow. 
{\bf Top Right:} The final \Halpha\ EW vs \boxfil\ colour distribution of the dust corrected \sample\ with $f=1$. Most galaxies lie at $([340]-[550])\sim0.6$, which corresponds to $\sim850$ Myr of age following the Salpeter IMF tracks. 
{\bf Bottom Left:} Similar to top left panel, but with a higher amount of extinction to nebular emission line regions compared ($\sim\times2$) to the continuum levels.
{\bf Bottom Right:} The final \Halpha\ EW vs \boxfil\ colour distribution of the dust corrected \sample\ with $f=2.27$. 
}
\label{fig:EW_with_dust_corrections}
\end{figure}

The form of the attenuation law of galaxies at $z>2$ show conflicting results between studies.  Observations from the Atacama Large Millimeter Array (ALMA)  have indicated the presence of galaxies with low infra-red (IR) luminosities suggesting galaxies with attenuation similar to the Small Magellanic Cloud \citep[SMC.,][]{Capak2015,Bouwens2016b}. 
\citet{Reddy2015} showed a SMC like attenuation curve for $z\sim2$ galaxies at $\lambda\gtrsim2500$\AA\ and a \citet{Calzetti2000} like attenuation curve for the shorter wavelengths.
However, \emph{HST} grism and SED fitting analysis of galaxies at $z\sim2-6$ has shown no deviation in the attenuation law derived by \citet{Calzetti2000} for local star-forming galaxies. 
Such conflicts are also apparent in simulation studies, where \citet{Mancini2016} showed evidence for SMC like attenuation with clumpy dust regions while \citet{Cullen2017} have shown that galaxies contain similar dust properties as inferred by \citet{Calzetti2000}. 

\begin{figure}
\centering
\includegraphics[trim=0 0 0 0 , clip, scale=0.7]{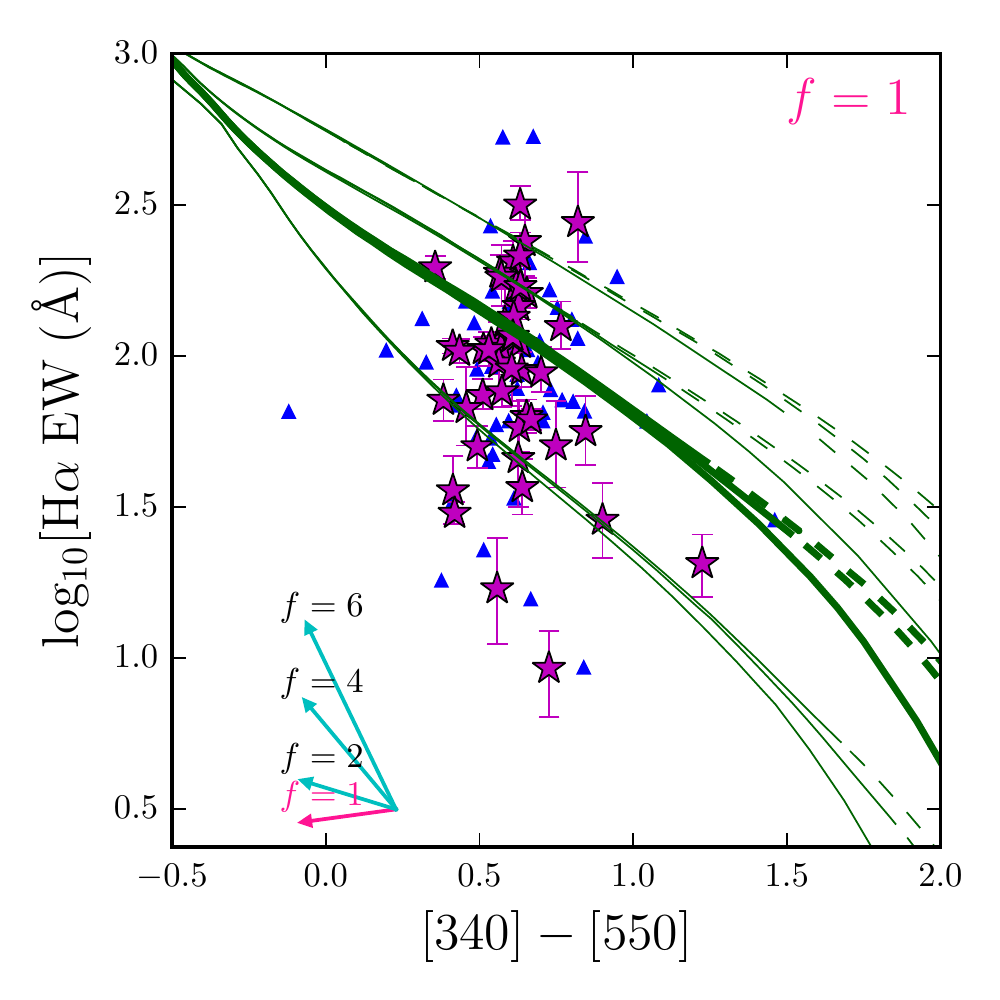}
\includegraphics[trim=0 0 0 0 , clip, scale=0.7]{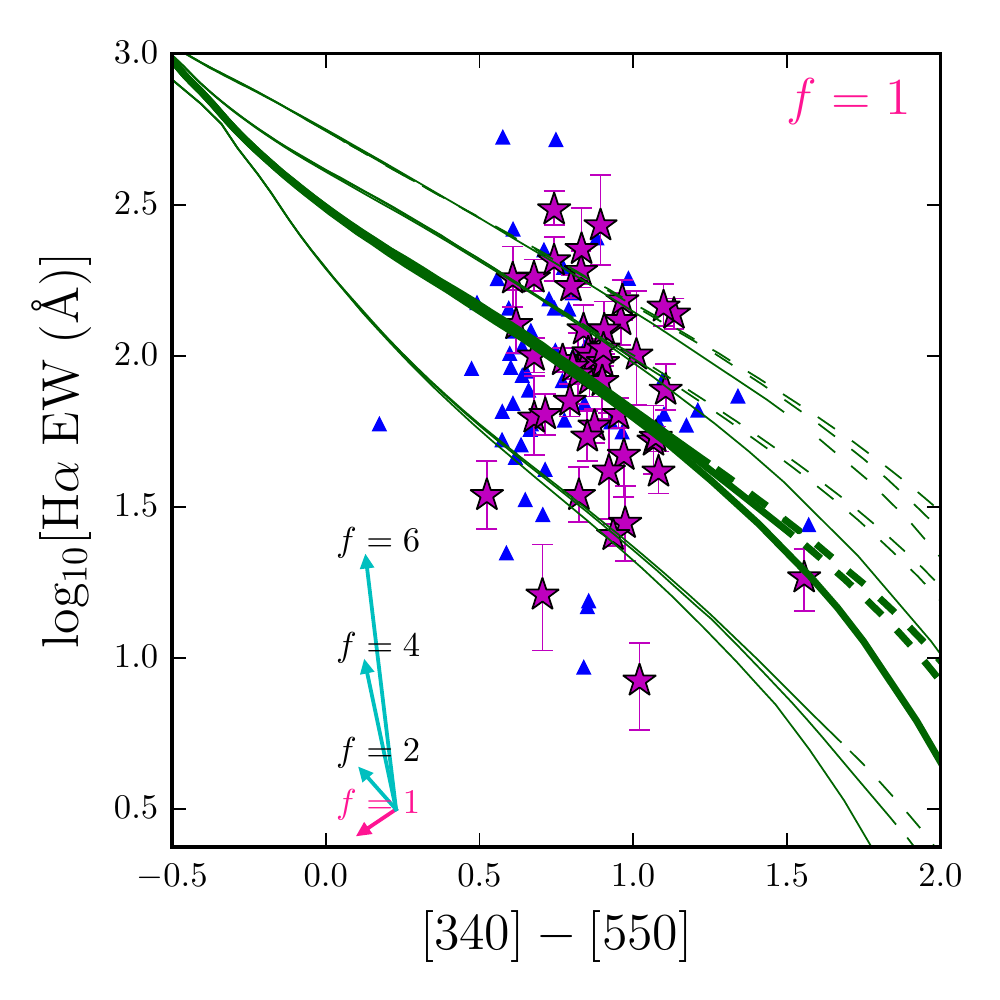}
\includegraphics[trim=0 0 0 0 , clip, scale=0.7]{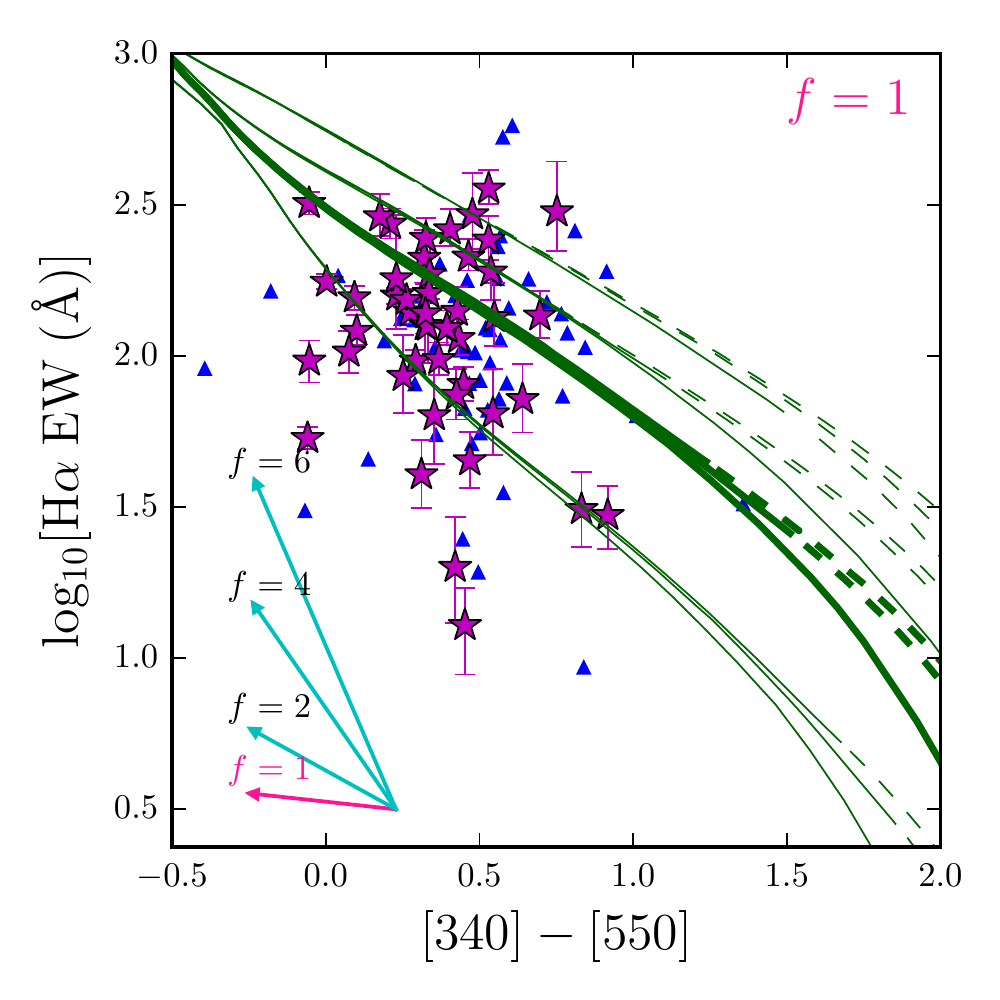}
\caption[Dust correction applied to the \sample\ using different dust laws.]{Here we show the distribution of the \sample\ in the \Halpha\ EW vs \boxfil\ colour parameter space with the \Halpha\ continuum and optical colour dust corrections applied following 
{\bf Top Left:} \cite{Calzetti2000} attenuation law, 
{\bf Top Right:} \cite{Pei1992} SMC attenuation law, and 
{\bf Bottom:} \cite{Reddy2015} attenuation law. 
In all panels \cite{Cardelli1989} attenuation law has been used to dust correct the nebular emission lines with equal amount of extinction applied to continuum and nebular emission line regions $(f=1)$.
The arrows in the bottom left corner show the dust vector for a galaxy with Av=0.5 but with varying \cite{Calzetti1994} factors, which is shown as \emph{f} next to each arrow. 
}
\label{fig:EW_with_dust_corrections_with_various_dust_laws}
\end{figure}

In order to understand the role of  dust laws in the \Halpha\ EW vs \boxfil\ colour parameter space we compare the results using other dust laws such as \citet{Pei1992} SMC dust law and \citet{Reddy2015} $z\sim2$ dust law to correct the stellar contributions (\Halpha\ continuum and optical colours).
A comparison between the distribution of galaxies obtained with different dust laws for a given $f$ are shown by Figure \ref{fig:EW_with_dust_corrections_with_various_dust_laws}. 
The fraction of galaxies with $\Delta$EW $>2\sigma$ from the $\Gamma=-1.35$ IMF track with $f=1 (f=2)$ dust corrections are $\sim20\% (\sim45\%),\sim35\% (\sim75\%),$ and $\sim15\% (\sim55\%)$ for \citet{Calzetti2000}, \citet{Pei1992} SMC, and \citet{Reddy2015} dust laws, respectively.  
However, we refrain from interpreting the differences in the distributions of the sample between the considered dust laws because the attenuation values used in the ZFIRE/ZFOURGE surveys have been derived from SED fitting by FAST using a \citet{Calzetti2000} dust law. 
Compared to the adopted dust law, the change in the value of $f$ has a stronger influence on the galaxies in our parameter space and can significantly affect the EW values, which is discussed further in Section \ref{sec:calzetti_factor}.

To investigate differences between our $z\sim2$ sample with HG08 $z\sim0$ sample, we derive dust corrections to the \gr\ colours. 
Using the following equations to apply dust corrections to g$_{0.1}$ and r$_{0.1}$ fluxes we recalculate the \gr\ colours for the \sample.   
\begin{subequations}
\begin{equation}
\label{eq:g dust corrected}
f(g_i)_{0.1} = f(g_{obs})_{0.1} \times 10^{0.4 \times 1.25 A_c(V)}
\end{equation}
\begin{equation}
\label{eq:r dust corrected}
f(r_i)_{0.1} = f(r_{obs})_{0.1} \times 10^{0.4 \times 0.96 A_c(V)}
\end{equation}
\end{subequations}

\begin{figure}
\centering
\includegraphics[scale=0.8]{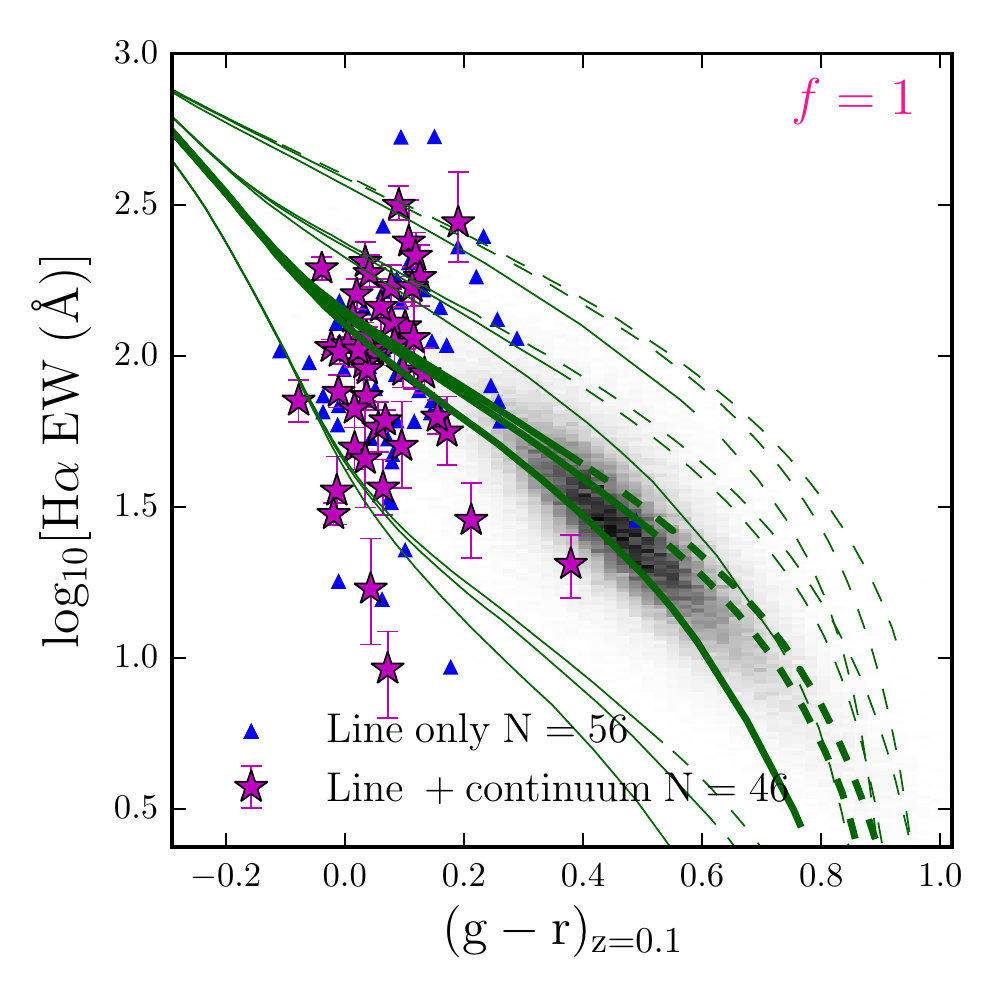}
\includegraphics[scale=0.8]{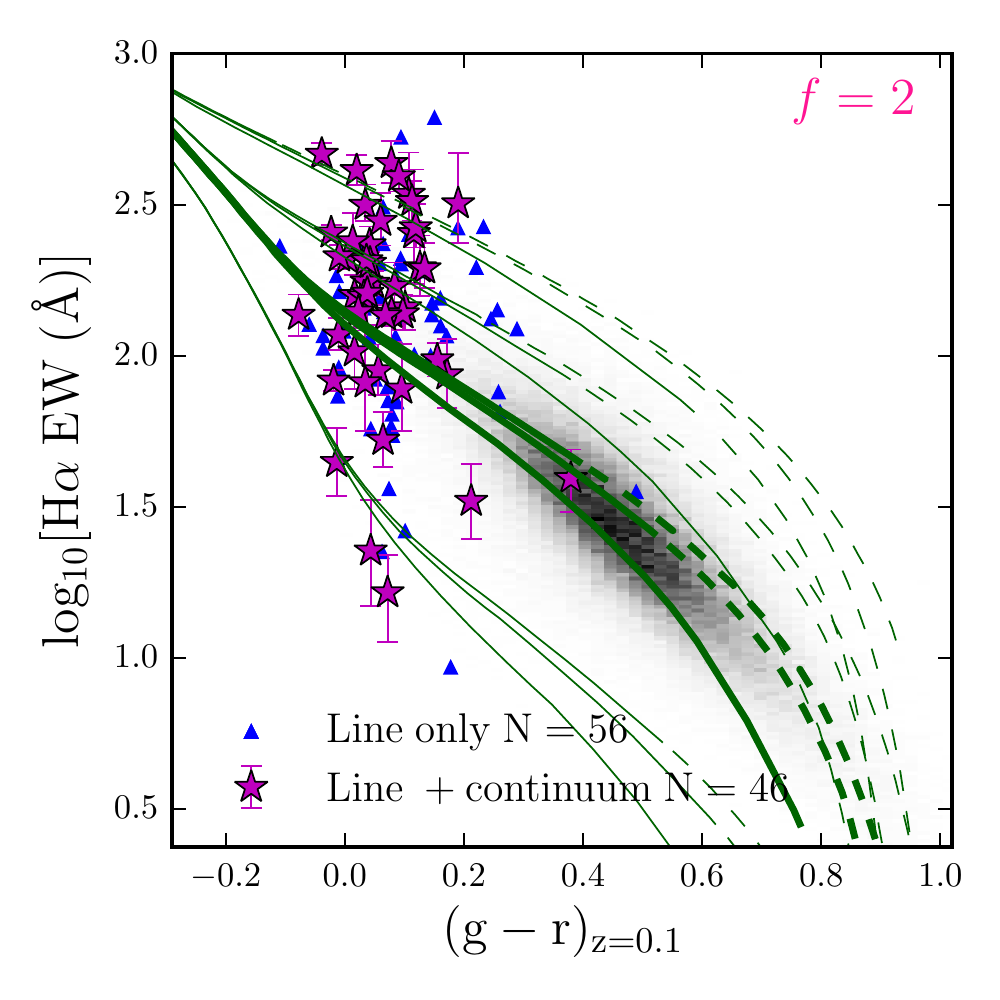}
\caption[ Comparison of \sample\ Hα EW and \gr\ colours with SDSS.]{Comparison of the \Halpha\ EW and \gr\ colours of the $z\sim2$ \sample\ with the HG08 $z\sim0$ sample. The HG08 sample is shown by the 2D grey histogram. The PEGASE models shown correspond to varying IMFs: from top to bottom $\Gamma=-0.5,-1.0,-1.35,\ \mathrm{and}\ -2.0$. Similar to Figure \ref{fig:EW_no_dust_corrections}, for each IMF we show three models with exponentially declining SFHs with varying p$_1$ values (from top to bottom p$_1$=1500 Myr, 1000 Myr, and 500 Myr). Model tracks at t$>3200$ Myr are shown by the dashed lines. 
{\bf Top:}  \sample\ with dust corrections applied with a $f=1$.
{\bf Bottom:} \sample\ with dust corrections applied with a $f=2$.
Note that HG08 uses a $f=2$ in their dust corrections. 
}
\label{fig:EW_HG08_comp}
\end{figure}

We show the \Halpha\ EW vs \gr\ colour comparison between ZFIRE and SDSS samples in Figure \ref{fig:EW_HG08_comp}. The dust corrections for the \sample\ has been performed using a $f=1$ and $f=2$. 
Similar to the \boxfil\ colour relationship, there is a significant presence of galaxies with extremely high \Halpha\ EW values and \around60\%of the galaxies lie above the Salpeter IMF track when dust corrections are applied with a $f=2$.
Furthermore, the $z\sim2$ sample shows much bluer colours compared to HG08 sample, which we attribute to the younger ages ($\sim850$ Myr inferred from tracks with a Salpeter IMF) and the higher SFRs of galaxies at $z\sim2$.

\begin{figure}
\centering
\includegraphics[scale=1.5]{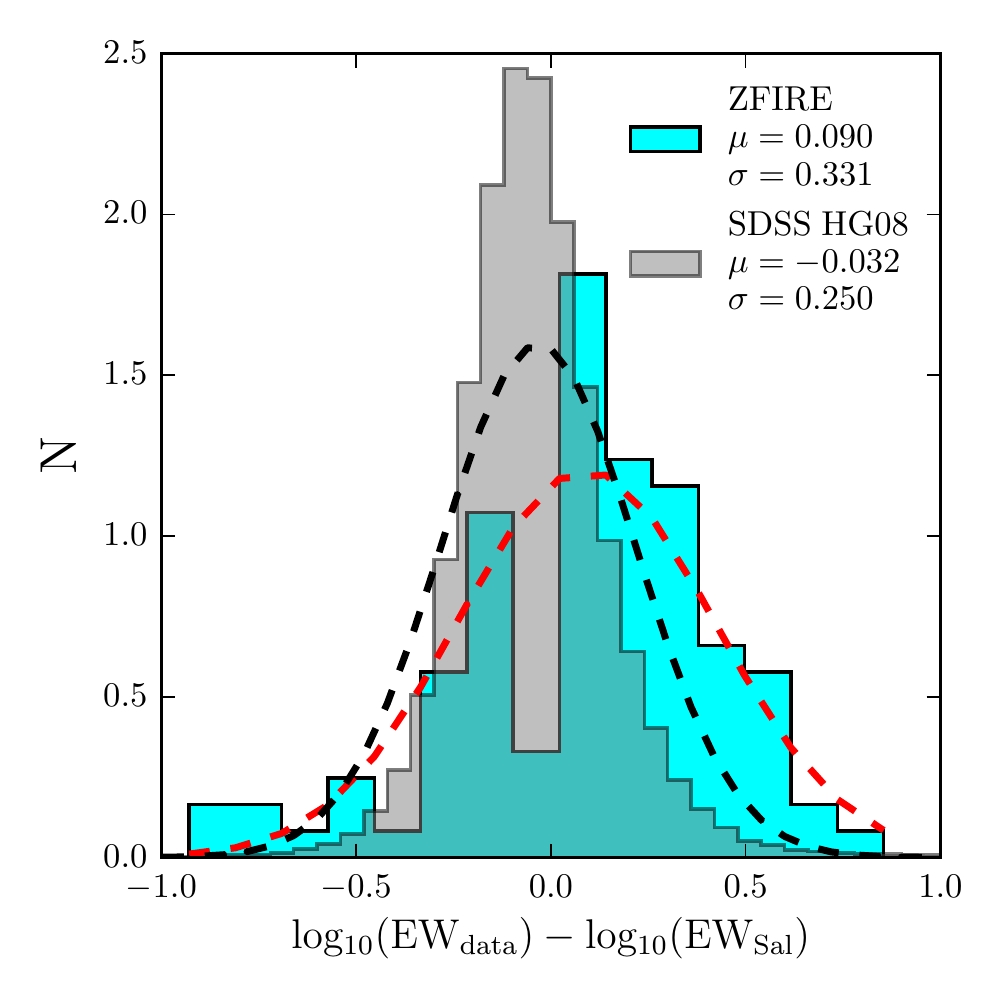}
\caption[Deviations of the observed \Halpha\ from the canonical Salpeter IMF expectations.]{Deviations of the observed \Halpha\ from the canonical Salpeter like IMF tracks in the \Halpha\ EW vs \gr\ colour space. We show the $z\sim0$ SDSS HG08 sample (grey--black) and the $z\sim2$ \sample (cyan--red). Both histograms are normalized to an integral sum of 1 and best-fitting Gaussian functions are overlaid. The parameters of the Gaussian functions are shown in the legend.
}
\label{fig:EW_deviations_from_salp}
\end{figure}

In Figure \ref{fig:EW_deviations_from_salp} we use the $\Gamma=-1.35$ IMF tracks to compute the deviation of observed \Halpha\ EW values from a canonical Salpeter like IMF. 
For each \gr\ galaxy colour we calculate the expected \Halpha\ EW using the standard PEGASE model computed using an exponential decaying SFH with a p$_1=1000$ Myr.  
We then calculate the deviation between the observed values to the expected values. Only the $f=2$ scenario is considered here to be consistent with the dust corrections applied by HG08.
Our results suggest that the ZFIRE sample exhibits a log-normal distribution with a mean and a standard deviation of 0.090 and 0.321 units, respectively.  Similarly for the HG08 sample, the values are distributed with a mean and a standard deviation of -0.032 and 0.250 units. 
Compared to HG08, the \sample\ shows a larger scatter and favours higher \Halpha\ EW values for a given Salpeter like IMF. 
A simple two sample K-S test for the \sample\ and HG08 gives a Ks statistic of 0.37 and a P value of $1.32\times10^{-12}$, which suggests that the two samples are distinctively different from each other. 
In subsequent sections, we further explore whether the differences between the $z\sim0$ and $z\sim2$ populations are driven by IMF change or other stellar population parameters.

\subsection{IMF dependence of extinction values}
\label{sec:further_dust}

Dust corrections applied to the \sample, as explained in Section \ref{sec:dust_corrections}, are derived from FAST \citep{Kriek2009} using best-fitting model SEDs to ZFOURGE photometric data. 
FAST uses a grid of SED template models to fit galaxy photometric data to derive the best-fitting redshift, metallicity, SFR, age, and Av  values for the galaxies via a $\chi^2$  fitting technique. Even though these derived properties may show degeneracy with each other (see \citet{Conroy2013} for a review), in general FAST successfully describes observed galaxy properties of deep photometric redshift surveys \citep{Whitaker2011,Skelton2014,Straatman2016}. FAST has a limited variety of stellar templates, and therefore, we cannot explore the effect of varying IMFs on the FAST derived extinction values.

In order to examine the role of IMF on derived extinction values, we compare the distribution of ZFIRE rest-frame UV and optical colours with PEGASE model galaxies. 
Following the same procedure used to derive the [340] and [550] filters, we design two boxcar filters centred at 1500\AA\ ([150]) and 2600\AA\ ([260]) with a length of 675\AA. The wavelength regime covered by these two filters approximately correspond to the B and I filters in the observed frame for galaxies at $z\sim2$ (further information is provided in Appendix \ref{sec:filter choice 150}). Therefore, K corrections are small and the computed values are robust.

\begin{figure}
\centering
\includegraphics[scale=0.8]{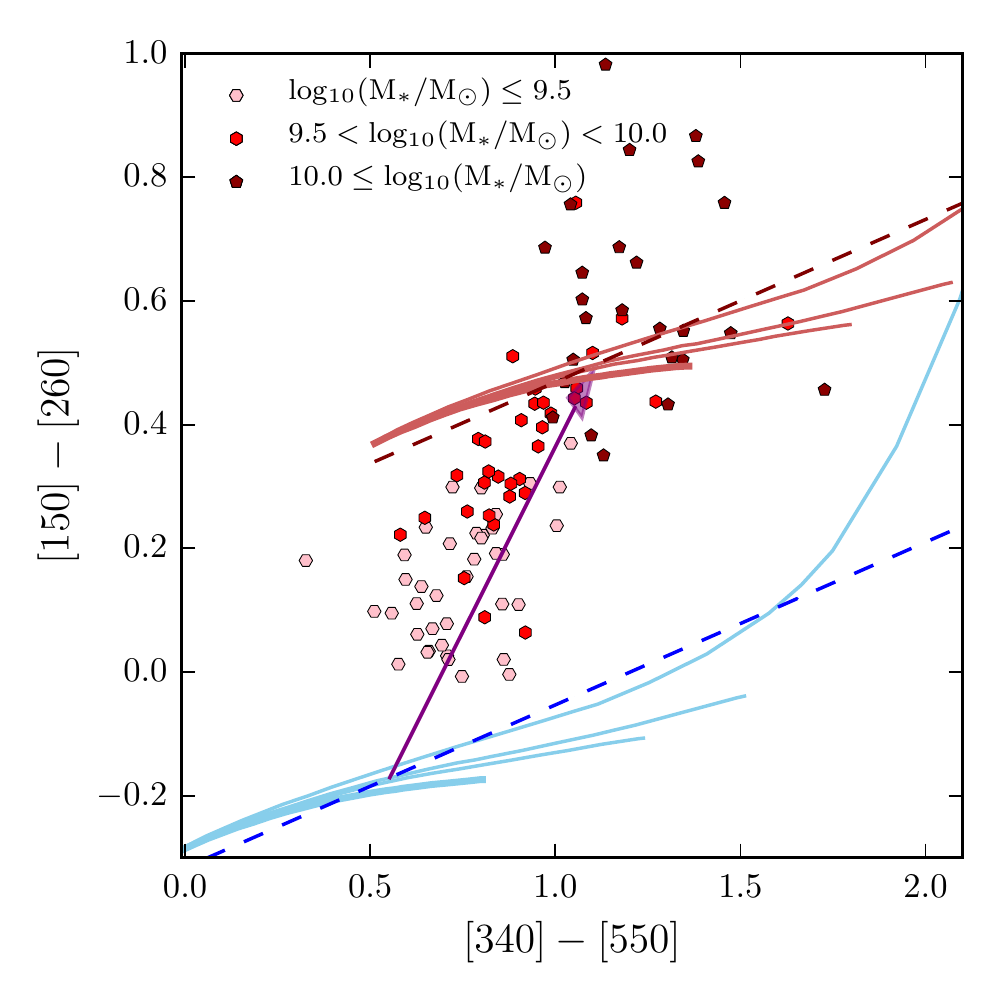}
\includegraphics[scale=0.8]{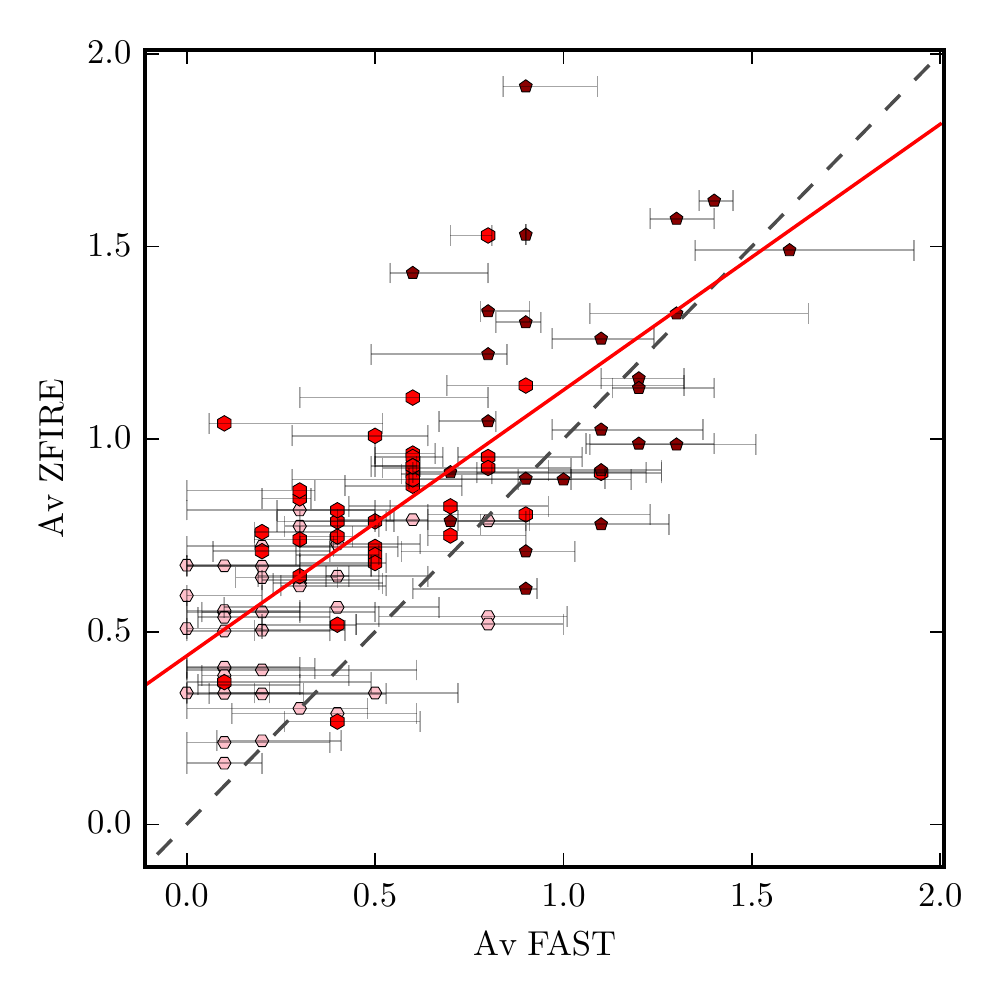}
\caption[Dust parametrisations to investigate IMF dependencies of dust extinction values.]{Dust parametrisations to investigate IMF dependencies of dust extinction values. Galaxies are divided to 3 mass bins. 
{\bf Top:} The dust content of the galaxies are parametrized using UV/optical colours. 
The cyan solid lines are PEGASE models for $\Gamma=-1.35$ IMF with varying SFHs. The dashed blue line is the best-fitting line for these models. 
The pink lines are similar to the cyan lines, but with an extinction of 1 mag. 
The magenta dashed line is the best-fitting line for these models with \Av=1. 
The purple arrow denotes the direction of the dust vector and connects the two best-fitting lines at time $t$.
{\bf Bottom:} Comparison between the extinction derived by Equation \ref{eq:Av_ZFIRE} with the extinction values derived by FAST. The error bars are the upper and lower 68th percentile of the \Av\ values compiled by FAST and the diagonal red line is the error weighted least squares fit to the data.
The diagonal dashed line is the \AvSED=\AvZFIRE\ line.
}
\label{fig:Av_ZFIRE}
\end{figure}

By binning galaxies in stellar mass, we find massive galaxies to be dusty than their less massive counterparts.
We show the distribution of our sample in the rest-frame UV vs rest-frame optical parameter space in Figure \ref{fig:Av_ZFIRE} (top panel).   PEGASE model galaxies with $\Gamma=-1.35$ and varying SFHs are shown by the solid model tracks.
When we apply a \Av=1 extinction, the models show a strong diagonal shift due to reddening of the colours in both axes.  For each set of tracks, we perform a best-fitting line to the varying SFH models.
The dust vector (shown by the arrow) joins the two best-fitting lines drawn to the models with \Av=0 and \Av=1 at time $t$. 
We define \AvZFIRE\ to be the correction needed for each individual galaxy to be brought down parallel to the dust vector to the best-fitting  line with \Av=0,  and is parametrized by the following equation:

\begin{equation}
\label{eq:Av_ZFIRE}
\begin{split}
A_{v}(\mathrm{ZF}) = -0.503\times ([340]-[550])\\
+ 1.914\times ([150]-[260])+ 0.607
\end{split}
\end{equation}

Our simple method of dust parametrization is similar to the technique used by FAST, which fits SED templates to the UV continuum to derive the extinction values.  
The \dustfil\ colour probes the UV continuum slope, which is ultra sensitive to dust, while the 
\boxfil\ probes the optical continuum slope which is less sensitive to dust. 
\Halpha\ emission does not fall within these filters, and hence, is not strongly sensitive to the SFR of the galaxies.

In the bottom panel of Figure \ref{fig:Av_ZFIRE}, we compare the derived extinction values from our method (\AvZFIRE) with the extinction values derived by FAST (\AvSED). Since \citet{Chabrier2003} IMF at $\mathrm{M_*>1M_\odot}$ is similar to the slope of Salpeter IMF ($\Gamma=-1.35$), the comparison is largely independent of the IMF. 
The median and \NMAD\ scatter of the Av values derived via FAST and our method is $\sim-0.3$ and $\sim-0.3$ respectively. Therefore, the values agree within $1\sigma$.
There is a systematic bias for \AvZFIRE\ to overestimate the extinction at lower \AvSED\ values and underestimate at higher \AvSED\ values. 
We attribute this residual pattern to age metallicity degeneracy, which is not considered in the derivation of \AvZFIRE.

The choice of IMF will affect dust corrections derived from UV photometry (using FAST or our empirical method) as there is a modest dependence of the rest-frame UV continuum slope on IMF for star-forming populations. We are primarily interested in IMF slopes shallower than Salpeter slope ($\Gamma> -1.35$) to explain our population of high \Halpha\ EW galaxies. 
For $\Gamma=-0.5$ we find the best-fitting model line in the top panel of Figure \ref{fig:Av_ZFIRE} shifts down by \around0.1 mag. 
This increases the magnitude of dust corrections 
and extends the arrows in Figure \ref{fig:EW_with_dust_corrections} to bluer colours and higher EWs and does not explain the presence of high \Halpha\ EW objects. For the purpose of comparing with our default hypothesis (Universal IMF with $\Gamma=-1.35$) we adopt the FAST derived dust corrections.

\subsection{Balmer decrements}
\label{sec:balmer_decrements}

Stellar attenuation values computed by fitting a slope to galaxy SEDs in UV, estimates the extinction of old stellar populations that primarily contributes to the galaxy continuum. 
Nebular emission lines originates from hot ionized gas around young and short-lived O and B stars. Given their short life-time ($\sim10-20$ Myr), O and B stars are not expected to move far from their birthplace (dusty clouds), thus,  the nebular emission-lines are expected to have high levels of extinction.
Next, we investigate the dust properties of the stars in different star-forming environments using the luminosity ratios of nebular hydrogen lines and observed UV colours.

Luminosity ratios of nebular hydrogen lines are insensitive to the underlying stellar population and IMF parameters for a fixed electron temperature \citep{Osterbrock1989}. 
These line ratios are governed by quantum mechanics, and therefore, can be used to probe the reddening of nebular emission lines and dust geometry under the assumption that ionized gas attenuation resembles that of the underlying stellar population.

With the recent development of sensitive NIR imagers and multi-object spectrographs, studies have now started to investigate the properties of dust at $z\sim2$ \citep{Shivaei2015, Reddy2015, deBarros2015}. These studies show conflicting results on the fraction of stellar to nebular attenuation of galaxies at $z\sim2$. 
Here we show  Balmer decrement results for a sub-sample of our \sample\ which shows SNR $>5$ detections for both \Halpha\ and \Hbeta. The data presented herein are a combination of data released by the ZFIRE data release \citet{Nanayakkara2016} and additional MOSFIRE observations carried out during January 2016. 
Our sample comprises of 42 galaxies with both \Halpha\ and \Hbeta\ emissions line detections with a SNR $>5$ and 35 galaxies are part of the \sample. Further details on \Hbeta\ detection properties are explained in Appendix \ref{sec:Balmer decrement extended}.

\begin{figure}
\centering
\includegraphics[scale=0.85]{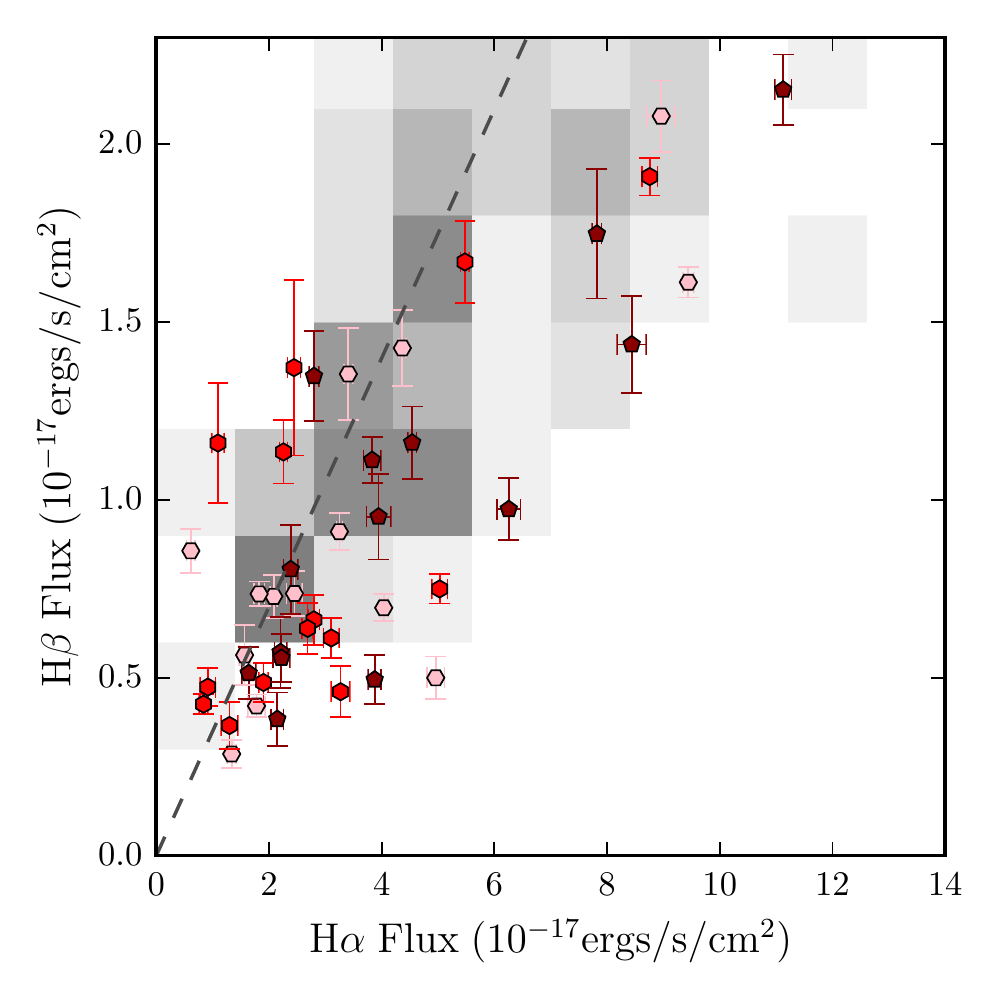}
\includegraphics[scale=0.85]{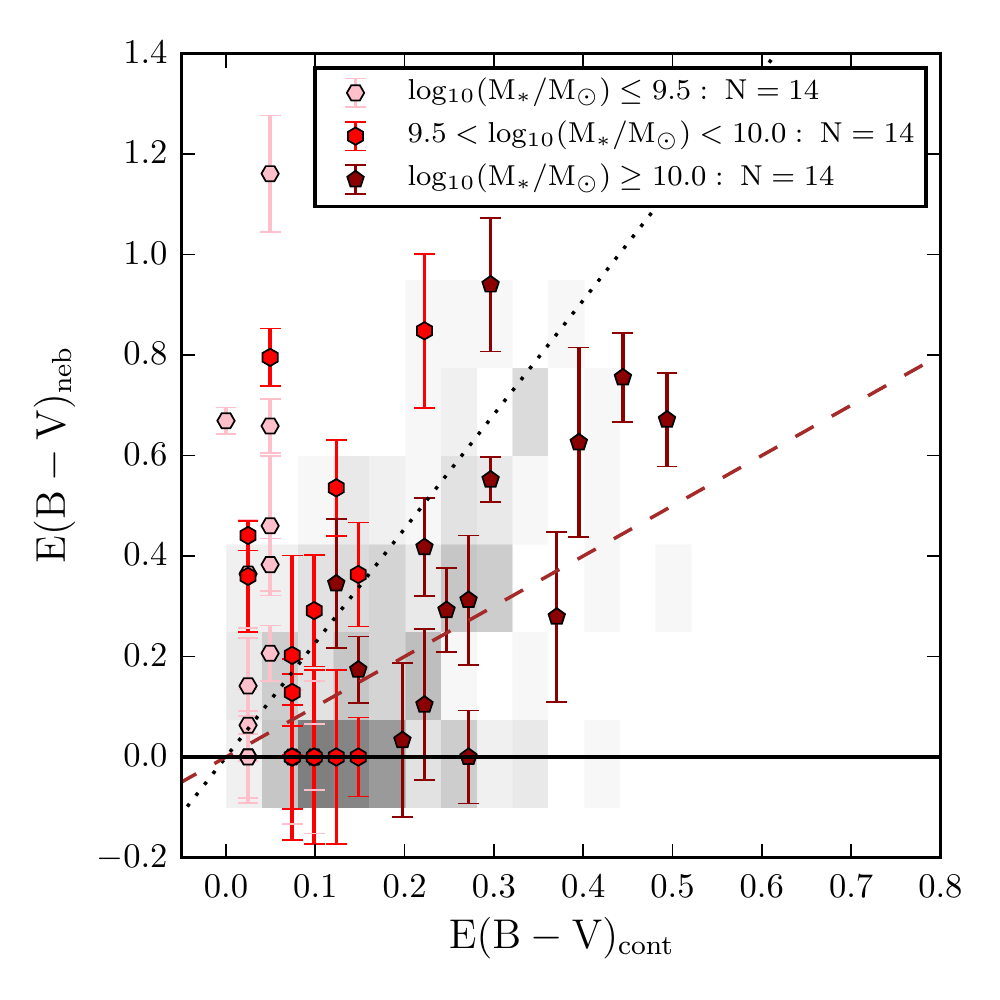}
\caption[Balmer decrement properties of the ZFIRE sample.]{Balmer decrement properties of the ZFIRE sample. Here we show the subset of \Halpha\ and \Hbeta\ detected galaxies in our \sample. The 2D density histogram shows the distribution of values from \citet{Reddy2015}
{\bf Top:} \Hbeta\ flux vs \Halpha\ flux measurements for the \sample.  
The individual galaxies are divided into three mass bins. 
The diagonal dashed line is the $f(H\alpha)=2.86\times f(H\beta)$ line which denotes the Balmer decrement for case B recombination.  
{\bf Bottom:} Comparison between SED derived extinction values with extinction computed from the Balmer decrement. The black dotted line is the $E(B-V)_{neb}=E(B-V)_{cont} /0.44$ relationship expected from \citet{Calzetti2000} to compute the extra extinction for nebular emission line regions. The brown dashed line is the $E(B-V)_{neb}=E(B-V)_{cont}$ line and the horizontal solid black line is the $E(B-V)_{neb}=0 $ line. Galaxies with $E(B-V)_{neb}$ $<0$ have been assigned a value of 0.
}
\label{fig:balmer_decrement}
\end{figure}

We show the \Hbeta\ flux vs \Halpha\ flux for our total ZFIRE galaxies Figure \ref{fig:balmer_decrement} (top panel). 
The diagonal dashed line of the left panel shows the Balmer decrement for Case B recombination models with \Halpha/\Hbeta = 2.86 \citep{Osterbrock1989}.
Galaxies that fall below this criteria are expected to have realistic dust models. 
In Figure \ref{fig:balmer_decrement} (bottom panel) we show the comparison between extinction computed for stars by FAST with the extinction computed for ionized gas regions using the Balmer decrement. The colour excess is computed from the Balmer decrement using:
\begin{equation}
\label{eq:balmer_dec}
E(B-V)_{neb} = \frac{2.5}{1.163} \times \mathrm{log}_{10} \left \{ \frac{f(H\alpha)}{2.86 \times f(H\beta)} \right \}
\end{equation}
The distribution of our sample in these panels is similar to \citet{Reddy2015} results as shown by the 2D density histogram. Therefore, both studies highlight the complicated dust properties of $z\sim2$ star-forming galaxy populations.

\subsection{The difference in extinction between stellar and nebular regions}
\label{sec:calzetti_factor}

In this section, we investigate how the differences in dust properties between stellar and ionized gas regions can affect the distribution of our galaxies in \Halpha\ EW vs \boxfil\ colour space. 
\cite{Calzetti2000} showed that, for $z\sim0$ star-forming galaxies, the nebular lines are $\sim\times2$ more attenuated than the continuum regions, but at $z\sim2$ studies show conflicting results \citep{Shivaei2015, Reddy2015, deBarros2015}. 
Using ZFIRE data in Figure \ref{fig:balmer_decrement}, we show that galaxies occupy a large range of $f$ values in our \Hbeta\ detected sample. We attribute the scatter in extinction to the properties of sight-lines of the nebular line regions.



The galaxies in our sample are at $z\sim2$ and are actively forming stars. Therefore, the dusty molecular gas regions are actively collapsing to sustain the ongoing star-formation. \citet{Straatman2017} showed that the velocity dispersion of the ZFIRE sample is $\mathrm{\lesssim100kms^{-1}}$. In a 1 Pc molecular cloud, stars will take $\gtrsim30$ Myr to move away from their parent birth clouds. Therefore, with active star-formation all massive stars and non-negligible fractions of smaller mass stars will be within these dusty regions, giving rise to varying extinction between nebular and stellar regions. Additionally, \citet{Reddy2015} showed SFR to be a dominant factor in determining the dust geometries within $z\sim2$ galaxy populations. 
Differences in the dust geometries in old and young stars regions that arise due to these reasons in ionizing clouds may result in non-uniform dust sight-lines for galaxies at $z\sim2$.

By varying the value of $f$ as a free parameter, we calculate the $f$ values required for our galaxies to be consistent with a universal IMF with slope $\Gamma=-1.35$. 
For each dust corrected \boxfil\ colour, we compute the \Halpha\ EW of the PEGASE $\Gamma=-1.35$ IMF track with p$_1=1000$ Myr. We then use the observed and required \Halpha\ EW values to compute the $f$ as follows:
\begin{equation}
\begin{split}
f =  \left \{ \frac{\mathrm{log}_{10}(\mathrm{H\alpha\ EW})_{\Gamma=-1.35} - \mathrm{log}_{10}(\mathrm{H\alpha\ EW})_{obs}}{0.44 \times 0.62 \times A_c(v)} 
+ \frac{0.82}{0.62}  \right \}  \\
\end{split}
\label{eq:varying_f}
\end{equation}
where log$_{10}(\mathrm{H\alpha\ EW})_{\Gamma=-1.35}$ is the \Halpha\ EW of the PEGASE model galaxy for dust corrected \boxfil\ colours of our sample and log$_{10}(\mathrm{H\alpha\ EW})_{obs}$ is the observed \Halpha\ EW. 

\begin{figure}
\centering
\includegraphics[scale=1.4]{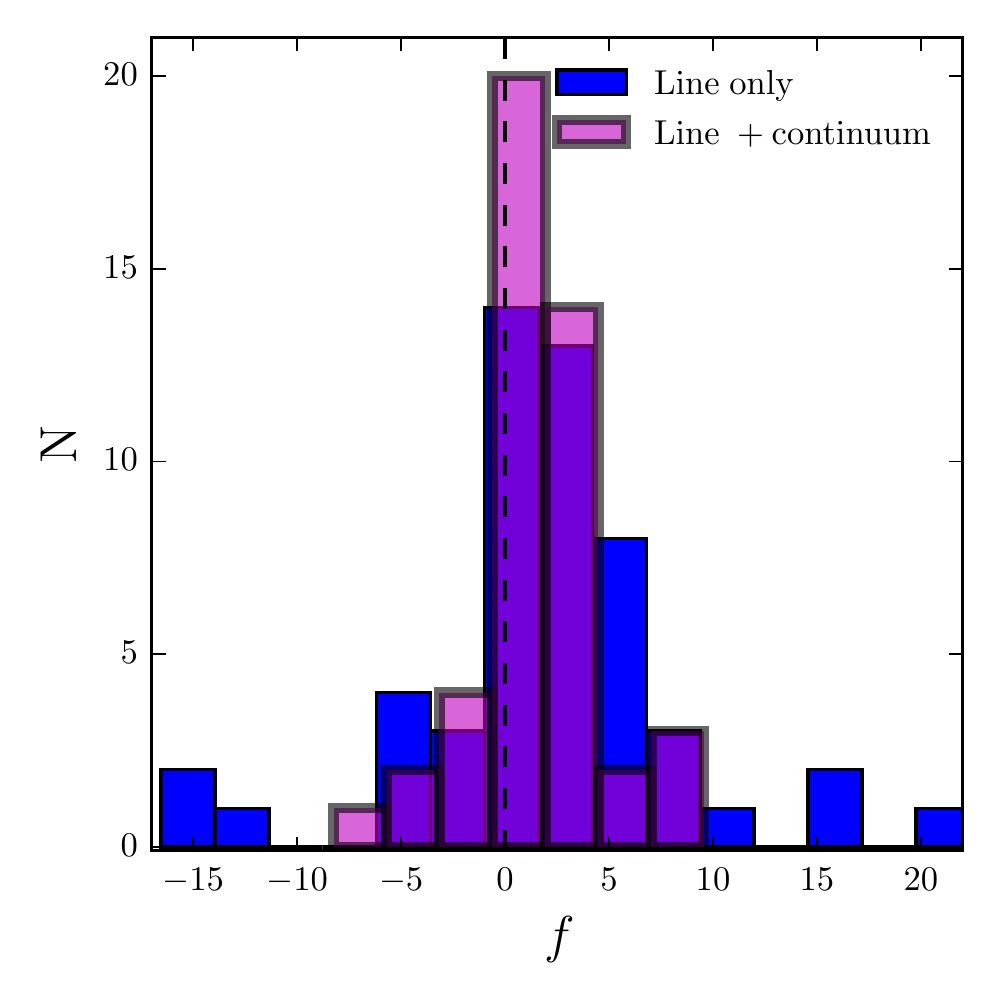}
\caption[The distribution of $f$ values required for galaxies in the \sample\ to agree with a $\Gamma=-1.35$ Salpeter like IMF. ]{The distribution of $f$ values required for galaxies in the \sample\ to agree with a $\Gamma=-1.35$ Salpeter like IMF. For each dust corrected \boxfil\ colour of the \sample\ galaxies, we use the corresponding \Halpha\ EW of the $\Gamma=-1.35$ PEGASE track with an exponentially declining SFH with a p$_1=1000$ Myr to compute the $f$ value required for the observed \Halpha\ EW to agree with the $\Gamma=-1.35$ IMF. The vertical dashed line is the f=0 line. 
}
\label{fig:varying_f}
\end{figure}

In Figure \ref{fig:varying_f}, we show the $f$ values required for our galaxies to agree with a universal IMF with a slope of $\Gamma=-1.35$. 
For the 46 continuum detected galaxies, $\sim30\%$  show $f<1$. It is extremely unlikely that galaxies at $z\sim2$ would have $f<1$, which suggests that ionizing dust clouds where the nebular emission lines originate from are less dusty than regions with old stellar populations.
Furthermore, galaxies that lie above the Salpeter track (log$_{10}$(\Halpha\ EW) $>2.2$) requires $f<1$ and therefore, even a varying $f$ hypothesis cannot account for the high EW galaxies. 
$\sim17\%$ of continuum detections have $f<0$ which is not physically feasible since it requires dust to have the opposite effect to attenuation. 
Therefore, we reject the hypothesis that varying $f$ values could explain the high \Halpha\ EWs of our galaxies.

\subsection{Observational bias}
\label{sec:observational_bias}

The \sample\ spans a large range of \Halpha\ EWs, suggesting a considerable variation in the sSFRs of the ZFIRE galaxies at $z\sim2$. High \Halpha\ EW can result due to two reasons:
\begin{enumerate}
\item High line flux: suggests a higher SFR in time scales of \around10 Myr.
\item Lower continuum level: suggests lower stellar mass for galaxies.
\end{enumerate}
These two scenarios should be considered together: i.e., a higher line flux with lower continuum level would suggest the galaxy to be going through an extreme star-formation phase. 
We investigate any detection bias that could explain our distribution of  \Halpha\ EWs.

\begin{figure}
\centering
\includegraphics[scale=0.90]{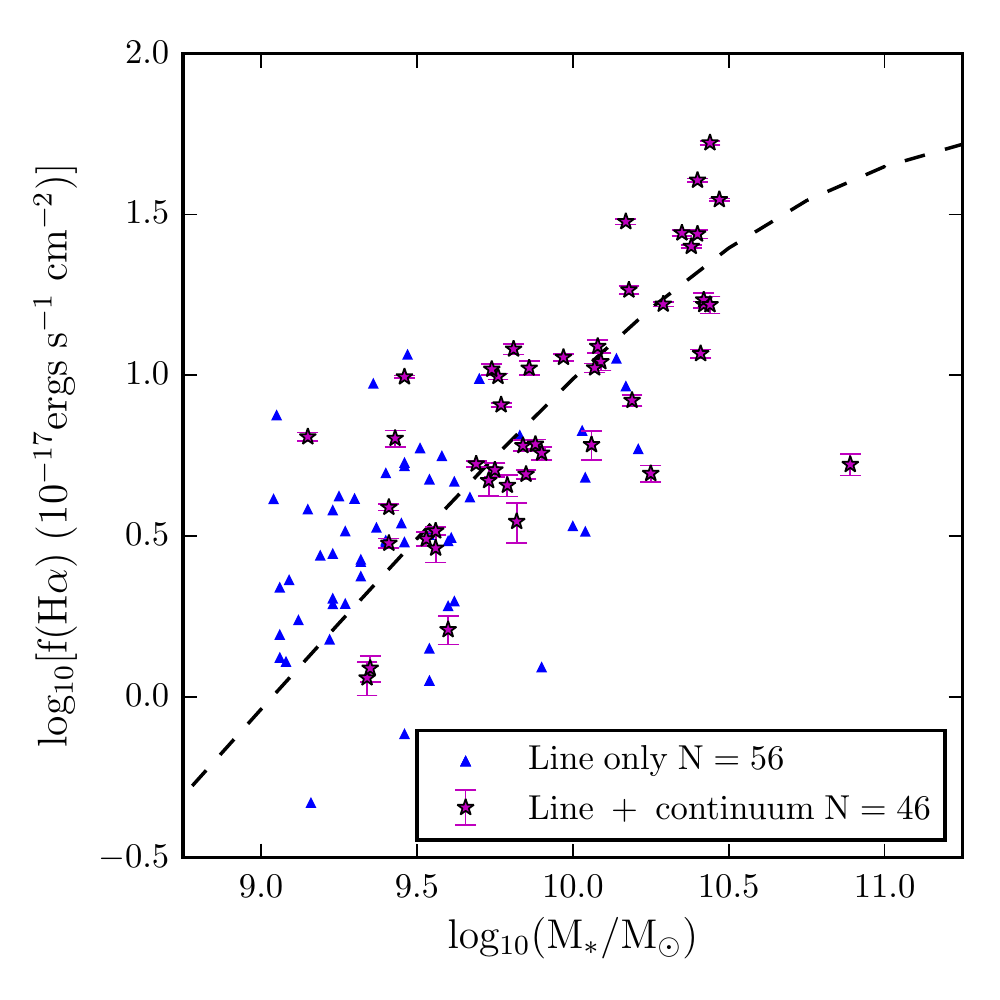}
\includegraphics[scale=0.95]{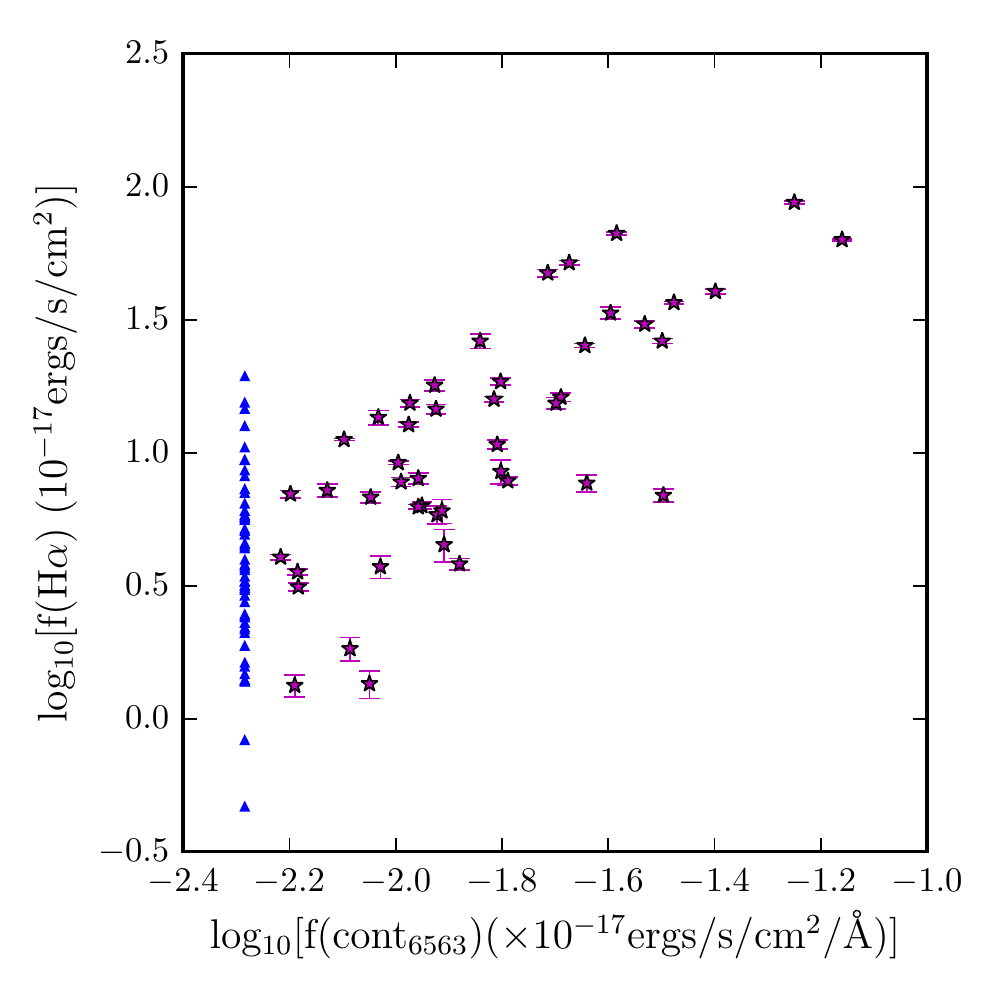}
\caption[Investigation of observable parameter/s that drive the high \Halpha\ EWs.]{ Here we investigate which observable parameter/s drive the high \Halpha\ EW values compared to $\Gamma=-1.35$ Salpeter like IMF expectations in \Halpha\ EW vs \boxfil\ colour space. 
{\bf Top:} The logarithmic \Halpha\ flux of the \sample\ as a function of stellar mass. The \Halpha\ flux has been dust corrected following Equation \ref{eq:Halpha dust corrected} with $f=1/0.44$. 
The black line is derived from the star-formation main sequence from \citet{Tomczak2014}, converted to \Halpha\ flux using \citet{Kennicutt1998} \Halpha\ star-formation law at $z=2.1$. 
{\bf Bottom:} The \Halpha\ flux vs the continuum level at 6563\AA. Both parameters (\Halpha\ flux as above, continuum level following Equation \ref{eq:cont dust corrected} ) have been corrected for dust extinction and are plot in logarithmic space. The \Halpha\ fluxes are $\sim2$ orders of magnitude brighter than the continuum levels. 
}
\label{fig:Ha_MS_and_cont}
\end{figure}

In \citet{Nanayakkara2016}, we show that the ZFIRE COSMOS K band detections are mass complete to $\mathrm{\log_{10}(M_*/M_\odot)\sim9.3}$.
In Figure \ref{fig:Ha_MS_and_cont} we show the distribution of the \Halpha\ flux and continuum levels of our sample.  It is evident from Figure \ref{fig:Ha_MS_and_cont} (top panel) that our galaxies evenly sample the star-forming main-sequence described by \citet{Tomczak2014} without significant bias towards extreme \Halpha\ flux values.
Therefore, we conclude that the \Halpha\ fluxes we detect are typical of star-forming galaxies at $z=2.1$. 

In Figure \ref{fig:Ha_MS_and_cont} (bottom panel), we compare our \Halpha\ flux values with the derived continuum levels. 
Continuum detected galaxies show continuum levels that are in the order of $\sim2$ mag smaller compared to the \Halpha\ fluxes, and therefore, the higher \Halpha\ EWs in our sample are primarily driven by the low continua. Note that our continuum detection level is $\sim-2.3$ log flux units. Therefore, for galaxies with only line detection, the difference between \Halpha\ flux and continuum level is much higher, which suggests much larger \Halpha\ EWs.

Several studies investigated the \Halpha\ EW of galaxies at higher redshifts ($z\geq1.5 $) \citep{Erb2006b,Shim2011,Fumagalli2012,Kashino2013,Stark2013,Masters2014,Sobral2014,Speagle2014,Marmol-Queralto2016,Rasappu2016} using SED fitting techniques and/or grism spectra. 
We find that our \Halpha\ EWs show good agreement with EWs expected at $z\sim2$ \citep{Marmol-Queralto2016} and conclude that our observed \Halpha\ EW values are typical of $z\sim2$ galaxies.

However, there are no studies that use high quality spectra to study the \Halpha\ EW at $z\sim2$. Even though our \Halpha\ fluxes and EWs are typical of $z\sim2$ galaxies, in the \Halpha\ EW vs \boxfil\ colour space, a large fraction of our galaxies show high EWs for a given \boxfil\ colour compared to the expectation by a Salpeter like IMF. 
Our high EWs are driven by lower continuum levels, for which we consider two possible explanations. 
\begin{enumerate}
\item Most galaxies have quenched their starburst phase in a time-scale of $\sim10$ Myr. Therefore, the old stellar populations are still being built up explaining the lack of continuum level from the older stars.
\item Stars are being formed continuously at $z\sim2$ with a higher fraction of high mass stars.
\end{enumerate}

In Section \ref{sec:star_bursts}, we investigate the effects of starbursts on our study to examine how probable it is for $\sim1/3$rd of our galaxies to have quenched their star-formation within a time-scale of  $\lesssim10$ Myr.

\section{Can star bursts explain the high \Halpha-EWs?}
\label{sec:star_bursts}

Galaxies at $z\sim2$ are at the peak of their star formation history \citep{Hopkins2006}. We expect these galaxies to be rapidly evolving with multiple stochastic star formation scenarios within their stellar populations \citep[eg.,][]{Pacifici2015}. If our sample consists of a significant population of starburst galaxies, it may cause significant systemic biases to our IMF analysis.

In this section, we investigate the effects of bursts on the SFHs of the galaxies. We study how the distribution of galaxies in \Halpha\ EW vs \boxfil\ colour space may be affected by such bursts and how we can mitigate their effects.  We demonstrate that our final conclusions are not affected by starbursts.


\subsection{Effects of starbursts}

A starburst event would abruptly increase the \Halpha\ EW of a galaxy within a very short time scale ($\lesssim5$ Myr). The increase in ionizing photons is driven by the extra presence of O and B stars during a starburst which increases the amount of Lyman continuum photons. Assuming that a constant factor of Lyman continuum photons get converted to \Halpha\ photons via multiple scattering events, we expect the number of \Halpha\ photons to increase as a proportion to the number of O and B stars. Furthermore, the increase of the O and B stars would drive the galaxy to be bluer causing the \boxfil\ colours to decrease.

The ability of a starburst to drive the points away from the monotonic Salpeter track is limited. The deviations are driven by the burst fraction, which we define as the burst strength divided by the length of the starburst. 
If the burst fraction is small, it has a small effect. However if it is very large it dominates both the \Halpha\ and the optical light, the older population is `masked', and it heads back towards the track albeit at a younger age, i.e. one is seeing the monotonic history of the burst component. The maximum deviation in our study occurs for burst mass fractions of 20--30\% occurring in time-scales of 100-200 Myr or fractions thereof, which can cause excursions of up to $\sim 1$ dex. However as we will see this only occurs for a short time.

\begin{figure}
\centering
\includegraphics[scale=1.1]{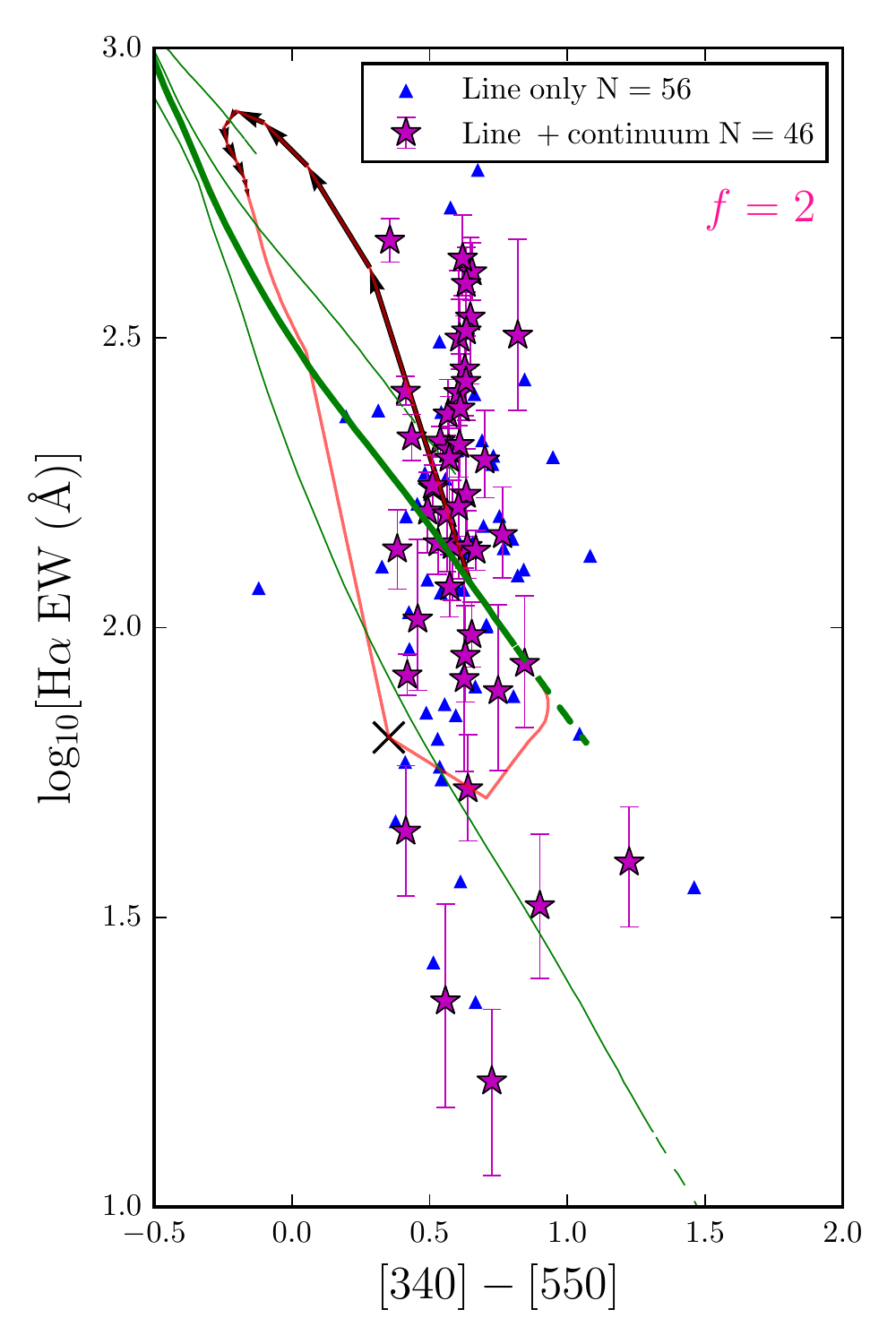}
\caption[The effect of a star burst on a PEGASE model galaxy track.]{The effect of a star burst on a PEGASE model galaxy track.   
The green tracks are computed with constant SFHs but with different IMFs. From top to bottom they have $\Gamma$ values of respectively $-0.5, -1.0, -1.35$, and $-2.0$. All solid tracks end \around 3.1 Gyr and the continuation up to 13 Gyr is shown by the dashed lines.
The track in red follows the constant SFH model with a IMF slope $\Gamma=-1.35$ (thick green track) up to t=1500 Myr. A burst with a duration of $t_b=100$ Myr is superimposed on the track at t=1500 Myr. 
The burst generates 30\% of the galaxies' total mass at \around3000 Myr. The black arrows superimposed on the tracks show the direction of the burst in the first 10 Myr and are plot every 1 Myr to distinctively demonstrate the short time scales in \Halpha\ EW evolution on the aftermath of a starburst. The cross denotes $t=1600$ Myr signalling the end of the burst. 
The \sample\ is dust corrected with a $f=2$ following prescriptions outlined in Section \ref{sec:dust_corrections}. The magenta stars show galaxies with continuum detections while solid blue triangles show galaxies only with \Halpha\ emission lines. 
}
\label{fig:burst_model_EW_BC}
\end{figure}

We show the effect of a starburst on a PEGASE model galaxy with a monotonic SFH in Figure \ref{fig:burst_model_EW_BC}.
A starburst with a time-scale of $\tau_b=100$ Myr and a strength $f_m=0.2$ (fraction of mass at $\sim3000$ Myr generated during the burst) is overlaid on the constant SFH model at time = 1500 Myr. The starburst drives the increase of \Halpha\ EW which occurs in a very short time scale. In Figure \ref{fig:burst_model_EW_BC} the galaxy deviates from the constant SFH track as soon as the burst occurs and reaches a maximum \Halpha\ EW within 4 Myr. 
At this point, the extremely high-mass stars made during the burst will start to leave the main sequence. This will increase the number of red-giant stars resulting in higher continuum level around the \Halpha\ emission line.
Therefore, the \Halpha\ EW starts to decrease slowly after $\sim4$ Myr. Once the burst stops the \Halpha\ EW drops rapidly to values lower than pre-burst levels.
The galaxy track will eventually join the $\Gamma=-1.35$ smooth SFH track at a later time than what is expected by a smooth SFH model.

We further investigate the effect of starbursts with smaller time-scales ($t_b<20$ Myr) and find that the evolution of \Halpha\ EW in the aftermath of the burst to be more extreme for similar $f_m$ values. This is driven by more intense star-formation required to generate the same amount of mass within $\sim1/10$th of the time scale. Since both \Halpha\ EW and \boxfil\ colours are a measure of sSFR we expect the evolution to strongly dependent on $f_m$ and $\tau_b$ of the burst and to be correlated with each other.

In our analysis, we do not consider star-bursts with time scales $>300$Myr. This is driven by the assumption that stars are formed within star-forming clumpy regions, and therefore time-scales of switching on and off star-bursts should occur in times comparable to the life-time of these clumps ($\sim10-200$Myr \citep[eg.,][]{Elmegreen2009b,NewmanS2012,Genzel2011}).

To consider effects of bursts, we adopt two complimentary approaches. First, we stack the data in stellar mass and \boxfil\ colour bins. Stellar populations are approximately additive and by stacking we smooth the stochastic SFHs in individual galaxies and also account the effect from galaxies with no \Halpha\ detections. Second, we use PEGASE to model starbursts to generate Monte Carlo simulations to predict the distribution of the galaxies in \Halpha\ EW vs \boxfil\ colour space. Using the simulations we investigate whether it is likely that the observed discrepancy is driven by starbursts and also double check whether the stacking of galaxies would generate smooth SFH models.

\subsection{Stacking}
\label{sec:EW_stacking}

In order to remove effects from stochastic SFHs of individual galaxies in our sample, we employ a spectral stacking technique. We first divide the galaxies into three mass and dust corrected \boxfil\ colour bins as follows.
\begin{itemize}
\item Mass bins: log$\mathrm{_{10}(M_*/M_\odot)}$ $\leq9.5$, $9.5<$ log$\mathrm{_{10}(M_*/M_\odot)} <10$, log$\mathrm{_{10}(M_*/M_\odot)}\geq10$ 
\item \boxfil\ colour bins:  (\boxfil) $\leq0.56$, $0.56<$ (\boxfil) $<0.65$, (\boxfil) $\geq0.65$
\end{itemize} 
We select a wavelength interval of \around1500\AA\ centred around the \Halpha\ emission for each spectra and mask out the sky lines with approximately $2\times$ the spectral resolution. 
In order to avoid systematic biases arisen from narrowing down the sampled wavelength region in the rest-frame, we instead redshift all spectra to a common $z=2.1$ around which most of the galaxies reside. 
We sum all the spectra at this redshift, in their respective bins.
The error spectra are stacked in quadrature following standard error propagation techniques. 

We mask out the nebular emission-line regions of the stacked spectra and use a sigma-clipping algorithm to fit a continuum (c1). 
The error in the continuum is assigned as the standard deviation of the continuum values of 1000 bootstrap iterations.

We visually inspect the stacked spectra to identify the \Halpha\ emission line profiles to calculate the integrated flux. Stacked \Halpha\ EW is calculated following equations \ref{eq:Ha_EW_obs} and \ref{eq:Ha_EW_rest}.

To estimate the error on the stack due to the stochastic variations between galaxies, we use a bootstrapping technique to calculate the error of the stacked \Halpha\ EW values.
We bootstrap galaxies with replacement in each bin to produce 1000 stacked spectra for each of which we calculate the \Halpha\ EW.
The standard deviation of the logarithmic EW values for each bin is considered as the error of the \Halpha\ EW of the stacked spectra. 
We expect the bootstrap errors to include stochastic variations in the SFHs between galaxies. If our sample comprises of galaxies undergoing extreme starbursts, the effects from such bursts should be quantified within these error limits. 

We stack the individual [340] and [550] fluxes of the galaxies in similar mass and \boxfil\ colour bins. The average extinction value of the galaxies in each bin is considered as the extinction  of the stacked spectra. 
We use this extinction value to dust correct the \Halpha\ EW and \boxfil\ colours of the stacked spectra, following recipes explained in Section \ref{sec:dust_corrections}.

\begin{figure}
\centering
\includegraphics[scale=0.7]{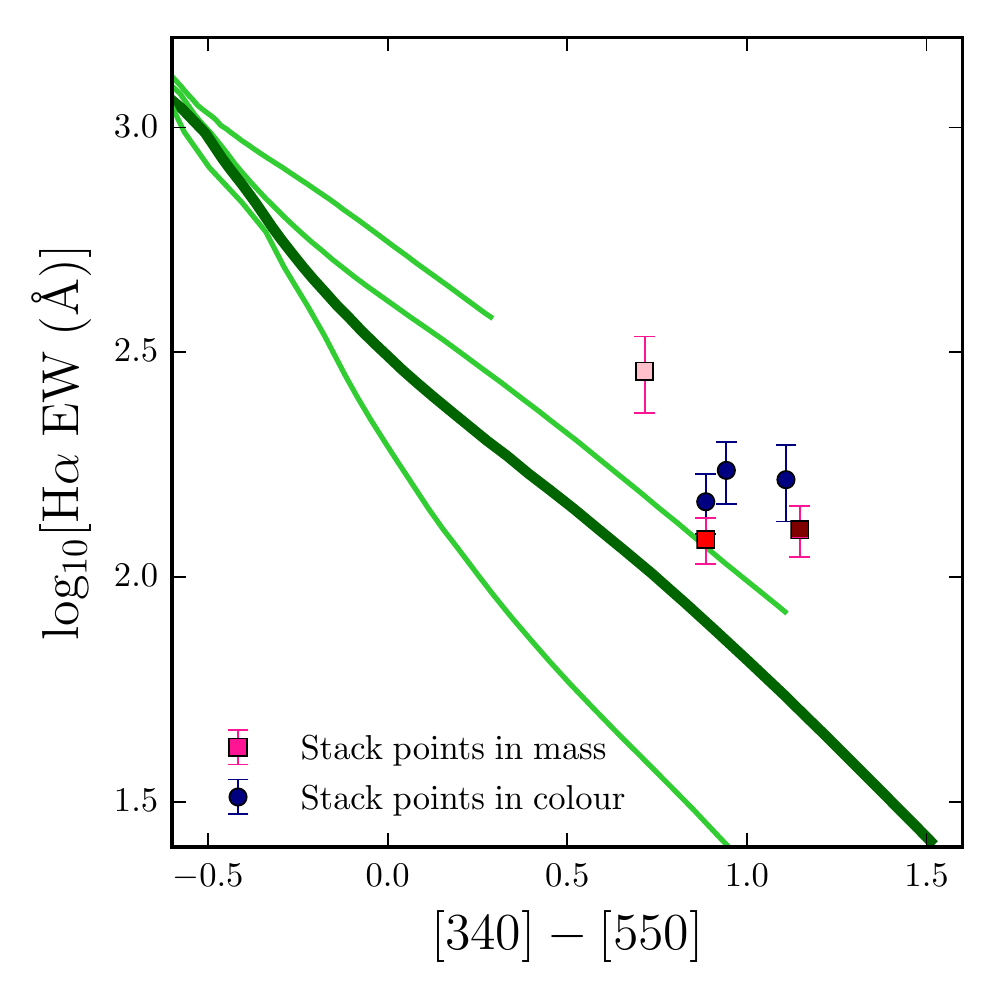}
\includegraphics[scale=0.7]{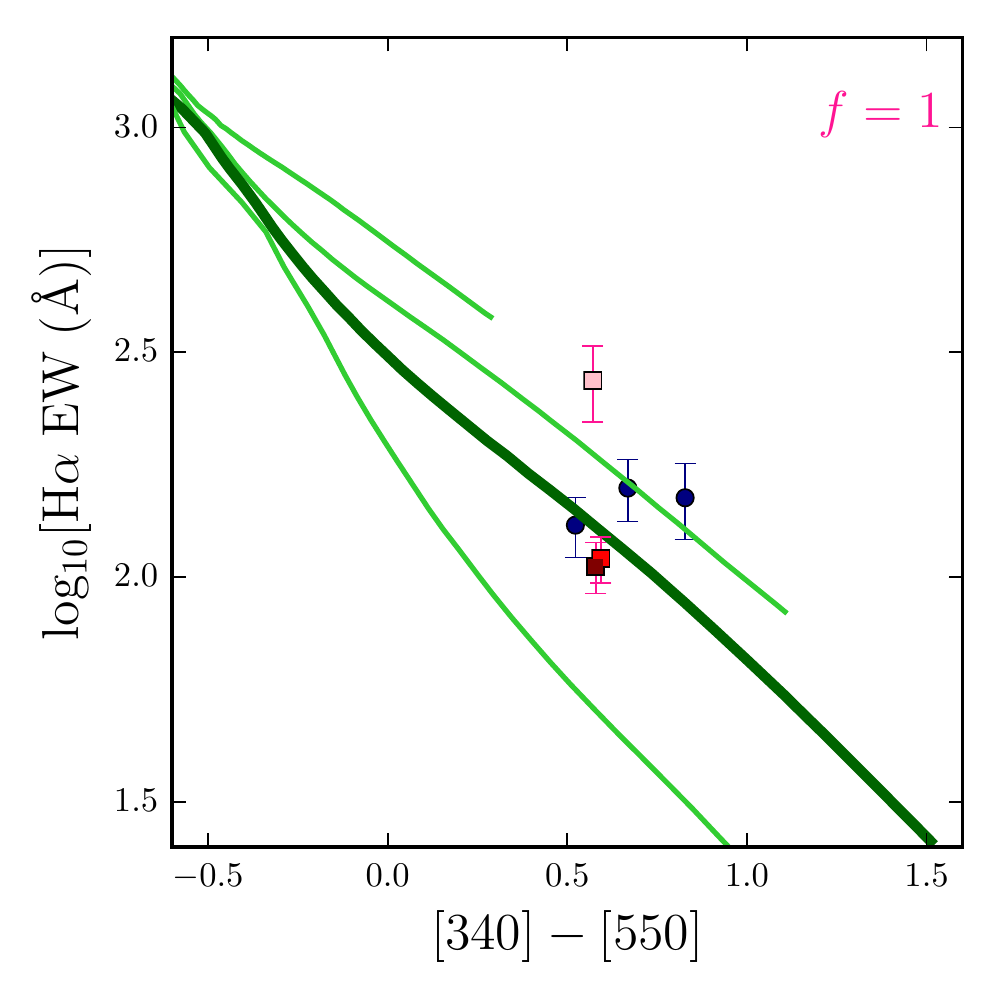}
\includegraphics[scale=0.7]{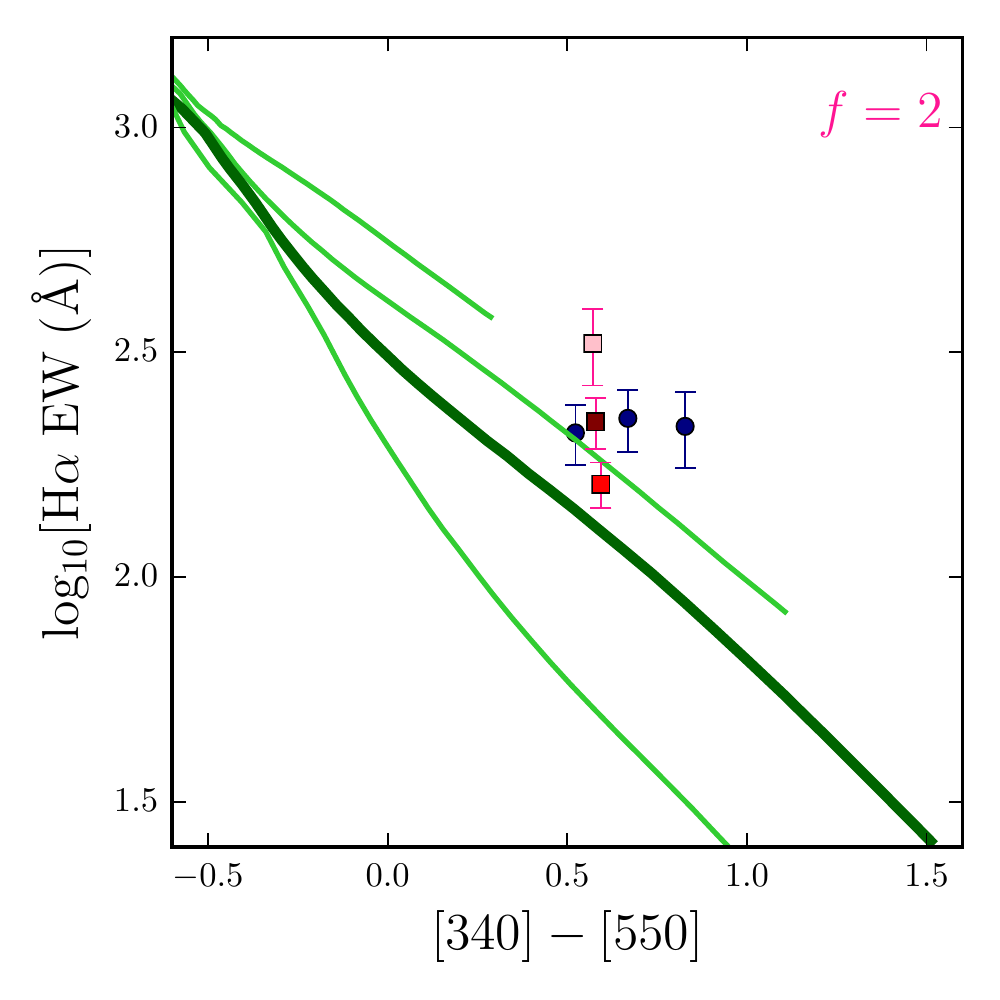}
\caption[The \Halpha\ EW vs \boxfil\ colour distribution of the stacked \sample.]{The \Halpha\ EW vs \boxfil\ colour distribution of the stacked \sample.  
The spectra are stacked in mass (squares) and \boxfil\ colour (circles) bins. The errors are from bootstrap re-sampling of the objects in each bin. The mass bins are colour coded where the higher masses have darker colours.
The tracks are SSP models computed from PEGASE with exponentially decaying SFHs of p$_1=1000$ Myr and varying IMFs. 
From top to bottom the tracks have $\Gamma$ values of respectively $-0.5, -1.0, -1.35$ (thick dark green line), and $-2.0$. 
All tracks end  \around3.1 Gyr ($z\sim2$).
{\bf Top Left:} Stacked galaxies before any dust corrections are applied. 
{\bf Top Right:} Stacked galaxies after dust corrections are applied following recipes outlined in Section \ref{sec:dust_corrections} with a $f=1$.
{\bf Bottom:} Similar to the centre panel but with a $f=2$. 
}
\label{fig:EW_stacked}
\end{figure}

Figure \ref{fig:EW_stacked} shows the distribution of the stacked spectra in \Halpha\ EW vs \boxfil\ colour space before and after dust corrections are applied. 
We consider dust corrections with $f=1$ and $f=2$.
For $f=1$ dust correction scenario the bluest colour bin and the medium and high mass stacked data points agree with the $\Gamma=-1.35$ track. However, the redder colours bins and lowest mass galaxies on average prefer shallower IMFs. 
With $f=2$ dust correction, the distribution of stacked data points even with bootstrap errors suggest values $\sim0.1-0.4$ dex above the Salpeter IMF, if interpreted as an IMF variation. 
In both scenarios, redder galaxies show larger deviation from the canonical Salpeter IMF. If starbursts drive the distribution of the galaxies, we expect the bluer galaxies on average to show larger deviation from the $\Gamma=-1.35$ tracks.

To further account for any detection bias arisen from \Halpha\ undetected galaxies, we use the ZFIRE COSMOS field K band targeted galaxies with no \Halpha\ emission detections to compute a continuum (c2) contribution to the stacked spectra. 
We use the photometric redshifts to select 37 galaxies within $1.90<z<2.66$, which is the redshift interval the \Halpha\ emission line falls within the MOSFIRE K band. 
Next we perform a cut to select galaxies with similar stellar masses ($9.04<\mathrm{log_{10}(M_\odot)}<10.90$) and \boxfil\ colours ($-0.12<$(\boxfil)$<1.46$) to the galaxies in the \sample. 
The final sample comprises of 21 galaxies which we use to stack the 1D spectra in mass and \boxfil\ colour bins.

\begin{figure}
\centering
\includegraphics[scale=1.0]{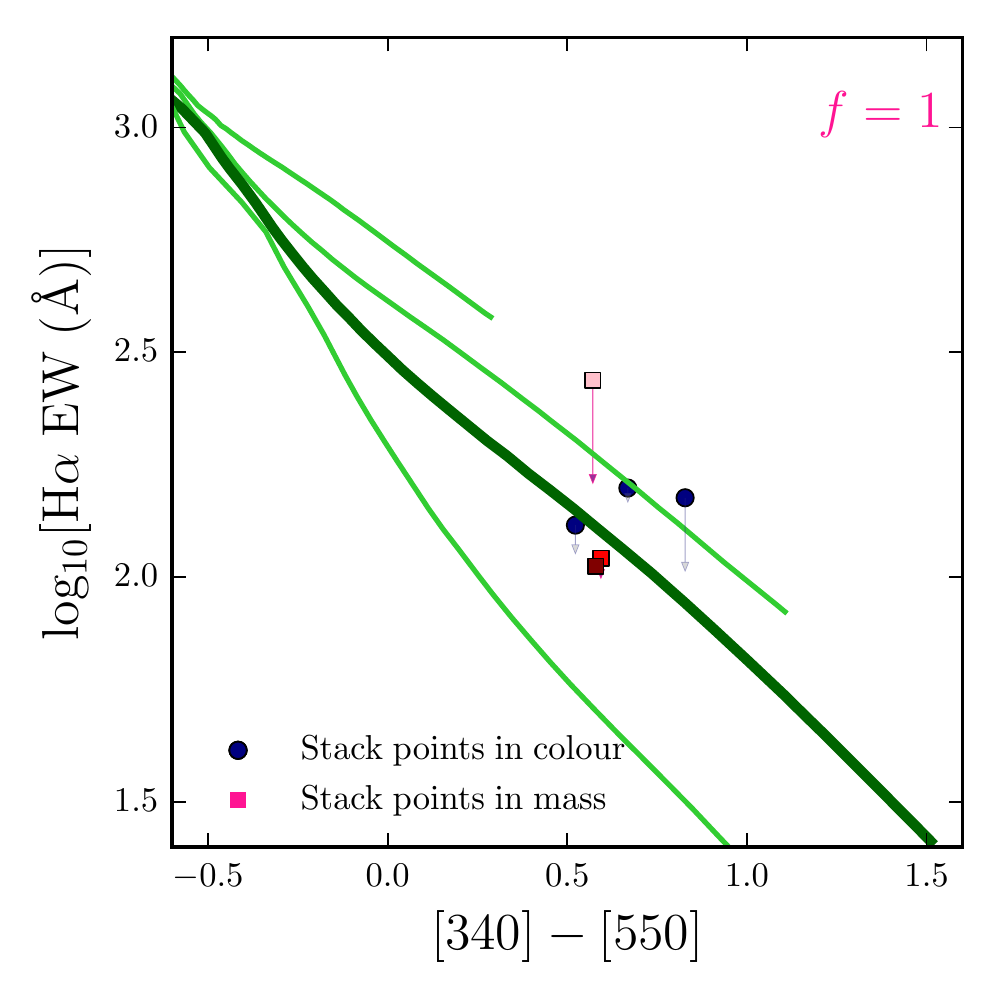}
\includegraphics[scale=1.0]{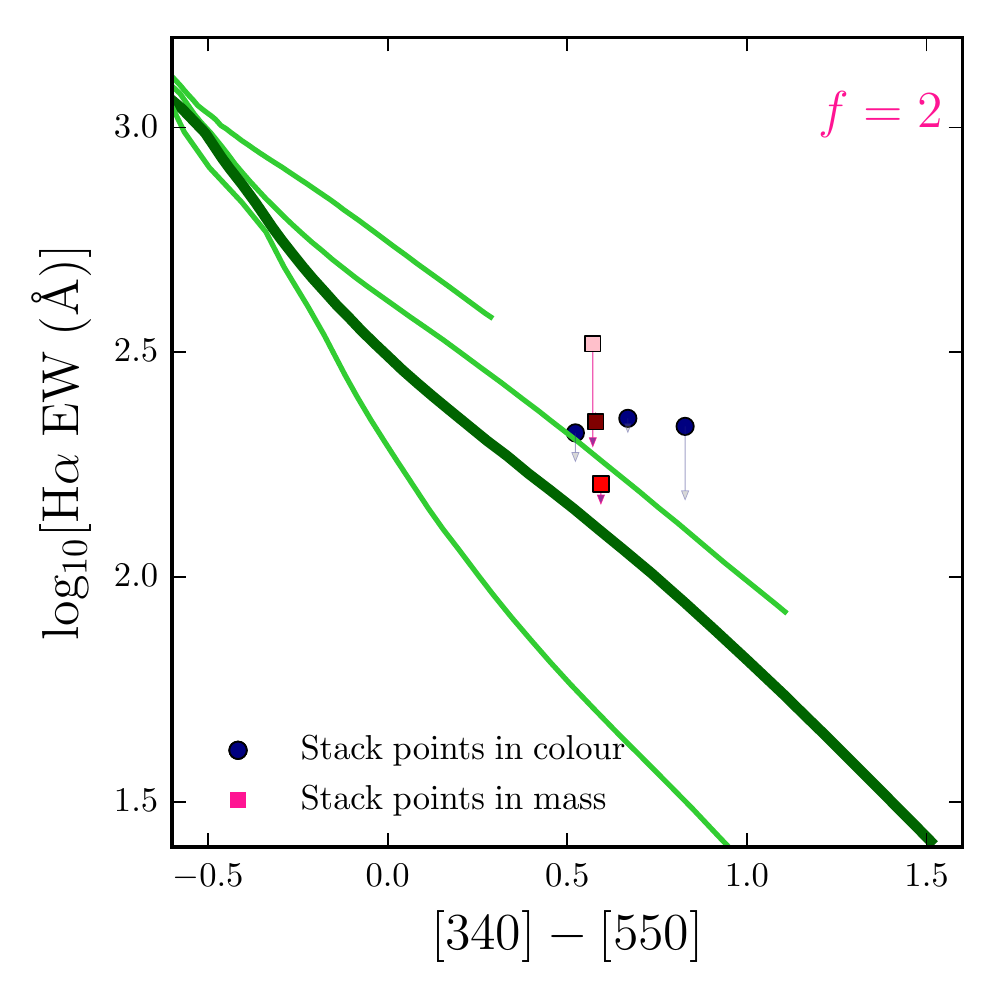}
\caption[Effect of considering the continuum contribution from the \Halpha\ non-detected galaxies to the mass and colour stacks of the \sample.]{ Here we show the effect of adding the continuum contribution (c2) of galaxies with no \Halpha\ detections to the \sample\ stacks. Galaxies are stacked in mass and colour bins and the arrows show the change in EW when the contribution from c2 is added to the data. All tracks shown are similar to Figure \ref{fig:EW_stacked}. 
{\bf Top:} Dust corrections applied with $f=1$. The errors for the data points are similar to Figure \ref{fig:EW_stacked} (centre panel).
{\bf Bottom:} Dust corrections applied with $f=2$. The errors for the data points are similar to Figure \ref{fig:EW_stacked} (right panel).
}
\label{fig:EW_c1c2_stacked}
\end{figure}

In order to stack the spectra, first we mask out the sky regions and assume that all galaxies are at a common $z=2.1$.
We mask out \Halpha\ and \NII\ emission line regions and fit a continuum similar to how c1 was derived. 
We add c1+c2 to re-calculate the \Halpha\ EW for each of the mass and colour bins. Since the continuum level is increased by the addition of c2, the \Halpha\ EW of the spectra reduces. We note that the highest mass bin contains no \Halpha\ undetected galaxies.
Figure \ref{fig:EW_c1c2_stacked} shows the change in stacked data points when the \Halpha\ non-detected continuum is considered with a dust correction of $f=1$ and $f=2$. 
The maximum deviation of the stacked \Halpha\ EW values is \around0.2 dex and the lowest mass and the reddest \boxfil\ colour bins show the largest deviation.  This is driven by the higher number of lower mass redder galaxies which have been targeted but not detected by the ZFIRE survey. 
The magnitude of the deviations are independent of the $f$ value used for the dust corrections and for both $f=1$ and $f=2$, the galaxies that show an excess of \Halpha\ EW compared to $\Gamma=1.35$ tracks still show an excess when the added c2 continuum contribution is considered.  
For $f=2$ dust corrections, even with considering the effect of non-detected continuum levels, majority of the stacked galaxies in our sample are significantly offset from the canonical Salpeter like IMF value.

\

\subsection{Simulations of starbursts}
\label{sec:simulations}

By employing spectral stacking and bootstrap techniques, we showed in Section \ref{sec:EW_stacking}, that our galaxies on average favour shallower IMFs than the universal Salpeter IMF.  
In this section, we use PEGASE SSP models to generate simulations with starbursts to calculate the likelihood for half of the \sample\ to be undergoing simultaneous starbursts at $z\sim2$. 
Furthermore, by randomly selecting galaxies from the simulation at random times, we stack the galaxies in mass and \boxfil\ colour bins to make comparisons with the stack properties of the \sample.

We empirically tune our burst parameters to produce the largest number of galaxies above the Salpeter track. 
A single starburst with time scales of $t_b\sim100-300$ Myr and with $f_m\sim0.1-0.3$ are overlaid on constant SFH models with the starburst occurring at any time between $0-3250$ Myr of the galaxies' lifetime. Simulation properties and the evolution of \Halpha\ EW and \boxfil\ colours during starbursts are discussed next.


\subsubsection{PEGASE simulations of starburst galaxies}
\label{sec:simulation_properties}

Here we describe the PEGASE simulations discussed in Section \ref{sec:simulations} to model the effects of starbursts in \Halpha\ EW vs \boxfil\ colour parameter space. 
We tune burst empirical parameters to maximize the number of high \Halpha\ EW objects.
We consider 4 scenarios in our simulations as shown in Table \ref{tab:simulation_param}.  For each scenario we model 100 galaxies and superimpose a single starburst on PEGASE model tracks with an IMF slope of $\Gamma=-1.35$ and a constant SFH. 

PEGASE model galaxies are generated from an initial gas reservoir of 1\msol. Therefore, we normalize the total mass generated by the constant SFH and the star burst to 1\msol\ in order to calculate the SFR for each time step. These values are used to calculate SSPs with a IMF slope of $\Gamma=-1.35$ and upper and lower mass cutoffs set at 0.5\msol\ and 120\msol\ respectively. We use a constant metallicity of 0.02  for all our simulations. The other parameters are kept similar at their default values as described in Section \ref{sec:PEGASE_models}. Following this recipe we generate the simulated galaxies with finer sampling of the time steps around the time of burst to better resolve the effects of the bursts.

\begin{figure}
\centering
\includegraphics[width=0.49\textwidth]{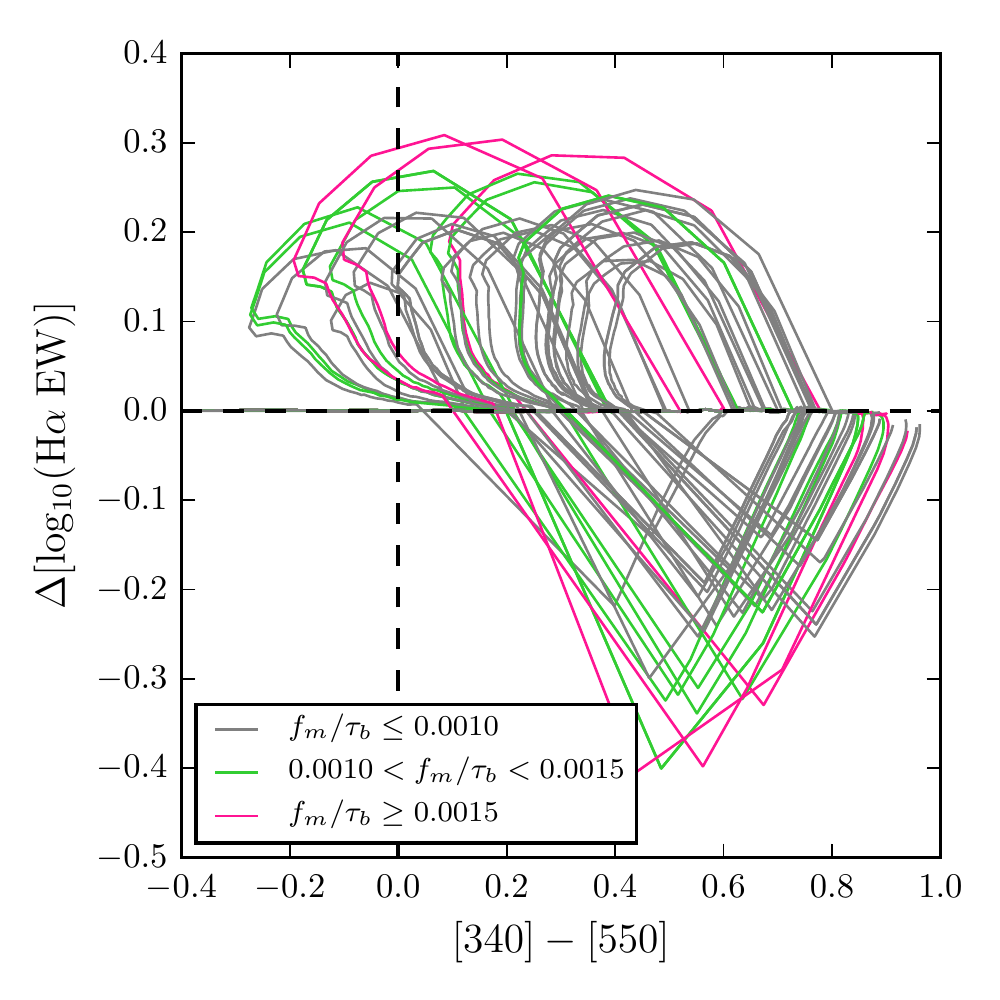}
\includegraphics[width=0.49\textwidth]{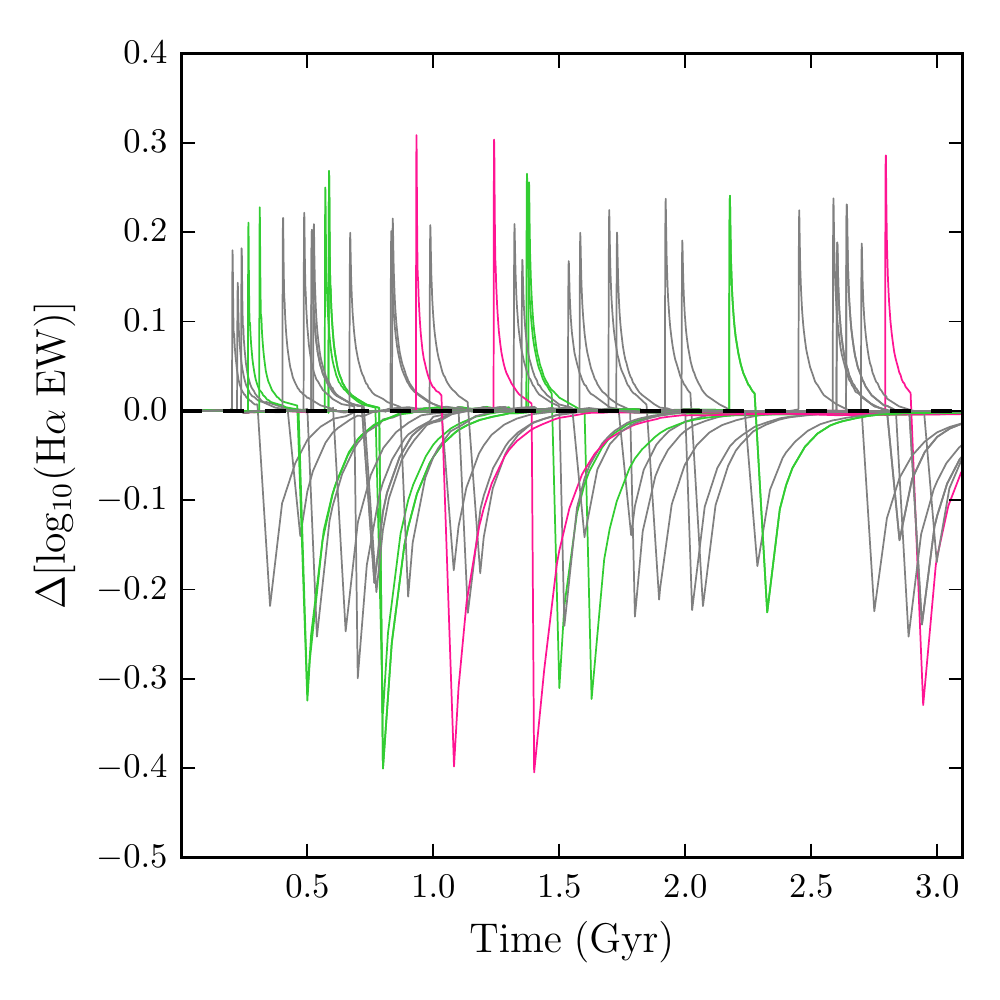}
\includegraphics[width=0.49\textwidth]{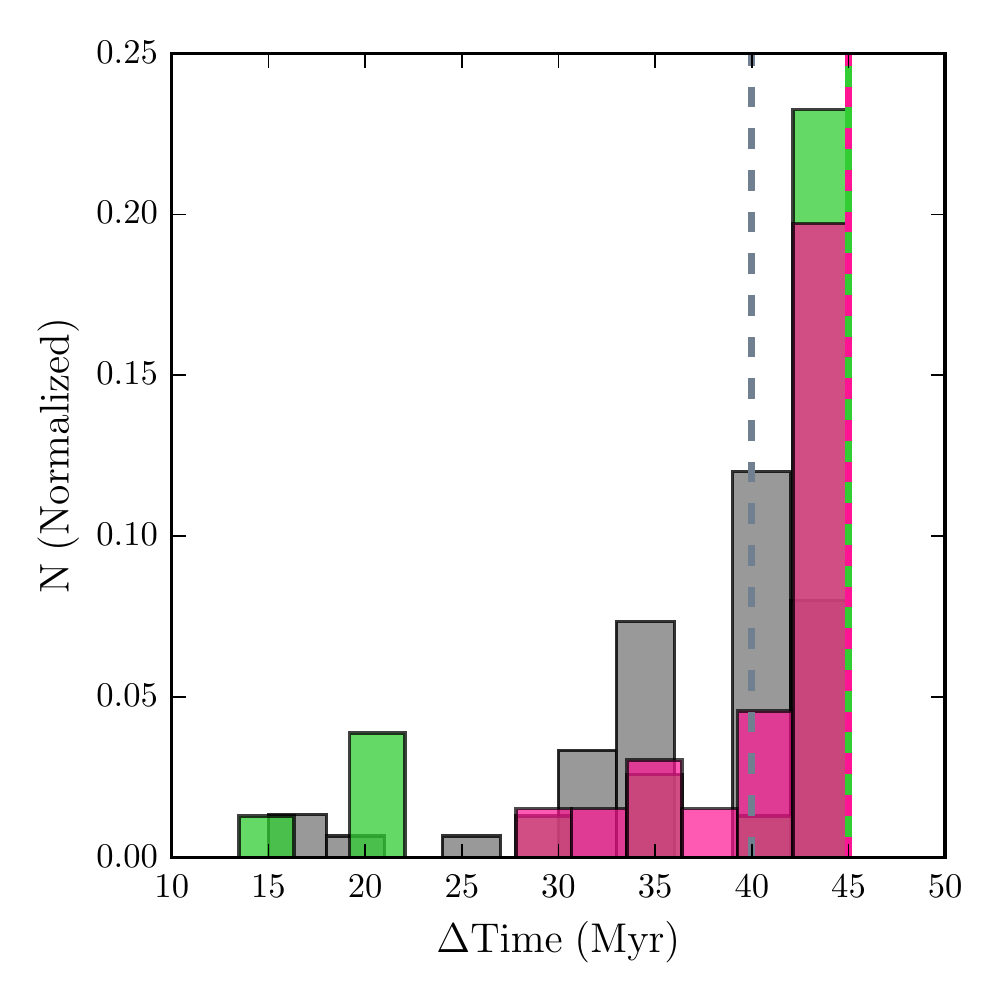}
\caption[EW deviation of galaxies during starbursts in the \Halpha\ EW vs \boxfil\ colour space.]
{ The \Halpha\ EW deviation of galaxies from our simulated sample (Table \ref{tab:simulation_param} Scenario 1). For each model galaxy, the deviations are calculated with respect to the $\Gamma=-1.35$ IMF constant SFH model at the same \boxfil\ colour. 
{\bf Top Left:} \Halpha\ EW deviations  as a function of \boxfil\ colour for 40 randomly selected galaxies. 
Galaxies are colour coded according to the fraction of stellar mass generated by the burst per unit time (measured in Myr$^{-1}$). The starbursts occur at random times, hence the galaxies deviate from the reference IMF track at random \boxfil\ colours. 
{\bf Top Right:} \Halpha\ EW deviations as a function of time.  The galaxy sample is similar to the left panel and are similarly colour coded according to the fraction of stellar mass generated by the burst per unit time.
{\bf Bottom:} The normalized histogram of the amount of time the starburst tracks stay at least 0.05 dex above the $\Gamma=-1.35$ smooth SFH model. All 100 galaxies in the simulation are shown here and have been colour coded according to the fraction of stellar mass generated by the burst per unit time. The vertical dashed lines show the median for each bin and are as follows: $\mu(f_m/\tau_b\leq0.0010)=40$ Myr, $\mu(0.0010<f_m/\tau_b<0.0015)=45$ Myr, and  $\mu(f_m/\tau_b\geq0.0015)=45$ Myr.
}
\label{fig:simulations_delta_EW_vs_col_time}
\end{figure}

We use the simulated galaxies from Scenario 1 (Table \ref{tab:simulation_param}) to further investigate the evolution of \Halpha\ EW and \boxfil\ colour of galaxies during a starburst phase. 
40 galaxies are chosen at random from the simulations in Figure \ref{fig:simulations_delta_EW_vs_col_time} to show the deviation of the \Halpha\ EW values from the $\Gamma=-1.35$ IMF track. In a smooth SFH, the \boxfil\ colours are correlated with the age of the galaxies due to stellar populations moving away from the main sequence making the galaxy redder with time. 
However, galaxies undergo bursts at random times resulting them to deviate from the smooth SFH track at random times. Galaxies with higher burst fractions per unit time ($f_m/\tau_b$), show larger deviations due to the extreme SFRs required to generate higher amount of mass within a shorter time period. 
The time scale galaxies populate above the reference IMF track is small (in the order of $\lesssim50$ Myr) compared to the total observable time window of the galaxies at $z\sim2 (\sim3$ Gyr). The \Halpha\ EW increase significantly soon after a burst within a short time scale (in order of few Myr) and decreases rapidly to be deficient in \Halpha\ EW compared to the reference constant SFH model. Afterwards, the \Halpha\ EW increases at a slower phase until the tracks join the smooth SFH model after the burst has past.

In Figure  \ref{fig:simulations_delta_EW_vs_col_time} (top right panel), we select all simulated galaxies to calculate the amount of time galaxies spend with higher \Halpha\ EW values ($0.05<\mathrm{\Delta[\log_{10}(H\alpha\ EW)]}$) compared to models with smooth SFHs. We bin the galaxies according to the burst fraction per unit time to find that there is no strong dependence of it on $\Delta$Time.

\subsubsection{Large bursts}
\label{sec:large_bursts}

Our final simulation grid contains 8337 possible time steps, which we use to randomly select galaxies within $2.0<z<2.5$ (similar to the time window where our observed sample lies) to perform a density distribution study and a stacking technique similar to the method described in Section \ref{sec:EW_stacking}.

\begin{figure}
\centering
\includegraphics[scale=0.5]{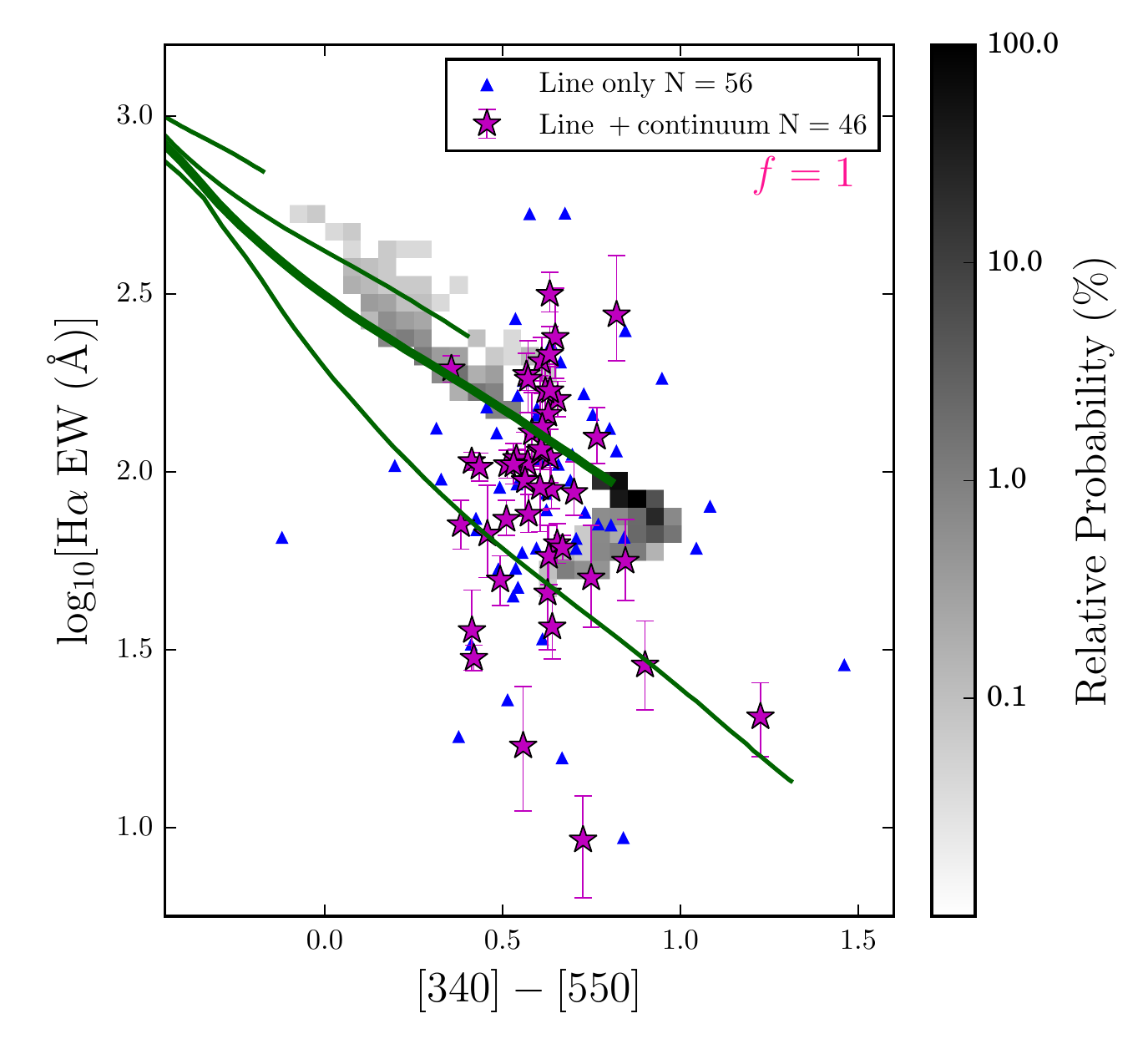}
\includegraphics[scale=0.5]{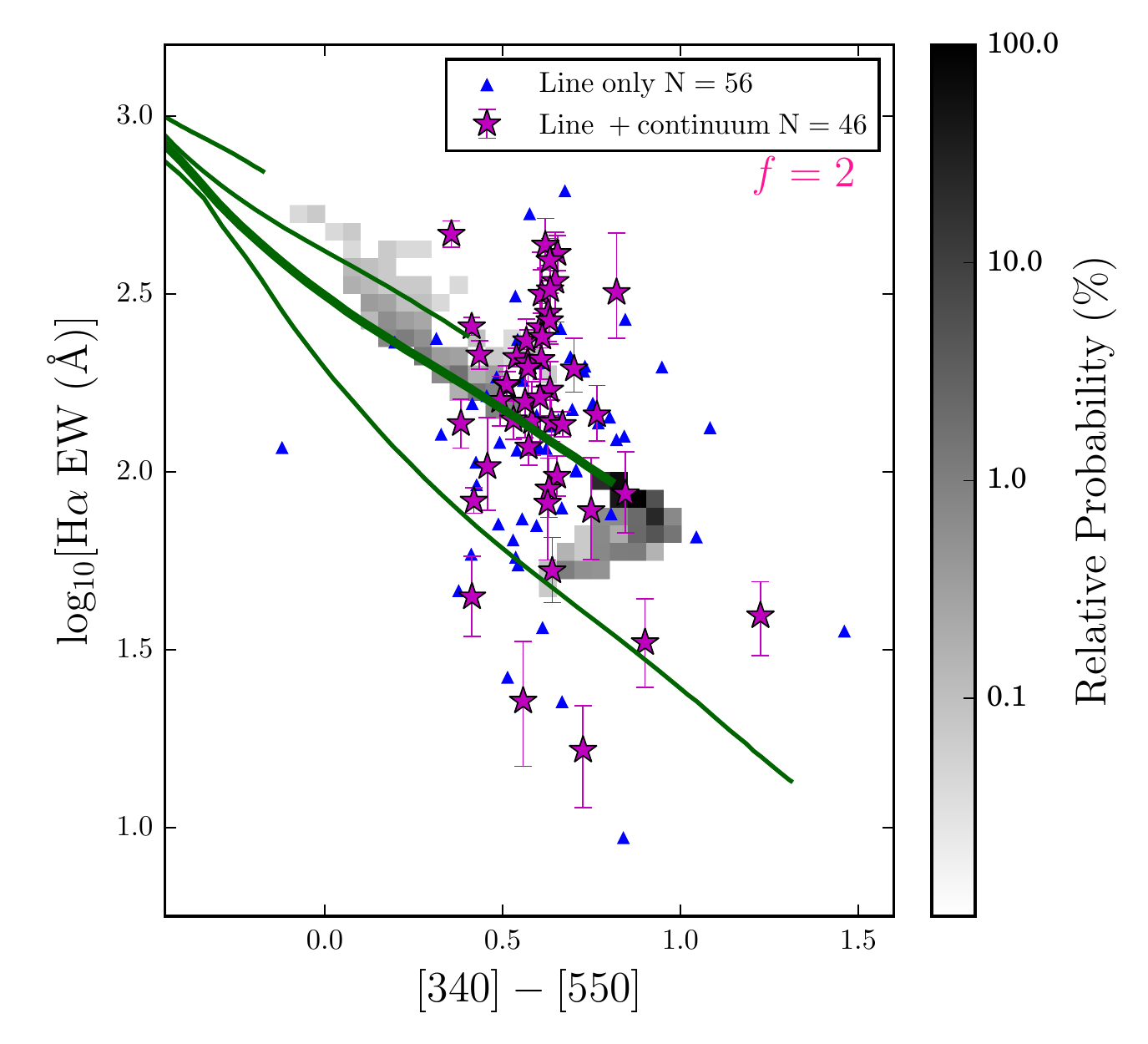}
\includegraphics[scale=0.5]{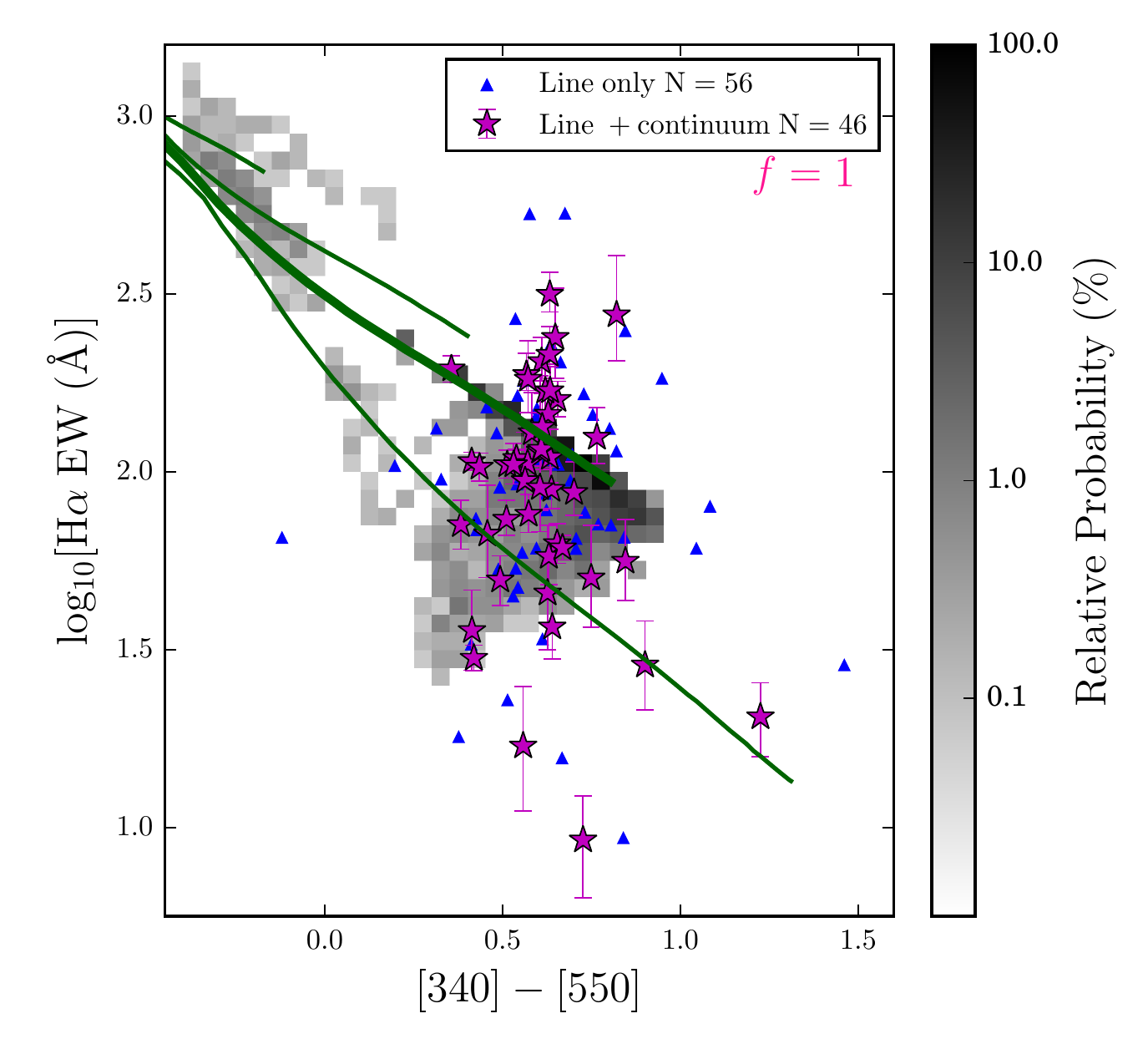}
\includegraphics[scale=0.5]{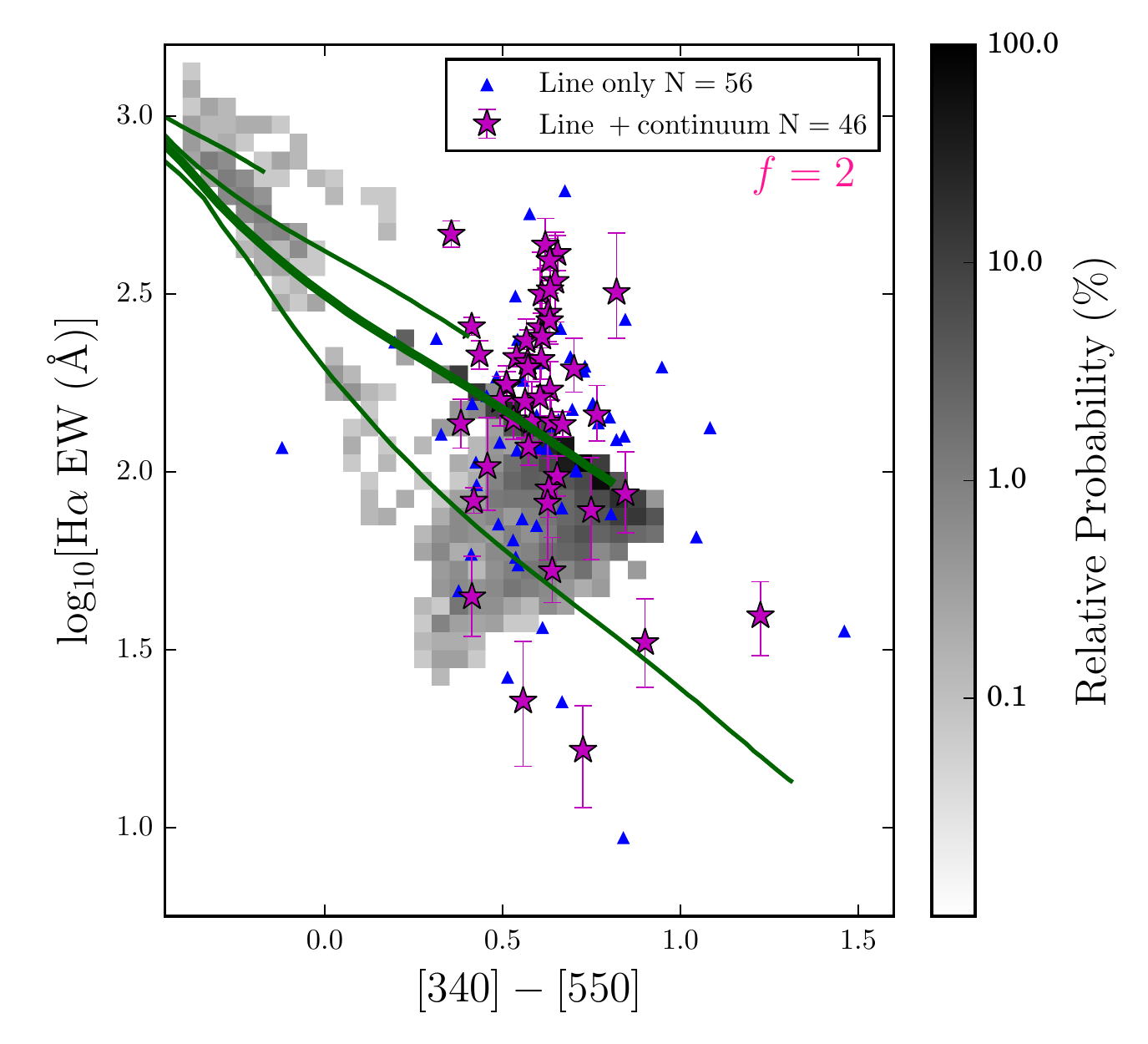}
\caption[Monte Carlo simulations of galaxies with starbursts.]{{\bf Top left:} Relative probability distribution of galaxies occupying the \Halpha\ EW vs \boxfil\ colour space. The density distribution is made from 10,000 galaxies chosen randomly from the simulated galaxies with $t_b=  100-300$ Myr and $f_m= 0.1-0.3$. A darker shade suggests a higher probability. The \sample\ is overlaid on the figure with dust corrections applied with a $f=1$ following prescriptions outlined in Section \ref{sec:dust_corrections}. The PEGASE tracks shown are similar to Figure \ref{fig:EW_stacked}. 
{\bf Top right:} Similar to left but the \sample\ has been dust corrected using a $f=2$.
{\bf Bottom:} Similar to the top panels, but galaxies are selected from a simulation with smaller $t_b$ ($10-20$ Myr) but similar $f_m$ ($0.1-0.3$) values constrained to a redshift window between $2.0<z<2.5$. 
}
\label{fig:simulation_density}
\end{figure}

To quantify the probability of starbursts dominating the scatter in the \Halpha\ EW vs \boxfil\ colour space, we select 10,000 galaxies randomly from the simulated sample to calculate the relative probability galaxies occupy in \Halpha\ EW vs \boxfil\ colour space. Figure \ref{fig:simulation_density} (top panels) shows the density distribution of the selected galaxy sample. 
The relative probability is calculated by normalizing the highest density bin to 100\%. To generate real values for the logarithmic densities we shift the distribution by 0.01 units. 
As evident from the figure, for both $f=1$ and $f=2$ dust corrections, there is a higher probability for galaxies to be sampled during the pre or post burst phase due to the very short time-scale the tracks take to reach a maximum \Halpha\ EW value during a starburst.
$\sim90\%$ of the galaxies in the \sample\ lie in regions with $\lesssim0.1\%$ probability. 
Therefore, we conclude that it is extremely unlikely that $\sim1/5$th ($f=1$) and $1/3$rd ($f=2$) of the galaxies in the \sample\ to be undergoing a starburst simultaneously and rule out the hypothesis that starbursts could explain the distribution of the \sample\ in the \Halpha\ EW vs \boxfil\ colour parameter space.

\begin{figure}
\centering
\includegraphics[scale=1.2]{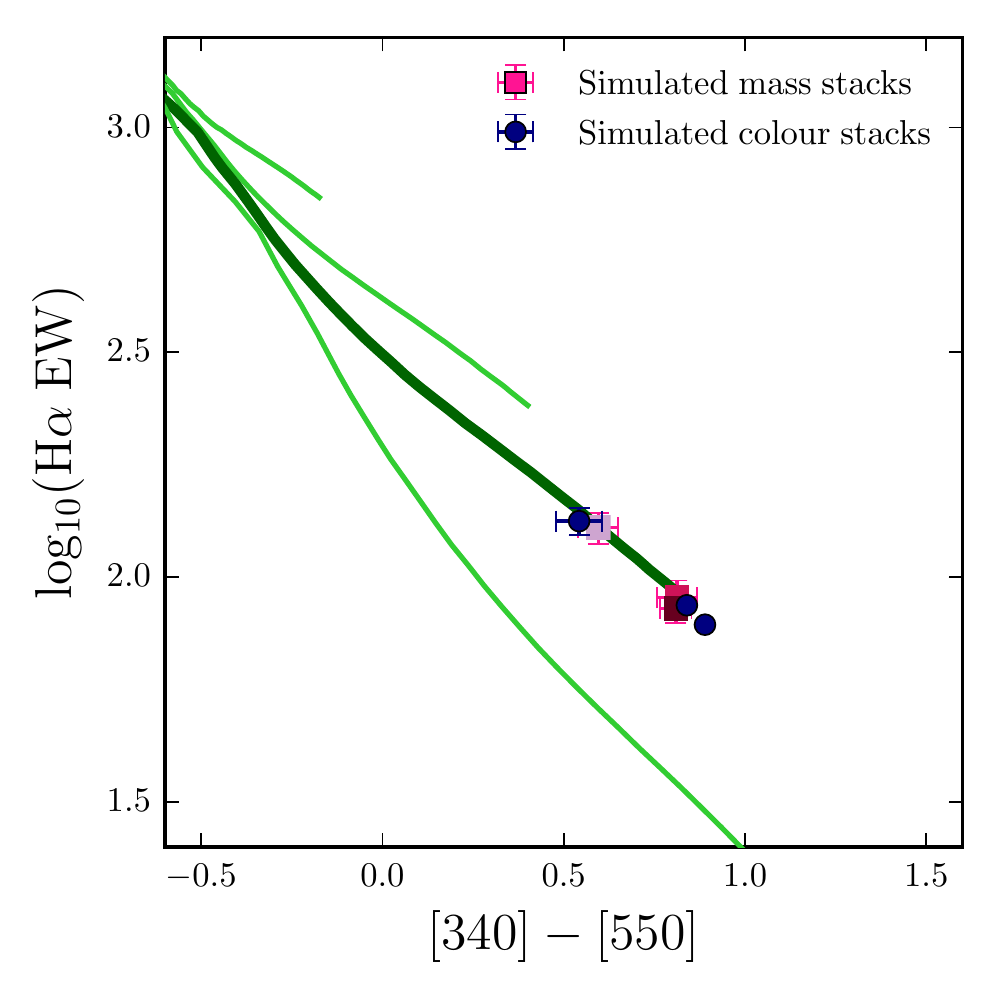}
\caption[Validation of our stacking analysis using PEGASE starburst simulations.]{\Halpha\ EW vs \boxfil\ colour distribution for stacked PEGASE galaxies used in the simulations in Figure \ref{fig:simulation_density}. Here we stack 100 randomly chosen galaxies from our simulated sample. The selected galaxies have constant SFHs with starbursts with varying strengths ($f_m=0.1-0.3$) and durations ($\tau_b=100-300$ Myr) overlaid at random times ($0-3250$ Myr) and have been stacked in stellar mass  and \boxfil\ colour bins. The errors of the stacks are computed using a bootstrap re-sampling. The PEGASE tracks shown have constant SFHs but varying IMFs.  From top to bottom the $\Gamma$ values of tracks are respectively $-0.5,-1.0, -1.35$, and -2.0. All tracks terminate at $t=3250$ Myr. 
}
\label{fig:simulation_stacks}
\end{figure}

In Figure \ref{fig:simulation_stacks}, we use our burst machinery to validate our stacking method.  
We select 100 galaxies randomly with replacement from the parent population of 100 galaxies. For each galaxy we select a random time to extract the galaxy spectra at the closest sampled time to retrieve stellar population parameters. 
We then stack the selected galaxies in stellar mass and \boxfil\ colour bins. The bins are generated in such a way that the selected galaxies are distributed evenly across the bins. 
We repeat the galaxy selection and stacking process 100 times to calculate bootstrap errors for the stacked data points. 

\newpage

\begin{deluxetable}{  c  r  r  r  r  c }
\tabletypesize{\normalsize}
\tablecaption{Summary of scenarios investigated in our starburst simulations.
\label{tab:simulation_param}}
\tablecolumns{5}
\tablewidth{0pt} 
\tablewidth{0pt}
\tablehead{
\colhead{Scenario} &
\colhead{Start of SFH (Myr)} &
\colhead{Time of burst (Myr)} &
\colhead{$\tau_b$ (Myr)} &
\colhead{$f_m$} &
}
\startdata
1 & 0       & 0--3250    & 100--300 & 0.10--0.30 \\
2 & 0--2500 & 2585--3250 &  10--30  & 0.01--0.03 \\
3 & 0--2500 & 2585--3250 &  10--30  & 0.10--0.30 \\
4 & 0--2500 & 2585--3250 & 100--300 & 0.01--0.03 \\
\enddata
\end{deluxetable}

\clearpage

Galaxies containing SFHs with bursts stacked in mass and \boxfil\ colour show similar distribution to galaxy tracks with constant SFHs. Even with large $t_b$ values, the time scale the tracks deviate significantly above the $\Gamma=-1.35$ is in the order of 1--5 Myr, and therefore it is extremely unlikely ($\lesssim5$ selected in the stacked sample of 100 galaxies) to preferentially select a large number of galaxies during this phase. 
Furthermore, stacked errors from bootstrap re-sampling do not deviate significantly from the $\Gamma=-1.35$ tracks. This further strengthens the point that repetitive sampling of galaxies does not yield stacks with higher \Halpha\ EW values for a given \boxfil\ colour.

\subsubsection{Smaller bursts}
\label{sec:small_bursts}

Galaxies at $z\sim2$ appear to be clumpy \citep[eg.,][]{Tadaki2013}. Therefore, it is possible for a single clump to have a high SFR that will add a significant contribution to the ionizing flux of a galaxy resulting in higher \Halpha\ EW values. To account for such scenarios, we perform starburst simulations with smaller burst time-scales ($t_b\sim10-30$ Myr) and large burst strengths ($f_m\sim0.1-0.3$) and allow the galaxies to commence their constant SFH at a random time between $0-2500$ Myr. We further constrain the bursts to redshifts between $2.0<z<2.5$ corresponding to a $\Delta t\sim650$ Myr, which is similar to the redshift distribution of our galaxies.

As described previously, we randomly select 10,000 galaxies from our simulated sample, but constraining the selection to the redshift window of $2.0<z<2.5$. We show the density distribution of our randomly selected sample with the observed \sample\ in Figure \ref{fig:simulation_density} (bottom panels). A large fraction of galaxies are now selected during the post-burst phase, thus with lower \Halpha\ EWs compared to the reference IMF track, specially with $f=1$ dust corrections. 
Since the star-bursts are now short lived but have to generate the same fraction of stellar masses as the longer lived bursts, the fraction of mass generated by the burst per unit time is extremely high. Therefore, changes in \Halpha\ EW and optical colour are much more drastic as a function of time and makes it further unlikely to select galaxies with high \Halpha\ EWs. 
We further test scenarios including smaller bursts within short time-scales (see Table \ref{tab:simulation_param}) and find the distribution of selected galaxies to be similar to \ref{fig:simulation_density} (top panels). 
We conclude that, even limiting the starbursts to a narrow redshift window, does not yield a distribution of galaxies that would explain our high \Halpha\ EW sample.


\section{Considering Other exotica}
\label{sec:other_exotica}

In the previous sections we have shown that the distribution of the \sample\ galaxies in the \Halpha\ EW vs \boxfil\ colour cannot solely be described by dust or starbursts within a universal IMF framework. In this section, we investigate whether other exotic parameters related to SSP models such as stellar rotations, binary stars, metallicity, and the high mass cutoff could influence the distribution of the galaxies in our parameters to impersonate a varying IMF.

\subsection{Stellar rotation}
\label{sec:stellar_rotation}

First, we consider effects of implementing stellar rotation in SSP models.
Rotating stellar models produce harder ionizing spectra with higher amounts of photons that are capable of ionizing Hydrogen. This is driven by rotationally induced larger Helium surface abundances and high luminosity of stars \citep{Leitherer2012}, which results in $\sim\times5$ higher ionizing photon output by massive O stars at solar metallicity \citep{Leitherer2014} and can be $\gtrsim\times10$ towards the end of the main sequence evolution \citep{Szecsi2015}. 
The minimum initial mass necessary to form W-R stars is also lowered by stellar rotation resulting in longer lived W-R stars \citep{Georgy2012} increasing the number of ionizing photons. 
Therefore, stars with rotation shows higher \Halpha\ fluxes compared to systems with no rotation, resulting in higher \Halpha\ EW values. 
Additionally, stellar rotation also leads to higher mass loss in stars, which results in bluer stars in the red supergiant phase \citep{Salasnich1999}. 
Furthermore, stellar models with rotation results in longer life-times by $10\%-20\%$ \citep[eg.,][]{Levesque2012,Leitherer2014}. This allows a larger build up of short lives O and B stars compared with similar IMF and SFH models with no rotation resulting in higher \Halpha\ flux values and bluer stellar populations.

S99 supports stellar tracks from the Geneva group (explained in detail in \citet{Ekstrom2012} and \citet{Georgy2013} and references therein), which allows the user to compute models with and without invoking stellar rotation. Models with stellar rotation assumes an initial stellar rotation velocity (\vini) of 40\% of the break up velocity of the zero-age main-sequence (\vcrit).

\citet{Leitherer2014} notes that, \vini$=0.4$\vcrit\ for stellar systems is of extreme nature and should be considered as an upper boundary for initial stellar rotation values. 
\vini\ is defined as the rotational velocity the star possess when it enter the zero-age main-sequence. Depending on stellar properties and interactions with other stars \citep[eg.,][]{deMink2013} the initial rotational velocity of the star will be regulated with time (see Figure 12 of \citet{Szecsi2015}, where the evolution of stellar rotation has been investigated as a function of time for models with different stellar masses and initial velocities). 

A realistic stellar population will contain a distribution of \vini/\vcrit\ values.  
\citet{Levesque2012} investigated galaxy models with 70\% of stars with stellar rotation following \vini=0.4\vcrit\ and 30\% with no stellar rotation, thus allowing more realistic conditions. 
They found that such models show  $\sim0.5$ dex less Hydrogen ionizing photons compared to a stellar population with all stars with \vini=0.4\vcrit\ stellar rotation and $\sim1.4$ dex higher number of Hydrogen ionizing photons compared to a stellar population with no stellar rotation.

The extent of stellar rotation required to describe observed properties of stellar populations is not well understood. 
Gravitational torques have been shown to prevent stars from rotating $>50\%$ of its breakup velocity during formation \citep{Lin2011}. 
\citet{Martins2013} showed that Geneva models with stellar rotation does not reproduce the distribution of massive, evolved stars accurately and requires less amounts of convective overshooting thus lowering the required \vini.
However, recent studies demonstrate the requirement for populations of stars with extreme rotation in low-metallicity scenarios to explain the origin of narrow He emission in galaxies \citep{Grafener2015,Szecsi2015} and long Gamma-ray bursts \citep{Woosley2006,Yoon2006}. 
Stellar populations of the Large Magellanic Cloud have shown to be distributed following a two peak rotational velocity distribution with $\sim50\%$ of galaxies rotating at $\sim20\%$ of their critical velocities while $\sim20\%$ of the population having near-critical velocities \citep{Ramirez-Agudelo2013}. 
Furthermore, populations of Be stars \citep{Secchi1866,Rivinius2013}, which are near-critically rotating main-sequence B stars observed in local stellar populations \citep{Lin2015,Yu2016,Bastian2017}, have shown evidence for the existence of rapidly rotating stars in massive stellar clusters \citep[eg.,][]{Bastian2017}.

We show the evolution of galaxy properties in the \Halpha\ EW vs \boxfil\ colour in Figure \ref{fig:rotation_and_binary} (top panels). 
Due to limitations in rotational stellar tracks, the metallicity of the stars are kept at $Z=0.014$, but the stellar atmospheres are kept at $Z=0.020$. 
Invoking stellar rotation increases the \Halpha\ EW by \around0.1 dex for similar IMFs and shows slightly bluer colours for a given time \emph{t}. Further analysis of the sub-components shows us that these changes in the \Halpha\ EW vs \boxfil\ colours are driven by the increase in \Halpha\ flux and bluer optical colours.

Implementing stellar rotation results in similar effects of a shallower IMF ($\Gamma>-1.35$), but the deviations are not sufficient to explain the $f=2$ dust corrected \sample\ within a universal IMF scenario. 
However, with $f=1$ dust corrections, only $\sim5\%$ of the sample lie above the $\Gamma=-1.35$ track with stellar rotation models with a majority of galaxies showing steeper IMFs.

Having a large fraction of stars with extreme rotation will lead to a higher number of ionizing photons and bluer colours and could potentially explain the high-EW objects in our sample. 
Sustaining such high rotation requires extremely low metallicities, which we further discuss in Section \ref{sec:model_Z}.  
Even though we expect the actual variation of the \Halpha\ EW and \boxfil\ colours due to stellar rotation at near-solar metallicity to be much smaller than what is shown in Figure \ref{fig:rotation_and_binary}, we cannot rule out extreme stellar rotation dominant in at low metallicities (Z$\sim0.002$). 
Therefore, extreme stellar rotation may provide one explanation independent of the IMF to describe the distribution of our galaxies in the \Halpha\ EW and \boxfil\ colour space (see Section \ref{sec:model_Z}). 
Furthermore, stellar rotation can introduce fundamental degeneracies to IMF determination which we discuss further in Section \ref{sec:ssp_issues}.

\subsection{Binary system evolution}
\label{sec:binaries}

We consider the effect of implementing the evolution of binary stellar systems on our study. All SSP models described thus far only considered single stellar populations, i.e. there were no interactions between stars in a stellar population. 
However, recent observational studies in our Galaxy have shown that \around50\% of massive O stars are in binary systems and that the environment may have a strong influence on the dynamical and/or stellar evolution \citep{Langer2012,Sana2012,Sana2013}. Only a minority of O stars would have undisturbed evolution leading to supernovae \citep{Leitherer2014}, thus introducing additional complexities to SSP models and strong implications for studies using these models to infer observed stellar properties. Furthermore, \citet{Steidel2016} demonstrated the necessity of invoking models with massive star binaries to fit rest-frame UV and optical features of star-forming galaxies at $z\sim2.5$.

\begin{figure}
\centering
\includegraphics[scale=0.70]{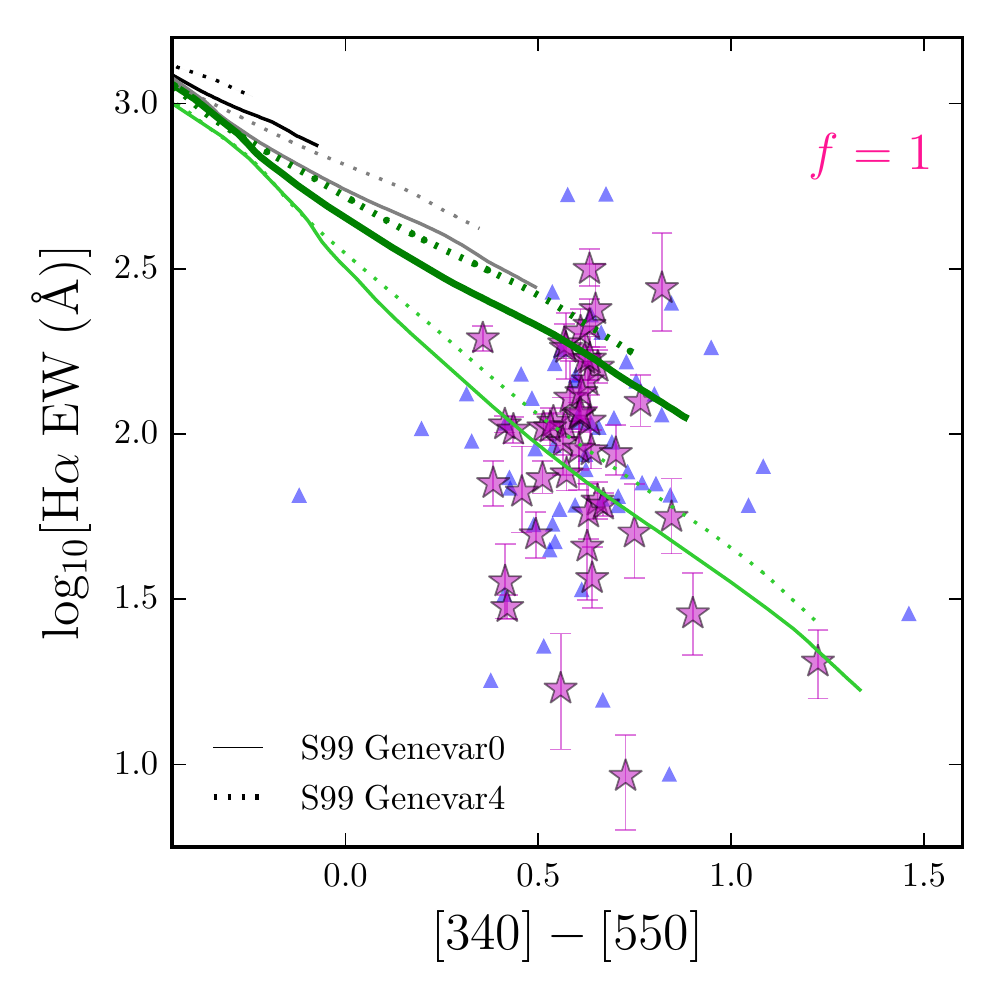}
\includegraphics[scale=0.70]{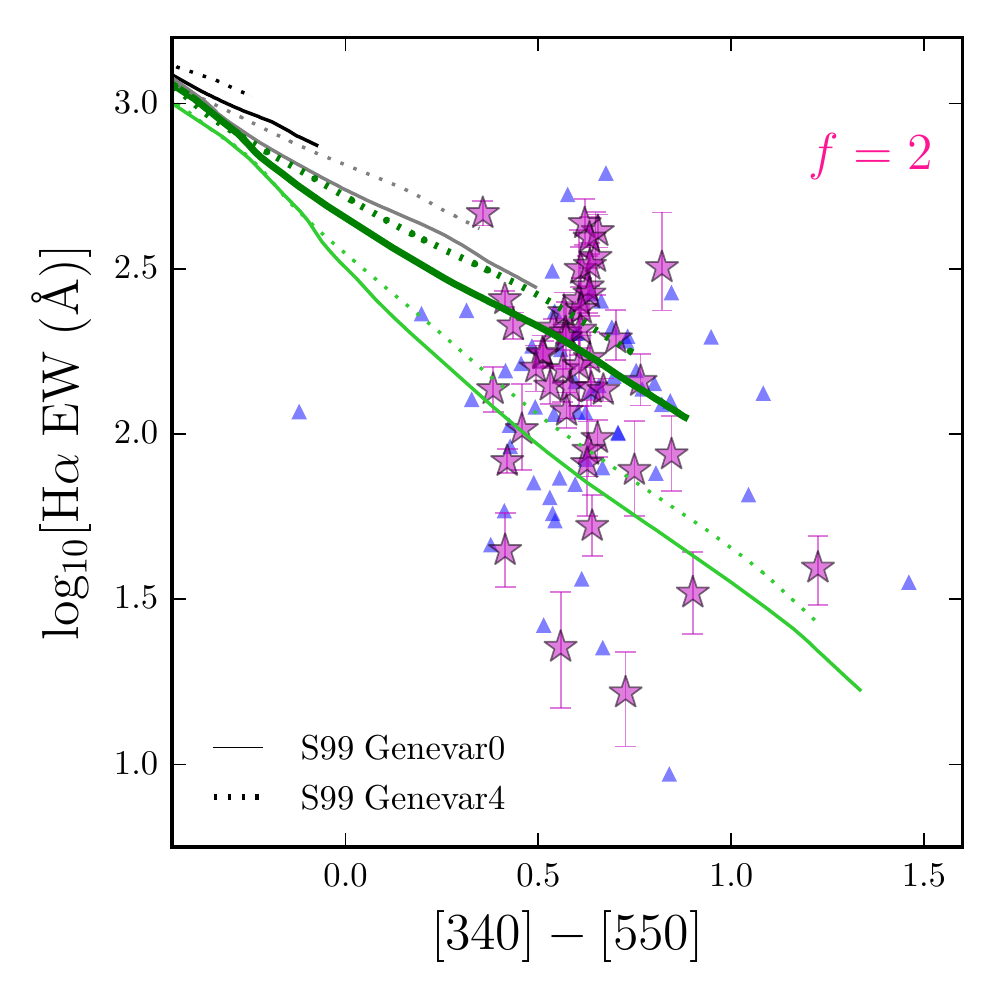}
\includegraphics[scale=0.70]{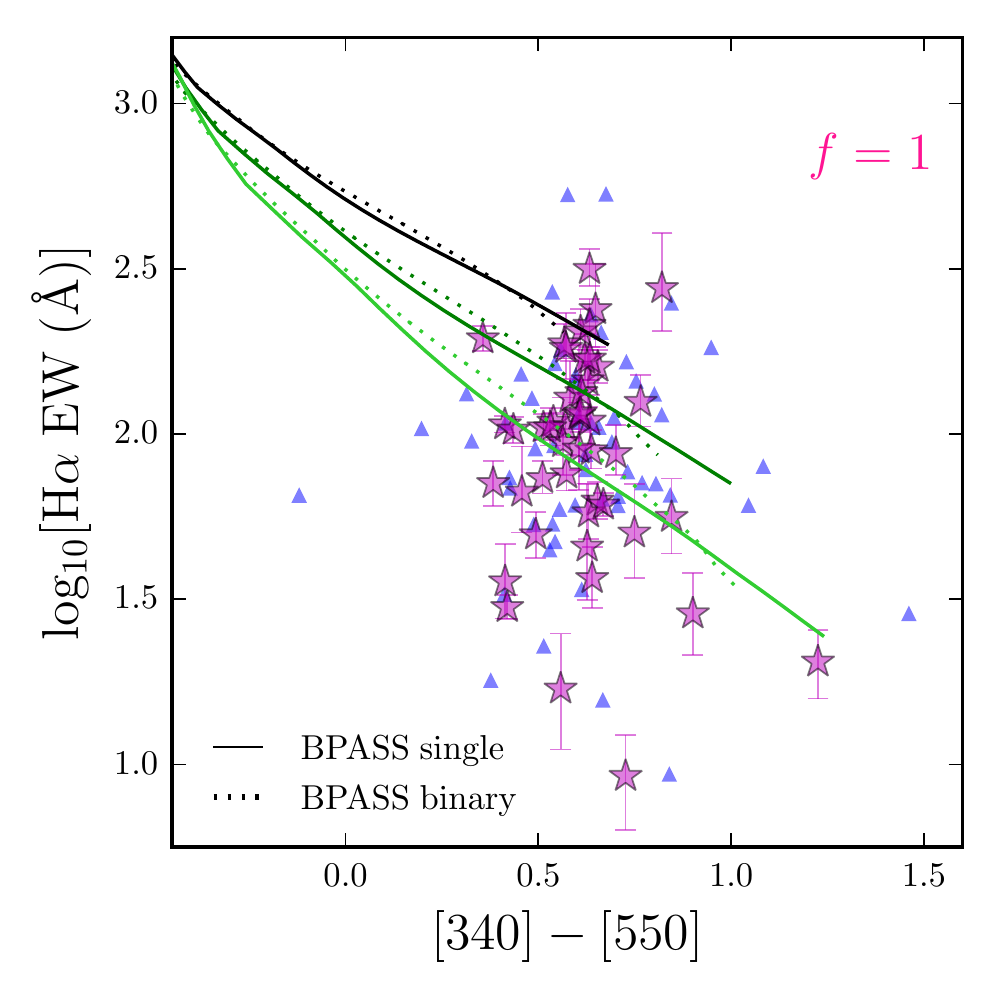}
\includegraphics[scale=0.70]{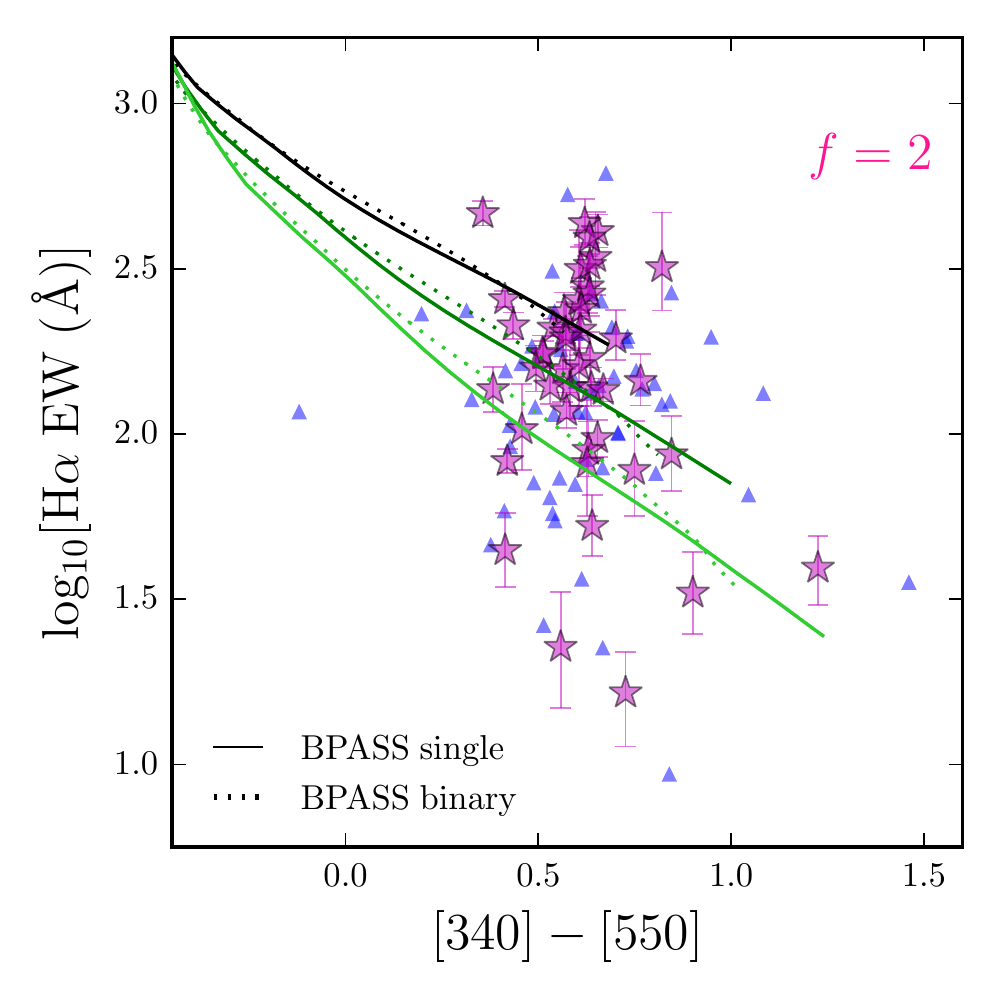}
\caption[Effects of stellar rotation and evolution of binary stars in the \Halpha\ EW vs \boxfil\ colour space.]{Effects of stellar rotation and evolution of binary stars in the \Halpha\ EW vs \boxfil\ colour space. 
{\bf Top left:} Here we show the evolution of  \Halpha\ EW vs \boxfil\ colours for Starburst99 SSP models using Geneva tracks with (dotted lines) and without (solid lines) stellar rotation. Each pair of tracks represent the same IMF and the IMFs plot are similar to Figure \ref{fig:EW_no_dust_corrections}. The overlaid \sample\ galaxies have been dust corrected using a $f=1$.
{\bf Top right:} Similar to the top left panel but the \sample  has been dust corrected using $f=2$.
{\bf Bottom left:} Here we show the effects of including the evolution of binary stellar systems in the \Halpha\ EW vs \boxfil\ colour space. The models shown here are from the \citet{Stanway2016} BPASSv2 models with Z=0.02. Each pair of tracks with same colour represent the same IMF with (dotted) and without (solid) considering the evolution of binary stellar systems. From top to bottom the IMFs plot are for $\Gamma$ values of$ -1.0, -1.35$ (Salpeter IMF), and $-1.70$. The overlaid \sample\ galaxies have been dust corrected using a $f=1$.
{\bf Bottom right:} Similar to the bottom left panel but the \sample\ has been dust corrected using $f=2$.}
\label{fig:rotation_and_binary}
\end{figure}

We use the BPASS v2.0 models \citep{Stanway2016} to investigate the effects of invoking stellar binary evolution in the \Halpha\ EW vs \boxfil\ colour space. The computed models have been released by the BPASS team only for a limited set of IMF models. We use IMF models with Z=0.02 and $\Gamma=-1.00, -1.35,\ \mathrm{and}\ -1.70$ with a lower and upper mass cutoff at 0.5\msol\ and 100\msol,respectively. 
The IMF slope for stellar masses between 0.1\msol$-$0.5\msol\ is kept at $\Gamma=-0.30$ for all the models. We remind the reader that stars with $M_*\lesssim0.5M_\odot$ have negligible effect on the \Halpha\ EW vs optical colour parameter space.
Figure \ref{fig:rotation_and_binary} (bottom panels) compares the effect of considering stellar binary system evolution in this parameter space. 
Binary rotation with simple prescriptions for stellar rotation slightly increases the \Halpha\ EW (max increase for a given time is \around0.2dex) and make galaxies look bluer for a given IMF at a time \emph{t}. These changes are more prominent for galaxies with steeper IMFs and are driven by the \Halpha\ flux and optical colours of the galaxies. Furthermore, unlike effects of rotation, we see a trend on which the steeper IMFs show larger changes (up to $\sim\times2$) in \Halpha\ flux and \boxfil\ colours  compared to shallower IMFs.

Due to higher ionizing flux and longer lifetimes of massive O type stars in binary systems, galaxies look bluer at an older age compared to what is predicted by single-star models \citep{Eldridge2016}. The increase in ionizing flux is driven by transfer of mass between stars causing rejuvenation, generation of massive stars via stellar mergers, and stripping of Hydrogen envelope to form more hot Helium or W-R stars. Mass transfer and mergers between stars also result in larger, bluer stars at later times contributing to the stellar population to be bluer. 
The change of \Halpha\ EW and \boxfil\ colours due to binary system evolution at Z=0.02 is not sufficient to explain the distribution of the \sample\ galaxies and is significantly smaller than the contribution from stellar rotation.

Note that BPASS single stellar evolutionary models do not consider any form of stellar rotation. 
BPASS binary models do consider stellar rotation, but only if a secondary star accretes material from a companion. In such scenarios at Z $>0.004$ the secondary star is spun up, fully mixed, and is rejuvenated resulting it to be a zero-age main-sequence star. 
However, it is assumed that the star is spun down quickly and stellar rotation is not considered for the rest of it's evolution \citep[][J.J. Elridge., private communication]{Eldridge2012a,Stanway2016}.
Since the current version of BPASS binary models does not consider aspects of stellar rotation in the context of reduction in surface gravity and the driving of extra-mixing beyond the expectations from the standard mixing-length theory as discussed in Section \ref{sec:stellar_rotation} (also see papers in the series by \citet{Meynet2000} and \citet{Potter2012}), comparisons between S99 Geneva models and BPASS cannot be performed to constrain the net effect of introducing stellar binary stellar evolution to SSP models that consider stellar rotation.

\subsection{Stellar metallicity}
\label{sec:model_Z}

\citet{Hoversten2008} showed that the evolution of galaxies \Halpha\ EW vs optical colour parameter space was largely independent of the metallicity of the galaxies. All PEGASE models in this thesis are generated using a fixed metallicity. 
We also tested models with consistent metallicity evolution and found to have no significant effect to the conclusions of this study. 
However, PEGASE models does not account for the increase in mass loss via stellar winds and increase in ionizing flux predicted in low metallicity scenarios, which are considered by models that include stellar rotation and binary interactions.

The lack of elements such as Fe that dominate the opacity in radiation-driven stellar winds, stellar interiors, and atmospheres in low metallicity stars results in the generation of higher amount of ionizing photons \citep[eg.,][]{Pauldrach1986,Vink2005,Steidel2016}.
Furthermore, at lower metallicities due to weaker stellar winds the mass loss is rate is low, thus most massive stars retain their luminosity and continue to shine for an extended time. 

When stellar rotation is introduced to single stellar population models, due to the higher fraction of W-R stars in higher metallicity environments, rotational stellar models with higher metallicity show a larger increment ($\Delta$EW) in ionization flux compared to the increment seen in lower metallicity models  \citep{Leitherer2014}. 
However, at $t<3100$ Myr low metallicity rotating stellar models on average show higher amount of ionising flux compared to higher metallicities.

When binary interactions are considered, the mass transfer between the binaries result in the increase of angular momentum of the stars causing an increase in stellar rotation \citep{deMink2013}. 
Additionally, at Z $\leq0.004$ if stars with \mass\ $>20$\msol\ has accreted $>5\%$ of its original mass, BPASS assumes that the star maintains its rapid rotation throughout its main-sequence lifetime \citep{Stanway2016}. 
This is driven by weaker stellar winds that allow the stars to maintain their rapid rotations for a prolonged period. 
Furthermore, rotationally induced mixing of stellar layers causes Hydrogen burning to be efficient resulting in rejuvenation of the main sequence stars. 
As we show in Section \ref{sec:stellar_rotation}, stellar rotation increases the production of ionizing photons and therefore lower metallicity systems with binary interactions show higher \Halpha\ EW values. 
Lower cooling efficiencies prominent in lower metallicity environments, also result in the stars to be bluer and brighter. 
Comparisons between S99 Geneva models with Z=0.002 and Z=0.014 suggest metallicity to have a prominent effect in increasing the \Halpha-EWs compared to stellar rotation. 
BPASS models also show metallicity effects to be prominent compared to effects by stellar rotation and binary interactions. 

Therefore, we conclude that within the scope of current stellar models, metallicity to be the prominent driver in increasing the \Halpha-EWs with stellar rotation and binary interactions contributing to a lesser degree.

In Figure \ref{fig:Z_and_high_mass} (top panels), we show the evolution of a $\Gamma=-1.35$ IMF constant SFH stellar tracks from BPASS with varying metallicities. The variation in metallicity in the \Halpha\ EW vs \boxfil\ colour is degenerated with IMF variation. 
Models with lower metallicities favour higher EWs and bluer colours compared to their higher metallicity counterparts.

Next, we explore whether gas phase metallicities computed for our galaxies \citep{Kacprzak2015,Kacprzak2016} suggest sufficiently low stellar metallicities to  produce ionising flux to explain the high-EW galaxies within a $\Gamma=-1.35$ IMF scenario. 
Converting gas-phase oxygen abundance to stellar iron abundance in high-z galaxies is nontrivial.
First, there are considerable systematic uncertainties in the gas phase metallicities measured using \NII/\Halpha\ ratios.
There are uncalibrated interrelations between ionization parameter, electron density and radiation field hardness at $z\sim2$ \citep{Kewley2013a}. 
For example, at a fixed metallicity, the \NII/\Halpha\ ratio can be enhanced by a lower ionization parameter or the presence of shocks \citep[eg.,][]{Yuan2012,Morales-Luis2013} and it is unknown whether the N/O ratio evolves with redshifts \citep{Steidel2014}.  From \citet{Kacprzak2015}, the gas-phase oxygen  abundance of our sample at $\mathrm{log10(M_*/M_\odot)=9.5}$ is $\sim0.5$ \zsol, however, the  systematic uncertainty can be  a factor of $\times$2 because of the unknown calibrations.  Because of this, we emphasize that metallicity can be compared reliably in a relative sense, but not yet on an absolute scale \citep{Kewley2008}.

Second,  there is limited knowledge on how iron abundances relative to $\alpha-$element (e.g, O, Mg, Si, S and Ca) abundance change over cosmic time and in different galactic environments \citep[eg.,][]{Wolfe2005,Kobayashi2006,Yuan2015}. 
In addition, there is a lack of consistency in  abundance scale used in   stellar atmosphere modelling, stellar evolutionary tracks and nebular models \citep{Nicholls2016}.    There are considerable variations  in the [O/Fe] ratios that are not well-calibrated at the low metallicity end.  
For example, at [Fe/H] $< - 1.0$ the extrapolated [O/Fe] ratio based on Milky Way data is  0.5 \citep{Nicholls2016}, with a $\sim0.3$ dex uncertainty in conversions of individual values \citep[eg.,][]{Stoll2013}.   
\citet{Steidel2016} argued an average [O/Fe] ratio of 0.74 for $z\sim2$ UV selected galaxies at oxygen nebular metallicity of $\sim0.5$ \zsol, suggesting a substantially lower stellar metallicity of [Fe/H] $\sim-1.0$.    
If we adopt the [O/Fe] ratio of \citet{Steidel2016},  then we would reach the same conclusion as \citet{Steidel2016} that our stellar abundance is [Fe/H] $\sim -1.0$.  In this case, we cannot completely rule out extremely low metallicity scenarios to explain the distribution of galaxies in the \Halpha\ EW vs \boxfil\ colour space. 
With $f=2$ ($f=1$) dust corrections between BPASS binary models with stellar metallicity of Z=0.02 to Z=0.002, the amount of objects that lie $2\sigma$ above the reference $\Gamma=-1.35$ track changes from $\sim40\%$ ($\sim9\%$) to $\sim9\%$ ($\sim2\%$).

Given all the uncertainties mentioned above, we think it is premature to convert our gas-phase oxygen  abundance  to stellar iron abundance and draw meaningful conclusions.
We further note that there are significant uncertainties in massive star evolution in SSP codes and the treatment of stellar rotation and binary stars, which we discuss further in Section \ref{sec:ssp_issues}.

\subsection{High mass cutoff}
\label{sec:mass_cutoff}

\begin{figure}
\centering
\includegraphics[scale=0.70]{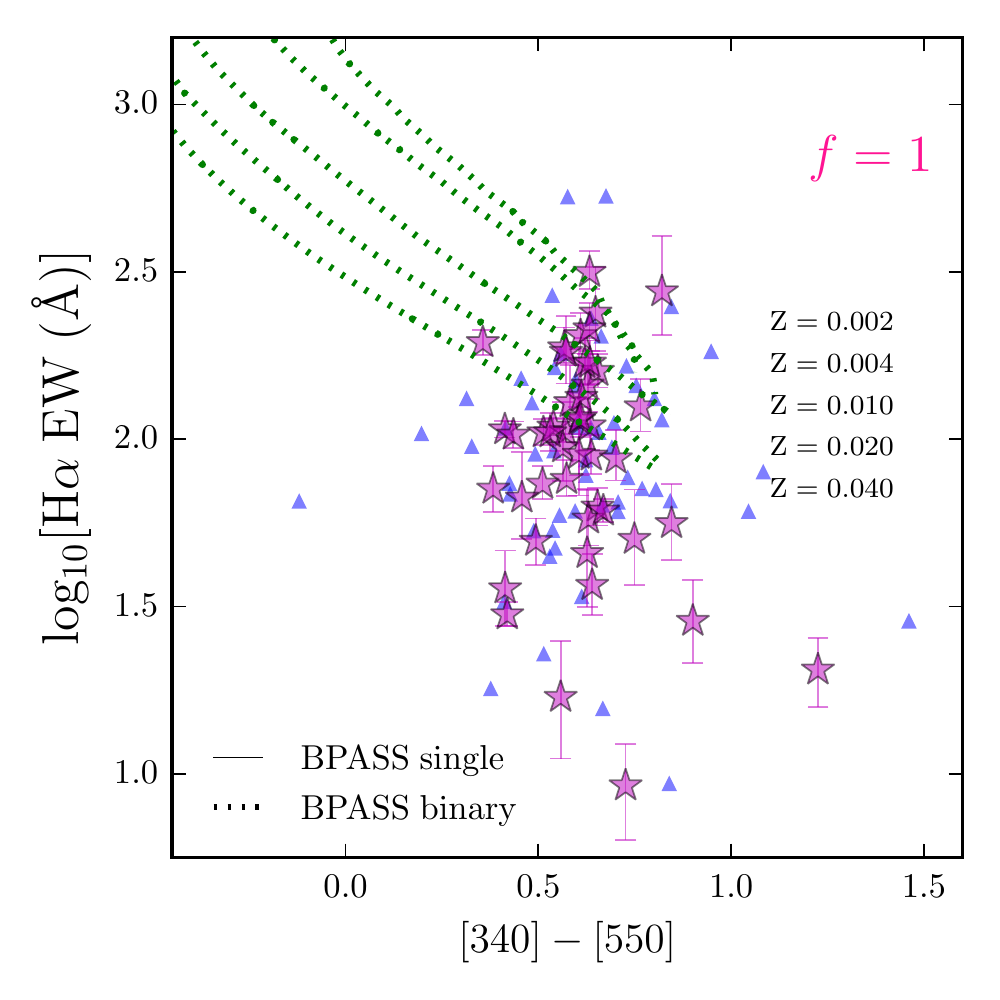}
\includegraphics[scale=0.70]{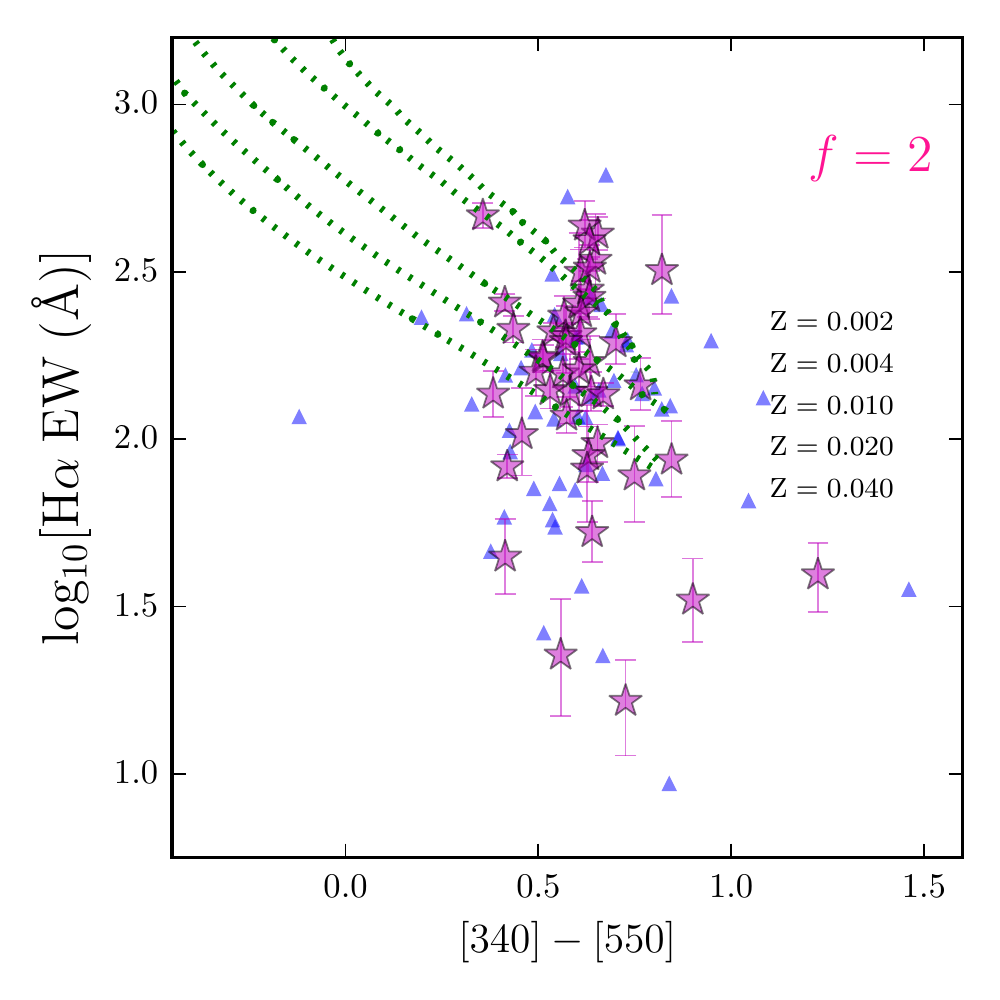}
\includegraphics[scale=0.70]{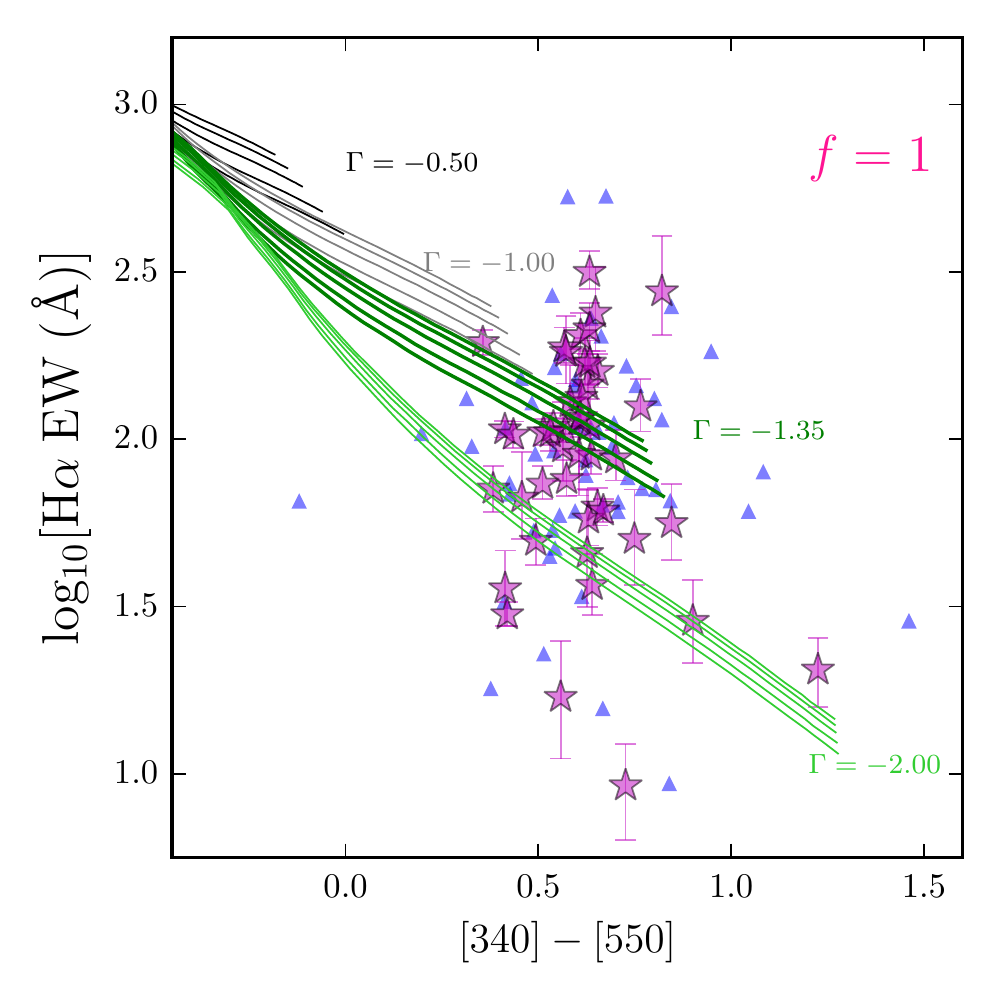}
\includegraphics[scale=0.70]{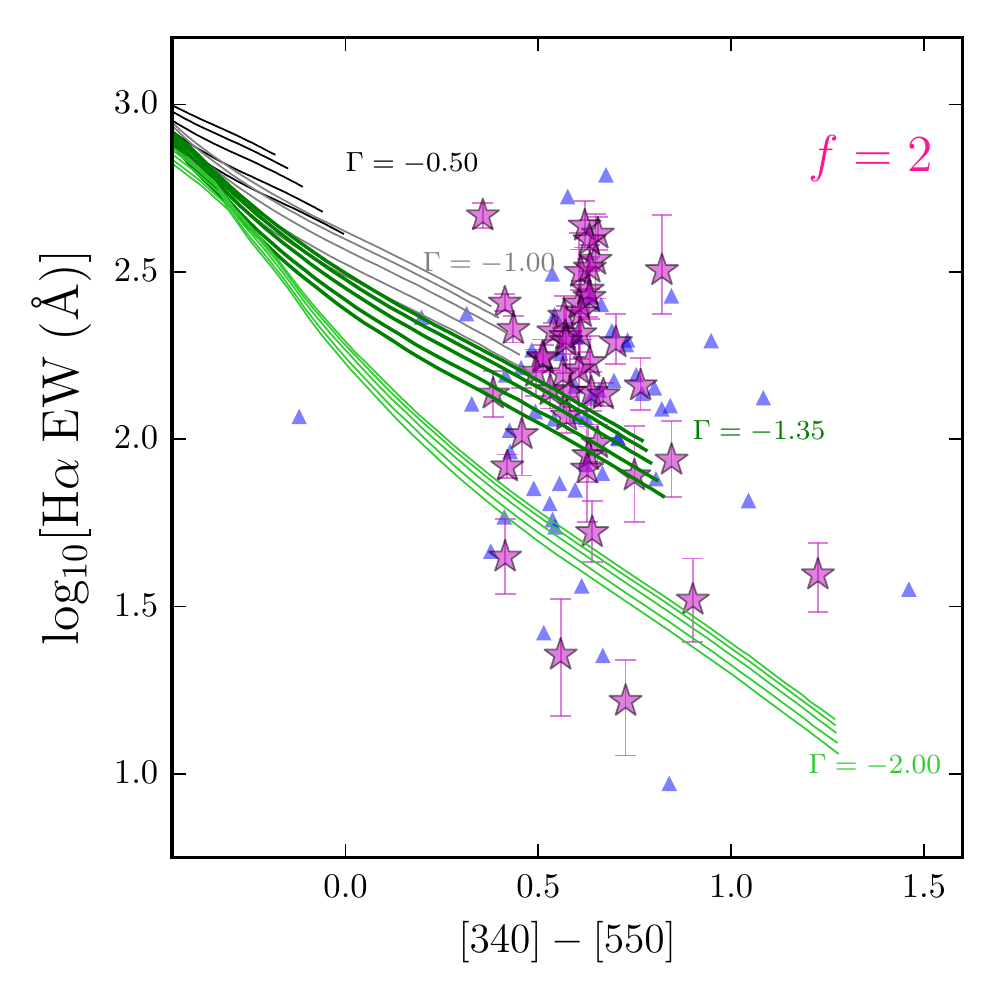}
\caption[Effects of metallicity and high mass cutoff of SSP models in the \Halpha\ EW vs \boxfil\ colour space.]{Effects of metallicity and  high mass cutoff of SSP models in the \Halpha\ EW vs \boxfil\ colour space. 
{\bf Top left:} Here we show the evolution of \Halpha\ EW vs \boxfil\ colours of the BPASS binary model galaxies with different metallicities. All BPASS models shown here have an IMF of slope $\Gamma=-1.35$ and a constant SFH. From top to bottom the metallicities of the tracks are Z=0.002, 0.004, 0.010, 0.020, and 0.040. The overlaid \sample\ has been dust corrected with a $f=1$.
{\bf Top right:} Similar to the top left panel but with the \sample\ dust corrected with a $f=2$.
{\bf Bottom left:}  Here we show the evolution of \Halpha\ EW vs \boxfil\ colours of PEGASE constant SFH models with varying IMFs and high mass cutoffs. From top to bottom each set of tracks with similar colour have IMF slopes $\Gamma=$ $-0.50, -1.00, -1.35$, and $-2.00$. Each set of IMFs are computed with varying high mass cutoffs. From top to bottom for each IMFs, the high mass cutoffs are respectively 120\msol, 110\msol, 100\msol, 90\msol, and 80\msol. The overlaid \sample\ has been dust corrected with a $f=1$.
{\bf Bottom right:} Similar to the bottom left panel but with the \sample\ dust corrected with a $f=2$.
}
\label{fig:Z_and_high_mass}
\end{figure}

\citet{Hoversten2008} showed that the high mass cutoff is degenerated with IMF in the \Halpha\ EW vs \boxfil\ colour. In Figure \ref{fig:Z_and_high_mass} (bottom panels) we show various IMF slopes with constant SFHs computed for varying values of high mass cutoff. The deviation between tracks with 80\msol\ and 120\msol\ high mass cutoff varies as a function of IMF slope. Shallower IMFs will have a larger effect when the high mass cutoff is increased due to the high number of stars that will populate the high mass regions.

The maximum deviation for the high mass cutoffs for the $\Gamma=-1.35$ tracks is 0.17 dex, which cannot describe the scatter we notice in \Halpha\ EWs of our sample. Furthermore, at $z\sim2$ we expect the molecular clouds forming the stars to be of low metallicity \citep{Kacprzak2016}, which favours the formation of high mass stars. Therefore, we require the high mass cutoff to increase, but we are limited by the maximum individual stellar mass allowed by PEGASE. BPASSv2 does allow stars up to 300\msol, however, we do not employ such high mass limits due to our poor understanding of evolution of massive stars. We conclude that it is extremely unlikely that the high mass cutoff to have a strong influence on the distribution of the \sample\ galaxies in the \Halpha\ EW vs \boxfil\ colour parameter space.

\section{Dependencies with other observables}
\label{sec:other_observables}

In this section, we investigate if the \Halpha\ EW vs \boxfil\ colour distribution show any relationship with environment, stellar mass, SFR, and metallicity of the galaxies.

ZFIRE surveyed the \citet{Spitler2012,Yuan2014} structure to great detail to probe the effects on environment on galaxy evolution. 
To date, there are 51 spectroscopically confirmed cluster candidates with ZFOURGE counterparts out of which 38 galaxies are included in our IMF analysis. The other 13 galaxies are removed from our analysis due to the following reasons: eight galaxies are flagged as AGN, two galaxies do not meet the $Q_z$ quality cut for our study, two galaxies give negative spectroscopic flux values and one object due to extreme sky line interference. 
We perform a 2-sample K-S test on the \Halpha\ EW values and \boxfil\ colours for the continuum detected cluster and field galaxies in our \sample\ and find the cluster and field samples to have similar parent properties. Therefore, we conclude that there are no strong environmental effects on the distribution of galaxies in our parameter space.


For the 22 continuum detected galaxies in common between \citet{Kacprzak2015} sample and \sample, we find no statistically significant differences between high and low metallicity samples for \Halpha\ EWs and \boxfil\ colours. 

\begin{landscape}

\begin{figure}
\centering
\includegraphics[scale=0.5]{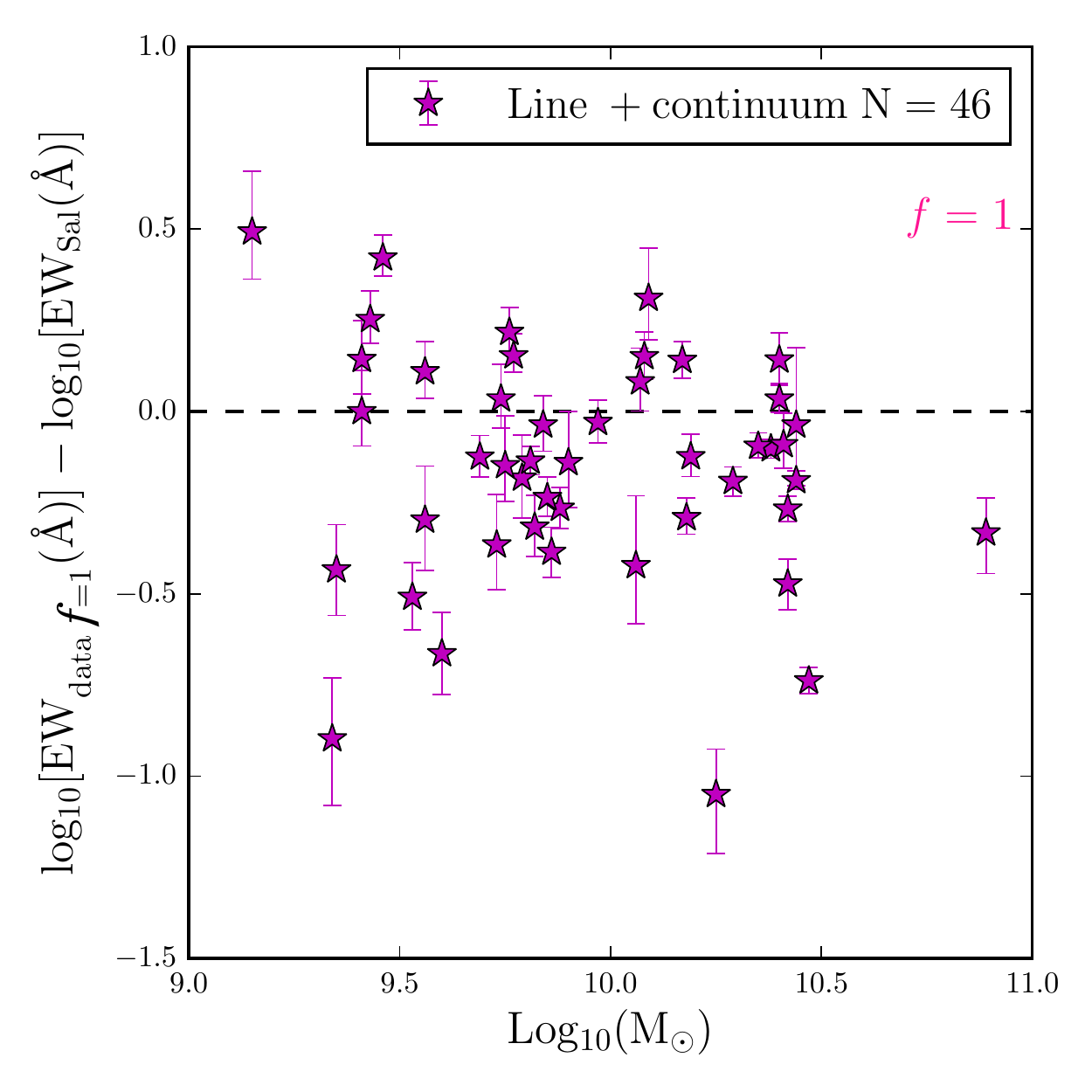}
\includegraphics[scale=0.5]{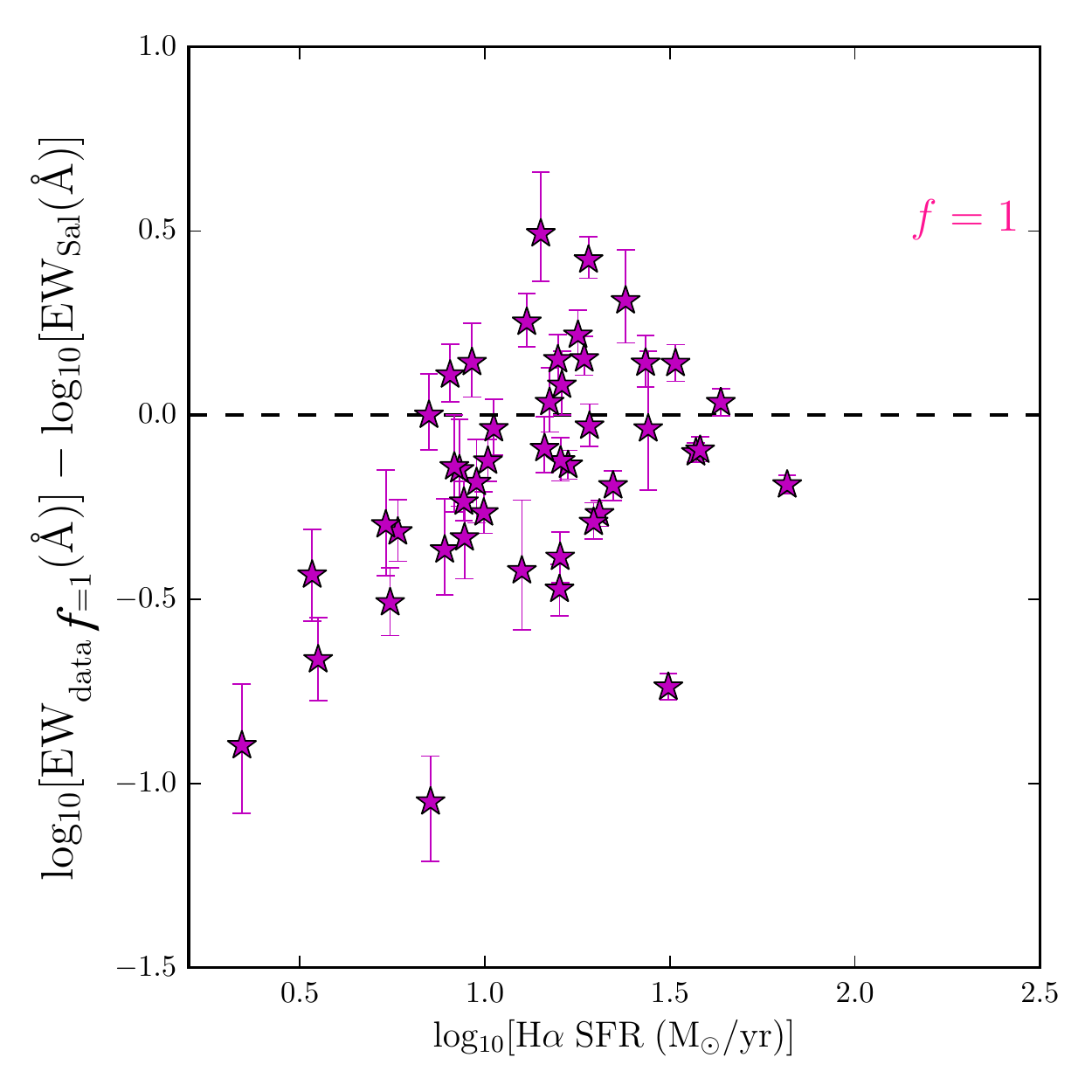}
\includegraphics[scale=0.5]{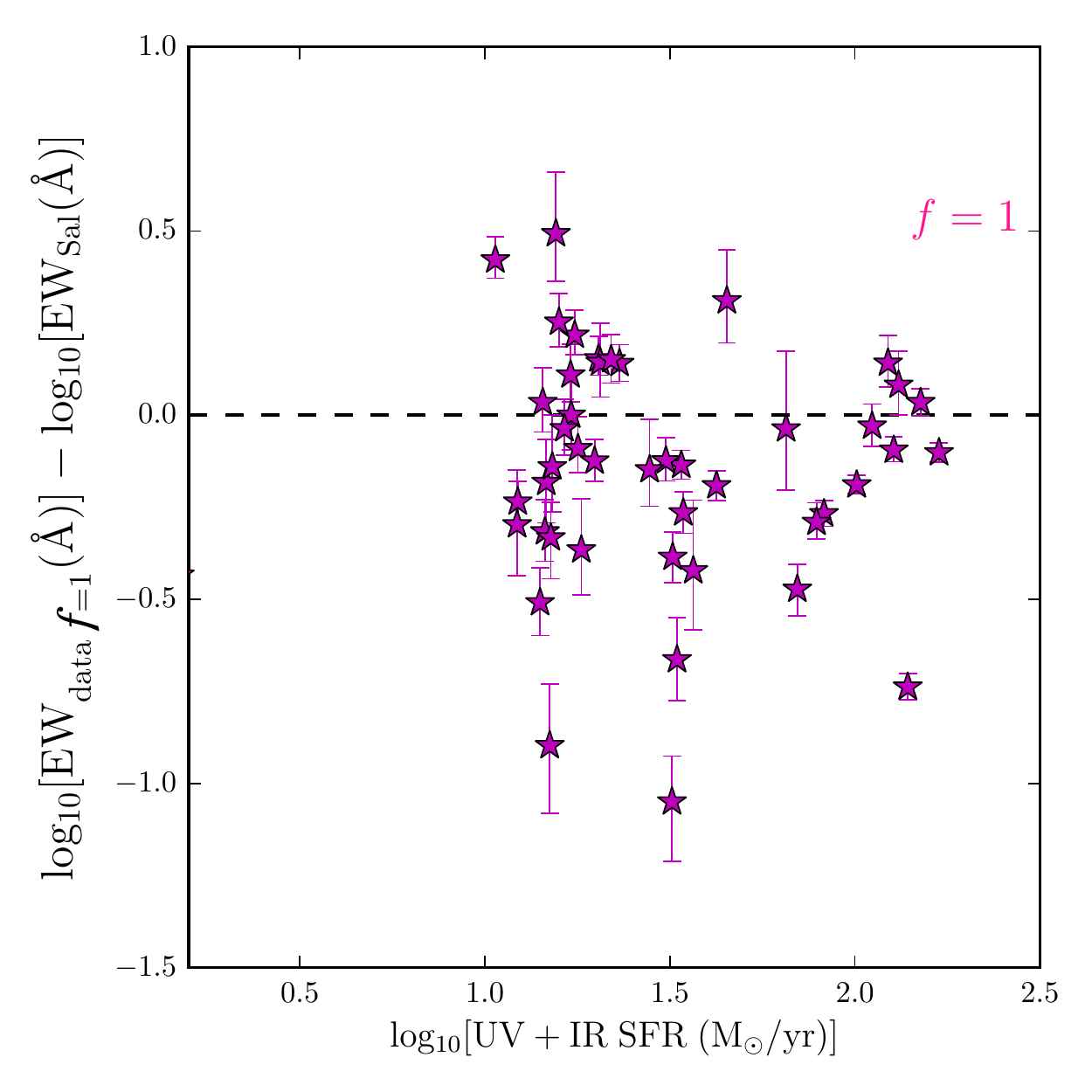}
\includegraphics[scale=0.5]{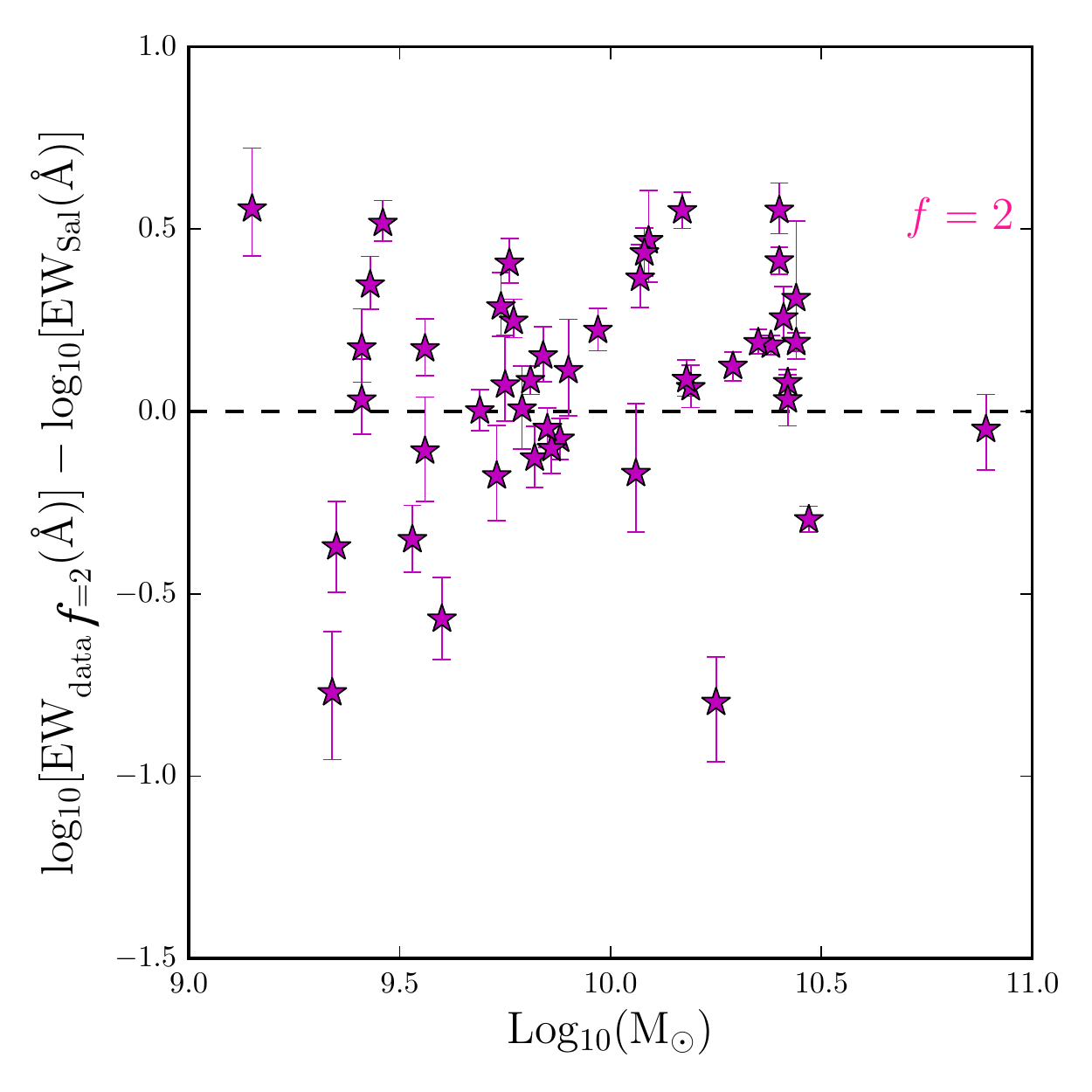}
\includegraphics[scale=0.5]{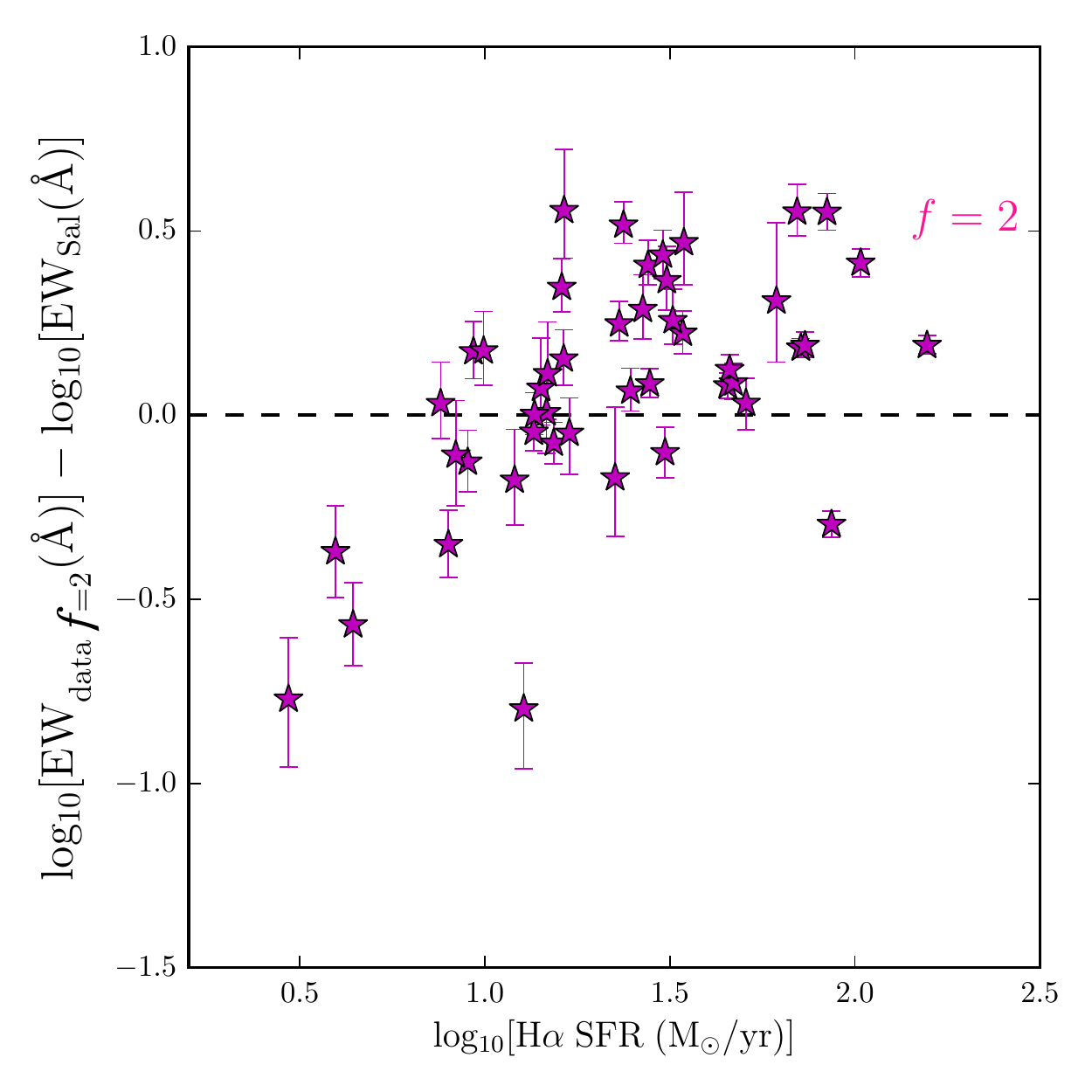}
\includegraphics[scale=0.5]{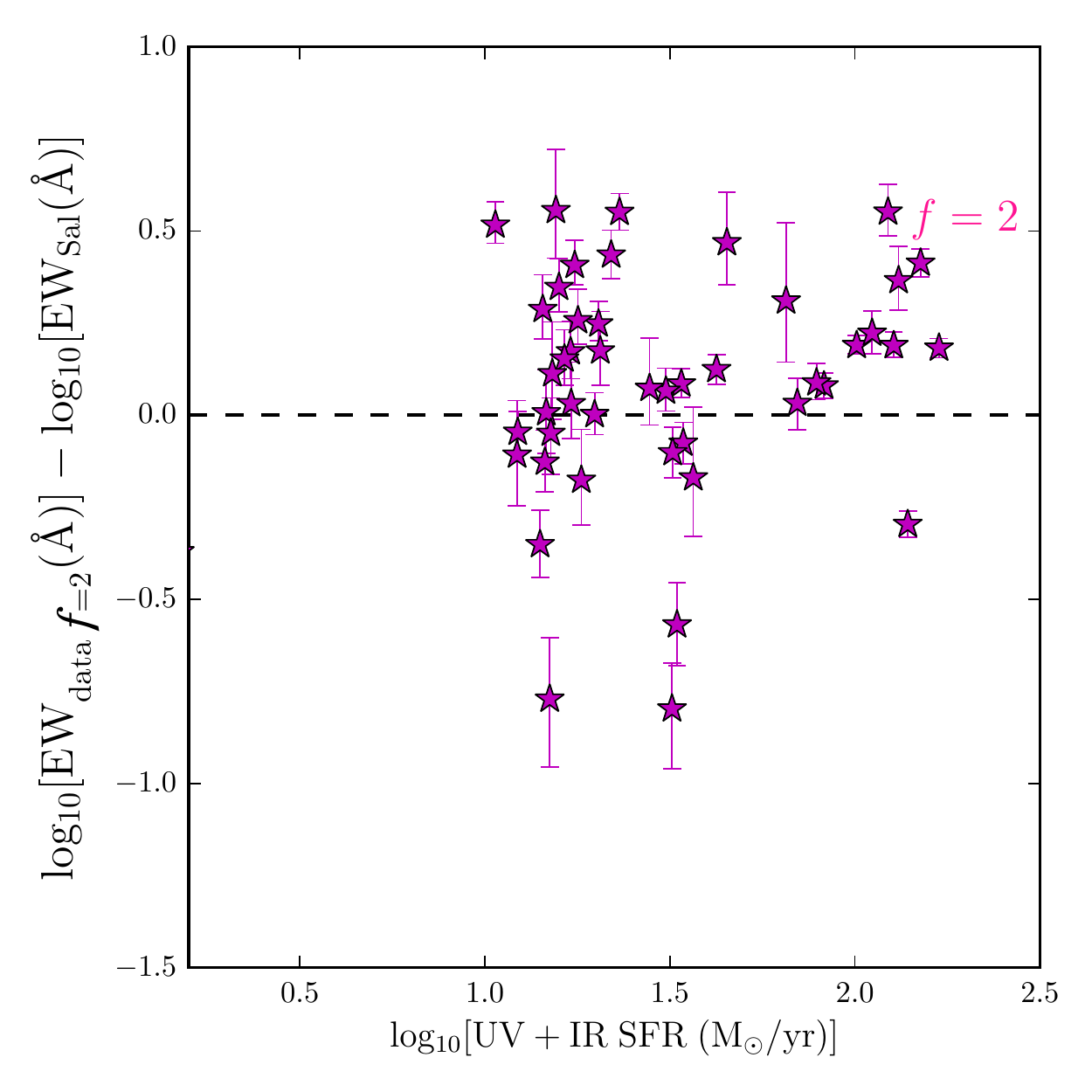}
\caption[EW excess of \sample\ from Salpeter IMF expectations.]{EW excess of the dust corrected (top panels $\rightarrow f=1$, bottom panels $\rightarrow f=2$) continuum detected sample  from a PEGASE track of $\Gamma=-1.35$ IMF slope with constant SFH.
{\bf Left panels:} EW excess as a function of stellar mass.
{\bf Centre panels:} EW excess as a function of \Halpha\ SFR. \Halpha\ SFR has been calculated using $f=1$ and $f=2$ in the top and bottom panels, respectively.
{\bf Right panels:} EW excess as a function of UV+IR SFR. 
In all panels the dashed line denotes y=0.
}
\label{fig:delta_EW_checks}
\end{figure}

\end{landscape}

We further use the Salpeter IMF tracks with constant SFHs to compare the EW excess of our continuum detected sample with stellar mass and SFR. 
In Figure \ref{fig:delta_EW_checks} (left panels) we show the EW excess as a function of stellar mass. We divide the sample into low mass ($\mathrm{log_{10}(M_*/M_\odot)<10.0}$) and high mass ($\mathrm{log_{10}(M_*/M_\odot)\geq10.0}$) bins and compute the scatter in EW excess to find that there is a greater tendency for low mass galaxies to show larger scatter in EW offsets.

Looking for IMF change as a function of SFR is inherently problematic, especially if SFR is itself computed from the \Halpha\ flux assuming a universal IMF. Nevertheless in order to compare with \citet{Gunawardhana2011} we show this in Figure \ref{fig:delta_EW_checks} (centre panels)  and confirm the trend they found of EW offset for higher ``SFR'' objects. However we refrain from interpreting this as a systematic trend for IMF variation. By using best-fitting SEDs from ZFOURGE, we compute the UV+IR SFRs \citep{Tomczak2014} and find that there is a greater tendency for low UV+IR SFR galaxies to show larger EW offsets, which is shown by Figure \ref{fig:delta_EW_checks} (right panels).

\section{Discussion}
\label{sec:discussion}

\subsection{Comparison with local studies}
\label{sec:HG08_comp}

Our study follows a method first outlined by \citet{Kennicutt1983} and later implemented on large data sets by \citet{Hoversten2008} and \citet{Gunawardhana2011} to study the IMF of star-forming galaxies.  
We find that, the distribution of \Halpha\ EWs and optical colour at $z\sim2$ to be unlikely to be driven by a sample of galaxies with a universal Salpeter like IMF. 
\citet{Hoversten2008} found a trend with galaxy luminosity with low luminosity galaxies in SDSS  favouring a steeper IMF and the highest luminosity ones showing a Salpeter slope. \citet{Gunawardhana2011} found a systematic variance in IMF as a function of SFR in GAMA galaxies with the highest-SFR galaxies lying above the Salpeter track. However, we note that the use of \gr\ colour by the $z\sim0$ studies may have given rise to additional complexities in the analysis, by introducing significant emission line contamination, and the use of SFR as a variable in IMF change is problematic as its calculation 
 depends on IMF and \Halpha\ luminosity.

Comparing our results with the local galaxies of HG08 show distinctive differences in the \Halpha\ EW vs \gr\ colour distribution.
Since galaxies at $z\sim2$ had only $\sim3.1$ Gyr to evolve, we observe  younger, bluer stellar populations giving rise to tighter \gr\ colours (distributed around 0.082 mag with a standard deviation of 0.085 mag). However, HG08 galaxy sample comprise of much redder colours with a larger scatter in \gr.  In a smooth star-formation scenario, we interpret the large scatter of the HG08 sample in \gr\ colour space to be driven by the large variety of ages of the galaxies.

Galaxies at $z\sim2$ show a large range in \Halpha\ EWs compared to $z\sim0$ results. In our analysis, we investigated several key factors that may contribute to the large scatter of \Halpha\ EW at $z\sim2$. 
Compared to $z\sim0$ galaxy populations, we expect galaxies at $z\sim2$ to be young, actively star-forming in various environments and physical conditions that may be distinctively different from local conditions. 
Therefore, effects such as starbursts may be prominent and dust properties may have significant variation, which can influence the observed \Halpha\ EW values. Galaxy mergers and multiple starburst phases in the evolutionary history of $z\sim0$ galaxies add additional layers of complexity. Furthermore, the presence of old stellar populations requires \Halpha\ absorption to be corrected, which we expect to be negligible at $z\sim2$. 
Due to the limited evolutionary time-scale at $z\sim2$ (only 3 Gyr), we consider most of these effects to have no significant influence on our analysis. 
However, we cannot completely rule out effects of dust sight-lines to our analysis, which we discuss further in Section \ref{sec:dust_discussion}.

The development of much advanced stellar tracks and greater understanding of stellar properties allow us to explore uncertainties related to stellar modelling that may significantly influence the observed parameters of galaxies at $z\sim2$.

\subsection{What do we really find?}

In Section \ref{sec:observational_bias}, we showed that our ZFIRE selected sample was not preferentially biased towards extremely high star-forming galaxies and that our mass-complete $z\sim 2$ sample is sensitive to quite low EWs. Since observational bias appears not to be the explanation we
can investigate physical factors that drive the difference in the  \sample\ from universal Salpeter like IMF scenarios in the \Halpha\ EW vs \boxfil\ colour plane.

\Halpha\ flux is a direct probe of the SFR on time scales of $\sim10$ Myr, the continuum level at 6563\AA\ provides an estimate of the mass of the old stellar populations, and therefore, \Halpha\ EW is a proxy for the sSFR. Similarly, for monotonic SFHs, the optical colours change smoothly with time, so the \boxfil\ colour 
is a second proxy for the sSFR, but with different IMF sensitivity. Of the two sSFR measures the \Halpha\ EW is the most sensitive to the highest mass stars,
so one way to express our result is to state that there is an excess of ionising photons (i.e. \Halpha) at a given SFR compared to a Salpeter-slope model. A similar result was found by \citet{Steidel2016} for a stacked galaxy spectrum, however the excess of ionising photons was attributed to the effects of binary stars.  

\citet{Shivaei2015,Shivaei2016} found that the \citet{Calzetti2000} dust law with a $f=1$ gives the best agreement between \Halpha\ derived SFRs and UV+IR derived SFRs. Given the excess in ionising photons that is observed by \Halpha\ flux is also transferred to higher UV flux and higher dust processed IR flux, such agreements are not in contradiction with our result.

In our sample, with $f=2$ dust corrections, \around50\%  galaxies have an excess of high mass stars for a given sSFR compared to the expectation by a Salpeter like IMF. 
By stacking galaxies in mass and \boxfil\ colour bins, we can average out stochastic variations in SFHs between galaxies. Our stacking results further confirmed that on average, for all masses and sSFR values, a universal IMF cannot produce the observed galaxy distribution in the \Halpha\ EW vs \boxfil\ colour space. 
We performed further analysis to understand other mechanisms that may drive this excess in \Halpha\ EW for a given sSFR.

\subsection{Dust and starbursts}
\label{sec:dust_discussion}

The dust extinction values in our analysis were derived using FAST, which uses underlying assumptions of IMF and SFH to produce best-fitting stellar parameters to galaxy observables. 
Our own analysis of dust showed SED derived extinction values from the UV slope to have a strong dependence on the assumed IMF. However for the purposes of testing consistency with a universal Salpeter-slope this suffices.

We further found that differential extinction in dust between the stellar continuum and nebular emission line regions can introduce significant scatter to galaxies in the \Halpha\ EW vs \boxfil\ colour parameter space. By analysing Balmer decrement values for a subset of galaxies in our sample, we found that, there was significant scatter in the relation between extinction of nebular and stellar continuum regions ($f$), which can be attributed to differences in dust sight-lines between galaxies. 
\citet{Reddy2015} showed that this scatter in extinction to be a function of \Halpha\ SFR, where galaxies with higher star-forming activity shows larger nebular extinction compared to galaxies with low SFRs. 
We test this by allowing $f$ values to vary as a free parameter in the \Halpha\ EW vs \boxfil\ colour space for each galaxy to force agreement with a universal Salpeter like IMF. Our results showed extreme values for the distribution of $f$, including unphysical negative values, suggesting that it is extremely unlikely that the scatter in the \Halpha\ EW vs \boxfil\ colour space is driven solely by the variation of $f$ values.

Starbursts in galaxies can introduce significant scatter in the \Halpha\ EW vs \boxfil\ colour space. We implemented a stacking procedure for the galaxies in mass and colour bins to remove stochastic SFHs of individual galaxies, treating them as an ensemble stellar population with a smooth SFH prior to $z\sim 2$. We found that, our stacks on average ($50\%$ of the $f=1$ stacks and $100\%$ of the $f=2$ stacks) favour shallower IMF slopes compared to the traditional $\Gamma =-1.35$ values from Salpeter. By performing Monte Carlo simulations of starbursts using PEGASE SSP models we found that time-scales of bursts makes it extremely unlikely for them to account for the galaxies which lie significantly above the $\Gamma=-1.35$ track.

\subsection{Dependencies on SSP models and stellar libraries}
\label{sec:ssp_issues}

We compared the evolution of model galaxies in the \Halpha\ EW vs \boxfil\ colours using PEGASE and Starburst99 SSP codes to conclude that the evolution of these parameters are largely independent of the SSP models used for a given stellar library. The \boxfil\ colours were designed in order to avoid strong emission lines regions in the rest-frame optical spectra which averts the need of complicated photo-ionization codes to generate nebular emission lines. \Halpha\ flux is generated using a constant value to convert Lyman continuum photons to \Halpha\ photons, which is similar between PEGASE and S99.

We found that, stellar libraries play a vital role in determining the evolutionary tracks of galaxies in the \Halpha\ EW vs \boxfil\ colour parameter space. Stellar libraries with rotation show higher amounts of ionizing flux which results in higher \Halpha\ EW values for a given \boxfil\ colour. 
\cite{Leitherer2014} further showed that rotation  leads to larger convective cores in stars increasing the total bolometric luminosity, which can mimic a shallower IMF. 
At \boxfil= 0.61, introducing stellar rotation via Geneva stellar tracks with Z=0.014 results in $\Delta\mathrm{log_{10}[EW (log_{10}(\AA))]\sim0.09}$. Therefore, we found that rotation cannot itself account for the scatter of our sample in \Halpha\ EW vs \boxfil\ colour parameter space at near solar metallicities.

Consideration of binary stellar systems is imperative to understand the stellar properties of $z\sim2$ galaxies \citep{Steidel2016}. 
However, added complexity arises due to angular momentum transfer during binary star interactions. This may influence the rotation of the galaxies and therefore it is necessary to consider the evolution of binary stars with detailed prescriptions of stellar rotation. 
Metallicity of the stars become important in such scenarios, which is a strong factor that regulates the evolution of stellar rotation. 
However, adding additional degrees of freedom for SSP models makes it harder to constrain their values, thus resulting in extra uncertainties \citep{Leitherer2014}. 
At \boxfil= 0.61, introducing the effect of binaries via BPASS models resulted in $\Delta\mathrm{log_{10}[EW (log_{10}(\AA))]\sim0.01}$. 
Comparing results between S99 (single stellar population stellar tracks with and without rotation) and BPASS (single and binary stellar tracks with rotation), we found stellar rotation to have a larger contribution to the $\Delta$EW compared to binaries. 
Direct comparisons require further work to investigate differences in the evolution of stellar systems between S99 and BPASS SSP codes.

We found low stellar metallicities (Z$\sim$0.002) to have a strong influence in increasing the  \Halpha\ EWs for a given \boxfil\ colour. 
At \boxfil= 0.61, reducing the metallicity of BPASS binary models from Z=0.02 to Z=0.002 resulted in  $\Delta\mathrm{log_{10}[EW (log_{10}(\AA))]\sim0.36}$. 
This was largely driven by the increase in the number of ionization photons in the stellar populations due to lower opacities, lower mass loss via stellar winds, and sustained stellar rotation. 
Interactions between stars also contribute to an increase in ionising flux. 
When considering the ionization energy generated by a stellar population, effects of stellar rotation is degenerated with the abundance of high mass stars (see Figure 16 of \citet{Szecsi2015}). 
Therefore, we cannot completely rule out effects of stars with extremely low metallicities to describe the distribution of our galaxies in the \Halpha\ EW vs \boxfil\ colour parameter space.
In Section \ref{sec:model_Z}, we provided a thorough analysis of the gas phase metallicities derived for the ZFIRE sample by \citet{Kacprzak2015} and \citet{Kacprzak2016} and inferred the metal abundances of stellar systems, which is a primary regulator of ionising photons. 
However, uncertainties in deriving gas phase abundances of elements via nebular emission line ratios (driven by our limited understanding of the ionization parameter at low metallicities), uncertainties in computing relative abundances of $\alpha$ elements in stellar systems, and our limited understanding on linking gas phase metallicities to stellar metallicities in $z\sim2$ stellar populations constrains our ability to distinguish between effects of metallicity and IMF.

\subsection{Case for the IMF}
\label{sec:IMF_discussion}

So far we have investigated various scenarios (summarised in Table \ref{tab:summary_table}) that could explain the distribution of the \sample\ galaxies in the \Halpha\ EW vs \boxfil\ colour without invoking changes in the IMF. However, none of the scenarios by itself could best describe the distribution of our galaxies.

The galaxies in our sample have stellar masses between $\log_{10}(M_*/M_\odot)=9-10$ and we expect these galaxies to grow in stellar mass during cosmic time to be galaxies with stellar masses of $\sim\log_{10}(M_*/M_\odot)=10-11$ at $z\sim0$ \citep{DeLucia2007,vanDokkum2013b,Genel2014,Papovich2015}. 
Recent studies of ETGs with physically motivated models have shown the possibility for a two phase star-formation \citep[eg.,][]{Ferreras2015}. Furthermore recent semi-analytic models have shown that a varying IMF  best reproduces observed galaxy chemical abundances of ETGs \citep[eg.,][and references therein]{Lacey2016,Fontanot2017}. According to these models, ETGs, during their starburst phases at high-redshift are expected to produce higher fraction of high mass stars (shallower IMFs). \citet{Gunawardhana2011} showed that $z\sim0$ star-forming galaxies also show an IMF dependence, where highly star-forming galaxies prefer shallower IMFs.

If we consider a varying IMF hypothesis, our results are consistent with a scenario where star-forming galaxies form stars with a high fraction of high mass stars compared to their local ETG counterparts. With lower metallicities and higher SFRs prominent at $z\sim2$, we expect the fragmentation of molecular clouds to favour the formation of larger stars due to lower cooling efficiencies and higher heating efficiencies due to radiation from the young massive stars \citep{Larson2005}. \citet{Krumholz2010} showed that radiation trapping prominent in high star-forming regions of dense gas surface density can also favour the formation of massive stars. 
If we allow the IMF to vary in our analysis, the distribution of the \sample\ in \Halpha\ EW vs \boxfil\ colour space can be explained, however values as shallow as $\Gamma=0.5$ could be required. This could be problematic for chemical evolution models and have implications to how galaxies form and evolve \citep{Romano2005}. 
We note that invoking extremely shallow IMFs can have a significant influence on the inferred evolution of the universe. Therefore, it is imperative to fully understand these observations and test alternate explanations.

\subsection{Effect of IMF variation on fundamental quantities} 

If the IMF does vary, we need to consider the potential effect on the basic parameters in our input ZFIRE survey, which were calculated using a Chabrier IMF \citep{Chabrier2003}. 
First we consider possible effects on the calculation of our rest frame \boxfil\ colours. This should not have a significant effect for several reasons: first we are only
using the spectral models as an interpolator, and by design we are interpolating only across a small redshift range. At $z=2.1$ the interpolated and observed colours
agree well as discussed in Appendix B. Second we note that the main effect is an increased scatter in the EW axis (Figure \ref{fig:EW_HG08_comp}), once dust corrected the colours are
quite tight.  Finally we note that at these young ages everything is quite blue, hence quite flat spectra are being interpolated.

Next is the effect on SFR and stellar mass, which have been used in many of the previous ZFIRE papers \citep{Yuan2014,Tran2015,Kacprzak2015,Alcorn2016,Kacprzak2016,Kewley2016,Nanayakkara2016}, and here in our own mass selection. To quantify the change
in mass we run PEGASE  for $\Gamma$ and constant SFH models and estimate the change in $R$-band mass-to-light ratio ($\simeq$K-band at
$z\simeq 2$) for ages 1--3 Gyr. We find for $-1.35<\Gamma<-0.5$ the change in mass-to-light is $<0.7$ dex, with shallower IMFs resulting in a lower stellar mass.
Thus we conclude that our stellar mass selection is only slightly effected by the possible IMF variations we have identified.

The effect is much more severe for \Halpha\ derived star-formation rates \citep{Tran2015,Tran2017} as these directly count the number of the most massive stars, a sensitivity we have exploited
in this paper to measure IMF. For $-1.35<\Gamma<-0.5$ the change is $\sim1.3$ dex. 
UV and far-IR derived SFRs are more complicated. The rest-frame UV is more sensitive to intermediate mass stars, at 1500\AA\ the change in flux is $\sim0.4$ dex
for $-1.35<\Gamma<-0.5$. The far-IR is from younger stars in deeper dust-enshrouded regions, at least in local galaxies \citep{Kennicutt1998}. It is common at high-redshift to use an indicator that combines UV and far-IR \citep[eg.,][]{Tomczak2014}. These are often calibrated using stellar population models with idealized SFHs, and traditional IMFs  and for a fixed dust mass the balance between UV and IR luminosities will depend on dust geometry, IMF, and SFH \citep{Kennicutt1998,Calzetti2013}. Therefore, IMF change could lead to difficulties in predicting the true underlying SFR of stellar populations.

\begin{landscape}
\begin{deluxetable}{  l  l  l  l  l   l   l  l   l  l l}
\tabletypesize{\scriptsize}
\tablecolumns{3}
\tablewidth{0pt} 
\tablecaption{Summary of scenarios investigated to explain the distribution of the \sample\ in the \Halpha\ EW vs \boxfil\ colour parameter space within a universal IMF framework.
\label{tab:summary_table}}
\tablehead{\colhead{Effect}                                    &
           \colhead{Section}                             	   &
           \colhead{SSP model}                             	   &
           \colhead{SFH}                             	   	   &
           \colhead{Z}                             	           &
           \multicolumn{2}{c}{Median($\Delta$EW)}              &
           \multicolumn{2}{c}{\%\tablenotemark{a}}     	   	   &
           \colhead{Conclusion}                                \\
           \colhead{}                                          & 
           \colhead{}                                          & 
           \colhead{}                                          & 
           \colhead{}                                          & 
           \colhead{}                                          &        
           \colhead{$f=1$}                                     &
           \colhead{$f=2$} 									   &
           \colhead{$f=1$}                                     &
           \colhead{$f=2$} 									   &
           \colhead{}                                          & 
           }
\startdata
& \\
{\bf Dust} 		& \ref{sec:dust} & PEGASE & Exp declining ($\tau=1000$Myr)	& 0.020	& -0.13 & 0.10  & 20\% & 46\% &  Unlikely\tablenotemark{b} \\
{\bf Observational bias} & \ref{sec:observational_bias} & PEGASE & Exp declining ($\tau=1000$Myr) & 0.020 & $--$ & $--$ & $--$ & $--$ &   Excluded \\ 
{\bf Star bursts} & \ref{sec:star_bursts} & PEGASE & Constant & 0.020 & $--$ & $--$ & $--$ & $--$ &   Excluded \\ 
{\bf Stellar rotation} & \ref{sec:stellar_rotation} & S99 & Constant & 0.020 & -0.39 & -0.15 & 2\% & 13\%  &  Probable\tablenotemark{c} \\
{\bf Binaries} & \ref{sec:binaries} & BPASS & Constant & 0.020 & -0.18 & 0.05 & 9\%  & 39\% &  Future work\tablenotemark{c} \\
\sidehead{\bf Metallicity}
Single stellar systems &  \ref{sec:model_Z} & BPASS & Constant & 0.020 & -0.17 & 0.06 & 13\% & 37\%  & Unlikely \\
(with rotation)		   &   					&  		& Constant & 0.010 & -0.29 & -0.06 & 4\% &  22\% & Probable \\ 
   					   &					&  		& Constant & 0.002 & -0.47 & -0.23 & 0\% &  6\%  & Probable \\

Binary stellar systems & \ref{sec:model_Z} & BPASS & Constant & 0.020 & -0.18  & 0.05 & 9\% & 39\% &  Unlikely\tablenotemark{c} \\
(with rotation)		   &  				   &  	   & Constant & 0.010 & -0.32 & -0.08 & 4\% & 22\% &  Probable\tablenotemark{c} \\
					   & 				   &  	   & Constant & 0.002 & -0.51 & -0.28 & 2\% & 9\% &   Probable\tablenotemark{c} \\
\sidehead{\bf High mass cutoff}
{80\msol} & \ref{sec:mass_cutoff} & PEGASE & Constant  & 0.020 & -0.01 & 0.22 & 28\% & 54\% &   Excluded\tablenotemark{c} \\
{120\msol} & \ref{sec:mass_cutoff} & PEGASE & Constant & 0.020 & -0.14 & 0.09 & 17\% & 39\% &   Excluded\tablenotemark{c} \\
\enddata
\tablenotetext{a}{The fraction of \sample\ galaxies with $>2\sigma$ positive deviations from the $\Gamma=-1.35$ tracks. 
}
\tablenotetext{b}{Even though we cannot exclude effects from various dust sight-lines, we demonstrated that effects from dust cannot explain the excess of high \Halpha\ EW galaxies.}
 \tablenotetext{c}{Conclusions driven within the bounds of current SSP models, however, more sophisticated models are required on stellar rotation, binary evolution, and high mass evolution to fully constrain the effects.}
\end{deluxetable}
\end{landscape}


\section{Summary \& Future Work}
\label{sec:summary}

We have used data from the ZFIRE survey along with multi-wavelength photometric data from ZFOURGE to study properties of a sample of star-forming galaxies in cluster and filed environments at $z\sim2$. By using the \Halpha\ EW and rest-frame optical colour of the galaxies we performed a thorough analysis to understand what physical properties could drive the distribution of galaxies in this parameter space. We have improved on earlier analysis by deriving synthetic rest-frame
filters that remove emission line contamination. We analysed effects from dust, starbursts, metallicity, stellar rotation, and binary stars in order to investigate whether the distribution of the \sample\ galaxies could be explained within a universal IMF framework.\\
 We found that:\\
\begin{itemize}
\item  \sample\ galaxies show a large range of \Halpha\ EW values, with $\sim1/3$rd of the sample showing extremely high values compared to expectation from a $\Gamma=-1.35$ Salpeter like IMF. Compared to the HG08 SDSS sample, galaxies at $z\sim2$ show bluer colours with a larger scatter in \Halpha\ EW values.

\item  The difference in extinction between nebular and stellar emission line regions ($f$) in galaxies can have a strong influence in determining the distribution of galaxies in the \Halpha\ EW vs \boxfil\ colour space. Our Balmer decrement studies for a sub-sample of galaxies showed a large scatter in $f$ values. However, we showed that considering $f$ value as a free parameter cannot describe the distribution of galaxies in the \Halpha\ EW vs \boxfil\ colour space.

\item  Starbursts can increase the \Halpha\ EW to extreme values providing an alternative explanation to IMF for a subset of our galaxies with high \Halpha\ EW values. By implementing a stacking technique to remove stochastic SFHs of individual galaxies we concluded that on average our \sample\ still shows higher \Halpha\ EW values for a given \boxfil\ colour compared to a $\Gamma=-1.35$ Salpeter like IMF. We further used Monte Carlo simulations to study time scales of starbursts to conclude that it was extremely unlikely that starbursts could explain the \Halpha\ EW vs \boxfil\ colour distribution of a large fraction of our galaxies. 

\item  Stellar rotation, binaries, and the high mass cutoff of SSP models could influence the distribution of galaxies in the \Halpha\ EW vs \boxfil\ colour parameter space. However, the individual effects of these were not sufficient to explain the distribution of the observed galaxies.

\item Considering multiple effects together can describe the galaxies in our parameter space. We showed that the fraction of galaxies above the $\Gamma=-1.35$ tracks reduces to $\sim5\%$ when considering stellar tracks with high initial rotations (\vini=0.4\vcrit) and equal dust extinction between nebular and stellar regions.

\item Including single or binary stars with stellar rotation in extreme low metallicity  scenarios can significantly increase the \Halpha\ EWs and is also one explanation to describe the distribution of our galaxies in \Halpha\ EW vs \boxfil\ colour parameter space. However, gas phase metallicity analysis of the ZFIRE sample by \citet{Kacprzak2015} and \citet{Kacprzak2016} rules our such low metallicities for our sample. 
We note that calibration of emission line ratios and differences between stellar and ionized gas metallicities at $z\sim2$ are uncertainties that may impact our inference about the stellar metallicity of our sample.  


\item  A non-universal high-mass IMF, varying between galaxies, could explain the distribution of galaxies in this parameter space. The \Halpha\ excess shows a broad trend with larger offsets for the less massive $z\sim 2$ galaxies.
We also confirm the same systematic trend in IMF slope with Chabrier-derived SFR as shown by \citet{Gunawardhana2011} but we refrain from interpreting this. 

\item Within the scope of our study, for $-1.35<\Gamma<-0.5$ the variation in high-mass IMF slope can lead to changes in mass-to-light ratios of up to $\sim0.7$ dex. Furthermore, ignoring calibration offsets we compute that \Halpha\ SFRs can show deviations up to $\sim1.3$ dex. 

 \end{itemize}

\chapter{Conclusions}
\label{chap:conclusions}

This thesis has focused on enhancing our understanding of galaxies at a time period when the universe was actively star-forming producing most of the currently observed baryonic matter (stellar matter). To understand the true nature of galaxies during such intense evolutionary times, it is imperative to probe galaxies in mass and/or magnitude complete samples. 
Photometry  and colours of galaxies provide us with vital clues about their fundamental properties. 
By observing galaxies using multi-wavelength photometry, astronomers derived SEDs to study galaxies in the distant universe. However, photometric studies do not give us the full story. There are inherent degeneracies associated with SED fitting techniques, coupled with our lack of understanding of the accuracy of model templates used in SED fitting for high-$z$ galaxies that may lead to complications in photometric studies. Furthermore, most inherent galaxy properties are probed via emission and absorption lines of nebular and stellar content of galaxies and even narrow band photometry cannot resolve such individual features. 
Therefore, at any redshift, spectroscopic observations of galaxies are vital to accurately understand fundamental galaxy properties.

Traditionally, spectroscopic surveys at $z\sim2$ have been biased towards high star-forming galaxies and thus have been unable to draw strong conclusions on galaxy evolution properties at high-$z$ by allowing direct comparisons with local galaxy populations. This bias has been driven by two main reasons.
\begin{enumerate}
\item Astronomers were unable to accurately select non-biased samples of galaxies from photometry for spectroscopic follow-up.
\item Most well know properties of galaxies in local galaxy populations are probed via rest-frame UV and optical emission/absorption lines that are redshifted to the NIR, thus requiring sensitive, multiplexed NIR detectors to observe galaxies, which were lacking in large telescopes.  
\end{enumerate}

This thesis take advantage of the new generation of NIR instruments and sophisticated SED fitting techniques to probe galaxy samples at $z\sim2$ to study:
\begin{enumerate}
\item The importance of spectroscopic redshifts in determining galaxy properties.
\item The stellar initial mass function of galaxies.  
\end{enumerate}

In Chapter \ref{chap:zfire_survey}, an introduction to the ZFIRE survey was provided. By taking advantage of high quality photometry from ZFOURGE, which used state-of-the-art medium-band imaging from FourStar and the wealth of public legacy data, ZFIRE survey was designed to probe regions of interest between $1.5<z<4.0$ with spectroscopy using the newly commissioned MOSFIRE instrument. 
Even though there were fundamental uncertainties with regard to IMFs, SFHs, dust parametrizations, metallicities, and nebular emission lines in the SED fits used by ZFOURGE, spectral features important to derive photometric redshifts occurred around the NIR wavelengths allowing high quality photometric redshifts to be determined. 
ZFOURGE being a mass complete survey at $z\sim2$ along with the exceptionally accurate photometric redshifts, enabled ZFIRE survey to select galaxies to lower stellar masses and low luminosities with very high accuracy.  A detailed description of the sample selection process, observing strategy, and data reduction methods were outlined in Chapter \ref{chap:zfire_survey}.

Chapter \ref{chap:spec_analysis} investigated basic properties of the observed galaxy samples of ZFIRE survey. Data for two legacy fields, COSMOS and UDS were analysed here to show the redshift distribution and completeness of the first ZFIRE data release. The COSMOS field was further analysed to investigate detection limits of the ZFIRE survey, assess the rest-frame UVJ completeness, and probe the spatial coverage of the $z=2.1$ density structure spectroscopically confirmed by \citet{Yuan2014}. 
Furthermore, photometric legacy surveys such as ZFOURGE, UKIDSS, NMBS, and 3DHST were used to compute photometric and grism redshift accuracies. It was evident at $z\sim2$, the deep imaging of ZFOURGE survey in the NIR medium band filters resulted in high quality photometric redshifts. 
However, even photometric redshift accuracies of $\sim1.5\%$ such as from ZFOURGE, was still not sufficient to probe gravitationally bound structures at $z\sim2$, thus asserting the need for spectroscopy. 
The high spectroscopic completeness of the ZFIRE detections in the K band COSMOS field at $z\sim2$ allowed me to study the importance of spectroscopic redshifts in determining galaxy properties via SED fitting techniques. 
Analysis of SFR and mass showed that they correlate with redshift error. SFR had a large scatter with redshift error, which was attributed to the strong dependence of UV flux in determining the SFR by SED fitting codes. The rest-frame UVJ selection of galaxies also showed a dependence in redshift, with (U$-$V) colours showing stronger dependence to redshift error. Therefore, spectroscopic redshifts provided greater accuracy in determining fundamental galaxy properties via SED fitting codes.

However, it is important to note the underlying assumptions of SED templates that may have significantly influenced derived galaxy properties. One such assumed parameter is the stellar IMF. 
Chapters \ref{chap:imf_observations} and \ref{chap:imf_analysis} discussed the analysis of the IMF of a sample of $z\sim2$ star-forming galaxies using the ZFIRE survey. 
By removing AGN and strong sky line contaminants  form the ZFIRE K band sample, the \Halpha\ EWs were derived for 102 galaxies, out of which 46 had strong continuum detections. 
This sample was shown to be a representative of $z\sim2$ star-forming galaxies that probe the SFR-Main Sequence without any bias towards extremely high star-forming galaxies. 
Using the \Halpha\ EW vs optical colour parameter space, which has been  used extensively to probe the IMF of $z\sim0$ star-forming populations, Chapter \ref{chap:imf_analysis} showed that galaxies at $z\sim2$ show a significant amount of scatter with 1/3rd of the population showing high \Halpha\ EW values compared to a Salpeter IMF for a given \boxfil\ colour. 
Multiple avenues were explored to explain the scatter in this parameter space, with change in IMF slope being one of them. Effects of dust sight lines were shown to have an effect on determining the scatter in this parameter space, however, by varying the factor that determined the difference in extinction between stellar and nebular regions, it was shown that dust sight-lines  could not explain the high \Halpha\ EW galaxies. 
Monte Carlo simulations of PEGASE models with star-bursts and stacking analysis of the \sample\ ruled out star-bursts to be a driver for the observed high \Halpha\ EW galaxies. Stellar rotation, binary stars, metallicity, and upper mass cutoff of the IMF were explored and ruled out to be viable to explain the observed population. 
Therefore, the analysis concluded that the change in IMF to be the strongest contender to explain the large scatter in \Halpha\ EW, specially for galaxies with high EWs compared to Salpeter IMF expectations.

The scatter in \Halpha\ EW vs optical colours for the \sample\ showed no statistically significant difference for the cluster and field sample. Metallicity, probed via \NII/\Halpha\ ratios showed no dependence on environments either. By computing the EW excess in comparison to a Salpeter IMF, it was found that low mass galaxies tend to show larger offsets compared to high mass galaxies. \Halpha\ SFRs showed a similar trend to \citet{Gunawardhana2011}, with high star-forming galaxies showing larger positive offsets in \Halpha\ EW from the Salpeter Slope compared to low star-forming galaxies. However, due to the inherent circular nature of probing systematic offsets of the IMF using the same proxy that is used to determine the SFR, this observed trend is difficult to interpret.

\begin{landscape}
\begin{figure}
\centering
\includegraphics[width=1.25\textwidth]{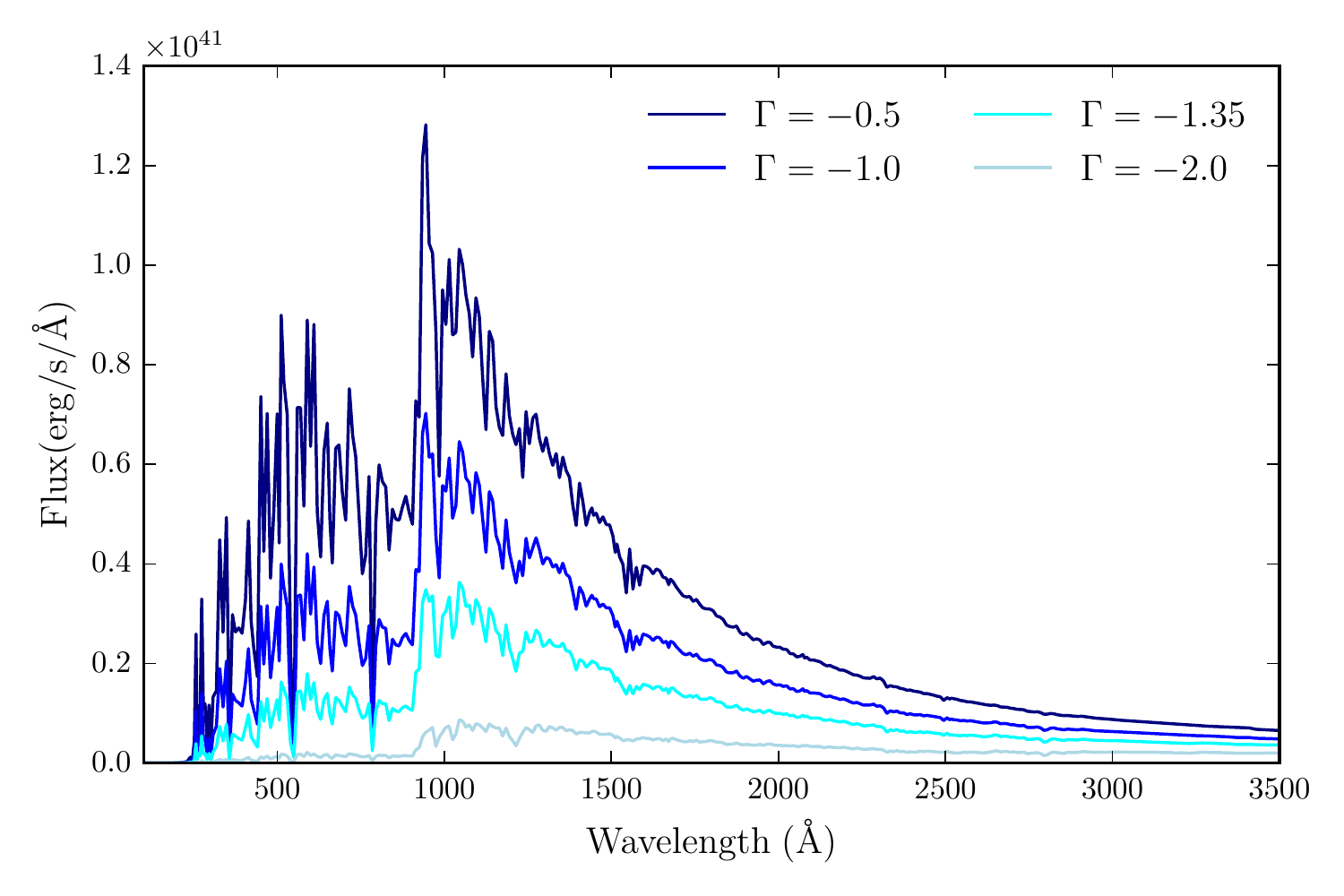} 
\caption[Rest-frame UV spectra from S99 models for varying IMFs.]{The change in rest-frame UV spectra in varying IMF scenarios. Four $\Gamma$ values are shown here, with the fraction of high mass stars decreasing from top to bottom. From top to bottom, $\Gamma$ values shown here are, $-0.5,-1.0,-1.35,$ and $-2.0$. Spectra are generated from S99 models using a constant SFH with genevar0 spectral library (stars with no rotation) and are plot at $t\sim3100$ Myr.	
}
\label{fig:s99_spectra}
\end{figure}
\end{landscape}

IMF change is an important topic as the IMF determines basic parameters such as stellar mass and star-formation rate, which are used to derive broad conclusions about galaxy evolution. 
An example set of rest-frame UV spectra are shown in Figure \ref{fig:s99_spectra}. As evident, for shallower IMFs, the rest-frame UV flux is extremely prominent compared to redder wavelengths. Therefore, change in IMF has significant influence to galaxy properties derived using UV spectral values.
By considering a change in IMF slope between $\Gamma=-1.35$ to $\Gamma=-0.5$, it was found that the mass-to-light ratio in the R band can change up to $\sim0.7$ dex, where galaxies with shallower IMFs will result in lower mass galaxies for an observed R band luminosity. \Halpha\ derived SFRs showed larger dependence in IMF slope, where changes were up to $\sim1.
3$ dex between $\Gamma=-1.35$ and $\Gamma=-0.5$. However, SFR measures are calibrated using an assumed IMF, and thus introduces added complexity if IMF is treated as a varying parameter.

What was observed by ZFIRE was a population of galaxies with high \Halpha\ equivalent widths, i.e. an excess of ionising photons for a given colour, with intermittent starbursts and alternate stellar population models ruled out as explanations. Such high-EW objects appear to become more common at high-redshift, for example similar observations have been reported at $z\sim4$ by multiple studies \citep[eg.,][]{Malhotra2002,Finkelstein2011b,McLinden2011,Hashimoto2013,Stark2013} and have even been invoked at $z>5$ as an explanation for cosmological re-ionisation \citep{Labbe2013,Labbe2015,Schenker2015_thesis,Stark2017}. It seems reasonable to hypothesize that the abundance of high-EW objects is evolving towards high-redshift and this study  observed this at $z\sim 2$.

Is  IMF change responsible? This currently seems to be the only explanation that is not ruled out, but it is not yet determined what would drive it to vary between individual galaxies. 
Since the IMF directly influences the cosmic star-formation history and the chemical evolution of the universe, any variations in the IMF should concurrently agree with observational values of such quantities. 
For example, any change in IMF should allow the cosmic star-formation history to be lower than the upper limits allowed by the diffuse supernovae neutrino background  and higher than the lower limits allowed by observed supernovae rates \citep{Hopkins2006}.

Recent hydrodynamical simulations and physically motivated models suggest that galaxies during their peak star-formation episodes produce stars with a preferential bias towards high mass stars \citep{Marks2012,Narayanan2012,Bekki2013,Narayanan2013,Weidner2013a,Chattopadhyay2015,Ferreras2015}.
However, caution is warranted when treating IMF as a variable parameter during cosmic evolution. 
In order to observationally confirm such models, further study is required to fully comprehend the stellar population parameters of the $z\sim2$ galaxies to resolve whether IMF is the main driver for the distribution of galaxies in the \Halpha\ EW vs rest-frame optical colour parameter space.

\section{Ideas for future work}

This thesis outlined a typical research project, one that concludes with more open questions than its beginning. However, astronomers now have access to sensitive multiplexed NIR instruments that can be used efficiently to probe galaxy populations at $z\gtrsim2$ via photometric and spectroscopic techniques.

The ZFOURGE survey has been a strong example in exhibiting the accomplishments of NIR $\mathrm{J_1,\ J_2,\ J_3,}$ ${H_l}$, and $\mathrm{H_s}$ medium band filters in determining high quality photometric redshifts up to $z\gtrsim4$. Prominent spectral features such as the D4000 break used by SED fitting techniques fall to the K band beyond this redshift, and therefore by breaking the classical broadband K filter in to medium bands will allow high quality photometric redshifts and to be probed up to $z\sim6$ and will enable astronomers to determine fundamental galaxy properties of these galaxies to high accuracy via SED fitting techniques.

It is imperative to acknowledge the underlying uncertainties that may contribute to errors in galaxy properties derived via SED fitting techniques, and therefore, future work should also consider the role of SED templates used in SED fitting techniques. Comparisons between empirically derived templates that are commonly used in SED fitting codes such as FAST, with physically motivated models, such as from MAGPHYS \citep{daCunha2008} will outline systematic effects in determining galaxy properties. Physically motivated models may lead to statistically robust values for dust, mass, SFRs, ages, and metallicities of galaxies. Furthermore, templates currently used in SED fitting codes does not include robust emission lines, which when included will provide more meaningful results in determining galaxy properties for active star-forming galaxies that are abundant in the high redshift universe.

Our analysis of stellar population properties of the \sample\ has shown that galaxies at $z\sim2$ are inherently different from local star-forming populations. Therefore, future spectroscopic surveys should probe galaxy properties via rest-frame UV and optical features along with integral field spectrography to determine the true nature of galaxies at these redshifts. Probing the nebular and stellar content of galaxies with the capability to spatially resolve them in statistically meaningful samples will allow better constrains to be made for fundamental galaxy properties such as IMF, mass, SFR, dust, metallicity, ISM conditions, gas fractions and will contribute to enhance our galaxy evolution models.

Future work on stellar models is warranted to include stellar rotation and binary stars as variable parameters along with sophisticated models for stellar rotation and further detailed modelling of higher mass stars in order to accurately determine stellar population properties of high-$z$ galaxies. Surveys such as Gaia \citep{Gaia2016} will be instrumental to identify stars with various physical properties in our Milky Way and will enable more robust stellar templates to be built.

Furthermore, future studies should investigate the IMF by probing direct stellar spectral signatures of galaxies, i.e., observations of rest-frame UV flux will provide vital observational signatures to indicate the presence of W-R stars, which contribute to excesses in ionized He photons and is a proxy for the presence of short lived massive stars, and therefore, can be used to further constrain the upper end IMF. 
Additionally, the UV to optical flux ratios can also be used to constrain the extreme upper end of the IMF \citep{Meurer2009} as well.

\begin{landscape}
\begin{figure}
\centering
\includegraphics[width=1.25\textwidth]{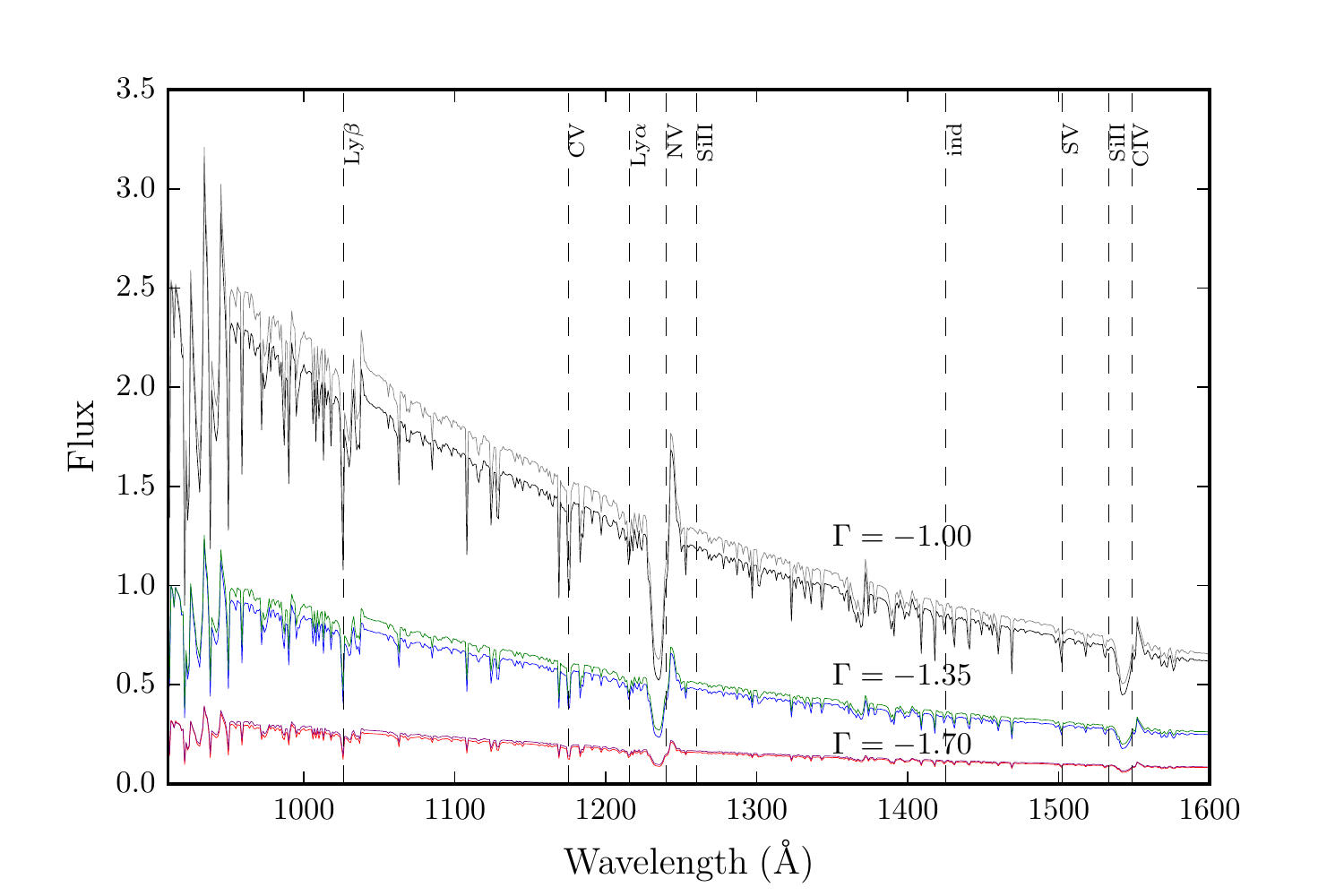} 
\caption[Rest-frame UV spectra of BPASS models with varying IMFs.]{Rest-frame UV spectra of BPASS models with varying IMFs. Three sets of  $\Gamma$ values are shown here, with the fraction of high mass stars decreasing from top to bottom. From top to bottom, $\Gamma$ values shown here are, $-1.0$ (black-grey), $-1.35$ (blue-green) , and $-1.70$ (red-purple). Spectra are generated from BPASS models using a constant SFH with a Z=0.002 and are plot at $t\sim3100$ Myr.	Spectra in black-blue-red are for single stellar populations, while spectra in grey-green-purple shows binary stellar populations. For a given IMF, binary stellar populations show a higher value in flux. All spectra are normalised to 912\AA\ of $\Gamma=-1.35$ single stellar population track. Some example rest-frame UV features are marked in the figure.
}
\label{fig:bpass_spectra}
\end{figure}
\end{landscape}

As shown by Figure \ref{fig:bpass_spectra}, stellar and ISM absorption features such as NV, CIV show stronger absorption and P Cigni profiles for shallower IMFs. 
The difference of these features between single and binary stellar populations are subtle, however, binary populations always show higher UV flux. The strength of the features vary as a function of stellar metallicity and therefore, adds further complexities when determining stellar population properties. 
Future work should consider a more thorough statistical analysis using all the broad-band colour information and multiple spectral diagnostics including simultaneous modelling of effects of
dust, starbursts, metallicity, stellar rotation, binary star evolution, and high mass cutoff of stellar systems together with systematic variances of the IMF.
Rest-frame optical line ratios such as \OIII, \Hbeta, \OII, and \SII\ will allow properties of the ionized gas to be determined to greater accuracy, while modelling of rest-frame UV and optical features together will allow stronger constraints to be made on the stellar and ionized gas properties of galaxies \citep{Steidel2016}. 
Through better understanding of the differences between stars and ionized gas in galaxy spectra, exotic stellar features such as stellar rotation, binaries, and stellar metallicity effects can be probed into greater detail. 
A new generation of stellar models are allowing many of these parameters to be varied and tested.

Additionally, flexible SED fitting codes such as Prospector-$\alpha$ \citep{Leja2017} has shown to be effective in predicting spectral properties of local galaxies purely from photometric data. Prospector-$\alpha$ takes advantage of FIR emissions and UV slope obtained through the photometry to fit non-parametric SFHs to galaxies via flexible stellar population synthesis \citep{Conroy2009,Conroy2010b} code. The accuracy of predicted emission line features for $z\sim2$ galaxies can be evaluated using spectra from surveys like ZFIRE and using different stellar libraries the effects of stellar rotation/binaries can be quantified to a greater depth.

Due to limitations in atmospheric windows, IR instruments can only probe limited redshift intervals to probe rest-frame optical nebular emission line properties. The launch of the \emph{James Webb Space Telescope} 
in 2018 will provide the opportunity to probe rest-frame optical and near-infrared stellar populations via high signal/noise absorption lines and will revolutionise our understanding of the processes of star-formation in the $z\sim 2$ universe.
Rest-frame NIR IMF sensitive features such as \NeIII/\NeII\ ratios (see \citet{Rigby2004}) can be probed at higher redshifts through JWST. Furthermore, \Halpha\ nebular emission lines can be probed beyond $z\gtrsim2.6$ and will be complemented by the large aperture and IFS capabilities of JWST, which will allow SFR/IMF along with dust properties (via \Halpha/\Hbeta) to be probed in sub-galactic scales. 
Furthermore, ALMA observations can be used to probe the molecular gas properties and obtain independent measurements of the SFR via emission lines such as \CII\ of these galaxy populations  \citep{Capak2015}. 
Therefore, combining observations of JWST with ALMA will provide stronger constraints to be made on star-formation conditions of the high redshift galaxy populations.

Local analogues can also be used to determine high-$z$ galaxy properties. Surveys such as DYnamics of Newly-Assembled Massive Objects \citep[DYNAMO.,][]{Fisher2017} has shown the existence of a highly selective population of local star-forming galaxies that resemble the properties of high-$z$ galaxies with extremely high SFR surface densities \citep{Fisher2017}, high gas fractions \citep{Fisher2014}, and high \Halpha\ velocity dispersions \citep{Green2010,Bassett2014}. 
The high resolution IFS data available in such surveys can be used to probe comparable properties to high-$z$ galaxies with many observational advantages such as significantly less exposure times, ability to probe with high angular and spatial resolutions to deeper surface brightness limits, and also has the inherent advantage of the optical wavelength regime, which has less observational constraints compared to the NIR \citep{Steidel2004}.

DYNAMO data could be used to probe the IMF via stellar population and dynamical analysis allowing stronger constrains on stellar population parameters. Furthermore, the high spatial resolution will allow the IMF to be determined within star-forming regions in sub-kPc scales enabling radial variations to be studied in gas phase metallicity, ionizing radiation field,  and ionization parameter and linking them with star formation rates and clump stellar masses enhancing the understanding of high-$z$ galaxy properties.

Studies of the stellar populations in local LSB galaxies can provide stronger constraints on galaxy formation models. LSBs generally live in low density environments and their \HI\ rich disks and the lack of molecular gas suggest extremely low interactions with their local environments. 
Furthermore, most LSBs show normal bulges suggesting normal SFHs in their past. 
Therefore, further detailed studies of LSBs can provide unique insight into galaxy and stellar formation and evolution models. Instruments such as DRAGONFLY \citep{Abraham2014} and KCWI \citep{Morrissey2012}, that can probe low surface brightness regions of galaxies will be crucial for such studies.

The future is looking bright for high-$z$ studies. JWST, along with the new next generation of 30m class telescopes will allow astronomers to probe galaxies in the deeper universe within the first 100 Myr of its birth. Robust statistical samples of galaxies at $z\gtrsim2$ that probes fainter magnitudes and lower stellar masses will further improve our understanding of galaxies in the low mass regime, how the first generation of galaxies were formed, the fuelling and conditions of star-formation, and answer the deep mysteries of the universe such as the existence of young mature massive galaxies in the early universe.

\bibliographystyle{apj}
\newpage
\bibliography{bibliography}
\addcontentsline{toc}{chapter}{Bibliography}


\appendix
\chapter{ZFIRE Survey: Calibrations and catalogue comparisions.}

\section{MOSFIRE calibrations}
\label{sec:MOSFIRE cals}

\subsection{Telluric Corrections}

\begin{figure}[h!]
\centering
\includegraphics[width=1.0\textwidth]{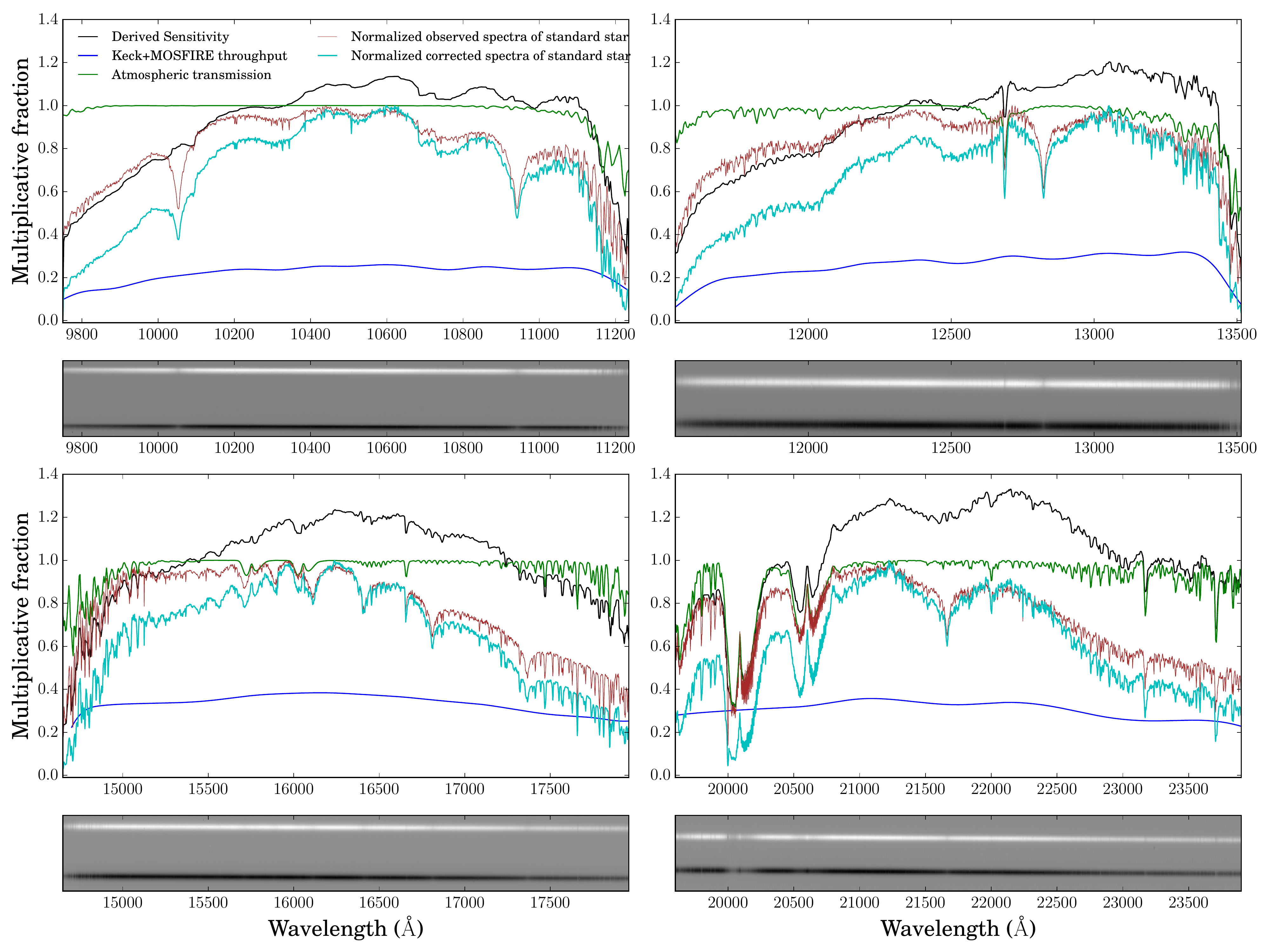}
\caption[Example set of derived sensitivity curves for MOSFIRE filters.]{Example set of derived sensitivity curves for MOSFIRE filters. From left to right, in the top panels we show the Y and J-bands and in the bottom panels we show the H and K bands. Pre-ship spectroscopic throughput for MOSFIRE is shown in blue. This takes into account the instrument response and the telescope throughput and \cite{McLean2012} shows that the these predictions agree extremely well with the measured values. The green line is the measured atmospheric transmission provided by the University of Hawai'i (private communication). 
The normalized spectra of the observed 1D standard stars before any corrections are applied are shown in brown. 
We remove the stellar atmospheric hydrogen lines and fit the spectra by a blackbody emission curve.
We use this derived spectra as a sensitivity curve (shown in black) and multiply our galaxy spectra by this to apply telluric corrections. 
We multiply the observed standard star spectra with the derived sensitivity curve to obtain a telluric corrected normalized standard star spectrum, which is shown in cyan. 
Each panel is accompanied with a 2D spectra of the standard star as given by the DRP. The black and white lines are the negative and positive images. Strong telluric features can be seen in regions where the intensity of the 2D spectra drops rapidly.
All 1D curves are normalized to a maximum value of 1.}
\label{fig:sensitivity}
\end{figure}

Additional figures related to the MOSFIRE data reduction process are shown in this section. 
Figure \ref{fig:sensitivity} shows an example set of derived sensitivity curves and the normalized 1D spectra applied to all observed bands. 

\clearpage

\subsection{Spectrophotometric Calibrations}

\begin{figure}
\centering
\includegraphics[width=1.0\textwidth]{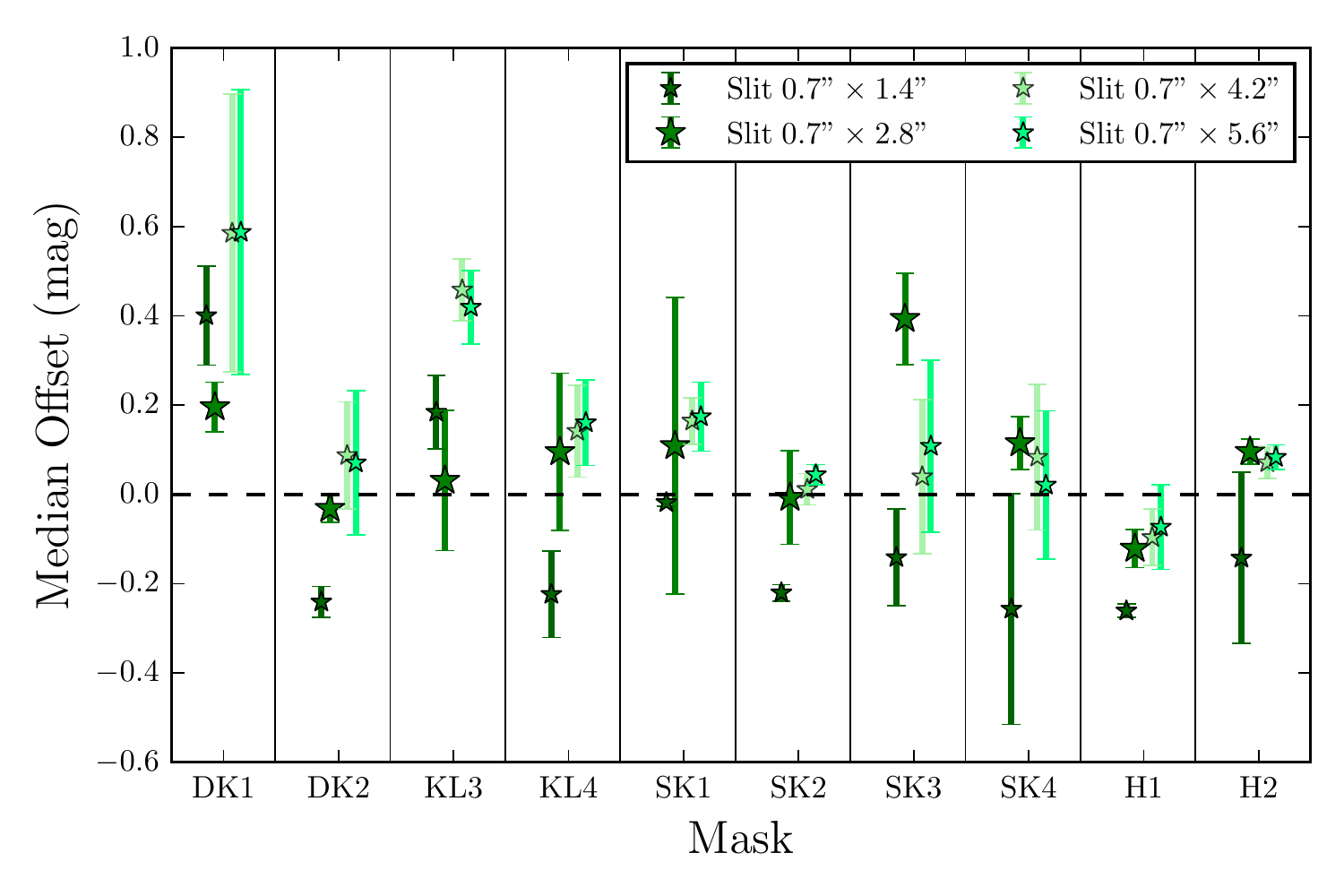}
\caption[The median offset values for different aperture sizes for the COSMOS field masks.]{The median offset values for different aperture sizes for the COSMOS field masks. This figure is similar to Figure \ref{fig:scaling_values} top panel, but shows the median offset values computed for all slit-box like aperture sizes considered in our spectrophotometric calibration process. 
Filter names correspond to the names in Table \ref{tab:observing_details}.
The green stars in different shades for a given mask relates to the median offset between spectroscopic magnitude of the objects in the mask to its photometric magnitude computed using ZFOURGE and HST imaging with varying aperture sizes. 
The errors are the \NMAD\ scatter of the median offsets calculated via bootstrap re-sampling of individual galaxies. 
The vertical lines are for visual purposes to show data points belonging to each mask. 
}
\label{fig:scaling_values_varying_slit_boxes}
\end{figure}

As mentioned in Section \ref{sec:sp calibration}, for the COSMOS field, we overlaid synthetic slit apertures with varying slit heights on the ZFOURGE imaging to count the integrated flux within each aperture. The main purpose of the process was to account for the light lost due to the finite slit size. 
Figure \ref{fig:scaling_values_varying_slit_boxes} shows the change of median offset values for varying aperture sizes for each of the COSMOS mask.  As is evident from the figure, when the slit height increases from $1''.4$ to $2''.8$, most of the light emitted by the galaxies is included within the slit aperture. For any slit height beyond that, there is no significant change to the integrated counts, thus suggesting the addition of noise. Driven by this reason, we choose the $0''.7\times2''.8$ slit size to perform the spectrophotometric calibrations.

\begin{figure}
\centering
\includegraphics[width=0.49\textwidth]{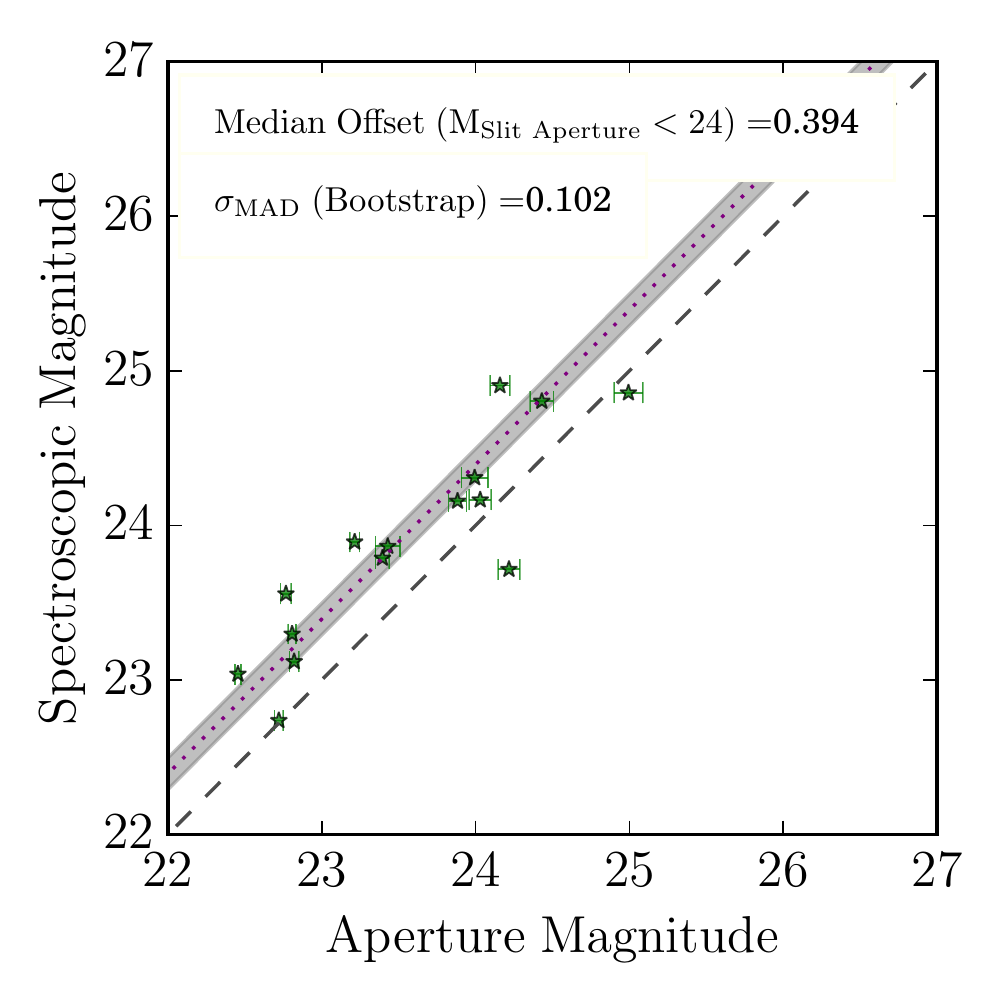}
\includegraphics[width=0.49\textwidth]{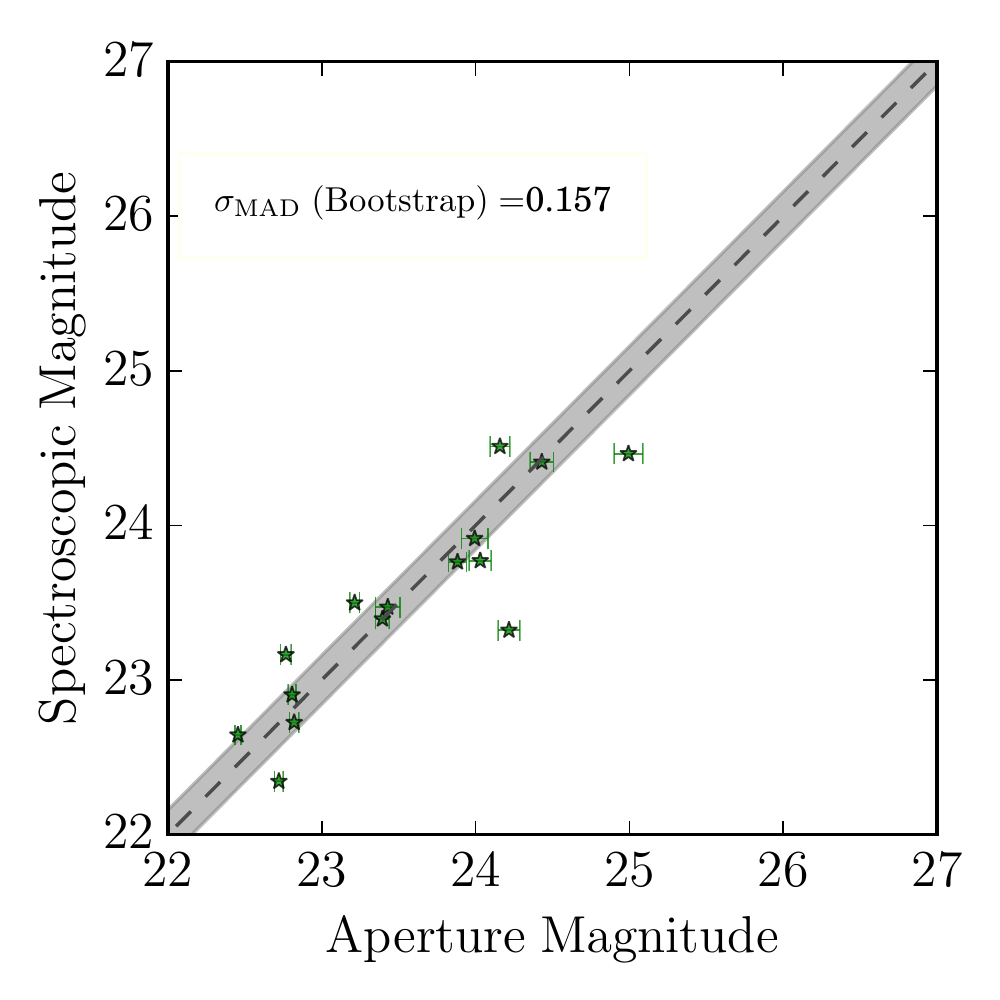}
\includegraphics[width=0.49\textwidth]{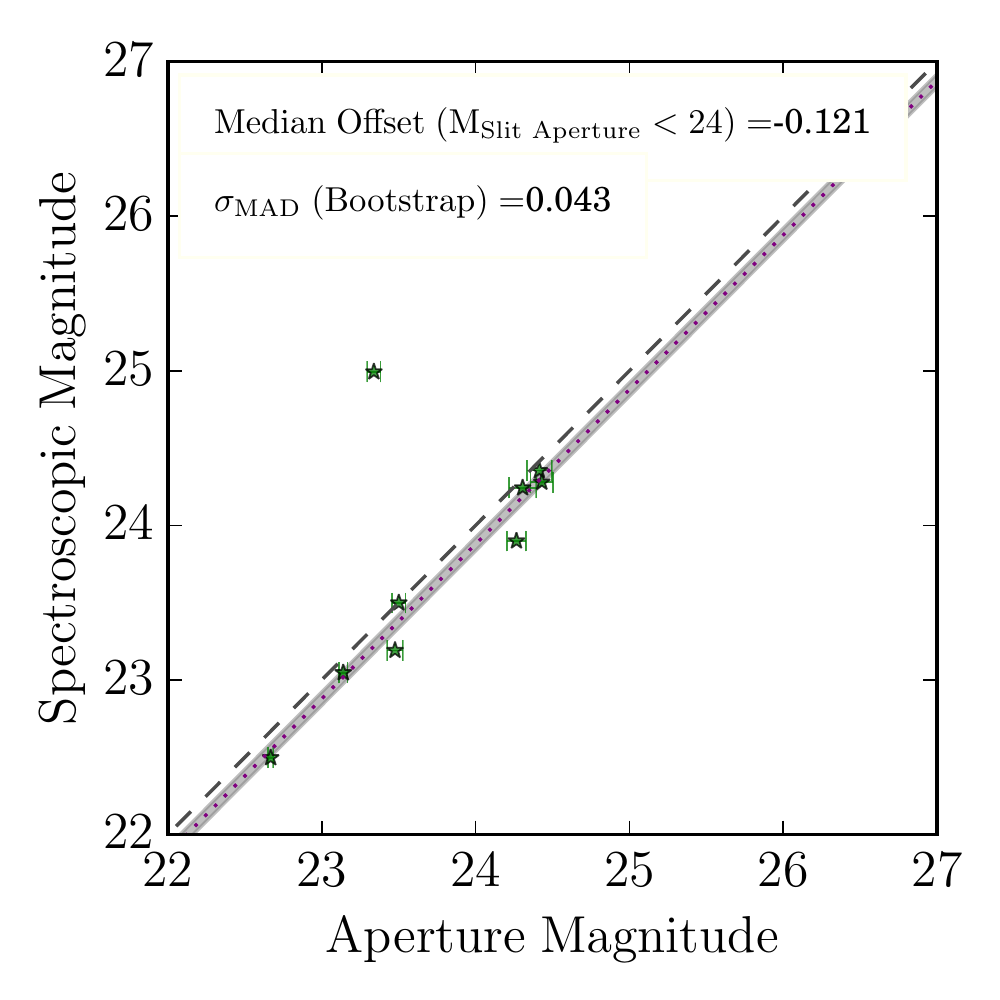}
\includegraphics[width=0.49\textwidth]{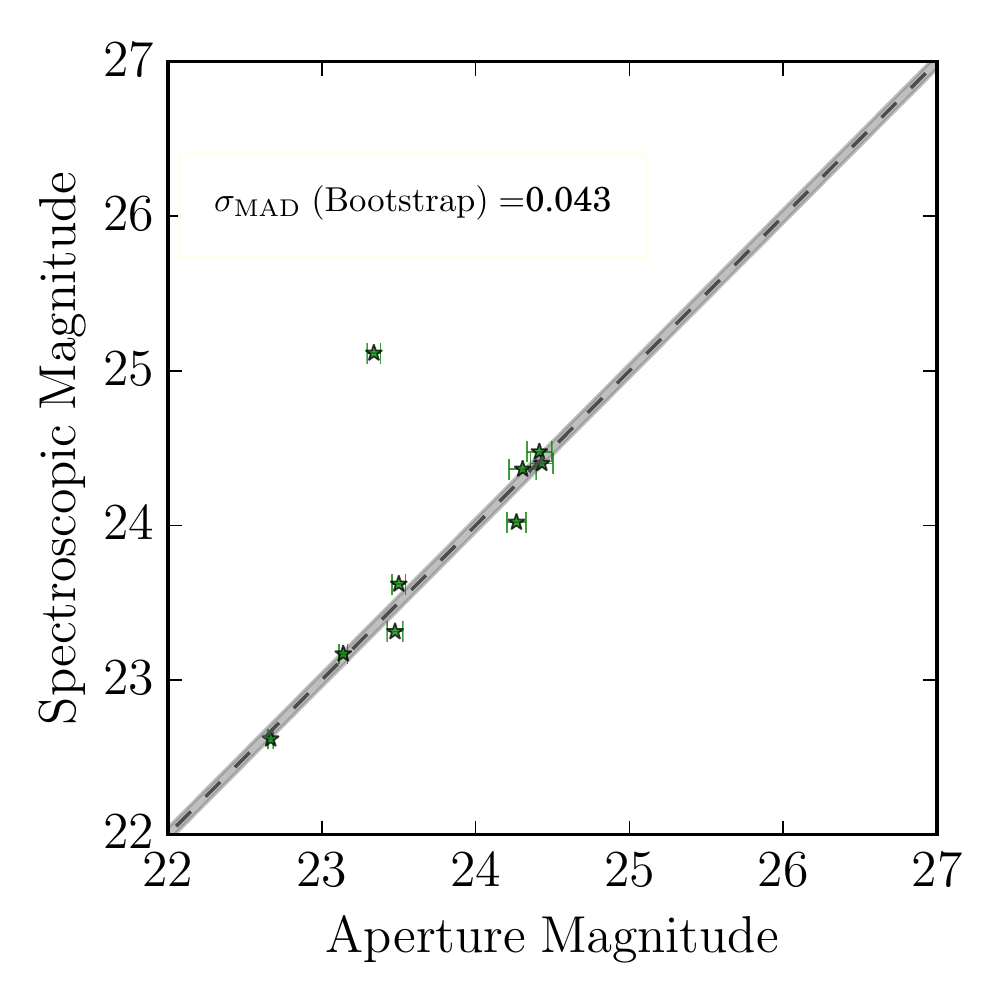}
\caption[Comparison between spectroscopically derived magnitude to the photometrically derived magnitude.]{Two example masks showing the comparison between spectroscopically derived magnitude to the photometrically derived magnitude using a $0''.7 \times 2''.8$ slit box. 
{\bf Top left:} K-band mask (KL3) before spectrophotometric calibration. The legend shows the median offset of galaxies with slit magnitude $<$24 and the corresponding bootstrap error.
{\bf Top right:} similar to left panel but after the spectrophotometric calibration has been applied. Since the scaling factor is now applied to the data, the median offset for galaxies with slit magnitude $<$24 is now 0. The inset shows the bootstrap error after the scaling is applied. This is considered to be the error of the spectrophotometric calibration process.
{\bf Bottom:} similar to the top panels but for a H-band mask (H1).
The grey shaded area in all the panels is the bootstrap error.
Error bars are from the ZFOURGE photometric catalogue. The flux monitor stars have been removed from the figure to focus the value range occupied by the galaxies. }
\label{fig:mask_scaling_example}
\end{figure}

We show the magnitude distribution of two example masks in Figure \ref{fig:mask_scaling_example}. Once a uniform scaling is applied to all the objects in a given mask, the agreement between the photometric slit-box magnitude and the spectroscopic magnitude increases.

\section{B: Differences between ZFOURGE versions}
\label{sec:ZFOURGE comparison}

Here we show the effect of minor changes between different versions of ZFOURGE catalogues. 
ZFIRE\ sample selection was performed using an internal data release intended for the ZFOURGE team (v2.1). In this version, detection maps were made from Ks band photometry from FourStar imaging.
The 5$\sigma$ depth for the public data release is Ks $\leq$25.3 in AB magnitude \citep{Straatman2016}, while the internal data released used  by \citet{Spitler2012} had a 5$\sigma$ depth  of 24.8 magnitude. 
All results shown in the paper, except for the photometric redshift analysis, are from v2.1. 

ZFOURGE COSMOS field has now been upgraded by combining the FourStar imaging with VISTA/K from UltraVISTA \citep{McCracken2012} to reach a 5$\sigma$  significance of 25.6 in AB magnitude (v3.1). This has increased the number of detected objects of the total COSMOS field by \around50\%. 

\begin{figure}[b]
\centering
\includegraphics[width=0.9\textwidth]{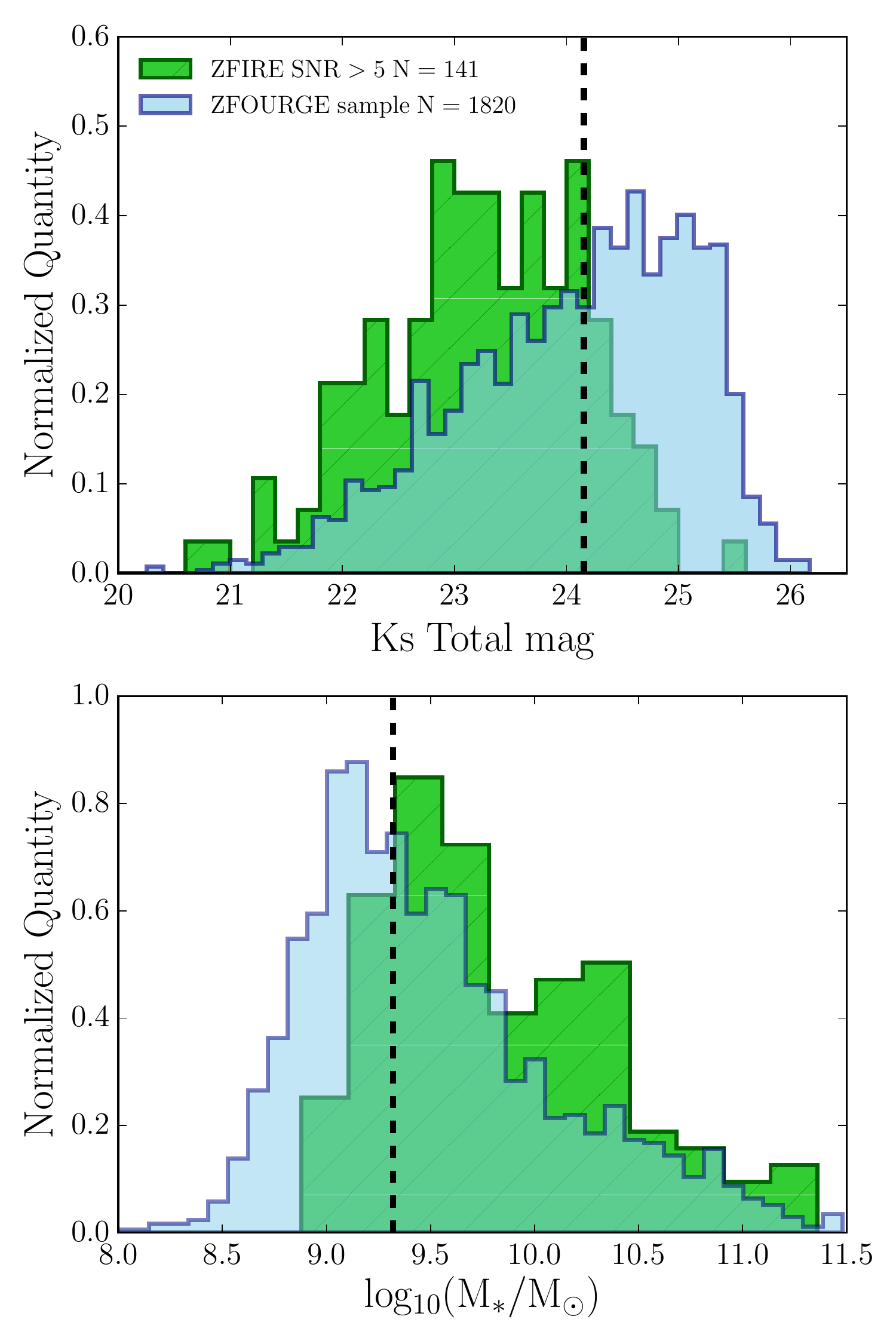}
\caption[The Ks magnitude and mass distribution of the ZFIRE K band sample compared with ZFOURGE public catalogue.]{The Ks magnitude and mass distribution of the $1.90<z<2.66$ galaxies from ZFOURGE overlaid on the ZFIRE\ detected sample.
This figure is similar to Figure \ref{fig:detection_limits}, but the ZFOURGE data has been replaced with the updated deeper ZFOURGE catalogue (v3.1) and shows all ZFOURGE and ZFIRE detected galaxies in this redshift bin (In Figure \ref{fig:detection_limits} the quiescent sample is removed to show only the red and blue star-forming galaxies). 
}
\label{fig:detection_limits_newcat}
\end{figure}

All ZFIRE\ galaxies identified by v2.1 of the catalogue are also identified with matching partners by v3.1. 
Figure \ref{fig:detection_limits_newcat} shows the distribution of the Ks magnitude and masses of the updated ZFOURGE catalogue in the redshift bin $1.90<z<2.66$. 
The 80\%-ile limit of ZFOURGE in this redshift bin increases by 0.4 magnitude to $Ks = 25.0$.
Similarly, the 80\% mass limit is \around$10^9$ \msol\ which is an increase of 0.2 dex in sensitivity.  
It is evident from the histograms that the significant increase of the detectable objects in this redshift bin has largely been driven by faint smaller mass galaxies. 
The 80\% limit for the ZFIRE-COSMOS galaxies is Ks=24.15 with the new catalogue. The change is due to the change of photometry between the two catalogues.

\begin{figure}
\centering
\includegraphics[width=1.0\textwidth]{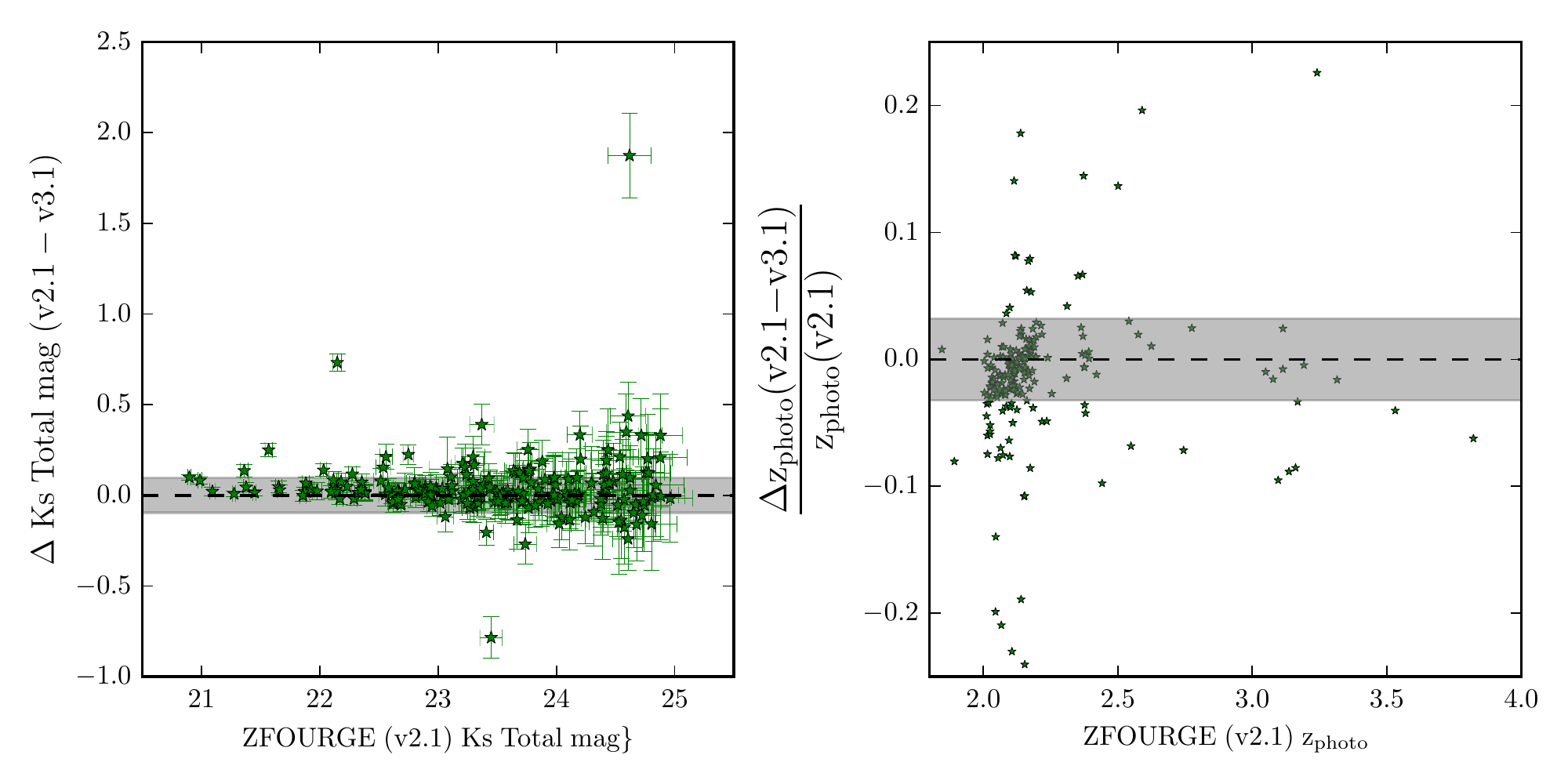}
\caption[Ks magnitude and the photometric redshift differences of ZFOURGE catalogues.]{Ks magnitude and the photometric redshift differences of ZFOURGE catalogues.
Only galaxies targeted by ZFIRE are shown.
{\bf Left:} the Ks band total magnitude difference between v2.1 and v3.1 of the ZFOURGE catalogues. 
{\bf Right:} the photometric redshift difference between v2.1 and v3.1 of the ZFOURGE catalogues. The grey shaded region denotes the \NMAD\ of the distribution. 
In both panels, the grey shaded region denotes the \NMAD\ of the distribution, which are respectively 0.09 magnitude and 0.03. 
}
\label{fig:cat_differences}
\end{figure}

Figure \ref{fig:cat_differences} shows the ZFOURGE catalogue differences between Ks total magnitude and the photometric redshift of the ZFIRE\ targeted galaxies. 
The Ks magnitude values may change due to the following reasons. 
\begin{enumerate}
\item The detection image is deeper and different, which causes subtle changes in the location and the extent of the galaxies. 
\item The zero point corrections applied to the data uses an improved method and therefore the corrections are different between the versions. 
\item The correction for the total flux is applied using the detection image, rather than the Ks image. Due to subtle changes mentioned in 1,  this leads to a different correction factor. 
\end{enumerate}
The \NMAD\ of the scatter for the Ks total magnitude is \around0.1 mag and is shown by the grey shaded region. There are few strong outliers. 
Two of the three catastrophic outliers are classified as dusty galaxies. One of them is close to a bright star and has an SNR of \around5 in v2.1. 
With the updated catalogue, the SNR has increased by \around30\% and therefore the new measurement is expected to be more robust.  
For the remaining galaxy, we see no obvious reason for the difference.

\begin{figure}
\centering
\includegraphics[width=1.0\textwidth]{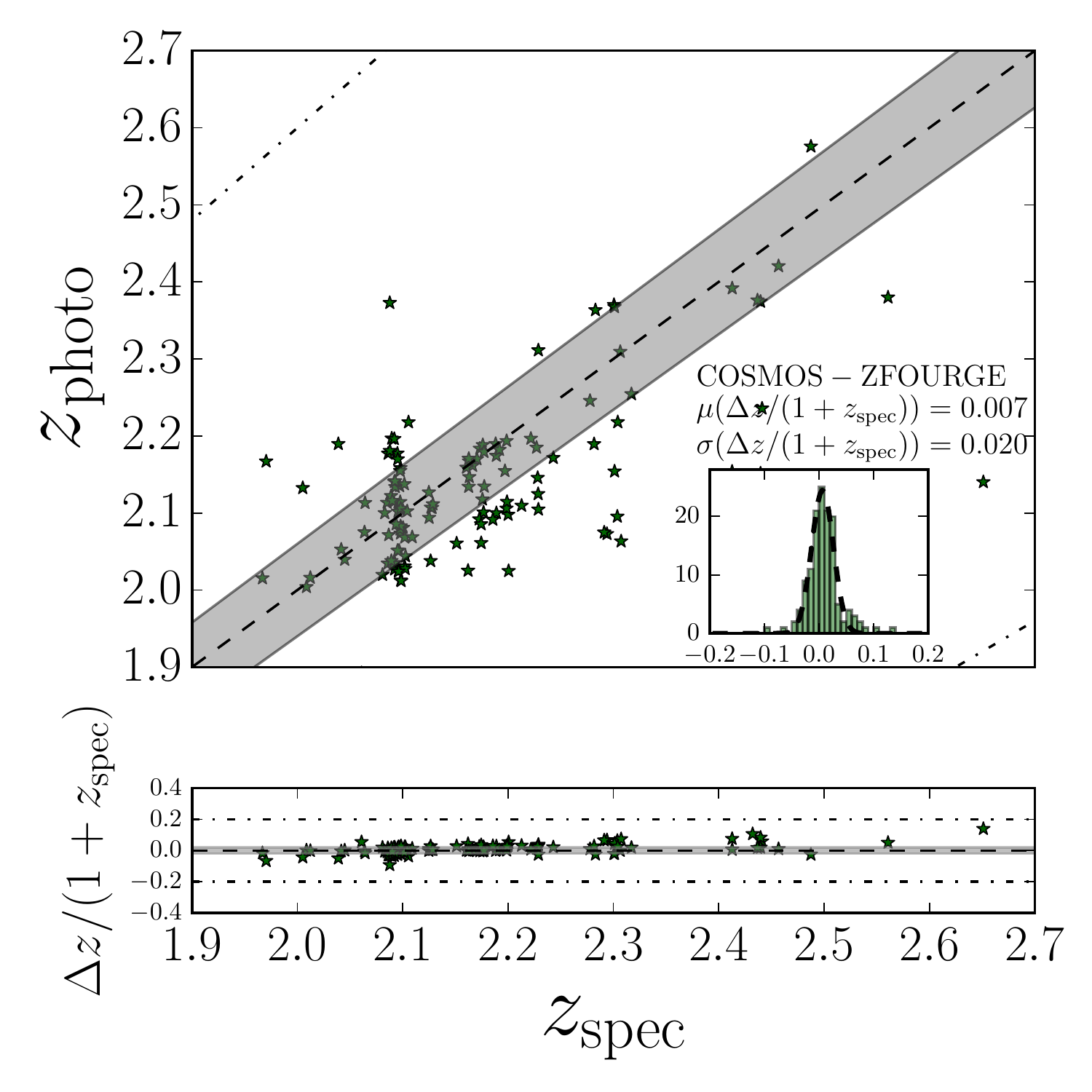}
\caption[Photometric and spectroscopic redshift comparison between ZFOURGE v2.1 and ZFIRE.]{Photometric and spectroscopic redshift comparison between ZFOURGE v2.1 and ZFIRE. 
This Figure is similar to Figure \ref{fig:specz_photoz_cosmos} with the exception of all photometric redshifts now being from v2.1 of the ZFOURGE catalogue. 
}
\label{fig:specz_photoz_newcat}
\end{figure}

Figure  \ref{fig:specz_photoz_newcat} shows the redshift comparison between ZFIRE spectroscopy and the v2.1 of the ZFOURGE catalogue. In v3.1, the photometric redshifts were updated by the introduction of high \Halpha\ equivalent width templates to EAZY and improved zero-point corrections to the photometric bands.
These changes along with the extra Ks depth have driven the increase in accuracy of the photometric redshifts from \around2.0\% in v2.1 to \around1.6\% in v3.1.

\chapter{Spectral fitting and AGN contamination}

\section{Robustness of continuum fit and \Halpha\ flux}
\label{sec:cont_Halpha_test}

The study presented in this paper relies on accurate computation of \Halpha\ flux and the underlying continuum level. In this section, we compare our computed continuum level and \Halpha\ flux values using two independent methods to investigate any systematic biases to our \Halpha\ EW values.

We examine the robustness of our continuum fit to the galaxies by using ZFOURGE imaging data to estimate a continuum level from photometry. 
By using a slit-box of size 0.7$''\times 2.8''$ overlaid on the $0.7''$ point spread function convolved FourStar Ks image, we calculate the photometric flux expected from the galaxy within the finite slit aperture.   Justification of this slit-size comes from the spectrophotometric calibration of the ZFIRE data which is explained in detail in \citet{Nanayakkara2016}. 
Since we remove slits that contain multiple galaxies within the selected aperture, only 38 continuum detected galaxies and 39 galaxies with continuum limits are used in this comparison. 
\begin{subequations}
We then convert the magnitude to $f_\lambda$ as follows:
\begin{equation}
f_\lambda = 10^{-0.4(mag +48.60)} \times \frac{3\times10^{-2}}{\lambda_{c}^2}\ \mathrm{erg/s/cm^2/\AA}
\end{equation}
where $\lambda_c$ is the central wavelength of the MOSFIRE K band which we set to 21757.5\AA. 
Next we compute the \Halpha\ flux contribution to $f_\lambda$ by using the photometric bandwidth of FOURSTAR Ks band ($\Delta \lambda_{FS}$=3300\AA). 
\begin{equation}
F_{H\alpha_{cont}} = \frac{F_{H\alpha}}{\Delta \lambda_{FS}}
\end{equation}
We then remove the \Halpha\ flux contribution to the photometric flux to compute the inferred continuum level from photometry as shown below:
\begin{equation}
F_{cont_{photo}} = f_\lambda-F_{H\alpha_{cont}}
\end{equation}
Since the \Halpha\ flux is the dominant emission line for star-forming non-AGN galaxies, we ignore any contributions from other nebular emission lines to the photometric continuum level. 
\end{subequations}
By using the medium band photometry to estimate the continuum level, we further validate the robustness of the spectrophotometric calibration of the ZFIRE data. We also note that our method of continuum estimation is somewhat similar to using the best-fit SEDs to estimate the continuum level, with the added advantage that it does not depend on the shape of the best-fit SED. 

\begin{figure}
\centering
\includegraphics[width=0.55\textwidth]{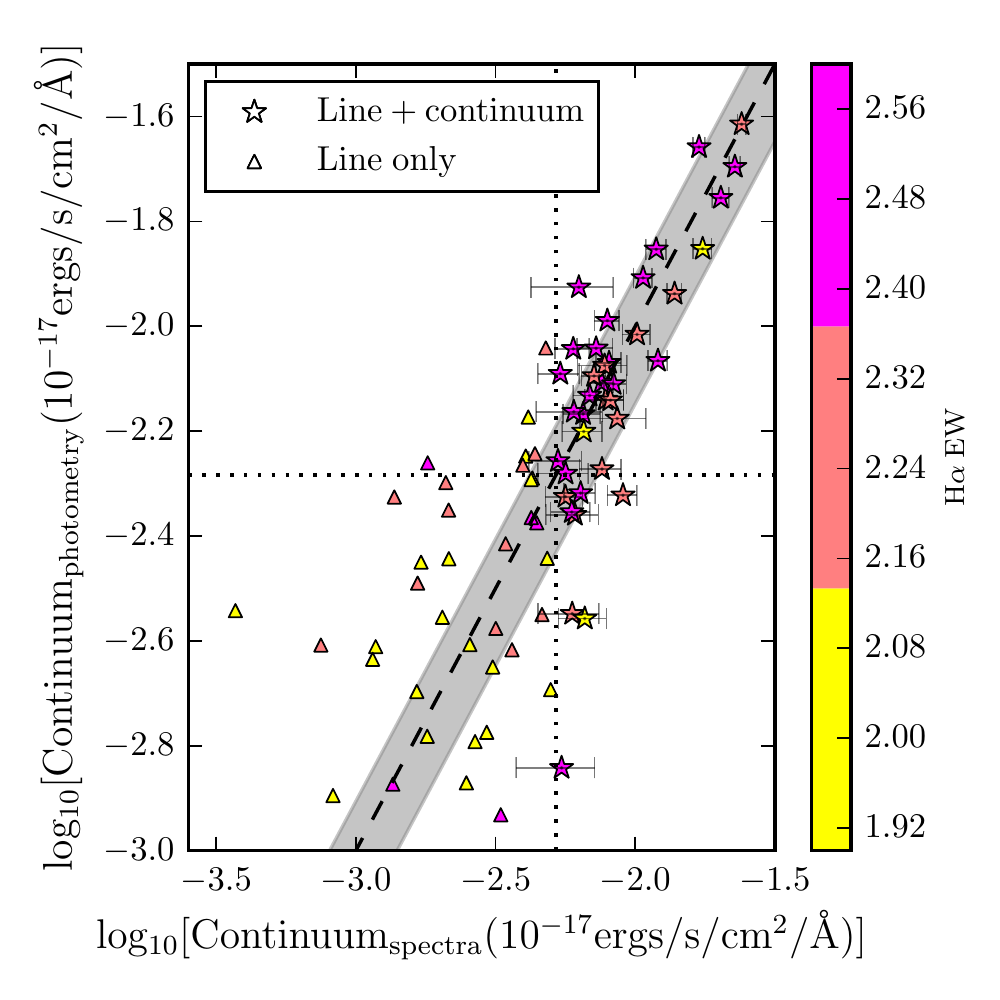}
\includegraphics[width=0.55\textwidth]{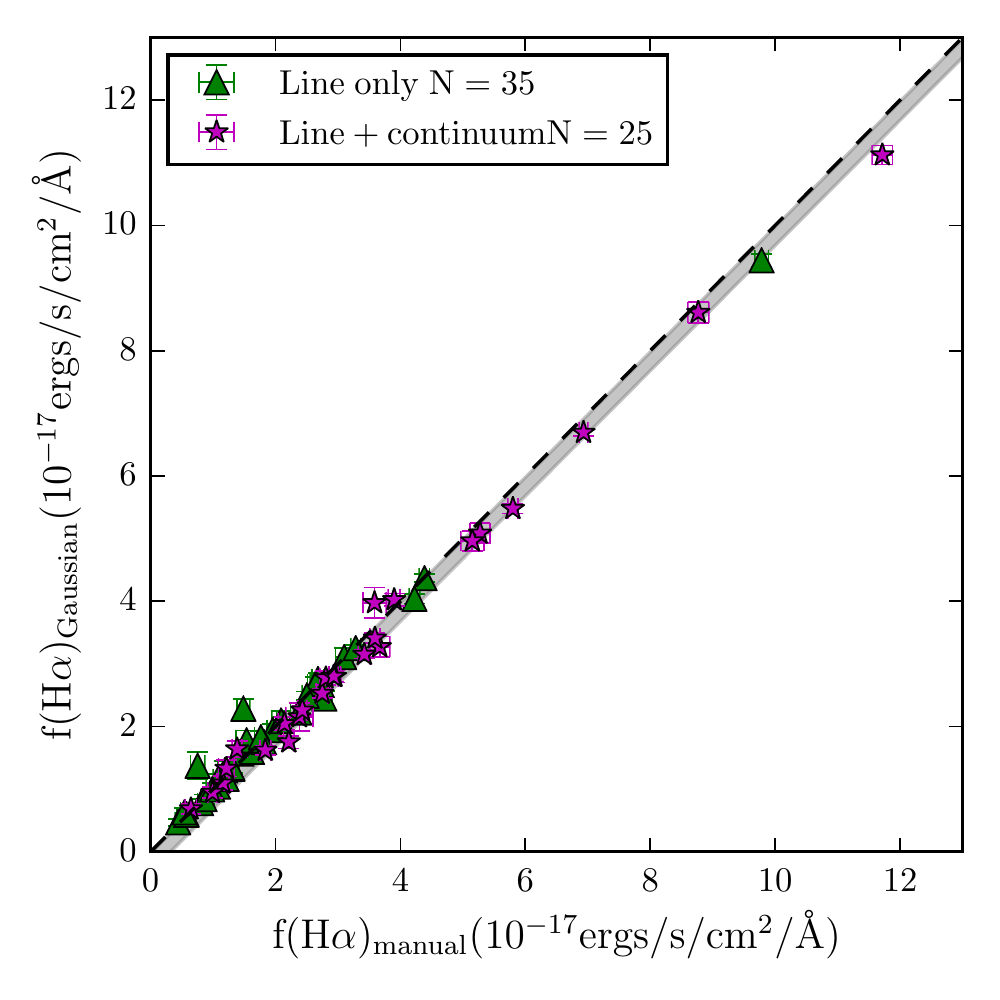}
\caption[Comparison of \Halpha\ flux and continuum level between spectroscopy and photometry.]
{{\bf Top:} We compare the continuum level derived from spectra with the expected continuum level from photometry. 
The stars resemble objects with robust continuum detections. The remaining sample is shown as triangles. The error bars for the continuum detections come from bootstrap re-sampling. The one to one line is shown as a black dashed line. 
The median from one to one deviation of the continuum detected objects is 0.06 logarithmic flux units with  \NMAD = 0.14 (shown by the grey shaded region). 
The horizontal and vertical dotted lines are the continuum detection levels. 
{\bf Bottom:} We compare the \Halpha\ flux values computed using Gaussian fits to that of visually identified limits for galaxies which show no strong sky contamination. 
The magenta stars resemble objects with robust continuum detections.  The remaining sample is shown as green triangles. The error values are from the integration of the error spectrum. The one to one line is shown as a black dashed line. 
The median one to one deviation of the continuum detected objects is 0.19 flux units with \NMAD\ = 0.20 (shown by the grey shaded region).
}
\label{fig:continuum_Halpha_checks}
\end{figure}

We compare this photometrically derived continuum level with the spectroscopic continuum level in Figure \ref{fig:continuum_Halpha_checks} (left panel). 
The median deviation of the detected continuum values are \around 0.026 in logarithmic flux values with a 1\NMAD scatter of 0.12, which leads us to conclude that the photometrically derived continuum values agree well with the spectroscopic continuum detections, thus confirming the robustness of our continuum calculations. 
We further note that the large scatter of galaxies below the continuum detection level is driven by the increased fraction of sky noise, which is expected and further confirms that we have robustly established the continuum detection limit. 
There is no strong dependence of the \Halpha\ EW on the continuum detection levels, which suggests that the \Halpha\ EW values are not purely driven by weak continuum levels. Further analysis of such detection biases are shown in Section \ref{sec:observational_bias}.

In order to test the robustness of the measured \Halpha\ flux, we compare the \Halpha\ fluxes 
between this study and the ZFIRE catalogue. Nebular line fluxes in the ZFIRE catalogue are measured by integrating a Gaussian fit to the emission lines \citep{Nanayakkara2016}. We follow a similar technique to calculate \Halpha\ fluxes for emission lines unless they show strong velocity structures.

It is vital to ascertain if Gaussian fits to emission lines would give drastically different \Halpha\ flux values compared to our visually integrated fluxes. 
In Figure  \ref{fig:continuum_Halpha_checks} (right panel), we show this comparison for \sample\ galaxies which do not have strong sky line residuals.  
For the continuum detections in the above subset, the median deviation between the manual limits and Gaussian fits is 0.19$\mathrm{\times 10^{-17} ergs/s/cm^2/A}$ with \NMAD= 0.20$\mathrm{\times 10^{-17} ergs/s/cm^2/A}$. 
Therefore,  \Halpha\ flux values agree with each other within error limits with minimal scatter. 
Single Gaussian fits would fail to describe \Halpha\ emission profiles of galaxies with strong rotations or galaxy that have undergone mergers. 
These features will require complicated multi-Gaussian fits to accurately provide the underlying \Halpha\ flux. All 3\NMAD\ outliers of continuum detections contain profiles that cannot be described using single Gaussian fits. 
Therefore, we expect the direct integration to be the most accurate method to calculate the \Halpha\ flux for galaxies with velocity structures because it is independent of the shape of the \Halpha\ emission. We note that all galaxies with sky line contamination shows profiles that are well described by single Gaussian fits. 

Due to above tests we are confident that neither the \Halpha\ flux or the continuum level calculations would give rise to systematic errors in our analysis. Therefore, we conclude that the \Halpha\ EW values derived for our continuum detected \sample\ are robust.

\section{AGN contamination to \Halpha\ flux}
\label{sec:AGN}

As described in Section \ref{sec:sample_selection}, we flag AGN of the ZFIRE sample following \citet{Coil2015} selection criteria. 
However, it is possible that weak AGN that are not flagged by our selection may still contaminate the \sample\ and contribute to higher \Halpha\ emission. 
In order to investigate effects from such AGN, we use \citet{Coil2015} selection and the measured \NII\ fluxes to compute upper limits to \Halpha\ fluxes required for the galaxies to be flagged as AGN as follows:
\begin{equation}
\label{eq:AGN_Ha_limit}
f(H\alpha)_{inf} < \frac{f([NII])}{0.316}
\end{equation} 
where f(\NII) is the measured \NII\ flux for our galaxies. 
We find that our measured \Halpha\ fluxes are $\sim\times2$ higher than the inferred \Halpha\ fluxes ($f(H\alpha)_{inf}$) computed using the above equation. 
Using the ratio of the measured and inferred \Halpha\ fluxes, we find the upper limit to the fraction of \Halpha\ photons produced by the strongest possible AGN that would not be flagged by the \citet{Coil2015} selection to be  $\sim0.4$. 
Therefore, if our sample is contaminated by weak sub-dominant AGN, we expect the AGN to be responsible for $\sim50\%$ of the observed \Halpha\ flux.

\chapter{Synthetic Stellar Population Models}

\section{Do SSP models give identical results?}
\label{sec:SSP comparision}

In order to investigate whether there is a strong dependence of the \Halpha\ EW and/or \gr\ colour evolution of model galaxies on the SSP models used, we compare the galaxy properties from PEGASE with that of Starburst99 \citep{Leitherer1999}. 
S99 models support the use of multiple stellar libraries. For this analysis we consider the Padova AGB stellar library that is an updated version of the \citet{Guiderdoni1988} stellar tracks that includes cold star parameters and thermally pulsating asymptotic giant branch (AGB) and post-AGB stars.

We compute PEGASE models using a constant SFR of $1\times10^{-4}$\msol/Myr with various $\Gamma$ values. PEGASE models are scale free and are generated using a baryonic mass reservoir normalized to 1\msol. Having higher SFRs in PEGASE will result in the SFR exceeding the maximal amount of SFR possible for the amount of gas available in the galaxy reservoir before reaching $z\sim2$. 
The other parameters of the PEGASE models are kept as mentioned previously.

S99 models employ a different approach to compute synthetic galaxy spectra where the SFR is not normalized and therefore the SFR should be kept at a reasonable level that allows the HR diagram to be populated with sufficient number of stars during the time steps the models are executed. For S99, we use a SFR of 1\msol using Padova AGB stellar libraries with a Z of 0.02, and similar IMFs to the PEGASE models. We do not change any other parameters in the S99 models from its default values. 
We state the S99 parameters below. \\
$\bullet$ Supernova cutoff mass is kept at 8\msol.\\
$\bullet$ Black hole cutoff mass is kept at 120\msol.\\
$\bullet$ Initial time is set to 0.01 Myr and time 1000 time steps are computed with logarithmic spacing up to 3100 Myr. \\
$\bullet$ We consider the full isochrone for mass interpolation. \\
$\bullet$ We leave the indices of the evolutionary tracks at 0. \\
$\bullet$ PAULDRACH/HILLIER option is used for the atmosphere for the low resolution spectra. \\
$\bullet$ Metallicity of the high resolution models are kept at 0.02.\\
$\bullet$ Solar library is used for the UV line spectrum.\\
$\bullet$ In order to generate spectral features in the NIR, we use microturbulent velocities of 3kms$^{-1}$ and solar abundance ratios for alpha-element to Fe ratio.

To account for the difference in the SFRs between the SSP models, we scale the PEGASE \Halpha\ flux and the corresponding continuum level to the 1\msol/yr value by multiplying by $10^{10}$. The scaling process assumes that the \Halpha\ luminosity $\propto$ SFR as shown by \citet{Kennicutt1983}. 
By interpolating the S99 models to the PEGASE time grid, we calculate the difference in the parameters between the two models for a given time.

All models are computed using a constant SFR. Since, the number of O and B stars that contribute strongly to the \Halpha\ flux is regenerated at a constant speed, the \Halpha\ flux reaches a constant value within a very short time scale and maintains this value. The lifetime of these O and B stars are in the order of $\sim10$ Myr and therefore, there is no effect from the accumulation of these stars to the \Halpha\ flux. \Halpha\ flux between the SSP models show good agreement with shallower IMFs showing larger \Halpha\ flux values.  
This is driven by the increase in the fraction of larger O and B stars, which contributes to the increase in ionizing photons to boost the \Halpha\ flux. The \Halpha\ flux generated by the two SSP models agree within \around0.03 dex. The discrepancy is slightly higher for steeper IMFs, perhaps driven by minor differences in the mass distribution of the stars in the SSP models.

The continuum level at 6563\AA\ also show good agreement between the SSP models and the differences are $\lesssim 0.1$ dex. Unlike \Halpha\ flux, the continuum level do not reach a constant value within 3 Gyr. This is driven by the larger lifetime of the A and G stars, which largely contributes to the galaxy continuum. 
The rate at which the continuum level increase is dependant on the IMF, where galaxies with steeper IMFs take longer times to reach a constant continuum level.
However, having a higher fraction of smaller stars eventually leads to a higher continuum level compared to a scenario with a shallower IMF. Since the fraction of A and G stars are higher in a steeper IMF, the higher continuum level is expected. PEGASE and S99 models follow different time scales for stellar evolution.  For a given IMF, PEGASE models evolves the continuum level faster to reach a higher value compared to the S99 models. The discrepancy between the models increases up to \around1500 Myr,  after which it decreases to reach a constant value.

\begin{figure}
\centering
\includegraphics[width=0.49\textwidth]{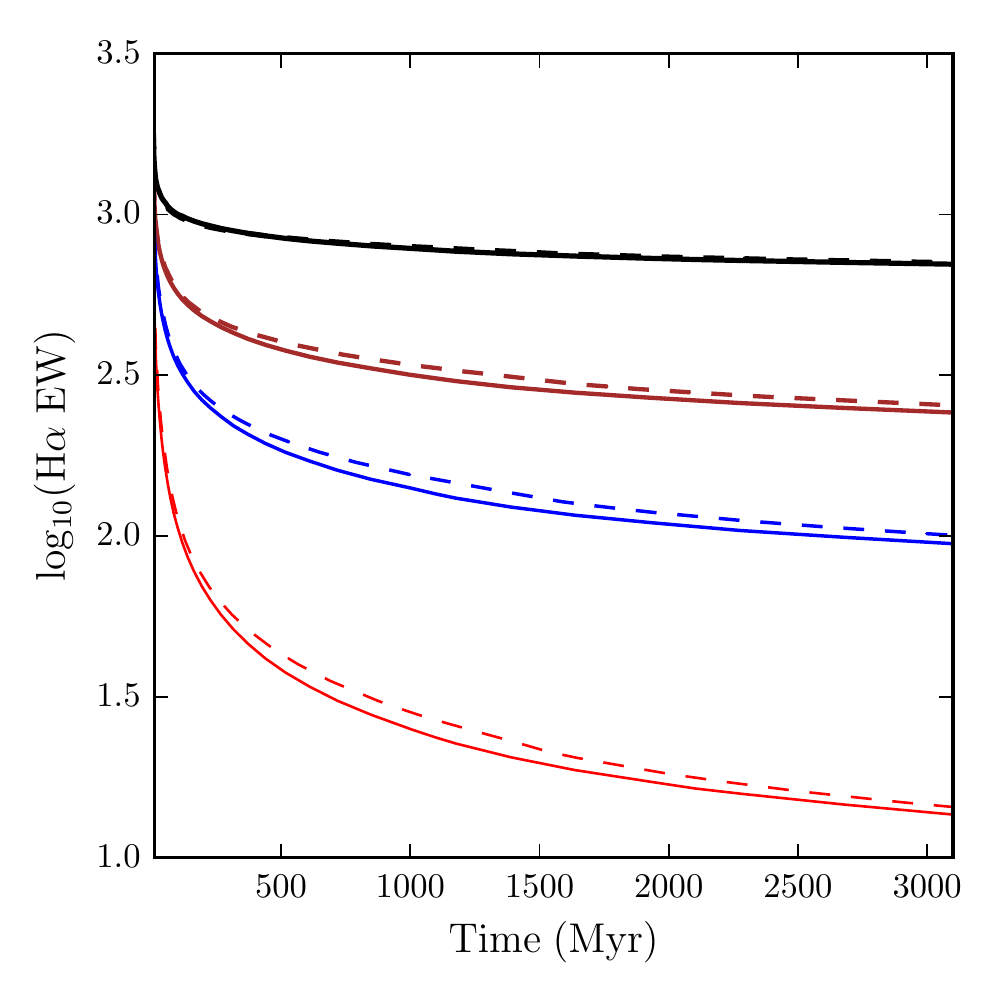}
\includegraphics[width=0.49\textwidth]{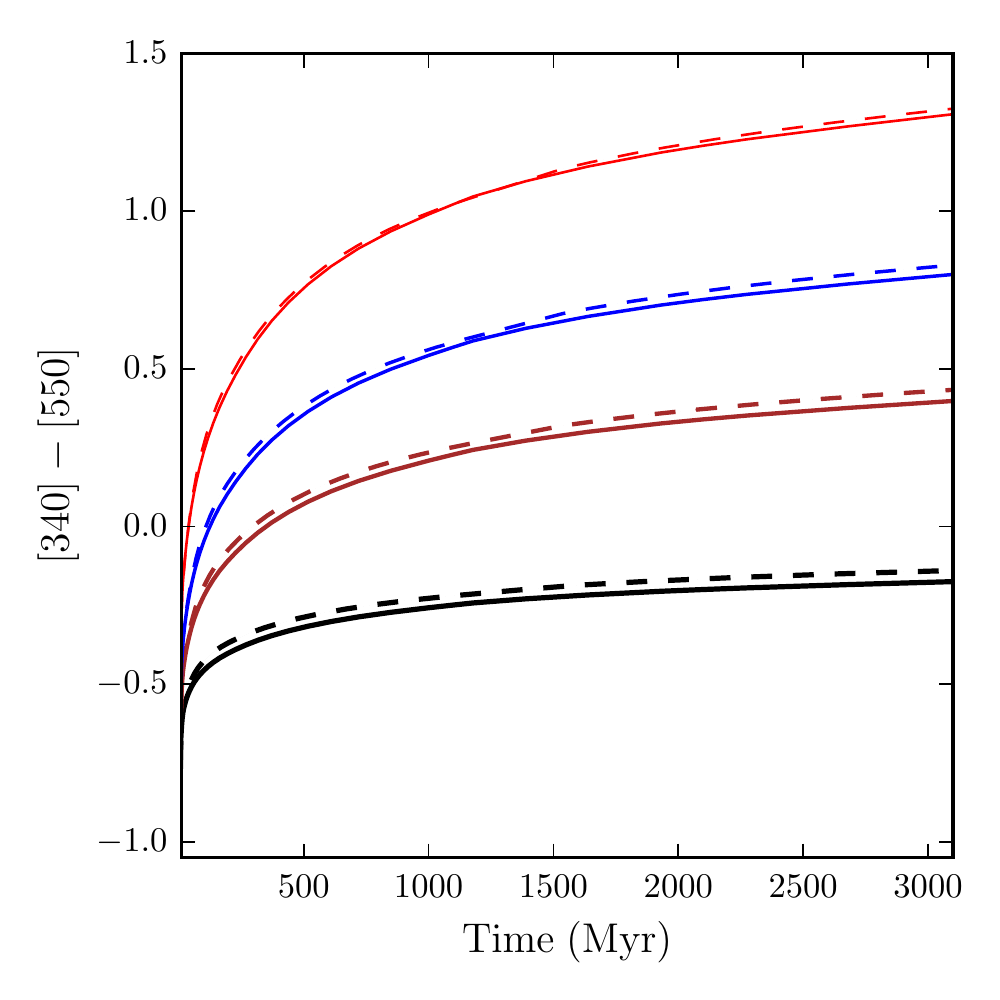}
\includegraphics[width=0.49\textwidth]{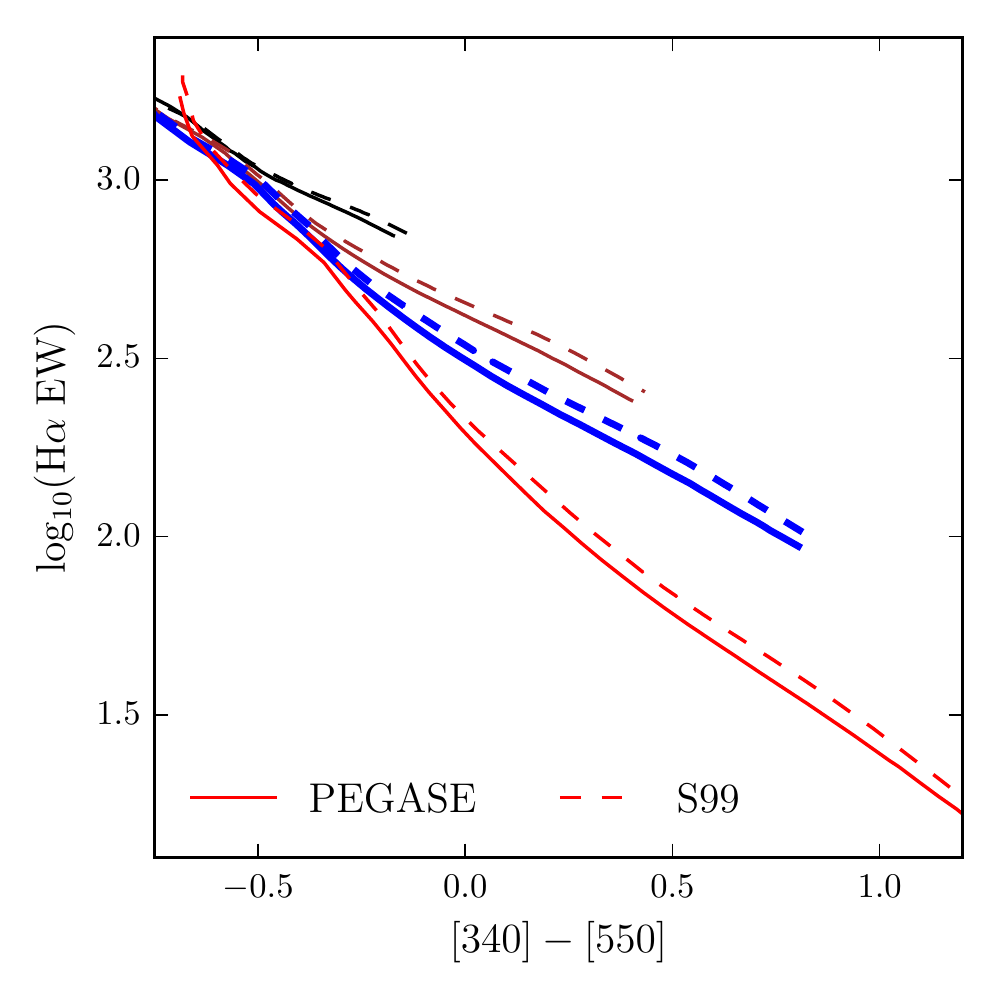}
\caption[Comparison of model parameters between PEGASE and Starburst99 (S99)]{Comparison of model parameters between PEGASE and Starburst99 (S99). We compare the evolution of \Halpha\ EW and \boxfil\ colours between PEGASE and S99  models. The S99 models are computed with similar to PEGASE using the updated Padova stellar tracks following prescriptions in Section \ref{sec:PEGASE_models}. Each SSP model has been computed using varying $\Gamma$ values of $-0.5$ (black), $-1.0$ (brown), $-1.35$ (blue), and $-2.0$ (red). The PEGASE models are shown as solid lines while S99 models are shown as dashed lines. 
{\bf Top Left:} The evolution of  \Halpha\ EW with time for PEGASE and S99 models. 
{\bf Top Right:} The evolution of PEGASE and S99 \boxfil\ colours with time. 
{\bf Bottom:} The evolution of PEGASE and S99 \Halpha\ EW and \boxfil\ colours. The wavelength coverage used for \boxfil\ colours do not include any strong emission lines and therefore the colours are independent of photo-ionization properties of the galaxies. 
}
\label{fig:PEGASE_S99_comp}
\end{figure}

The change of the \Halpha\ EW between the two SSP models (shown by \ref{fig:PEGASE_S99_comp} left panel) is driven by the differences in the \Halpha\ flux and the continuum level. Both the models behave similarly by decreasing \Halpha\ EW with time. 
Shallower IMFs show higher EW values driven by higher \Halpha\ flux and lower continuum values and the shape of the $\Delta$EW function is driven by the differences in the continuum evolution.

Furthermore, in Figure \ref{fig:PEGASE_S99_comp} (centre panel) we investigate the evolution of \boxfil\ colours derived between PEGASE and S99. 
Since the wavelngth regime covered by [340] and [550] colours do not include emission lines, a direct \boxfil\ colour comparison between S99 models (with no emission lines) with PEGASE models (with emission lines) is possible. 
Models with different IMFs show distinctive differences between the derived \boxfil\ colours. Driven by the excess of higher mass blue O and B stars, galaxies with shallower IMFs show bluer colours compared to galaxies with steeper IMFs for a given time.
Steeper IMFs show a better agreement between the two SSP models. Both, PEAGSE and S99 use the same stellar tracks from the Padova group and  therefore, we attribute the differences between the SSP models to differences in methods used by PEAGSE and S99 to produce the composite stellar populations.

In Figure \ref{fig:PEGASE_S99_comp} (right panel), we compare the evolution of \Halpha\ EW with \boxfil\ colours for PEGASE and S99 models. 
Following on close agreement between the evolution of \Halpha\ EW and \boxfil\ colours between the two SSP models, in the \Halpha\ EW vs \boxfil\ colour plane galaxies from both  PEGASE and S99  show similar evolution. Therefore, our conclusions in this study are not affected by the choice of SSP model (PEGASE or S99) but we note that stellar libraries do play a more prominent role, which we discuss in detail in Section \ref{sec:other_exotica}.

\chapter{Dust properties of $z\sim2$ galaxies}
\label{chap:extended_dust}

\section{Derivation of box-car filters for IMF and dust analysis}
\label{sec:box car filters}

\subsection{The choice of 340 and 550 filters}
\label{sec:filter choice 340}

\begin{figure}
\centering
\includegraphics[width=1.0\textwidth]{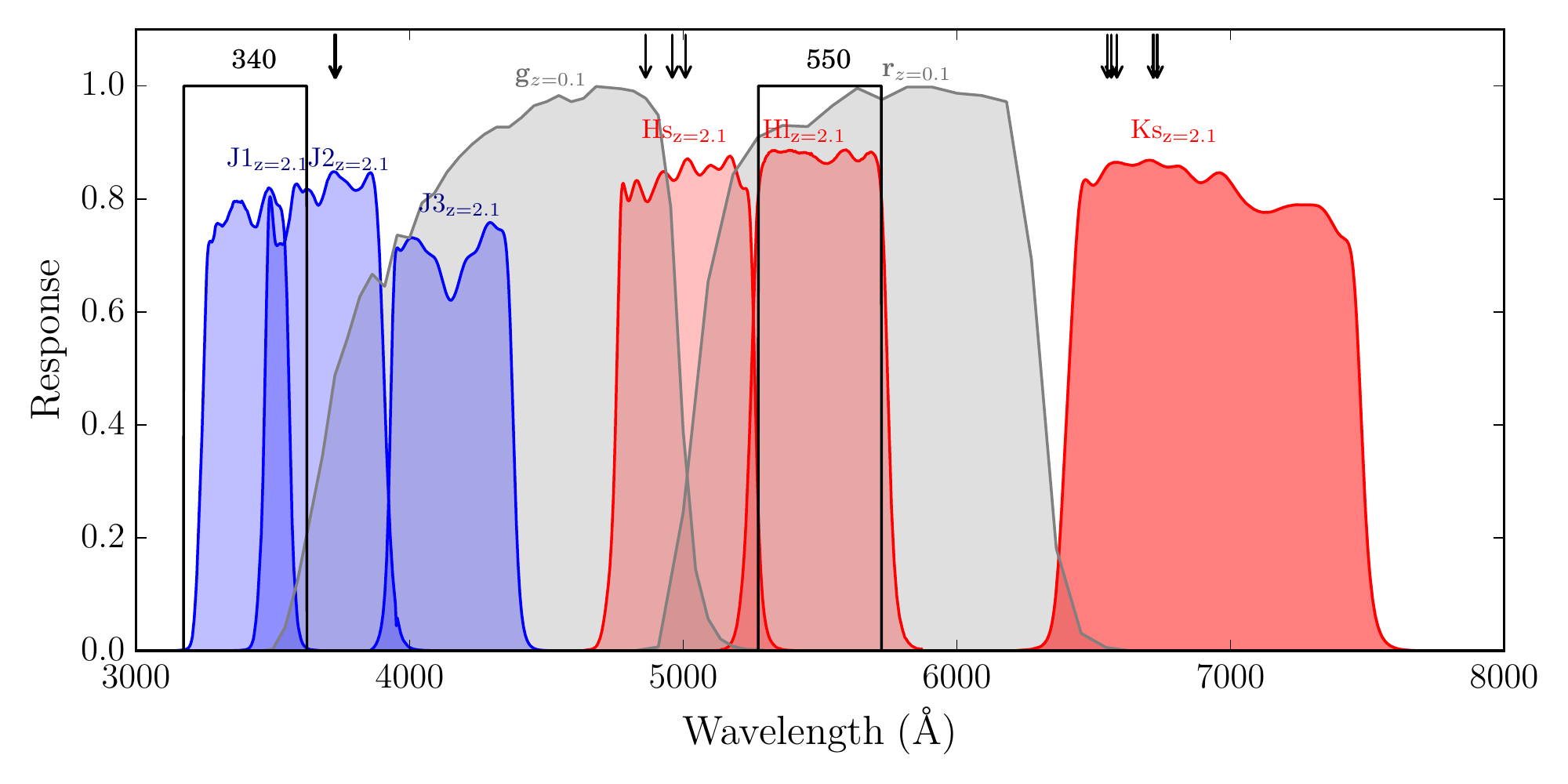}
\includegraphics[width=1.0\textwidth]{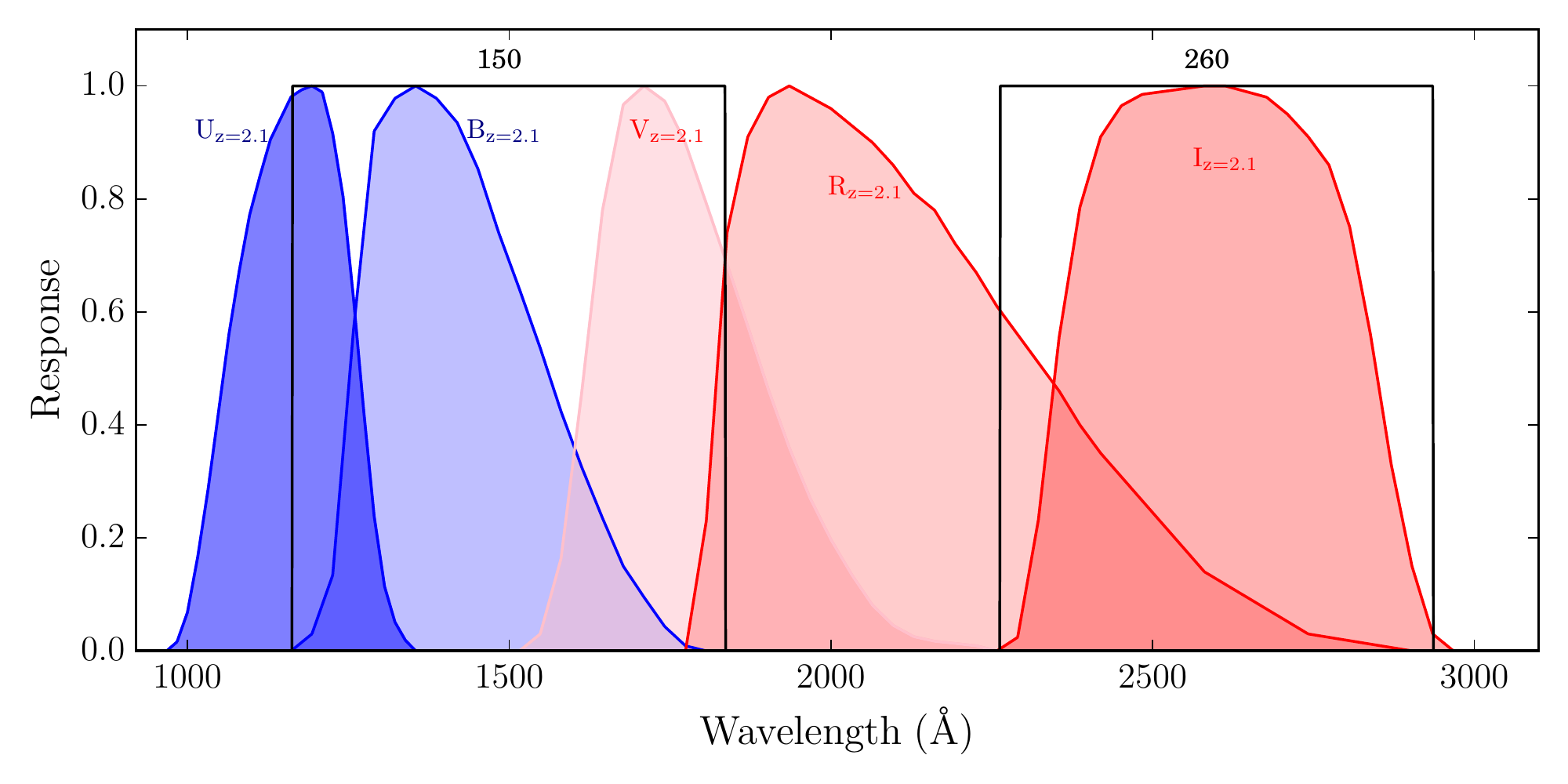}
\caption[The wavelength coverage of the synthetic filters used in our analysis.]
{{\bf Top:} The wavelength coverage of the [340] and [550] filters compared with the wavelength coverage of rest-frame FourStar NIR filters for a galaxy at $z=2.1$. 
We also show the wavelength coverage of the $g_{z=0.1}$ and $r_{z=0.1}$ filters used by the HG08 analysis. The arrows denote locations of strong emission lines. 
{\bf Bottom:} The wavelength coverage of the [150] and [340] filters along with  \citet{Bessell1990} filters de-redshifted from $z=2.1$. 
}
\label{fig:filter_coverage}
\end{figure}

Due to the strong dependence of nebular emission line properties in the \gr\ colour regime, we shift our analysis to synthetic box car optical filters that avoid regions of strong emission lines. Figure \ref{fig:filter_coverage} shows the wavelength coverage of our purpose built [340] and [550] box car filters along with the wavelength coverage of the FourStar filters in the rest-frame of a galaxy at $z=2.1$. It is evident from the figure that $\mathrm{J1_{z=2.1}}$, $\mathrm{J3_{z=2.1}}$, and $\mathrm{Hl_{z=2.1}}$ filters avoid wavelengths with strong emission lines. 
We choose the median wavelength of $\mathrm{J1_{z=2.1}}$ (3400\AA) and $\mathrm{Hl_{z=2.1}}$ (5500\AA) filters to develop  box-car filters with a with a wavelength coverage of 4500\AA. 
These box-car filters are used to compute optical colours for the ZFIRE-IMF sample. 


\subsection{The choice of 150 and 260 filters}
\label{sec:filter choice 150}

To be consistent with our IMF analysis, which employs the \boxfil\ colours, we compute UV filters  employing a similar technique described in Appendix \ref{sec:filter choice 340}. 
\citet{Bessell1990} B and I filters which samples the optical wavelength regime are chosen for this purpose. For a typical galaxy at $z=2.1$, these filters sample the UV wavelength regime. 
Therefore, by dividing the wavelength coverage of the B and I filters by 3.1, we define a filter set that samples the UV region in the rest-frame of a galaxy at $z=2.1$

We then define two box-car filters that has similar wavelength coverage of $\sim6700$\AA\ to the blue-shifted B and I filters. The bluer filter is centred at 1500\AA\ while the redder filter is centred at 2600\AA, both with a width of 673\AA. We name these filters [150] and [260] respectively, and are used in our analysis to investigate the IMF dependence of the dust parameters derived by FAST.

\subsection{Comparison between observed colours and EAZY derived rest-frame colours}
\label{sec:EAZY colour comparision}

\begin{figure}
\centering
\includegraphics[width=0.55\textwidth]{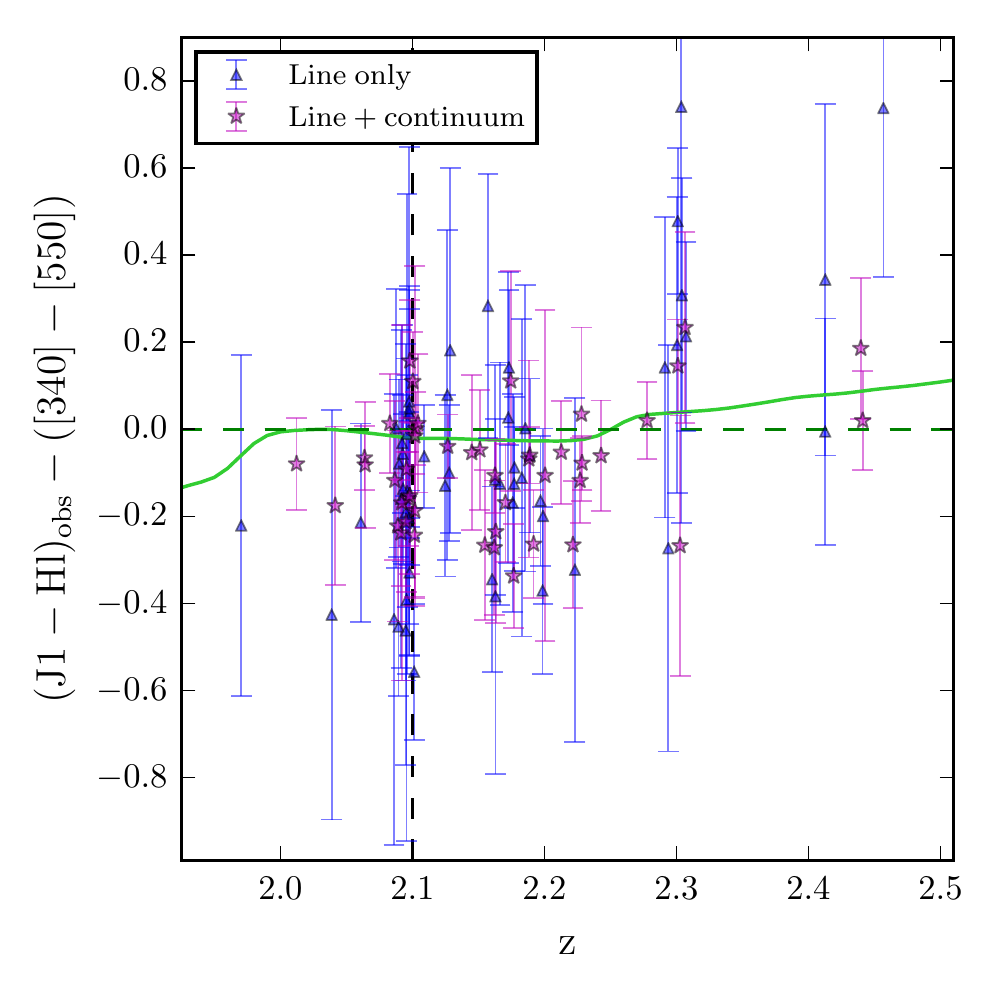}
\includegraphics[width=0.55\textwidth]{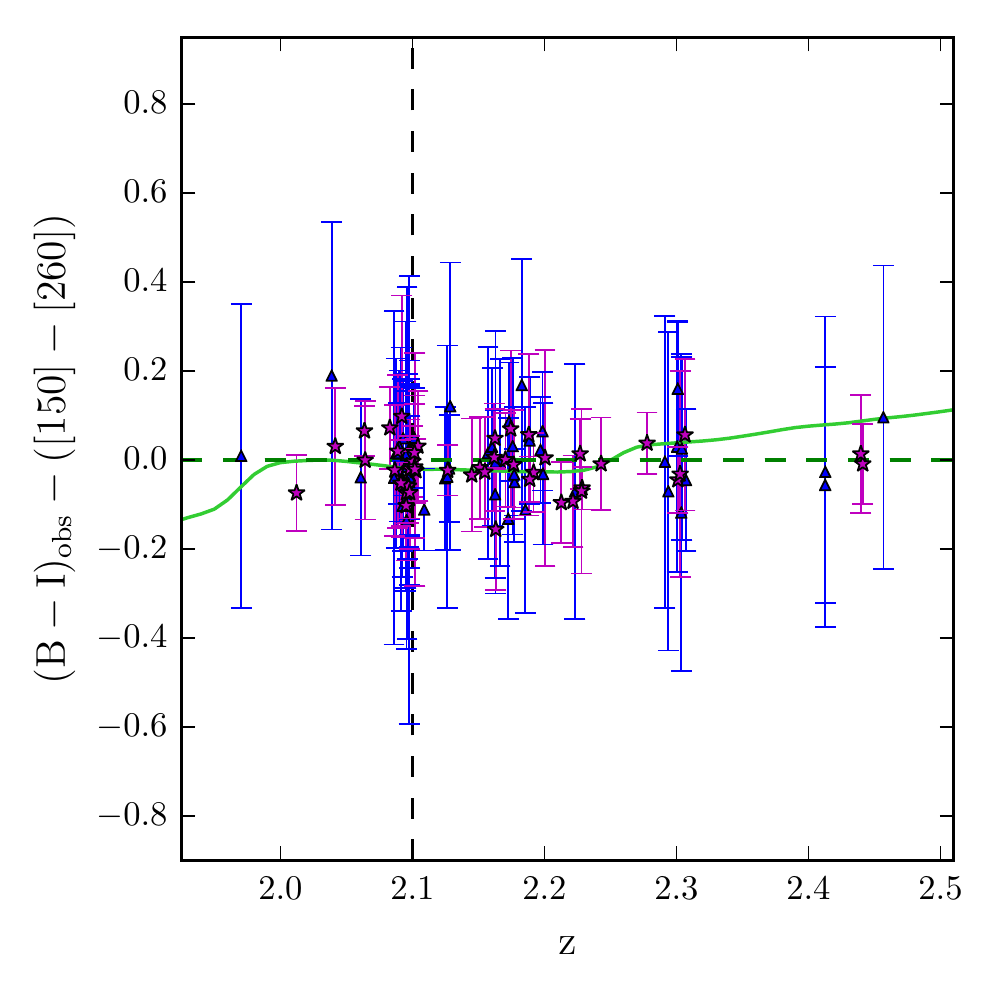}
\caption[The comparsion between observed FourStar colours and SED derived rest-frame colours.]{ The colour difference between observed colours and rest-frame colours derived from best-fitting SED templates from EAZY. Rest-frame colours are computed in such a way that the wavelength coverage in the observed frame at $z=2.1$ is approximately similar to the wavelength coverage of the rest-frame filters at $z=0$.
{\bf Top:} The colour difference between the observed (J1$-$Hl) colours and the \boxfil\ colours of the galaxies in our IMF sample. Galaxies with continuum detections are shown as magenta stars while galaxies with only \Halpha\ emission are shown as blue triangles. The error bars are from the errors in J1 and Hl filters from the ZFOURGE survey photometry. The green line shows the evolution of the colour difference of a PEGASE model galaxy. The vertical dashed line denotes $z=2.1$, which is the redshift used to de-redshift the NIR filters in order to compute rest-frame colours. The horizontal dashed line shown is the $\Delta(colour)=0$ line. Galaxies lying on this line shows no difference between the observed colours and the rest-frame colours derived via EAZY. 
{\bf Bottom:}  Similar to left but for (B$-$I) and \dustfil\ colours. 
}
\label{fig:rest_frame_colour_comparision}
\end{figure}

We use the observed FourStar J1 and Hl fluxes and the best-fitting SED fits of our \sample\ to test the robustness of the observed colours with the EAZY derived rest-frame colours. 
In Figure \ref{fig:rest_frame_colour_comparision} (left panel), we show the differences between the observed (J1$-$Hl) colours and the rest-frame \boxfil\ colours  (as described in Appendix \ref{sec:filter choice 340}) computed from the best-fitting EAZY SED templates.
We compare J1$-$Hl with \boxfil\ colours and expect them to approximately agree by construction at $z=2.1$.

Using a PEGASE model spectrum, we compare the difference with $z$ of J1$-$Hl with \boxfil\ colours with what we expect from SED templates. Lines go through zero at $z=2.1$ as we expect. 
The model spectrum is extracted at $t = 3100$ Myr from a galaxy with an exponentially declining SFH with a p$_1$ = 1000 Myr, $\gamma=-1.35$ IMF, and no metallicity evolution. 
We use the model spectrum to compute the \boxfil\ colour.  We then make a grid of redshifts between $z=1.8$ to $z=2.7$ with $\Delta z = 0.01$ and redshift the wavelength of the model spectra for redshifts in this grid by multiplying the wavelength by $(1+z)$. For each redshift we compute the (J1$-$Hl) colours and since we only investigate the colour difference, there is no need to consider the redshift dimming or K corrections etc.

Since the rest-frame filters assume that the galaxies are at $z=2.1$, we expect the observed colours and rest-frame colours to agree at this redshift. Figure \ref{fig:rest_frame_colour_comparision} (left panel) shows that for the model galaxy this expectation holds with a maximum deviation of \around$\pm0.1$ mag in colour difference between $z=1.8$ to $z=2.7$. Galaxies in the \sample\ shows a much larger deviation of \around$\pm0.5$ mag, which we attribute to errors in photometry as evident from the large error bars. 
Furthermore, zero point corrections in the SED fitting techniques can give rise to additional systematic variations.

A similar analysis is performed on the (B$-$I) colours using the observed fluxes from the ZFOURGE survey and \dustfil\ colours (as described in Appendix \ref{sec:filter choice 150}) on the best-fitting EAZY SED templates. The same PEGASE model galaxy used for the (J1$-$Hl) comparison is used to derive the (B$-$I) colours by red-shifting the spectra to redshifts between $z=1.8$ to $z=2.7$. Figure \ref{fig:rest_frame_colour_comparision} (right panel) shows the comparison between observed  and EAZY derived colours along with the $ideal$ expectation computed from PEGASE spectra. Due to the intrinsic shape of the SED, the redshift evolution of $\Delta$(B$-$I) is opposite to that of $\Delta$(J1$-$Hl).

\section{Nebular extinction properties of ZFIRE $z\sim2$ sample}

\subsection{\Hbeta\ detection properties}
\label{sec:Balmer decrement extended}

\begin{figure}
\centering
\includegraphics[width=0.6\textwidth]{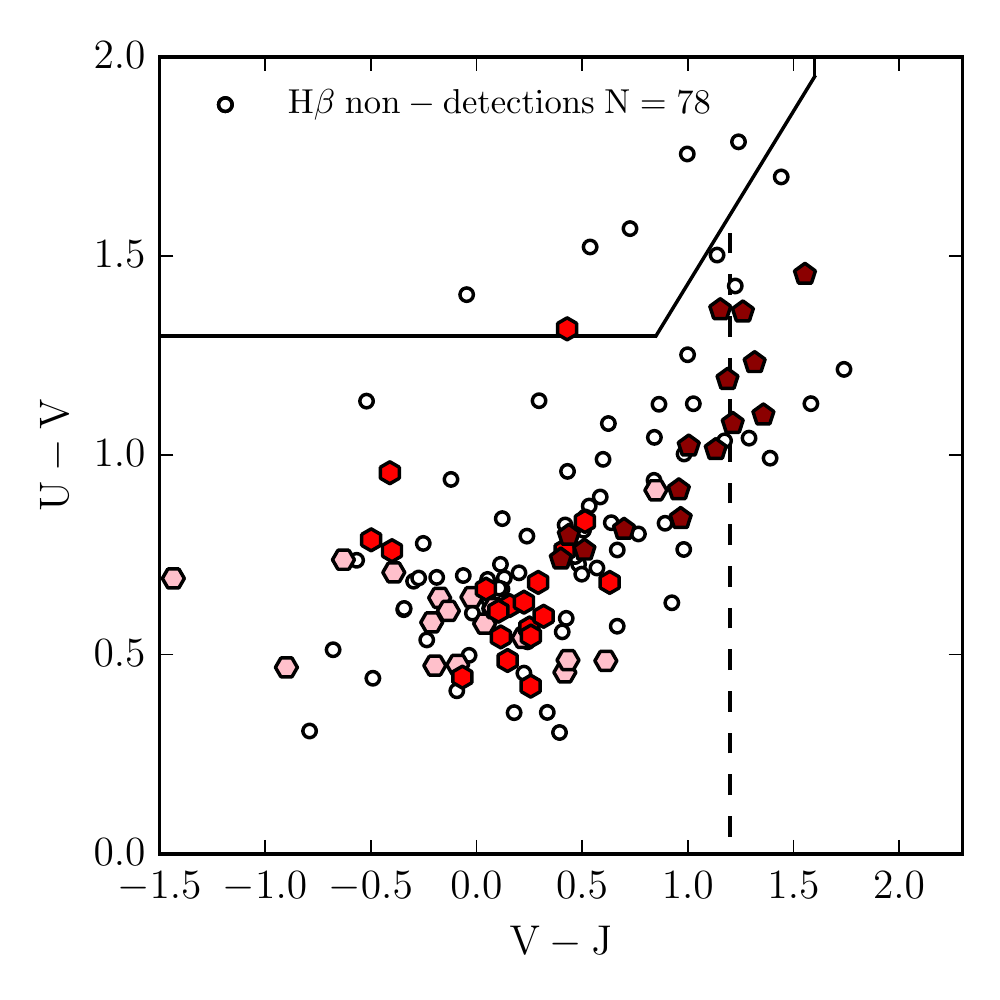}
\includegraphics[width=0.6\textwidth]{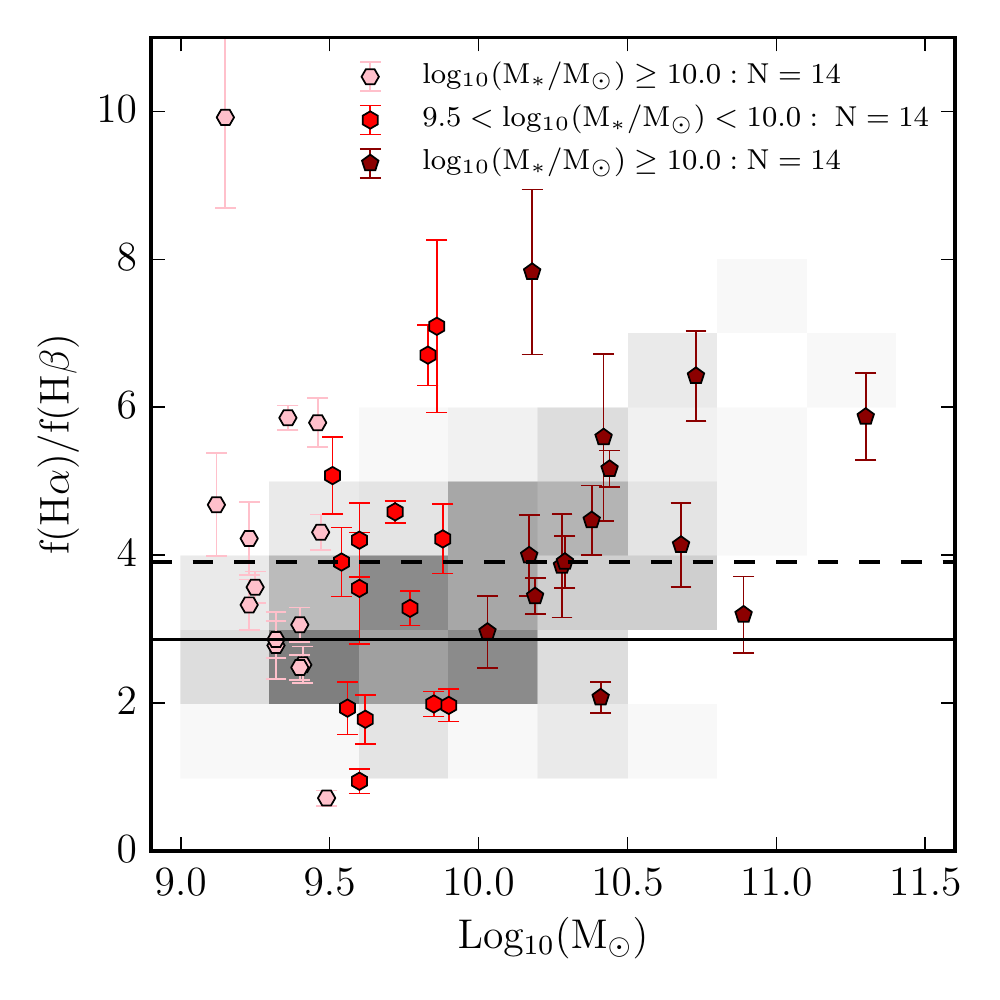}
\caption[\Hbeta\ detection properties of the ZFIRE sample.]{
{\bf Top:} Rest-frame UVJ diagram of the ZFIRE \Hbeta\ targeted sample.  Galaxies shown by green filled symbols have been detected in \Hbeta\ with SNR$>5 $ while black open circles show \Hbeta\ non-detected galaxies. Note that not all galaxies with \Hbeta\ detections have been targeted in K band. The rest-frame colours have been derived using spectroscopic redshifts where available. 
{\bf Bottom:} Balmer decrement vs mass of the ZFIRE sample. 
The black dashed horizontal line is the median Balmer decrement value (3.9) of the sample and the black solid line is the Balmer decrement = 2.86 limit from Case B recombination at $T=10,000$ K \citep{Osterbrock1989}. 
All masses are derived from SED fitting by FAST. The 2D density histogram shows the distribution of values from \citet{Reddy2015}. 
In both panels, the \Hbeta\ detected sample is colour coded according to their stellar mass. 
}
\label{fig:balmer_decrement_properties}
\end{figure}

Figure \ref{fig:balmer_decrement_properties} (left panel) shows the UVJ diagram \citep{Spitler2014} of the ZFIRE  \Hbeta\ targeted and detected sample.
Rest frame UVJ analysis  shows that our \Hbeta\ detected sample is a reasonably representative subset of our star forming galaxies.


Figure \ref{fig:balmer_decrement_properties} (right panel) shows Balmer decrement values as a function of stellar mass. We calculate the median Balmer decrement for our sample to be 3.9. Using a least-squares polynomial fit to the data we find that galaxies with higher mass are biased towards high Balmer decrement values. There are 9 galaxies with Balmer decrements below 2.86, which are below the theoretical minimum value for case B recombination at $T=10,000$ K \citep{Osterbrock1989}.

\newpage

\subsection{Derivation of the dust corrections to the \sample}
\label{sec:dust_derivation}

 In this section, we show how we used the \citet{Calzetti2000} and \citet{Cardelli1989} attenuation laws to derive extinction values for the \sample. 

We first calculate the starburst reddening curve at $0.6563\mu$m using the following equation:  
\begin{subequations}
\begin{equation}
\label{eq:starburst_curve_calzetti_IR}
k'(\lambda) = 2.659(-1.857+ \frac{1.040}{\lambda}) +R'_v
\end{equation}
where $\lambda$ is in $\mu m$. This equation is only valid for wavelengths between $0.63\mu m<\lambda<2.2\mu m$. 
Following \cite{Calzetti2000} the total attenuation ($R^{'}_{v}$) is set to 4.05. 
We use the derived value for the reddening curve to calculate the attenuation of the continuum at $0.6563\mu$m\ ($A_c(0.6563)$).
\begin{equation}
\label{eq:cont attenuation}
A_{c}(0.6563) = k'(0.6563) \times \frac{A_v}{R'_v} = 0.82A_{c}(V)
\end{equation}
\end{subequations}
Next we use the \cite{Cardelli1989} prescription to calculate the attenuation of the nebular emission lines. This law is valid for both diffuse and dense regions of the ISM and therefore we expect it to provide a reasonable approximation to the ISM of galaxies at $z\sim2$. 
We use the following equations to evaluate the extinction curve at 6563\AA. 
\begin{subequations}
\begin{equation}
x = 1/\lambda
\end{equation}
where $\lambda$ is in $\mu$m and is between $\mathrm{1.1\mu m^{-1} \leq x \leq 3.3 \mu m^{-1}}$. 
Wavelength dependent values $a(x)$ and $b(x) $are defined as follows:
\begin{equation}
\begin{split}
a(x) = 1 + (0.17699 \times y) - (0.50447\times(y^2)) \\
- (0.02427\times(y^3))+ (0.72085\times(y^4)) + \\
(0.01979\times(y^5))-(0.77530\times(y^6))+ \\
(0.32999\times(y^7))
\end{split}
\end{equation}
\begin{equation}
\begin{split}
b(x) = (1.41338*y)+(2.28305*(y^2))+(1.07233*(y^3))-\\
(5.38434*(y^4))-(0.62251*(y^5))+\\
(5.30260*(y^6))-(2.09002*(y^7))
\end{split}
\end{equation}  
where $y = x-1.82$. 
Using $a(x)$ and $b(x)$ values, the attenuation of the nebular emission line at 0.6563$\mu m$\ ($A_{n}(0.6563)$)  can be expressed as follows: 
\begin{equation}
\label{eq:A_n}
A_{n}(0.6563) = A_{n}[a(0.6563^{-1}) + \frac{b(0.6563^{-1})}{R''_v}] = 0.82A_{n}(V)
\end{equation}
$R^{''}_{v}$ is set to 3.1 following \cite{Cardelli1989}. 
\end{subequations}
Colour excess is defined as:
\begin{equation}
\label{eq:colour excess}
E(B-V) = A(V) /R_v
\end{equation}
\cite{Calzetti1994} shows that at $z\sim0$ newly formed hot ionizing stars reside in dustier regions of a galaxy compared to old stellar populations. Ionizing stars mainly contribute to the nebular emission lines while the old stellar populations contribute the stellar continuum of a galaxy. Therefore, they find that nebular emission lines of a galaxy to be $\sim2$ times more dust attenuated than the stellar continuum. Here, we denote this correction factor as $f$. Using $n$ and $c$ subscripts to denote the nebular and continuum parts respectively,
\begin{equation}
\label{eq:cal fac}
E_n(B-V) = f \times E_c(B-V)
\end{equation}
Substituting equation \ref{eq:colour excess} to Equation \ref{eq:A_n}:
\begin{subequations}
\begin{equation}
A_{n}(0.6563) = 0.82 \times\ R''_v \times E_n(B-V)
\end{equation}
Using Equation \ref{eq:cal fac}:
\begin{equation}
A_{n}(0.6563) = 0.82 \times R''_v \times f \times E_c(B-V)	
\end{equation}
$E_c(B-V)$ is computed using the \citet{Calzetti2000} dust law,
\begin{equation}
A_{n}(0.6563) = 0.82 \times f \times A_c(V) \frac{R''_v}{R'_v}	
\end{equation}
\end{subequations}
Therefore, we express the dust corrected nebular line ($f_i$(\Halpha)) and continuum flux ($f_i$(cont)) as follows:
\begin{subequations}
\begin{equation}
\label{eq:Halpha dust corrected}
f_i(H\alpha)= f_{obs}(H\alpha) \times 10^{0.4(0.62\times f\times A_{c}(V))}
\end{equation}
\begin{equation}
\label{eq:cont dust corrected}
f_i(cont)= f_{obs}(cont) \times 10^{0.4(0.82\times A_{c}(V))}
\end{equation}
where the subscript $obs$ refers to the observed quantity while $i$ refers to the intrinsic quantity. 
\end{subequations}
Since $EW_i =f_i(H\alpha)/f_i(cont))$, finally the dust corrected \Halpha\ EW can be expressed as follows:
\begin{equation}
\label{eg:EW dust corrected}
\log_{10}(EW_i) = \log_{10}(EW_{obs}) + 0.4A_c(V)(0.62f-0.82)
\end{equation}
\\

Next we consider the dust correction for the $z=0.1$ optical colours. Using \cite{Calzetti2000} attenuation law we calculate the starburst reddening curve for these wavelengths using the following equation: 
\begin{equation}
\label{eq:starburst_curve_calzetti_optical}
k'(\lambda) = 2.659(-2.156+ \frac{1.509}{\lambda} - \frac{0.198}{\lambda^2} +\frac{0.011}{\lambda^3}) +R'_v
\end{equation}
This equation is different from Equation \ref{eq:starburst_curve_calzetti_IR}, since this is valid for more bluer wavelengths between $0.12\mu m<\lambda<0.63\mu m$. 
Similar to Equation \ref{eq:cont attenuation}, we work out the attenuation for median wavelengths of the [340] and [550] filters (by definition the filter medians are respectively at $0.34\mu$m and $0.55\mu$m) as follows:. 
\begin{subequations}
\begin{equation}
\label{eq:BC340 dust corrected_extended}
f([340]) = f([340]_{obs}) \times 10^{0.4 \times 1.56 A_c(V)}
\end{equation}
\begin{equation}
\label{eq:BC550 dust corrected_extended}
f([550]) = f([550]_{obs}) \times 10^{0.4 \times 1.00 A_c(V)}
\end{equation}
\end{subequations}
Dust corrected fluxes are used to recalculate the \boxfil\ colours.

The median wavelengths of the g$_{0.1}$ and r$_{0.1}$ filters are respectively at 0.44$\mu m$ and 0.57$\mu m$. Similar to [340] and [550] filters, we use Equation \ref{eq:starburst_curve_calzetti_optical} to calculate the attenuation for g$_{0.1}$ and r$_{0.1}$ filters as follows:
\begin{subequations}
\begin{equation}
\label{eq:g_dust_corrected_derivation}
f(g_i)_{0.1} = f(g_{obs})_{0.1} \times 10^{0.4 \times 1.25 A_c(V)}
\end{equation}
\begin{equation}
\label{eq:r_dust_corrected_derivation}
f(r_i)_{0.1} = f(r_{obs})_{0.1} \times 10^{0.4 \times 0.96 A_c(V)}
\end{equation}
\end{subequations}


\end{document}